%% file: Thesis.tex
\documentclass[sotoncolour,openany]{uosthesis}      
\graphicspath{{Figures/}}   
\usepackage[numbers]{natbib}            
\usepackage{bibentry}          
\nobibliography*               
\usepackage{attrib}            
\hypersetup{colorlinks=true}   
\input{Definitions}            

\usepackage{xcolor}
\usepackage{textcomp}
\usepackage{cleveref}
\usepackage[sumlimits,tbtags]{amsmath}

\department  {School of Mathematical Sciences}
\DEPARTMENT  {\MakeUppercase{\deptname}}
\group       {Gravity Group}
\GROUP       {\MakeUppercase{\groupname}}
\faculty     {Faculty of Social Sciences}
\FACULTY     {\MakeUppercase{\facname}}
\title      {Dissipation and turbulence in general relativistic hydrodynamics}

\authors    {Thomas Celora} 
\addresses  {\groupname\\\deptname\\\univname}
\date       {\today}

\orcidid{0000-0002-6515-3644}
\subject    {}

\begin{document}
\pagenumbering{gobble} 
\copyrightDeclaration{} 
\frontmatter
\maketitle
\begin{abstract}
Hydrodynamics is one of the oldest research areas in physics, with applications across all macroscopic scales in the Universe.
Despite the long history of successes, however, fluid modelling still presents severe conceptual and computational challenges.
Not surprisingly, the hurdles become even more formidable for relativistic flows, and new issues come to the fore too. 
This work is concerned with advancing multi-fluid models in General Relativity, and in particular focuses on the modelling of dissipative fluids and turbulent flows. 
Such models are required for an accurate description of neutron star phenomenology, and binary neutron star mergers in particular. 
In fact, the advent of multi-messenger astronomy---started with the first detection of a binary neutron star coalescence in 2017---offers exciting prospects for exploring the extreme physics at play during such cosmic fireworks. 

In this work we first focus on modelling dissipative fluids in relativity, and explore the arguably unique model that is ideally suited for describing dissipative multi-fluids in General Relativity. 
Modelling single fluids in relativity is already a hard task, but for neutron stars it is easy to argue that we need to understand even more complicated settings: the presence of superfluid/superconducting mixtures, for example, means that we need to go beyond single-fluid descriptions. 
We then consider turbulent flows and focus on how to perform ``filtering'' in a curved spacetime setting.
We do so as most recent turbulent models in a Newtonian setting are based on the notion of spatial filtering. 
As the same strategy is beginning to be applied in numerical relativity, we focus on the foundational underpinnings and propose a novel scheme for carrying out filtering, ensuring consistency with the tenets of General Relativity.
Finally, we discuss two applications of relevance for binary neutron star mergers. We focus on the modelling of ($\beta$-)reactions in neutron star simulations, and provide a discussion of the magneto-rotational instability that is suited to highly dynamical environments like mergers. We focus on these two problems as reactions are expected to source the dominant dissipative contribution to the overall dynamics, while the magneto-rotational instability is considered crucial for sustaining the development of turbulence in mergers. 
\end{abstract}
\setcounter{tocdepth}{3}
\tableofcontents
\listoffigures*
\addtolom{Material Name e.g Map}
\addtolom{Material Name e.g CD}
\addtolom{Test Material}
\authorshipdeclaration{
\begin{itemize}
    \item[--] \citet{ActionDissip1st},
    \item[--] \citet{fibrLES},
    \item[--] \citet{dMHD31},
    \item[--] \citet{BV_in_sim}.
    \item[--] \citet{MRIpaper}.
\end{itemize}
}
\acknowledgements{
Even though at times it feels like yesterday, almost four years have passed since I first set foot in Southampton. 
All the experiences I have had during this time have massively shaped the person I have become, and if I can proudly stand where I am today I owe it to the great people I shared parts of this journey with. 
Since the list is too long to acknowledge everyone personally, I will only thank here those that deserve special mentions and without whom this would not have been possible. 

First of all, I am hugely indebted to my supervisors Nils Andersson and Ian Hakwe, and to Greg Comer---a.k.a. the fantastic trio. 
I have indeed been extremely lucky to have had not two, but three superb mentors. 
You have been an endless source of inspiration and guidance, and I did my best to learn as much as I could from you. Working with giants like you has been an immense pleasure and honour, and I hope our collaboration will continue for a long time. 

I must also express my sincere thanks to the members of the Gravity Group, particularly those working on neutron star physics. I am honoured to have been part of such a group of brilliant scientists, and grateful for all I have learnt from you. 

Next, I want to thank my mother Anna. You were the first one to show me how fascinating physics is, and I strongly believe you were the real trigger for everything that came later. I can never repay you for all the sacrifices you have made, which have allowed me to be here today. I must also thank my brother and sisters Agostino, Eleonora and Giuditta---rigorously in alphabetic order. I truly came to appreciate during my time here how much you all mean to me. Knowing that I can always count on you is priceless. 

My friends also played a key role in this endeavour. Both the new ones I met in Southampton, and those who have known me for longer. Without you, I do not think I would have made it this far.

Last, but by no means least, I wish to thank my partner in crime Elisa. 
You are the one that has been closest to me throughout the emotional roller coaster of these years. 
I know I have been difficult way too often, and I cannot find the words to express how grateful I am for your love, support and faith in me. 
You should know that this achievement is also yours, and I look forward with indescribable excitement to the next steps of our journey. 
}


\include{Notation}
\mainmatter


\include{Parts/Introduction}

\part{Dissipative (multi-)fluids in General Relativity}\label{part1}
\include{Parts/DissipativeHydro/part1macro}

\part{A covariant approach to large-eddy filtering}\label{part2}
\include{Parts/TurbulenceAndLES/part2macro}

\part{Binary neutron-star merger applications}\label{part3}
\include{Parts/BNSapplications/part3macro}

\part{Conclusions}\label{part4}
\include{Parts/Conclusions}

\appendix
\include{Parts/Appendix}

\backmatter
\bibliographystyle{abbrvnat}
\bibliography{Thesis.bib}

\end{document}

%% file: Definitions.tex
\newcommand{\BibTeX}{{\rm B\kern-.05em{\sc i\kern-.025em b}\kern-.08em T\kern-.1667em\lower.7ex\hbox{E}\kern-.125emX}}

\def\A{{\mathcal A}}
\def\B{{\mathcal B}}
\def\C{{\mathcal C}}
\def\D{{\mathcal D}}
\def\E{{\mathcal E}}
\def\F{{\mathcal F}}
\def\L{{\mathcal L}}
\def\M{{\mathcal M}}
\def\N{{\mathcal N}}
\def\O{{\mathcal O}}
\def\Q{{\mathcal Q}}

\def\S{{\mathcal S}}

\newcommand{\Lie}{\ensuremath{\mathcal{L}}}

\def\veps{{\varepsilon}}
\def\eps{{\epsilon}}
\def\om{{\omega}}
\def\Om{{\Omega}}
\def\varp{{\varphi}}

\def\b{{\rm b}}
\def\c{{\rm c}}
\def\d{{\rm d}}
\def\e{{\rm e}}
\def\f{{\rm f}}
\def\h{{\rm h}}
\def\l{{\rm l}}
\def\n{{\rm n}}
\def\o{{\rm o}}
\def\p{{\rm p}}
\def\q{{\rm q}}
\def\s{{\rm s}}
\def\u{{\rm u}}
\def\x{{\mathrm{x}}}
\def\y{{\mathrm{y}}}

\def\z{{\mathrm{z}}}

\def\ha{{\hat a}}
\def\hb{{\hat b}}
\def\hc{{\hat c}}
\def\hd{{\hat d}}
\def\hk{\hat{k}}
\def\hi{{\hat i}}
\def\hj{{\hat j}}
\def\hk{{\hat k}}
\def\hm{{\hat m}}

\def\hZ{{\hat 0}}
\def\hU{{\hat 1}}
\def\hD{{\hat 2}}
\def\hT{{\hat 3}}

\newcommand{\wh}{\hat{\omega}}
\newcommand{\whd}{\hat{\omega}_{\Delta}}

\def\fint{\int_{\cal M} {\rm d}^4 x \sqrt{- g} \;} 
\def\tintp{\int_{\partial {\cal M}_+} {\rm d}^3 x \sqrt{h_+} \;}
\def\tintm{\int_{\partial {\cal M}_-} {\rm d}^3 x \sqrt{h_-} \;}
\def\tintL{\int_{\partial {\cal M}_L} {\rm d}^3 x \sqrt{- h_L} \;}

\newcommand{\la}{{\langle}}
\newcommand{\ra}{{\rangle}}

\def\ta{{\tilde{a}}}

\def\tn{{\tilde{n}}}
\def\tp{{\tilde{p}}}
\def\ts{{\tilde{s}}}
\def\tu{{\tilde{u}}}

\def\tB{{\tilde{B}}}
\def\tE{{\tilde{E}}}

\def\tJ{{\tilde{J}}}
\def\tT{{\tilde{T}}}
\def\tY{{\tilde{Y}}}

\def\tbeta{{\tilde{\beta}}}
\def\teps{{\tilde{\varepsilon}}}
\def\tsig{{\tilde{\sigma}}}
\def\ttheta{{\tilde{\theta}}}
\def\tvort{{\tilde{\omega}}}
\def\tmu{{\tilde{\mu}}}
\def\tnu{{\tilde{\nu}}}

\def\tperp{{\tilde{\perp}}}

\def\pw{e^{ik_dx^d}}

\newcommand{\pd}[2]{\partial_{#1}{#2}}

\newcommand{\parallelsum}{\mathrel{/\mkern-5mu/}}

\renewcommand{\vec}[1]{\mathbf{#1}}

\newcommand{\gvec}[1]{\boldsymbol{#1}}

\newcommand{\comment}[1]{}

\usepackage{color}
\usepackage{cancel}
\usepackage{physics}
\usepackage{xfrac}


%% file: Notation.tex
\chapter*{Notation}\label{ch:notation}
\markboth{Notation}{Notation}

\addcontentsline{toc}{chapter}{\nameref{ch:notation}}

\paragraph{General conventions and units.}\phantom{A} \\
In this thesis we work with geometric units, that is 
\begin{align*}
    G &= 6.674 \times 10^{-11} \,\text{m}^2\, \text{kg}^{-1} \,\text{s}^{-2} = 1 \;, \\
    c &= 2.997 \times 10^{8}\, \text{m}\, \text{s}^{-1} = 1 \;,
\end{align*}
where $G$ is Newton's gravitational constant and $c$ is the speed of light in vacuum. 

We use the ``mostly-plus'' signature for the metric, so that the flat line element in Minkowski coordinates $\{t,x,y,z\}$ takes the form 
\begin{equation*}
    \d s^2 = -\d t^2 + \d x^2 + \d y^2 + \d z^2 \;.
\end{equation*}
As such, time-like four vectors have negative norm. 

Throughout this document we mainly work using explicit components notation for tensors.
Space-time indices are denoted by ``early'' Latin characters $a,b,c,\dots$ and take values from 0 to 3, while spatial indices are denoted with ``late'' Latin characters $i,j,k\dots$ and take values from 1 to 3.
We reserve the Latin characters $\x,\y,\z$ to denote different chemical species/components. In a multi-component system made of neutrons ($\n$) and protons ($\p$) for example, the chemical index takes values $\x = \n , \p$.
Capital Latin characters $A,B,C,\dots$ will be used to indicate coordinates on the matter spaces only. 
Finally, we distinguish indices with respect to a coordinate basis from those relative to an orthonormal basis or tetrad. 
The latter are denoted with an additional ``hat'' symbol on top, such as $v^{\hat a}$.

We make use of the Einstein summation convention, where repeated indices (one contravariant or ``up'' and one covariant or ``down'') imply summation. For example 
\begin{equation*}
    v_a v^a = \sum_{a = 0}^3 v_a v^a  \;,
\end{equation*}
where $v^a$ is an arbitrary four vector. 
Note that the Einstein summation convention does not apply to the chemical indices $\x,\y,\z$. Similarly, we will not distinguish between ``up or down'' chemical indices. 

Indices enclosed in round or square brackets denote, respectively, symmetrization and anti-symmetrization. 
If one (or more) of the indices within such brackets has ``straight lines'' to its left and right, the (anti-)symmetrization does not apply to it.  
For example, given a generic tensor $A^{abc}$ 
\begin{equation*}
    A^{a(bc)} = \frac{1}{2}\left( A^{abc} + A^{acb}\right) \;, \quad A^{a[bc]} = \frac{1}{2}\left( A^{abc} - A^{acb}\right) \;, \quad A^{(a|b|c)} = \frac{1}{2}\left( A^{abc} + A^{cba}\right) \;.
\end{equation*}

\paragraph{Derivatives and forms.}\phantom{A} \\
Throughout this document we make use of different notions of derivatives of a tensor. 
As usual, partial and covariant derivatives are denoted with $\partial$ and $\nabla$ respectively. 
Working in a coordinate basis, the tangent space is spanned by the basis vectors
\begin{equation*}
    \partial_a = \pdv{}{x^a} \;,
\end{equation*}
and the covariant derivative of a generic tensor $T^{ab\dots}_{\;\;\;cd\dots}$ takes the form
\begin{equation*}
\begin{split}
    \nabla_e T^{ab\dots}_{\;\;\;cd\dots}  = \partial_e T^{ab\dots}_{\;\;\;cd\dots} &+ \Gamma^a_{\;fe} T^{fb\dots}_{\;\;\;cd\dots} + \Gamma^b_{\;ef} T^{af\dots}_{\;\;\;cd\dots} + \dots \\
    &- \Gamma^f_{\;ec} T^{ab\dots}_{\;\;\;fd\dots} - \Gamma^d_{\;ed} T^{ab\dots}_{\;\;\;cf\dots} -\dots\;.
\end{split}
\end{equation*}
Clearly, this reduces to partial derivative for scalar quantities. 
The connection coefficients $\Gamma^a_{\;bc}$ are taken as Christoffel symbols and are determined by the metric $g_{ab}$ and its first derivatives as
\begin{equation*}
	\Gamma^{a}_{\;bc} = \frac{1}{2}g^{ad}\left(\partial_b g_{dc} + \partial_c g_{bd} - \partial_d g_{bc}\right) \;.
\end{equation*}
This means that, in particular, the connection is compatible with the metric $\nabla_a g_{bc} = 0 $ and the connection is symmetric (i.e. torsion free) $\Gamma^a_{\;bc} = \Gamma^a_{\;(bc)}$.

An orthonormal basis $e_{\hat a}$ is linked to a coordinate one through the ``matrix'' $e^{a}_{\hat b}$ (or its inverse $e^{\hat a}_b$)  so that 
\begin{equation*}
	\partial_a = e_a ^{\hat b} \,e_{\hat b} \;.
\end{equation*}
The covariant derivative of a tensor $T^{\hat a\hat b\dots}_{\;\;\hat c\hat d\dots}$  in a tetrad basis takes the form 
\begin{equation*}
\begin{split}
	\nabla_e T^{\hat a\hat b\dots}_{\;\;\;\hat c\hat d\dots} = \partial_e T^{\hat a\hat b\dots}_{\;\;\;\hat c\hat d\dots} &+ \om_{e\;\hat f}^{\;\hat a}T^{\hat f\hat b\dots}_{\;\;\;\hat c\hat d\dots} + \om_{e\;\hat f}^{\;\hat b}T^{\hat a\hat f\dots}_{\;\;\;\hat c\hat d\dots} + \dots  \\ 
    &-\om_{e\;\hat c}^{\;\hat f}T^{\hat a\hat b\dots}_{\;\;\;\hat f\hat d\dots} -\om_{e\;\hat f}^{\;\hat d}T^{\hat a\hat b\dots}_{\;\;\;\hat c\hat f\dots}  -\dots \;,
\end{split}
\end{equation*}
where the spin coefficients $\om_{a\;\hat b}^{\;\hat c}$ can be obtained from the connection coefficients via 
\begin{equation*}
	\om_{a\;\hat b}^{\;\hat c} = e_e^{\hat c} \;e^d_{\hat b}\;\Gamma^e_{\;ad} - e^d_{\hat b} \;\partial_a e_d^{\hat c} \;.
\end{equation*}

We also make use of the notion of Lie derivative. Given a tensor $T^{ab\dots}_{\;\;cd\dots}$and a vector $v^a$, the Lie derivative of $T^{ab\dots}_{\;\;cd\dots}$ along $v^a$ is given by
\begin{equation*}
\begin{split}
	\L_v T^{ab\dots}_{\;\;\;cd\dots} = v^d \partial_d T^{ab\dots}_{\;\;\;cd\dots} &- (\partial_e v^a) T^{eb\dots}_{\;\;\;cd\dots} - (\partial_e v^b) T^{ae\dots}_{\;\;\;cd\dots} - \dots \\
    &+(\partial_c v^e) T^{ab\dots}_{\;\;ed\;\dots} + (\partial_d v^e) T^{ab\dots}_{\;\;\;ce\dots} + \dots \;.
\end{split}
\end{equation*}
Whilst the Lie derivative does not require a metric or a connection to be defined, it is often convenient to rewrite the expression above in a form that is manifestly covariant. 
This can be obtained by substituting $\partial\to\nabla$ in the expression above, provided the connection is symmetric (the case of interest here as we work with Christoffel symbols). 

We also make use of differential (p-)forms, namely covariant tensors that are totally anti-symmetric. Given a p-form $n$ we write it in components notation using the ``natural basis''
\begin{equation*}
	n = \frac{1}{p!} n_{i_1\,\dots\,i_p} dx^{i_1}\wedge \dots \wedge dx^{i_p}\;,
\end{equation*}
where $\wedge$ is the exterior product and $n_{i_1\,\dots\,i_p}$ is totally anti-symmetric in its indices. 
The exterior derivative $d$ is an operation that takes a p-form as input and outputs a (p+1)-form. 
In components notation this is defined as 
\begin{equation*}
	dn_{i_1\,\dots\,i_{p+1}} = (p+1) \partial_{[i_1}n_{i_2\,\dots\,i_{p+1}]} \;.
\end{equation*}
In a similar fashion as for the Lie derivative, the exterior derivative of a p-form can be written in a manifestly covariant way by substituting $\partial \to \nabla$ in the expression above provided we work with a symmetric connection.

The Hodge-dual operation, denoted by $*$, takes a p-form as input and outputs a (n-p)-form, where $n$ is the dimension of the manifold the p-form lives in. 
In component notation,
\begin{equation*}
	(*n)_{i_1\,\dots\,i_q} = \veps_{i_1\,\dots\,i_q}^{\phantom{i_1\,\dots\,i_q}j_1\,\dots\,i_p}	n_{j_1\,\dots\,j_p} \;, \quad q = n- p
\end{equation*}
where $\veps$ is the spacetime volume form, defined as
\begin{equation*}
	\veps_{abcd} = \sqrt{-g}[a\,b\,c\,d] \;, \quad \veps_{ab}^{\;\;\;cd} = \veps_{abef}g^{ce}g^{df}
\end{equation*}
with $[a\,b\,c\,d]$ denoting the Levi-Civita symbol. 

\paragraph{Riemann, Ricci and Einstein tensor.}\phantom{A} \\
The space-time curvature is encoded in the Riemann tensor, namely the space-time is flat provided the Riemann tensor vanishes. 
Given a generic co-vector (or 1-form) $w_a$, the Riemann tensor is defined as
\begin{equation*}
    [\nabla_a,\nabla_b]w_c = \left(\nabla_a \nabla_b - \nabla_b \nabla_a \right)w_c = - R^d_{\;cab}w_d \;.
\end{equation*}
In components notation,
\begin{equation*}
    R^{a}_{\;bcd} = \partial_c \Gamma^a_{\;bd} - \partial_d \Gamma^a_{\;bc} + \Gamma^e_{\;db} \Gamma^a_{\;ce} - \Gamma^e_{\;bc} \Gamma^a_{\;de} \;.
\end{equation*}
The Ricci tensor, Ricci scalar and Einstein tensor are obtained from the Riemann tensor as
\begin{equation*}
    R_{ab} = R^c_{\;acb} \;, \quad R = R^a_{\;a} \;, \quad   G_{ab} = R_{ab} - \frac{1}{2}Rg_{ab} \;.
\end{equation*}

\paragraph{Projections and velocity gradients decomposition.}\phantom{A} \\
Throughout this document we often use parallel and perpendicular projection operators with respect to an observer. 
Given a time-like unit vector $u^a$, these are defined as 
\begin{equation*}
    \parallelsum^a_b = -u^a u_b \;, \quad \perp^a_b = g^a_b + u^au_b \;. 
\end{equation*}

We also make frequent use of the observer four velocity gradients decomposition. This is given by
\begin{equation*}
    \nabla_au_b = - a_b u_a + \om_{ab} + \sigma_{ab} + \frac{1}{3}\theta \perp_{ab}
\end{equation*}
where the acceleration and vorticity, shear and expansion rate are defined (respectively) as 
\begin{align*}
    a^b &= u^a \nabla_a u^b \;, \\
    \om_{ab} & = \perp^c_{[a}\perp^d_{b]}\nabla_{c}u_d \;, \\
    \sigma_{ab} & =  \perp^c_{(a}\perp^d_{b)}\nabla_{c}u_d - \frac{1}{3}\theta \perp_{ab} \;, \\
    \theta &= \nabla_a u^a = \perp^a_b \nabla_a u^b  \;.
\end{align*}

Finally, the notion of Hodge-dual in the subspace orthogonal with respect to some observer $u^a$ four velocity is frequently used in this work.  
This is defined in terms of the space-time volume form $\veps_{abcd}$ as 
\begin{equation*}
    \veps^{\u}_{abc} = \veps_{dabc}u^d \;.
\end{equation*}
Note that, for notational clarity, we will often drop the $\u$-label and write the object as $\veps_{abc}$ instead. 
We do so whenever there is no risk of confusion. 

Any variation from (or addition to) the notational conventions discussed here will be made explicit throughout the document. 

%% file: Parts/Introduction.tex
\chapter{Introduction} \label{Chapter:Introduction}

\section{Motivation}

Fluid behaviour is relevant across all macroscopic scales in the Universe, from the interior of biological cells and the cardiovascular apparatus all the way up to planetary systems, stars, galaxies and beyond. This, together with the fact that fluids are vital to human survival, is arguably the reason why hydrodynamics---namely the study and modelling of fluid behaviour---is one of the oldest research areas in physics. 
And yet, despite fluids having attracted the attention of numerous scientists and engineers over the years, hydrodynamic modelling still presents severe conceptual and computational challenges. 
This is demonstrated, for example, by the ``Navier-Stokes existence and smoothness problem'' Millennium Prize, and/or the fact that turbulent flows---namely, the vast majority of real-life flows---are extremely costly to simulate.
Such challenges are ever more formidable when we couple hydrodynamics to relativity, meaning that we consider fluids flowing at velocities commensurate to the speed of light and, at the same time, immersed in a strong gravitational field. 
The challenges added by the (successful) marriage of relativity and hydrodynamics are, not surprisingly, both conceptual and practical. 
To give one example, the most intuitive and natural extensions to relativity of the Navier-Stokes equations---further discussed in this thesis---are well-known to give rise to problems \cite{HiscockInsta}. 
On the practical side, simulations of relativistic fluids are intimidating by their complexity and computational cost. 

As a small side-step before carrying on, let us state clearly that throughout this thesis we consider Einstein's General Relativity as the theory of gravity, even though we know that this cannot be the ``ultimate theory''. 
The indications of this are, in fact both theoretical and observational.
On the one hand we know General Relativity breaks down on the quantum scales.
On the other hand, we may view the need of including mysterious (and predominant) dark energy/matter contents in our cosmological models---required to match observational evidences such as the accelerated expansion of the Universe---as an indication that something is missing in our understanding. 
Nonetheless we will here take a pragmatic/conservative approach and consider General Relativity as the ``correct'' theory of gravity. 
This is well-motivated considering that General Relativity has passed, over the years, weak and strong-field tests with flying colours (see, for example, the recent review by \citet{WillGRtests}).

Given the tremendous challenges posed by the modelling of relativistic fluids, it is perfectly logical to wonder why we should care.
The obvious reason is that there exist real systems whose description/modelling requires relativistic hydrodynamics. 
The most intriguing ones, possibly from a biased perspective, are neutron stars. 
Neutron stars form, quite dramatically, as result of a dying star that was born sufficiently massive---namely with mass $ 8 M_\odot \lesssim M \lesssim 30  M_\odot$, where $M_\odot$ is the mass of our Sun \cite{ShapiroTeukolsky83,NCompstarWB,anderssonGWbook}.
Once it has emptied its nuclear fuel reservoir, such a massive star undergoes an extremely powerful explosion, known as ``core-collapse supernova'', that leaves behind a proto-neutron star. 
In essence, neutron stars are among the most exotic objects in the Universe: they contain roughly the same mass as the Sun squeezed into a $10$ km radius sphere.
Their extreme compactness, second only to black holes, means that General Relativity is an absolute must for accurately modelling neutron stars' phenomenology. 
At the same time, the extreme densities reached inside neutron stars (a few times that of an atomic nucleus) make them ideal laboratories to probe otherwise inaccessible physical regimes. 
In particular, realistic modelling of neutron star phenomenology---and its ``validation'' through simulations and observations---allow us to set constraints on the elusive equation of state encoding matter properties at such extreme densities, and test our understanding of gravity in the strong field regime at the same time. 

When it comes to neutron star astrophysics, it is almost impossible not to mention the spectacular detection of 
$17^{\text{th}}$ August $2017$---hence dubbed $GW170817$. 
For the first time, the Laser Interferometer Gravitational-wave Observatory (LIGO) and Virgo teams captured the faint gravitational wave signal emitted during a binary neutron star inspiral \cite{LIGOScientific:2017vwq}.
This detection came roughly two years after the very first gravitational wave detection accomplished by the LIGO team on $14^{\text{th}}$ September $2015$ \cite{GW150914}. 
In this first remarkable event---known as $GW150914$---the gravitational waves were generated by two black holes inspiralling and merging, and travelled for about $1.3$ billion years (at the speed of light) before reaching the detectors on Earth.
Despite this earlier detection and the many more binary black hole merger detections that followed\footnote{See \href{https://www.ligo.org/detections.php}{https://www.ligo.org/detections.php} for an up to date catalogue.}, $GW170817$ truly represents a milestone for astrophysics. 
The gravitational wave event was soon followed by the detection of a post-merger electromagnetic counterpart \cite{LIGOScientific:2017ync} thus marking the beginning of a new ``multi-messenger era''\footnote{In truth, it was the FERMI gamma ray telescope that sent out the trigger alert to LIGO and other telescopes. A large noise glitch in one of the LIGO detectors at the same time as the real signal, in fact prevented LIGO from sending out the alert first---as the gravitational wave signal arrived in the detectors roughly two seconds before the gamma ray burst detection. However, after having cleaned the data, LIGO confirmed the detection in gravitational waves of the first binary neutron star merger.}.
Moreover, this event also confirmed the long-standing paradigm that much of the heavy elements in our Universe form precisely during such spectacular cosmic fireworks \cite{Kasen+2017,Pian+2017}.  

Whilst it is undeniable that $GW170817$ represents an astonishing achievement, it has also demonstrated the range of exciting features we can explore with neutron star mergers \cite{Bauswein_2017,Rezzolla_2018,Margalit_2017,RuizPRD18,ShibataPRD19}.
Neutron star mergers provide an opportunity to explore many extremes of physics, from the state of matter beyond the nuclear saturation density (the elusive equation of state \cite{Oertel+EoSrev,BurgioFantinaEoSRev,VidanaEoSRev,Lattimer2021})  to the formation of a long-lived merger remnant (likely a black hole \cite{ShibataHotokezakaRev2019,BernuzziRemnant,BaiottiRezzollaRev}). 
Matter outflows and the associated rapid nuclear reactions determine the (hopefully) observable kilonova signature \cite{MetzgerKilo} and the twisting of the stars' magnetic field may help collimate an emerging jet and hence explain observed short gamma-ray bursts \cite{Paczynski,CiolfiRev}. 

Current gravitational-wave facilities, however, are only sensitive to the signal produced in the inspiral phase (during which the orbit shrinks due to energy lost in gravitational waves until the neutron stars touch and merge), even though much of the interesting dynamics happens at merger and during the post merger phase (at the end of which the system eventually settles down to either a black hole or neutron star). 
The violent merger dynamics is affected by the details of the transport properties of matter at such extreme densities and temperatures \cite{AlfordPRL2018}, and hence crucial to set tight constraints on the still-baffling equation of state.
For this reason, new and more sensitive facilities are currently at the planning stage---such as the Einstein Telescope \cite{ETScienceCase} and the Cosmic Explorer \cite{CEbeyondLIGO,CEScienceCase} in the context of gravitational wave detectors---and expected to come online in the $2030$s.
With more sensitive instruments coming on-stream in the future, higher precision observations are anticipated and one may hope to extract more detailed information about the involved physics.
These extremely exciting prospects constitute the motivation for the work presented in this thesis. 

\section{Outline}

As present (and future) binary neutron star merger detections offer a wealth of opportunities to explore several extremes of physics, it is important to have an understanding of the different ingredients required to fully realize this potential---this will also provide further context for the results discussed within this thesis. 

We obviously need sensitive detectors/telescopes (of all sorts) and, for the purpose of this thesis, it will suffice to say that efforts in this direction are under way. 
We obviously also need sophisticated tools to analyze the data, and extract useful physics information from a merger detection. 
Focusing on the gravitational aspects, for example, we need to contrast the gravitational wave signal to waveform templates. 
To construct the templates, we need accurate theoretical modelling of the entire merger process, from inspiral to ringdown \cite{DamourEOB}.
Analytical models (based on the so-called Post-Newtonian expansion \cite{PoissonWill}) can accurately describe the inspiral phase, but they prove inadequate as the orbital separation of the two neutron stars (or black holes) approaches a few times their radii. 
To model the actual merger and post-merger phase we must rely on numerical simulations \cite{RezzollaZanotti,shibataNR}.
This means that, as mergers are highly dynamical events, the full suite of Einstein equations have to be solved together with a suitable description of matter. 
Current simulations tend to involve an evolution of the ideal fluid equations, possibly augmented with a relatively simple scheme (or closure) to account for electromagnetic effects and (neutrino) radiation. 
Although these models already provide a formidable computational task, it is easy to argue that an accurate description of the physics involved in a merger would require even more complicated theoretical models. This is ever more true given the expected advancements in detector technology and sensitivity.

When it comes to neutron stars, it can be easily argued that a fluid model capable of accounting for all the relevant physics should involve at least four distinct components \cite{livrev}. 
We expect, in fact, to find superfluid neutrons and superconducting protons as we go deep enough into the crust and core of a neutron star \cite{Haensel+NS,ChamelLRcrust}, meaning that neutrons and protons flow independently from each other. 
We also expect a flow of electrons, coupled to the protons due to electromagnetic interactions, and finally a heat flux/entropy flow. 
This means that sophisticated multi-fluid models constitute an important part of the story, and a substantial part of this thesis is dedicated to discussing them.
Moreover, as the violent dynamics of a merger is expected to push matter out of (local) thermodynamic equilibrium \cite{PeteThermal,PeteGWreactions}, we also need to include dissipation in the models we would like to simulate. 
At the same time, however, while there is no doubt that multifluid models are crucial for modelling the phenomenology of neutron stars\footnote{Pulsar glitches, for example, are considered smoking gun evidence for the presence of superfluid phases in the interiors of neutron star \cite{HaskellMelatosRev,Marco+PierreBook}.}, the question arises as to whether, say, superconductivity in the core will impact merger and post-merger dynamics. 
We may expect, for example, that the heating generated by the two neutron stars smashing against each other is going to melt the crust and push matter above the relevant critical temperature. In this respect, however, current simulations do not provide a definite answer, and some suggest that parts of the core could remain cold enough for superfluidity/superconductivity to play a role \cite{PeteThermal}. Obviously, the presence of superfluid/superconducting phases would change the underlying dissipative mechanisms.
In essence, given the expected sensitivity of future detectors, we are urged to check whether a more realistic modelling of neutron star matter can leave a measurable imprint in the detected signals.

In developing such complicated theoretical models, however, we need to bear in mind the extreme computational costs of simulating them, forcing us to consider simplifications whenever possible.
The limitations associated with actual numerical implementations become ever more crucial if we consider that we expect turbulence to develop in mergers due to known fluid instabilities \cite{Lesieur,mcdonough}. In fact, the modelling of turbulence is an extremely complicated and subtle business already at the Newtonian level. 
In essence, we would like to make sure that the underlying physical features are faithfully represented by our theoretical models and, at the same time, make sure that we ``get the physics right'' in our numerical simulations. This is clearly a very challenging issue. 
As a final point, it is also crucial to keep in mind that we need to make sure that the physics we hope to explore is cleanly associated with particular observational features.

In this thesis we will discuss recent theoretical advancements in (electromagnetic) multi-fluid modelling in General Relativity, motivated by applications to neutron star astrophysics and neutron star mergers in particular.
The introduction so far already suggests that there are a number of interconnected aspects to be kept in mind, from dissipation to turbulence. 
Given the complexity of the systems we aim to model, it is natural to consider each of these different aspects ``one at a time''. 
As such, the thesis is divided into three different parts, each of which is somewhat self-contained and could be read almost independently from the rest. 
Nonetheless, the discussions provided in the different parts complement each other to form a coherent story. 

The first part of this thesis is devoted to the modelling of dissipative multi-fluids. 
Modelling dissipation in relativity is challenging already for single fluid models, and has baffled physicists for quite some time. 
Several ideas and prescriptions are currently on the market, and some of them are fairly recent. 
We will therefore start the first part by reviewing in \cref{ch:DissipationLiteratureReview} the different modelling strategies. 
The discussion will highlight, in particular, that most of the present strategies do not seem to allow for an ``easy'' extension to multi-fluid models, which is presumably required for neutron stars.
We will then continue in \cref{ch:Linearizing} focusing on the (arguably) unique strategy that is clearly suited for this.
In particular we will study the close to thermodynamic equilibrium regime of an action based variational model for dissipative multi-fluids. 

The second part of this thesis is devoted to discussing some of the issues that arise when modelling turbulence in relativity.
Turbulent models in Newtonian hydrodynamics often involve some notion of filtering/averaging, and such strategies are becoming fashionable also in the relativity community, with impressive successes \cite{PalenzuelaBampinBNS,Aguilera-Miret-Universality}. 
However, whilst there has been significant numerical effort going into extending the Newtonian logic to relativity, the foundational underpinnings for these strategies are not as well explored as one may like. 
As such, after a brief introduction to hydrodynamic turbulence, we will focus in \cref{ch:LES} on discussing the foundational issues that arise in extending the Newtonian logic to the relativistic setting. 
This brings us to propose a novel scheme for carrying out filtering for relativistic fluids.
Whilst the analysis in \cref{ch:LES} focuses on single-fluid models, in the following \cref{ch:LESMHD} we will discuss the first steps towards extending such a scheme to multi-fluids.
Even though at first sight this second part may seem relatively independent from the previous one, it is rather evident that the two are linked. 
A first indication of this comes from the well-known fact that energy and momentum transport are enhanced in turbulent flows.
Furthermore, filtering an ideal fluid model---that is a model that does not account for dissipative effects---inevitably introduces additional terms in the equations that are akin to ``effective dissipative terms'', as we demonstrate in \cref{ch:LES}. 

While the first two parts of this thesis are rather theoretical---although the discussion in \cref{part2} brings to the fore the key role played by computational limitations---we will continue in \cref{part3} by considering two relevant binary neutron star merger applications. 
In particular, in \cref{ch:BVinSIM} we will focus on modelling reaction-sourced bulk viscosity for neutron star simulations---as reactions are expected to source the dominant dissipative mechanism in mergers.
The topic is relatively well-explored from a ``theoretical'' perspective \cite{AlfordBulk10,AlfordHarris18,AlfordHarrisDamping18,AlfordHarutyunuanSedrakian19,AlfordHarutyunuanSedrakian21}---but still numerically challenging---so that our analysis aims to establish how the inevitable ``limitations'' of a numerical simulation (in terms of resolution) enter the discussion. 
In particular, we also explore the link to (or conflict with) strategies for dealing with turbulent flows. 
We then continue our journey focusing in \cref{ch:MRI} on the magneto-rotational instability (MRI), which is considered a key mechanism for developing/sustaining turbulence in the outer envelopes of merger remnants. 
Crucially, the analysis is framed in a way suited for highly dynamical environments such as mergers. 
We provide our concluding remarks and comment on future works in \cref{part4}.

Supplemental material is provided in appendices. 
In \cref{app:FermiCoord} we discuss the Fermi coordinates as these play a key role in \cref{ch:LES}.
Appendix \ref{app:MultiScale} provides additional information about the multi-scale methods used in \cref{ch:BVinSIM}, while in \cref{app:compOSE} we make explicit contact with quantities that can be computed from standard equation of state tables as collected, for example, in the compOSE database \cite{compOSE}.
In \cref{app:MRIlocal} we link the analysis of \cref{ch:MRI} to the usual MRI results/criteria. 
Finally, in \cref{app:RH} we discuss the Routh-Hurwitz criterion, which is going to be used both in \cref{ch:LES,ch:MRI}.

\section{Relativistic perfect fluids}\label{sec:PerfectFluids}

As an appetizer before we begin the main part of our journey, let us cover some background material that is relevant to all the following parts.
Quite naturally, dissipative fluid models build on the notion of perfect fluids, namely fluids where dissipative effects can be neglected. 
As for turbulence, we expect it to develop in fluid flows where inertial effects prevail over viscous/dissipative ones. 
As such, we need to understand how to model (relativistic) ideal fluids before we can meaningfully start talking about dissipation and turbulence.
Excellent introductions to relativistic ideal fluids can be found in several textbooks/reviews---such as, for example, \cite{LandauFLuidMechanics,WeinbergGravitation,GravitationMTW,friedman2013rotating,RezzollaZanotti,livrev}---and the material is often covered in introductions to General Relativity. 
Nonetheless, we take this as an opportunity to set the stage for the analysis presented in the following chapters.

Let us begin by writing down the Einstein field equations (in geometric units) 
\begin{equation}
    G^{ab} = 8\pi T^{ab} \;,
\end{equation}
where $G^{ab}$ is the Einstein tensor associated with the spacetime metric, while $T^{ab}$ is the stress-energy-momentum tensor associated with the matter content within the spacetime. 
In essence, these equations prescribe how the matter/energy content curves the spacetime and, in turn, the spacetime metric/curvature dictates how (test, freely-falling) particles move in spacetime. 
An important property of the Einstein tensor is that it is divergence free, meaning that 
\begin{equation}
    \nabla_a G^{ab} = 0 \;,
\end{equation}
which follows from its definition and the second Bianchi identity satisfied by the Riemann tensor. 
This means that, as a consequence of the Einstein field equations, we also have 
\begin{equation}\label{eq:TabCons}
    \nabla_a T^{ab} = 0\;, 
\end{equation}
namely, energy and momentum are locally conserved. 
From a field-theory perspective, the conservation of the stress-energy-momentum tensor is associated with diffeomorphism invariance---and hence is analogous to conservation laws obeyed by the Noether currents whenever the field-theory has some underlying symmetry \cite{maggioreQFT,weinbergQFT}.
Whilst the distinction between equations of motion and conserved Noether currents becomes important in the context of multi-fluid modelling---and we will come back to stress this issue---we here follow the ``tradition'' and refer to \cref{eq:TabCons} as the equations of motion of hydrodynamics. 

It is immediately clear, however, that in order for \cref{eq:TabCons} to yield equations we can work with (or simulate), we need to specify the stress-energy-momentum tensor---otherwise our fluid model is somewhat ``empty''. 
To do so, we first note that the stress-energy-momentum tensor can be decomposed---just like any other tensor---into parts that are, algebraically, parallel or orthogonal with respect to some observer $U^a$.\footnote{By observer we here mean a normalized time-like unit vector. The concept will be further developed in the following chapters.}
In particular, we can decompose the stress-energy-momentum tensor as 
\begin{equation}
    T^{ab} = \E U^a U^b + 2 \Q^{(a} U^{b)} + \S^{ab}
\end{equation}
where 
\begin{subequations}
\begin{equation}
    \E = T^{ab}U_a U_b \;, \quad \Q^a = \perp^a_b T^{bc}U_c \;, \quad \S^{ab} = \perp^{ac}\perp^{bd}T_{cd} \;.
\end{equation}
\end{subequations}
The different terms are then interpreted as follows: $\E$ is the energy density measured by the observer, $\Q^a$ is the spatial momentum flux measured by the observer (and, by definition, satisfies $\Q^a U_a = 0$) while $\S^{ab}$ encodes the spatial stresses (and is obviously a symmetric tensor). 
Whilst this decomposition is useful, it is just an algebraic decomposition. 
As such, we need to do some more work. 

In order to arrive at the perfect fluid equations we now introduce the concept of locally co-moving observer, which we denote with $u^a$ in order to distinguish it from the generic one introduced above. 
We assume that the particles constituting the fluid collide sufficiently often (using a classical mechanics language) that we can meaningfully talk about a mean velocity field associated with the particles' collective motion. 
The co-moving observer is then defined as the observer that moves with the mean flow.
Next, we further assume that the frequent collisions are random in nature, so that an observer moving with the mean flow would observe an isotropic distribution of particles.
This then brings us to consider the most generic stress-energy-momentum tensor that is consistent with such a notion of isotropy.
In essence, we will require every quantity in the stress-energy-momentum tensor decomposition to be invariant with respect to spatial rotations (spatial with respect to the co-moving observer $u^a$).

We can rephrase the last assumption in a more formal way, by saying that we retain only those parts of $T^{ab}$---which belongs to a rank-two tensorial representation of the Lorentz group\footnote{In fact, we can always introduce a non-coordinate basis/tetrad such that the metric is in Minkowski form at each point in spacetime. 
Using such a basis, any tensor becomes a Lorentz tensor.
This has the additional advantage that it shows how in General Relativity we take the Poincaré group---namely the Lorentz group augmented with space-time translations---and require it to be a local/gauge symmetry of the theory.
Invariance with respect to gauged space-time translations is nothing but invariance with respect to diffeomorphisms. See, for example, \citet{GRCarroll}.} $SO(3,1)$---that transform as scalars under the rotation group $SO(3)$. 
As the rotation group is a sub-group of the Lorentz one, the decomposition of different representations of the Lorentz group with respect to the rotation group are well known. 
In particular, a symmetric rank-two (Lorentz) tensor is made up of two scalars, one vector, and a spin-2 part with respect to the rotation group.
The assumption of isotropy means that we take the vectorial and spin-2 part to vanish.
However, let us here leave the formalities aside and proceed with a more intuitive discussion.
We begin with the energy density, which is obviously a scalar under the Lorentz group. 
As such, the energy density is also a scalar with respect to rotations and we can keep it. 
Next, we observe that the momentum flux is a spatial vector living in the subspace orthogonal to the observer four-vector $u^a$. 
This means that a non vanishing momentum flux vector would identify a preferred direction in the subspace orthogonal to $u^a$. 
In order to not break isotropy, then, we need to assume that it vanishes.  
Finally, we focus on the spatial stresses and note these can be further decomposed as 
\begin{equation}
    \S^{ab} = \S \perp^{ab} + \S^{\langle ab\rangle} \;, \quad \text{where } \S = \S^a_{\;a} \;,\; \S^{\langle ab\rangle} = \S^{ab} - \frac{1}{3}g^{ab} \S \;.
\end{equation}
In short, $\S$ is the trace and $\S^{\langle ab\rangle}$ is the (symmetric) trace-free part of the stress tensor. 
The trace is a Lorentz scalar so that the same logic as for the energy density applies.
On the other hand, the trace-free part describes anisotropic stresses and hence must vanish in isotropic fluids.
We conclude that the stress-energy-momentum tensor of an ideal fluid takes the form
\begin{equation}
    T^{ab} = \veps u^a u^b + p ( g^{ab} + u^a u^b )\;.
\end{equation}
Note that we have here re-labelled the energy density and isotropic stresses as $\veps$ and $p$ (respectively), consistent with the standard notation. 
The isotropic stresses are then identified with the (equilibrium) thermodynamic pressure, usually given in the form of an equation of state (either in analytical form or provided as a table of data). 

Since we now have a prescription for the stress-energy-momentum tensor, we can use it to make the energy-momentum conservation law more explicit.
Working out the parallel and orthogonal projections of \cref{eq:TabCons} we obtain\footnote{It should be obvious by now that we here mean parallel and orthogonal with respect to the comoving observer.}
\begin{subequations}\label{eq:IdealFluidConsLaw}
\begin{align}
     &u^a \nabla_a \veps + (\veps + p) \nabla_a u^a = 0 \;, \\
     &(p + \veps) a^b + \perp^{bc} \nabla_c p = 0 \;.
\end{align}
\end{subequations}
The first relation is, intuitively, the familiar energy density balance law.
It says, in fact, that the time derivative of the energy density is proportional to the expansion rate of the fluid (with a negative sign). 
If the fluid undergoes compression (negative expansion rate), the energy density will increase because i) the same energy is now stored in a smaller volume ii) of the ``$p dV$''-work the fluid element does on its surroundings.
The second equation is instead the relativistic Euler equation: it says that the fluid is accelerated in such a way to minimize pressure gradients, where the pre-factor in front of the four-acceleration $a^b$ accounts for pressure contributions to the total mass-energy of the fluid. 

We can think of the two equations in \eqref{eq:IdealFluidConsLaw} as evolution equations for the energy density and the four velocity, respectively\footnote{That we only need three equations to fully determine the four velocity is due to the fact that a fourth component can be obtained from the constraint $u^a u_a = -1$ provided the other three are known.}.
The system would then be closed provided we have an equation of state of the form $p(\veps)$. 
Whilst this is a valid equation of state for, say, a fluid made of radiation, we would normally work with a two-parameter (or more) equation of state. For example, we can think of it as a function $p = p (n,\veps)$, where $n$ is the particle number density\footnote{In non-relativistic fluids, the pressure is often given as function of the mass density instead.}. 
This means that we need an additional evolution equation for $n$. 
As the particle number density is typically associated with baryons, we are then led to write a continuity equation 
\begin{equation}\label{eq:particleNumberContinuity}
    \nabla_a (n u^a) =0 \Longrightarrow u^a \nabla_a n + n \nabla_a u^a = 0 \;.
\end{equation}
Note that the particle number density continuity equation plays the role of the mass continuity equation in non-relativistic fluids (see, e.g. \cite{ShapiroTeukolsky83}).
Taken together, \cref{eq:particleNumberContinuity,eq:IdealFluidConsLaw} form a closed system of equations\footnote{We also note that more work is required to write the same equations in a form suitable for numerical implementations \cite{RezzollaZanotti,shibataNR,Gourgoulhon3+1}. 
However, this would require us to cover additional background, which would be beyond the scope of this thesis anyway.}.

We conclude this introduction to relativistic perfect fluids showing why this model truly describes an ideal fluid. 
To do so, however, we need to talk a little bit about (equilibrium) thermodynamics. 
Let us first of all consider the first law of thermodynamics written in terms of quantities per unit volume (see, for example, \cite{livrev})
\begin{equation}
    \d\veps = T\d s + \mu\d n  \;, \quad T = \left(\pdv{\veps}{s}\right)_{n} \;, \quad  \quad \mu = \left(\pdv{\veps}{n}\right)_{s} 
\end{equation}
where $\mu\;,T$ are the chemical potential and temperature, while $s$ in the entropy density---and $n,\,\veps$ are the particle number and energy densities as before.
Similarly, the ``Euler relation'' is 
\begin{equation}
    p + \veps = T s + \mu n \;.
\end{equation}
In essence, what this shows is that we can think of the equation of state as some relation that allows us to express the various thermodynamic quantities in terms of our two favourite ones. 
Quite naturally from a theory perspective, we are then led to choose these as the energy and particle number densities. 
In particular, we can think of the entropy density as $s = s(n,\veps)$. 
As such, we can derive an evolution equation for the entropy density starting from the relevant equations for the energy and particle densities. 
It is then easy to show that
\begin{multline}\label{eq:EntropyAdvection}
    T \nabla_a (s u^a ) = T u^a \nabla_a s + Ts \nabla_a u^a = u^a \nabla_a \veps - \mu u^a \nabla_a n + \left(p + \veps - \mu n\right)\nabla_a u^a \\
    = \left( u^a \nabla_a \veps + (p+\veps) \nabla_a u^a \right)- \mu\left( u^a \nabla_a n + n\nabla_a u^a \right) = 0 \;,   
\end{multline}
where we have used the thermodynamic relations above as well as the continuity equations for the energy and particle number densities.
In essence, this equation says that the entropy-current $su^a$ satisfies a continuity equation as well.
As thermodynamics dictates that dissipation is associated with an increase in entropy, the model truly represents an ideal fluid.

%% file: Parts/DissipativeHydro/part1macro.tex
\input{Parts/DissipativeHydro/LiteratureReview}
\input{Parts/DissipativeHydro/Linearizing}

%% file: Parts/DissipativeHydro/LiteratureReview.tex
\chapter{Dissipative fluid models in General Relativity: an overview} \label{ch:DissipationLiteratureReview}

The first part of this thesis focuses on modelling dissipative fluids in (general) relativity, which is a thorny issue that has kept physicists busy for quite some time. 
The covariant nature of General Relativity highlights the central role played by the reference frame used to describe a physical system. 
At the same time, the evolution of dissipative fluids must be consistent with thermodynamic principles and the arrow of time associated with the second law. 
As such, it is clear that the marriage of General Relativity and thermodynamics poses interesting foundational questions, so that it should not come a surprise that a number of authors have made significant contributions over the years\footnote{Since the list of authors is indeed very long (and growing), we find it difficult to do justice to all of them here. Instead, we will acknowledge as we go along the most important contributions (for our purposes) made by the various authors, and point to \cite{RezzollaZanotti,livrev,RomatschkeRFL,Bemfica2020} for further references.}.

Here we present an overview of the most important developments and results obtained over the past seventy years. 
The aim is not to draw a complete picture but only to introduce the main ideas and set the stage for what comes next. 
We will, in fact, continue in the following \cref{ch:Linearizing} by focusing on the action-based dissipative framework of \citet{2015CQAnderssonComer}.
We do so as this is the only framework currently on the market that is suited for describing multi-fluid systems.

\section{Non-equilibrium thermodynamics}

As hydrodynamics is a macroscopic theory, fluid models must be complemented with some information about the microphysics of the system they aim to describe. 
For the ideal fluid case, for instance, such information may be the chemical potentials of the fluid constituents. 
These can be obtained within the equilibrium thermodynamics framework. 
However, if we move to the non-ideal case, new effects like viscosity or heat conductivity start to play a role in the macroscopic dynamics. 
To take these effects into account, one needs to provide hydrodynamic models with new information, typically in the form of transport coefficients. 
As these describe out of equilibrium processes, we start with a short discussion of non-equilibrium thermodynamics.
Note that the analysis in this section is Newtonian. We will discuss how to re-phrase the same ideas in a relativistic setting later. 

\subsection{Linear irreversible thermodynamics}\label{subsec:LIT}
Classical equilibrium thermodynamics is a macroscopic theory that aims at describing the observed properties of a many-particle system---at thermodynamic equilibrium--- in terms of a finite set of macroscopic variables (like energy $E$, volume $V$ and particle's number $N$). It does not describe, however, the evolution of a system towards such equilibrium states, nor the damping of statistical fluctuations around it. These kind of phenomena are the realm of irreversible thermodynamics. 

A first fundamental result in the study of irreversible processes is due to the pioneering work of \citet{onsager31:_symmetry}, and it is now customary to refer to it as ``Linear Irreversible Thermodynamics'' (LIT). To sketch the main ideas of LIT, let us consider the entropy of a system out of (but close to) equilibrium, assuming that the entropy $S$ is function of some state variables $A_1,\,A_2,\,A_3 \dots$ The key assumption of LIT is that the number of variables necessary to specify the out of equilibrium state of the system is the same as in equilibrium. Then we can write\footnote{In this section we will use the greek letters $\alpha,\,\beta\dots$ to label the dissipative processes that lead to some entropy production. The Einstein summation convention does not apply to them.} 
\begin{equation}\label{eq:OnsagerEntropy}
    S= S_{\e\q} - \frac{1}{2}\sum_{\beta,\gamma}G_{\beta\gamma}\alpha_\beta\alpha_\gamma \;,
\end{equation}
where $\alpha_\beta$ is the deviation from the equilibrium value of the corresponding variable $A_{\beta}$. If we then assume the \textit{Onsager regression hypothesis}---which states that spontaneous fluctuations of the system decay with the same evolution law as the perturbations caused by external forces---we can write 
\begin{equation}\label{eq:OnsagerRegression}
    \frac{d\alpha_\beta}{dt} = -\sum_\gamma M_{\beta\gamma}\alpha_\gamma \;,
\end{equation}
where the matrix $\boldsymbol{M}$ describes the decay of spontaneous fluctuations. The entropy production in turn is\footnote{Let us note for clarity that the thermodynamic quantities in this section are global and refer to the entire system under consideration. As a consequence, all equilibrium quantities are constant in time. This applies also to $G_{\beta \gamma}$ defined in \cref{eq:OnsagerEntropy}, which is evaluated at equilibrium.}
\begin{equation}
    \frac{d S}{dt} =- \sum_{\beta,\gamma}  G_{\beta\gamma}\alpha_\beta\frac{d\alpha_\gamma}{dt} = \sum_\gamma J_\gamma X_\gamma \;,
\end{equation}
where we have introduced the \textit{thermodynamic forces} $X_\gamma=- \sum_\beta  G_{\beta\gamma}\alpha_\beta$ and \textit{thermodynamic fluxes} $J_\gamma = d\alpha_\gamma/dt$. By means of \cref{eq:OnsagerRegression} we can rewrite the entropy production rate as 
\begin{equation}\label{eq:OnsagerEntropyProduction}
	\Gamma_s = \frac{dS}{dt} = \sum_{\beta\gamma}L_{\beta\gamma}X_\beta X_\gamma \;,
\end{equation}
where the matrix $\boldsymbol{L}= \boldsymbol{M}\cdot\boldsymbol{G}^{-1}$.  The second law of thermodynamics then implies that the symmetric part of $\boldsymbol{L}$ must be a positive semi-definite matrix. 
While in his original discussion Onsager only considered thermodynamic variables that are even under time reversal, Casimir later extended the analysis to include also odd variables (e.g. the magnetic field) \cite{Casimir45}, demonstrating the so-called Onsager-Casimir relations: 
\begin{equation}
	L_{\beta\gamma} = \veps_\beta \veps_\gamma L_{\gamma\beta} \;,
\end{equation}
where $\veps_\beta$ is the parity of the variable $\alpha_\beta$ under time reversal. 

\subsection{Causality and extended irreversible thermodynamics}\label{ch1subsec:EITCattaneo}

Despite LIT being successful and largely used in many different contexts, it has serious drawbacks that are disturbing already at the Newtonian level, and unacceptable at the relativistic one. For instance, the standard Fourier law for the heat-flux $q$ (here given in 1+1 dimensions) $q= - \kappa \partial_x T$ can be deduced within the LIT program\footnote{Here $\kappa$ is the heat conductivity, while $T$ denotes the temperature.}. When the Fourier law is coupled to the balance law for the energy density $\veps$ of a system at rest $\pd{t}{\veps} = -\pd{x}{q}$, it yields the heat equation 
\begin{equation}
    \pdv{T}{t} - D\pdv[2]{T}{x} = 0 \;,
\end{equation}
where we introduced the heat diffusivity $D$, defined as $D = \kappa/ c_v$ where $c_v$ is the heat capacity at constant volume. 
As a result, a temperature perturbation will propagate with infinite speed\footnote{To better understand this statement it is sufficient to calculate the kernel of the heat equation---that is, a Gaussian.
As a result, an initial temperature perturbation localized at the origin will result after some finite time in a non-zero disturbance in the temperature field at all distances from the origin.}. 
This is obviously against our intuition as we would expect thermal and viscous disturbances to propagate with a speed of the order of, say, the mean molecular speed.
Whilst this problem may be ignored for non-relativistic applications where typical speeds are much smaller than the speed of light $c$, it comes back to bite us in a special/general relativistic setting. 
The Extended Irreversible Thermodynamics (EIT) program is built in order to address the issues of LIT keeping the analysis at a thermodynamical level. We now sketch the main ideas of EIT and point to \cite{EIT} for an exhaustive discussion.

The key assumption of EIT is that some (the majority) of the microscopic degrees of freedom rapidly decay towards their equilibrium values while others will do so on longer timescales. The causality issues of LIT are solved in the EIT paradigm by enlarging the set of variables that describe the out-of-equilibrium state of the system---through the inclusion of slowly decaying variables\footnote{This idea was first proposed by \citet{Muller:1967zza}.}. At the level of a phenomenological macroscopic theory, the additional variables can be chosen as the thermodynamic fluxes introduced above. 
The reason for this can be seen by going back to \cref{eq:OnsagerEntropyProduction}. 
On the one hand, the entropy production rate---a central quantity in the Onsager LIT program---must be quadratic to retain consistency with the Second law of thermodynamics. On the other hand, the out-of-equilibrium entropy is in general a function of a larger set of variables---compared with the equilibrium one. 
If we then expand the entropy up to second order, like in \cref{eq:OnsagerEntropy,eq:OnsagerEntropyProduction}, the dependence on these additional variables must be taken into account. This is exactly the ``missing piece'' in Onsager LIT, and the ultimate reason for its non-causal predictions. To better understand this, let us consider the simple example of a Newtonian fluid with heat-flux. 

Let us start by assuming the (generalized) entropy density  $s$ of the system is a function of the  energy density as usual and one additional non-equilibrium variable, the heat-flux $q$:
\begin{equation}
    s = s(\veps,q) \;.
\end{equation}
As a result, its differential reads
\begin{equation}
    	ds = \theta^{-1} d\veps + adq  \Longrightarrow
		\pdv{s}{t} = \theta^{-1} \pdv{\veps}{t} + a\pdv{q}{t} \;.
\end{equation}
where both $\theta$---which is a generalized temperature\footnote{By generalized temperature we here mean a notion of temperature valid also out of equilibrium, noting that this notion is not unique \cite{EIT}.}---and $a$ are functions of $(\veps,q)$.  Next, since in equilibrium there is no heat-flux and the entropy is maximized, we can expand the generalized entropy as $s(\veps,q)=s_\e(\veps) -\frac{1}{2} s_2(\veps)q^2$ and write 
\begin{subequations}
\begin{align}
	a &= \left(\frac{\partial s}{\partial q}\right)_\veps =- s_2q \;,\\
    \theta^{-1} &= \left(\frac{\partial s}{\partial u}\right)_{q} = \frac{1}{T} -\frac{1}{2} \dv{s_2}{\veps}q^2 \;.    
\end{align}
\end{subequations}
The net result is that we can now compute $\pd{t}{s}$ and write it in the form of a balance equation as 
\begin{equation}
    \pdv{s}{t} = - \pdv{j_s}{x} + \Gamma_\s \;, 
\end{equation}
where the entropy current $j_\s$ and the entropy production rate are
\begin{subequations}
\begin{align}
    j_\s &= \frac{1}{T} q\;,\\ 
    \Gamma_\s &= q \left[ \pdv{\left(T^{-1}\right)}{x}  - s_2\, \pdv{q}{t}\right]  \;,
\end{align}
\end{subequations}
and we have used the energy conservation law as above.
We can then proceed exactly as in the LIT program by writing the entropy production rate as the product of thermodynamic forces and fluxes; namely $\Gamma_\s = q X_q$. Assuming a phenomenological law for the force $X_\q = \mu (\veps)q$ we obtain the following law for the heat-flux\footnote{Where $\tau = s_2 / \mu$ and $\kappa = (\mu T^2)^{-1}$, although this is not really crucial for the present discussion.}
\begin{equation}\label{eq:CattaneoLaw}
	\tau \pdv{q}{t} + q = - \kappa \pdv{T}{x} \;.
\end{equation}
In conclusion, by simply enlarging the set of state variables used to described an out-of-equilibrium system---and then following the same strategy as in the Onsager LIT program---we obtained the Maxwell-Cattaneo law for the heat-flux (\cref{eq:CattaneoLaw}). 
We can then immediately use the last equation and the energy balance law to get an evolution equation for the temperature of the telegrapher type: 
\begin{equation}
	\tau \pdv[2]{T}{t} + \pdv{T}{t} - D\,\pdv[2]{T}{x} = 0 \;.
\end{equation}
This equation is hyperbolic and propagates temperature disturbances with a finite speed, its solution becoming indistinguishable from that of the heat-equation at late times. 

\section{Traditional strategies for dissipative fluids}\label{sec:TraditionalStrategies}

Having discussed the basic ideas behind the EIT paradigm, we now proceed to describe the traditional strategies to model dissipative fluids in relativity. We will first focus on the so-called first-order theories---the Eckart-Landau-Lifshitz models---and then on the M{\"u}ller-Israel-Stewart theory, following the presentation in \cite{livrev}. We conclude with a brief discussion on the more abstract divergence-type theories. 

\subsection{Eckart-Landau-Lifshitz models}\label{subsec:LandauEckart}

With the term ``first-order'' theories one typically refers to the dissipative models introduced by \citet{Eckart} and \citet{LandauFLuidMechanics}. They represent the most simple and natural generalization of the Newtonian Fourier-Navier-Stokes equations. These two models are essentially the same in terms of physics content---the difference being in the observer chosen to ``measure'' the various quantities---so we follow \cite{livrev} and present them as one. Let us note that, however, the mathematical properties of the equations in the two models are not the same---and this impacts on their stability and causality properties.  

In the Landau-Eckart models, the equations of motion are given by simple conservation laws\footnote{These equations constitute the ``matter-sector'' of the theory, so that in General Relativity they must be coupled with Einstein equations.}
\begin{subequations}\label{eq:BalanceEquations}
\begin{align}
    &\nabla_an^a = 0 \;, \\
    &\nabla_a T^{ab} = 0 \;,
\end{align}
\end{subequations}
where $n^a$ is the particle flux (or conserved baryon current) and $T^{ab}$ is the total stress-energy-momentum tensor. 
In order to account for dissipative effects the stress-energy-momentum tensor is decomposed as 
\begin{equation}\label{eq:Stress-energyDissipativeDecomposition}
	T^{ab} = \underbrace{ (\veps + p)u^au^b + p g^{ab}}_{\text{ideal terms}} + \underbrace{\chi\perp^{ab} + 2q^{(a}u^{b)} + \chi^{ab}}_{\text{dissipative terms}}\;,
\end{equation}
where we have separated the terms that would be present also in the ideal case from the rest.
Making contact with the discussion in \cref{sec:PerfectFluids}, we have split $\S = p + \chi$ and renamed $S^{\langle ab\rangle} = \chi^{ab}$.
In practice, the stress-energy-momentum tensor of some reference equilibrium state is augmented by the dissipative fluxes introduced above. In Eckart-Landau-Lifshitz theories (as well as the M{\"u}ller-Israel-Stewart model discussed later) the pressure and energy density of the fluid are \underline{assumed} to be equal to the reference equilibrium values\footnote{Or better, the reference equilibrium is defined as the one having the same energy and pressure, and vanishing thermodynamic fluxes.}. The terms $q^a,\,\chi,\,\chi^{ab}$ represent dissipative fluxes, respectively the heat-flux, the bulk-viscous scalar and the shear-viscous tensor. These additional terms are required to satisfy the following algebraic constraints
\begin{subequations}
\begin{align}
	u^aq_a = \chi^a_a &= 0\;, \\
    u^a\chi_{ab} &= 0 \;, \\
    \chi_{[ab]} &= 0 \;.    
\end{align}
\end{subequations}
At the end of the day, \cref{eq:Stress-energyDissipativeDecomposition} is just the same algebraic decomposition we encountered in \cref{sec:PerfectFluids}.
In a similar fashion, the the particle flux $n^a$ is linked to the observer 4-velocity as
\begin{equation}
	n^a = nu^a + \nu^a \;,
\end{equation}
where the diffusion vector $\nu^a$  is required to be orthogonal to the observer 4-velocity $u^a\nu_a=0$. Hence it is proportional to the relative velocity between the observer and the particle flux. The Eckart or Landau-Lifschitz frame are obtained by choosing the observer to be respectively the matter frame or the momentum frame---practically this means setting either $\nu^a$ or $q^a$ to zero in the equations above. 

Because of the additional quantities in the stress-energy-momentum tensor decomposition,  \cref{eq:BalanceEquations} are under-determined. 
In first-order theories, the system is closed by introducing an entropy flux $s^a$, which is assumed to be a linear combination of all the available vectors as 
\begin{equation}\label{eq:EntropyFluxLinearOrder}
    s^a = su^a + \beta q^a - \lambda \nu^a \;, 
	\end{equation}
with so-far unspecified coefficients.
We also have 
\begin{equation}\label{eq:LEentropydensity}
	-u_as^a = s \;,
\end{equation}
so that $s = s(\veps, n)$ in \cref{eq:EntropyFluxLinearOrder} is the entropy density (as measured by the chosen observer) and it is assumed to satisfy an equilibrium thermodynamic relation  
\begin{equation}\label{eq:LITEntropyFunctional}
	n\nabla_ax_s = \frac{1}{T}\nabla_a\veps -\frac{p+\veps}{nT}\nabla_an \;,
\end{equation}
where $x_s = s/n$ is the entropy per particle. 
By means of the last equation and the energy conservation law $u_b\nabla_aT^{ab} = 0$ we can write the entropy production rate as
\begin{equation}
\begin{split}
	\nabla_as^a &= q^a \Big( \nabla_a\beta -\frac{1}{T} u^b\nabla_bu_a\Big) + \Big(\frac{1}{T} -\beta\Big) \nabla_aq^a \\
    &- \Big( x_s + \lambda - \frac{p +\veps}{nT}\Big) \nabla_a\nu^a -\nu^a\nabla_a\lambda - \frac{\chi}{T} \nabla_au^a -\frac{\chi^{ab}}{T} \nabla_au_b \;.
\end{split} 
\end{equation}
To ensure consistency with the thermodynamic second law we first set
\begin{subequations}
\begin{align}
	\beta &= \frac{1}{T} \;,\\
    \lambda &= \frac{1}{nT} \bigg(p + \veps - sT\bigg) = \frac{\mu}{T}\;.
\end{align}
\end{subequations}
In practice, we identify the parameters introduced in \cref{eq:EntropyFluxLinearOrder} with the (inverse) temperature and the Gibbs free-energy per particle. Then, the second law is guaranteed to hold in the simplest possible way---often referred to as the ``natural way''---by guessing the following constitutive relations for the dissipative fluxes
\begin{subequations}\label{eq:1stOrderConstitutiveEqs}
\begin{align}
    q^a &= -\kappa\,T\,\perp^{ab}\Big(\frac{1}{T}\nabla_bT + u^c\nabla_cu_b \Big) \;, \\
    \nu^a &= -\sigma T^2 \perp^{ab}\nabla_b\lambda \;, \\
    \chi &= -\zeta \theta \;, \\
    \chi^{ab} &= - \eta \sigma^{ab} \;,
\end{align}
\end{subequations}
where $\kappa$ is the heat conductivity (as before) and we introduced the diffusion coefficient $\sigma$ and the bulk-and shear viscosity coefficients $\zeta,\eta$. 
Moreover, let us stress that $\theta$ here is the expansion rate of the observer (and not a generalized temperature), and similarly $\sigma^{ab}$ is the shear tensor.  
We can now rewrite the entropy production rate as 
\begin{equation}
	\nabla_as^a = \frac{q^aq_a}{\kappa T} + \frac{\chi^2}{\eta T} + \frac{\nu^a\nu_a}{\sigma T^2} + \frac{\chi^{ab}\chi_{ab}}{\eta T} \ge 0 \;.
\end{equation}
The system of equation is now determined and we just need to make sure that the transport coefficients---determined by microphysics--- introduced above are all positive.  

Let us conclude with a few comments on these first-order models. First of all, despite being quite natural and intuitive, these models have been shown to suffer from stability and causality issues. This means that if we set the system to deviate slightly from thermodynamic equilibrium, the deviations from it may grow rapidly in time (see \cite{HiscockInsta}). Also, they give rise to non-causal behaviour. To better understand this it is sufficient to go back to \cref{eq:1stOrderConstitutiveEqs} and observe that, for instance, the heat equation is nothing more than a relativistic version of the usual Fourier law.  The ultimate reason for this lies in the use of an equilibrium entropy (see \cref{eq:LEentropydensity,eq:LITEntropyFunctional}). In the parlance of the previous section this is an LIT description---hence, is not surprising that the resulting equations are non-causal. 

More recent works (see \cite{Bemfica2018,KovtunStable,HoultKovtun2020} for instance) have shown that one can build first-order theories that respects the stability/causality requirements. Because such models are motivated within a field-theory perspective, we will discuss these theories in \cref{sec:HydrodynamicFieldTheory}.

\subsection{M{\"u}ller-Israel-Stewart models}\label{subsec:MIS}

The take-home message from \cref{subsec:LandauEckart} is that the most intuitive and simple way of including dissipative effects in (general) relativistic fluid models leads to problematic results. Inspired by the work of \citet{Muller:1967zza} and supported by relativistic kinetic theory, Israel and Stewart proposed an extension of the first order theories to overcome their stability and causality flaws \cite{ISRAEL1976,IsraelStewart79,IsraelStewart79bis}.

The logic of these ``second-order theories'' does not differ much from first-order ones. The main difference is in the ansatz for the entropy flux. In second-order theories the entropy flux is expanded up to second order in the dissipative quantities\footnote{It is quite common to write down the equations of the second-order theories in the Eckart frame, namely set the diffusion vector $\nu^a$ to vanish. As this simplifies a little the expressions that follow, we will stick to this convention.} (compare with \cref{eq:EntropyFluxLinearOrder})
\begin{equation}\label{eq:EntropyFluxSecondOrder}
	s^a = su^a + \frac{1}{T}q^a -\frac{1}{2T}\Big(\beta_0 \chi^2 + \beta_1 q^bq_b + \beta_2 \chi_{bc}\chi^{bc}\Big) u^a +  \alpha_0\frac{\chi q^a}{T} + \alpha_1 \frac{\chi^{ab}q_b}{T} \;.
\end{equation}
As a result, the number of unknown parameters---that have to be determined by microphysical calculations---is now larger. It is also interesting to note that the entropy measured in the Eckart frame is now
\begin{equation}
	 - u^as_a = s -\frac{1}{2T} \Big(\beta_0 \chi^2 + \beta_1 q^bq_b + \beta_2 \chi_{bc}\chi^{bc}\Big)  \;.
\end{equation}
Here $s$ is, as before, the entropy density of the fluid at equilibrium---so that \cref{eq:LITEntropyFunctional} still holds. It is clear that the ansatz in \cref{eq:EntropyFluxSecondOrder} is consistent with the EIT paradigm as second order combinations of the dissipative terms enter the formula for the out-of-equilibrium entropy of the system\footnote{Actually, it may be more correct to view/present the EIT program as an attempt to systematize the ideas behind Müller-Israel-Stewart theory.}. Also, because the entropy is maximized at equilibrium, we get ``for free'' the following constraints on (some of) the additional parameters: $\beta_0,\,\beta_1,\,\beta_2\ge0$. 

The strategy then follows the logic of first-order theories. Making use of the equation of motion $\nabla_aT^{ab}=0$ and \cref{eq:LITEntropyFunctional} we arrive at 
\begin{equation}
\begin{split}
	\nabla_as^a = -\frac{1}{T} \chi \Bigg[ &\theta +\beta_0 u^a\nabla_a\chi -\alpha_0\nabla_aq^a -\gamma_0 Tq^a\nabla_a\bigg(\frac{\alpha_0}{T}\bigg) +\frac{\chi T}{2}\nabla_a\bigg(\frac{\beta_0u^a}{T}\bigg) \Bigg] \\
    -\frac{1}{T} q^a\Bigg[&\frac{1}{T}\nabla_aT +a_a + \beta_1 u^b\nabla_bq^a - \alpha_0\nabla_a\chi -\alpha_1\nabla_b\chi^{b}_{\,a} + \\
    &+\frac{T}{2}q_a\nabla_b\bigg(\frac{\beta_1 u^b}{T}\bigg) -(1-\gamma_0) \chi T \nabla_a\bigg(\frac{\alpha_0}{T}\bigg)  - (1-\gamma_1)T\chi^{b}_{\,a} \nabla_b\bigg(\frac{\alpha_1}{T}\bigg) \Bigg] \\
    -\frac{1}{T}\chi^{ab}\Bigg[&\nabla_au_b +\beta_2 u^c\nabla_c\chi_{ab} -\alpha_1\nabla_aq_b +\frac{T}{2}\chi_{ab}\nabla_c\bigg(\frac{\beta_2u^c}{T}\bigg) - \gamma_1 Tq_a\nabla_b\bigg(\frac{\alpha_1}{T}\bigg)\Bigg] \;.
\end{split}
\end{equation}
Let us note that, following \cite{HiscockLindblom1983} we included two additional parameters $\gamma_0,\,\gamma_1$ because of the freedom we have in distributing the mixed quadratic terms. Consistency with the thermodynamic second law may then be enforced as in the previous subsection by assuming the entropy production rate is a sum of quadratic terms. This in turn yields the equation of motion for the dissipative fluxes and the system of equations is now determined. Not surprisingly, they look like the general relativistic version of the Cattaneo laws
\begin{subequations}
\begin{align}
    \tau_\b \dot\chi + \chi &= -\zeta[\dots]  \;,\\ 
    \tau_\s \dot\chi_{ab} + \chi_{ab} &= -2\eta[\dots] \;, \\
    \tau_\h \dot q_a + q^a &= - \kappa T\perp^{ab}[\dots]_b \;,
\end{align}
\end{subequations}
where we have introduced three different relaxation timescales $\tau_\b,\,\tau_\s,\,\tau_\h$---which can be related to the parameters introduced above---and the ``dot'' stands for the proper time derivative associated with the Eckart frame 4-velocity: $\dot A = u^a\nabla_aA$. 
We refer to, for example, \cite{livrev,RezzollaZanotti} for explicit expressions of the terms we omitted  in square brackets. 
We nonetheless anticipate that they include, as one may expect, the first order forces---namely the thermodynamic forces of the first order theories, such as the expansion rate for the bulk-viscous scalar---and additional terms quadratic in the fluxes themselves. 

The M{\"u}ller-Israel-Stewart (MIS) model has been proven to overcome the stability and causality issues of the first-order theories \cite{HiscockLindblom1983}: it possesses stable equilibrium states, and deviations from these states propagate causally\footnote{This can be intuitively understood from the Cattaneo-type form of the equations for the fluxes.}. However, a number of other issues remain to be addressed. 

First, the M\"uller-Israel-Stewart model is based on an implicit expansion in deviations away from thermal equilibrium, and stability/causality are guaranteed only for the linearized system of equations. The non-linear behaviour is a completely different game, and not well-explored. One exception is the analysis of \citet{Hiscock88NonlinearPathologies}, which explores the presence of non-linear pathologies (in an extremely simplified case), even though this relates to such an extreme regime that it may not be relevant for any physical or astrophysical application. 
When it comes to causality in the non-linear regime, whilst there have been recent progress for the bulk-viscous case \cite{Bemfica19MIScausal}, a definite answer is still missing. 
Second, from a field-theory perspective, the ``second-order'' expansion of the MIS model cannot be considered complete. Even though the dissipative terms are based on kinetic theory, the model contains only squares of first-order ``thermodynamic fluxes'' (as in the sense of \citet{onsager31:_symmetry})  in all possible combinations. Last, and maybe most importantly, the equations of motion are obtained from the conservation of the total stress-energy-momentum tensor of the system, and it is not clear how to extend the model to multi-fluid systems.

\subsection{Liu-M{\"u}ller-Ruggeri and divergence-type theories}\label{subsec:divergence-type}

As we have discussed above, the MIS model represents a significant improvement over the first order theories. Nonetheless, there is another open question regarding the MIS model we have not discussed, and we will briefly touch upon it now. Local well-posedness and strong hyperbolicity are not firmly established\footnote{The bulk-viscous MIS model has been shown to be weakly hyperbolic only recently in \cite{Bemfica19MIScausal}.}. Motivated by this and the quest for a theory with more solid mathematical foundations, \citet{RETMullerRuggeri} have proposed a class of models (then slightly generalized by Geroch and Lindblom \cite{GerochLindblom1990,GerochLindblom1991}) known as divergence-type theories (see \citet{RezzollaZanotti} and the review by \citet{Salazar2019} for more details). This class of theories is based on three fundamental principles: i) Principle of Relativity ii) Maximum Entropy Principle iii) Hyperbolicity. In practice, the formal hydrodynamic equations are 
\begin{subequations}\label{eq:divergene-type}
\begin{align}
	&\nabla_a n^a = 0 \;, \\
    &\nabla_a T^{ab} = 0 \;,\\
    &\nabla_a A^{abc} = I^{bc} \;,
\end{align}
\end{subequations}
where $A^{abc} = A^{a(bc)},\,A^{ab}_{\;\,b}=0,\,I^{bc}=I^{(bc)},\,I^a_{\,a}=0$ and the system is closed assuming $A^{abc},\,I^{bc}$ are functions of $n^a,\,T^{ab}$. The first two equations represent the conservation of particle flux and stress-energy-momentum tensor as before, while $A^{abc}$ is supposed to be the third moment of the one particle distribution function of some underlying relativistic kinetic theory (see \cite{Lindquist1966,ThorneMoments,Cercignani}). Similarly $I^{bc}$ should represent an approximation to the second moment of the collisional integral of some underlying kinetic theory model. In addition, this class of theories must be completed by an entropy current $s^a$, again considered as function of  $n^a,\,T^{ab}$. 

The number of theories that can be formulated as in \cref{eq:divergene-type} is obviously quite large, and one can see, for example, that the set contains both the Eckart-Landau-Lifschitz and the Müller-Israel-Stewart models. 
It is fair to say, however, that the number of acceptable theories can be somewhat reduced using constraints coming from the principles stated above---although to do this in practice one often assumes $A^{abc}$ and $I^{bc}$ are linear in the dissipative fluxes. 
While the final aim is that of constructing a framework in which discussing issues like stability, causality and hyperbolicity becomes relatively simple, this is (to the best of our knowledge) very much a work in progress. 

Before we move on, it is worth mentioning an important result that was derived independently by \citet{Geroch1995Relaxation} and \citet{Lindblom1996Relaxation}. They argued that fluid states predicted by the causal divergence-type theories decay on very short timescales---of the order of the characteristic timescale of microscopic particle interactions---to ones that are well-described by the Eckart-Landau-Lifschitz model. 
In essence, the results suggest that while the first-order theories are non causal, unstable and not well-posed---and hence problematic for numerical applications---their physical predictions would be practically/experimentally indistinguishable from those of second order ones. 

\section{Variational models}\label{sec:VariationalModels}

In this section we review variational approaches to model dissipative multi-fluids in General Relativity. Most of these models are built on extending the action-based model for non dissipative multi-fluids first championed by \citet{TaubActionPrinciple} and then developed by Carter and collaborators (see \citet{CarterNoto,comer93:_hamil_multi_con,comer94:_hamil_sf,CarterLanglois} and \citet{livrev} for an up-to-date and pedagogical review). We start with a summary of the variational principle for non dissipative multi-fluids. 

\subsection{Non dissipative multi-fluids models}\label{subsec:VariationalNonDissipativeModels}

To model general relativistic multi-fluids systems we start from an action of the form 
\begin{equation}
	S = \int \d^4x \sqrt{-g} \big(R + \Lambda\big) \;,
\end{equation}
where $R$ is the Ricci scalar and the so-called ``master function'' $\Lambda$ accounts for the matter content of the theory. To model a multi-fluid system with different chemical species (or constituents) labelled by $\x,\,\y\dots$, we take the master function $\Lambda$ to depend on the particle fluxes $n_\x^a$. Assuming the system to be isotropic, the master function is considered as a function of all the possible scalars that can be constructed from the fluxes $n_\x^a$ and the spacetime metric: $\Lambda=\Lambda(n_\x^2, n_{\x\y}^2)$, where $n_\x^2 = - n_\x^a n^\x_a$ and $n_{\x\y}^2 = -n_\x^a n^\y_a$. Then, performing the variation of the master function we obtain\footnote{Hereafter we ignore boundary terms unless they become relevant to the discussion.} 
\begin{equation}\label{eq:UnconstrainedVariation}
	\delta \big(\sqrt{-g} \Lambda\big) = \sqrt{-g} \left[ \sum_\x \mu^\x_a \delta n_\x^a + \frac{1}{2}\left(\Lambda g^{ab} + \sum_\x n_\x^a \mu_\x^b\right)\delta g_{ab}\right] \;,
\end{equation}
where we have introduced the particle four-momenta 
\begin{equation}
	\mu^\x_a = \frac{\partial \Lambda}{\partial n_\x^a}= \B^\x n^\x_a + \sum_{\y\neq\x} \A^{\x\y}n^\y_a \;,
	\label{mudef}
\end{equation}
and
\begin{equation}
    	\B^\x = -2\frac{\partial\Lambda}{ \partial n_\x^2} \;,
\end{equation}
while the entrainment coefficients are defined as
\begin{equation}
    \A^{\x\y} = - \frac{\partial\Lambda}{ \partial n_{\x\y}^2} \;.
    \end{equation}
By inspecting \cref{eq:UnconstrainedVariation} we immediately learn two things. First, we see that the variational approach automatically accounts for the entrainment effect. Roughly speaking, entrainment is a non-dissipative interaction between the species and causes a species' four-momentum $\mu^\x_a$ to be misaligned with its respective particle flux $n^a_\x$. Entrainment was first recognized as an important dynamical effect in superfluid mixtures by \citet{AndreevBashkin} and, from a microphysical perspective, is akin to the notion of effective masses (gained by the electrons, say, when they move past a ion lattice). 
Second, we learn that an unconstrained variation on the fluxes gives rise to trivial equations of motion. In fact, the fluid equations that follow from \cref{eq:UnconstrainedVariation} are $\mu^\x_a = 0$. This is a  well-established result: to obtain non-trivial fluid equations of motion from a Lagrangian, the variation of the particle fluxes must be constrained \cite{SchutzSorkin1977,CarterNoto,2004Prix}.

\begin{figure}
   \centering
   \includegraphics[width=0.8\linewidth]{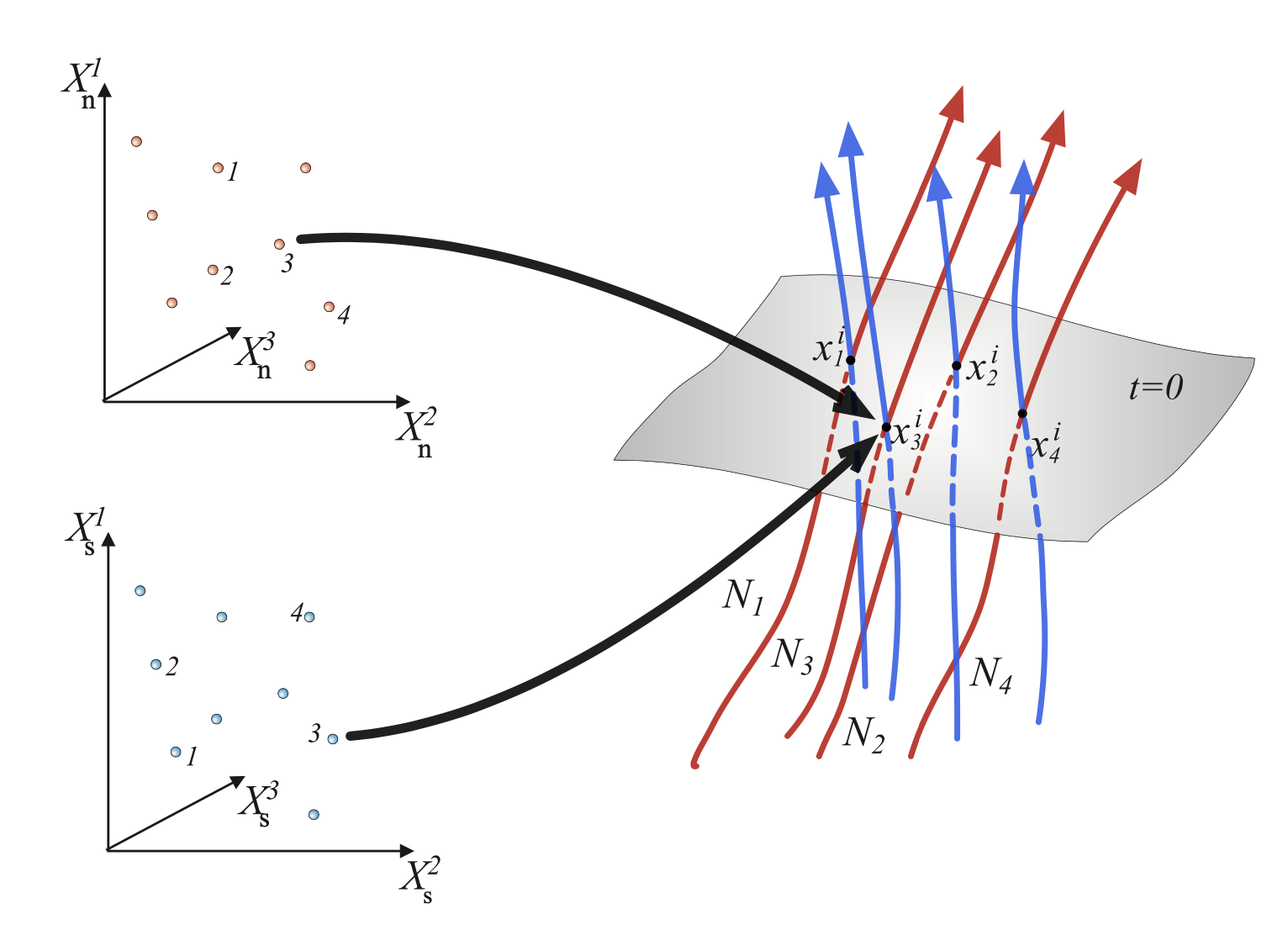}
   \caption{The pull-back from a point in the $\x^{th}$ matter space to the corresponding spacetime worldline. The points in matter space are labelled by $X^A_\x$ with $A= 1,2,3$. Figure taken from \citet{livrev}.}
   \label{fig:Pull-back}
\end{figure}

A particularly elegant way of imposing the relevant constraint involves introducing the matter space, defined by identifying each currents worldline as a single point \cite{CarterQuintana72}, see \cref{fig:Pull-back} for an illustration of the idea. For each fluid, the matter space is a three-dimensional manifold, so that when we introduce a set of coordinates $X_\x^A$ on, say, the $\x$-fluid's matter space, we attach a ``name'', or label, to each fluid element. Because the entire worldline of each fluid element is mapped to a single matter space point,  it is clear that the fluid element's label $X^A_\x$, now considered as a collection of three scalars on spacetime, takes the same value at each point on the worldline. After assigning a label to each fluid element worldline, we can use the linear map $\Psi^A_{\x\,a}$, defined as
\begin{equation}
	\Psi^A_{\x\,a}\doteq \frac{\partial X_\x^A}{\partial x^a}\;, 
\end{equation}
to push-forward (pull-back) vectors (co-vectors) between spacetime and the matter spaces. This is important because we can associate with each of the particle fluxes $n^a_\x$ a three-form $n^\x_{abc}$ by the standard Hodge-dual procedure: 
\begin{equation}
	n_\x^a = \frac{1}{3!} \veps^{bcda}\,n^\x_{bcd} \;,\quad n^\x_{abc} = \veps_{eabc}\,n_\x^e \;.
\end{equation}
Now we can assume that the spacetime three-form $n^\x_{abc}$ is obtained by pulling back a corresponding matter space three-form, to be denoted $n^\x_{ABC}$; namely,
\begin{equation}
	n^\x_{abc} = \Psi^A_{\x\,[a}\Psi^B_{\x\,b}\Psi^C_{\x\,c]}n^\x_{ABC} \;,
\end{equation}
where, as usual, straight brackets indicate anti-symmetrization (and round ones symmetrization). Similarly, upon applying the Hodge-dual to the four-momentum $\mu^\x_a$, we can push-forward with the map and identify a matter space momentum ``three-form'' $\mu^{ABC}_\x$ via
\begin{subequations}
\begin{align}
    	\mu_\x^{abc} &=\veps^{dabc}\,\mu^\x_d	\;,\\
    \mu_\x^{ABC} &= \Psi^A_{\x\,[a}\Psi^B_{\x\,b}\Psi^C_{\x\,c]} \,\mu_\x^{abc}\;.
\end{align}
\end{subequations}
The main idea of the convective variational principle is to obtain the particle flux variation $\delta n_\x^a$ by first varying the matter-space three-form and then working backwards. 

Generally speaking, there are two ways of tracking changes in a fluid system---Eulerian and Lagrangian. The first, to be denoted by a $\delta$, measures changes in the fluid at fixed spacetime coordinates. The second, to be denoted $\Delta_\x$, measures changes following the motion of fluid elements. Locally, the two are related through the Lie derivative along some displacement vector field $\xi^a_\x$ as\footnote{We note that this relation between Lagrangian and Eulerian variation works only to first order in the perturbation fields $\xi^a_\x$, see \citet{Friedman1978a} for further details.}
\begin{equation}
\Delta_\x = \delta + \mathcal{L}_{\xi_\x} \; ,
\end{equation}
where $\mathcal{L}_{\xi_\x}$ is the Lie derivative with respect to $\xi^a_\x$. Because the label $X^A_\x$ of a fluid element is fixed, we can assert
\begin{equation}
	\Delta_\x X^A_{\x} = 0 
	\; \Longrightarrow \; \delta X^A_{\x} = - \mathcal{L}_{\xi_\x} X^A_{\x} = - \Psi^A_{\x\,a} \xi^a_\x \;.
\end{equation}
Now, it is easy to show that the particle flux variation $\delta n^a_\x$ is (see \cite{livrev}) 
\begin{equation}\label{eq:ConvectiveVariation4CurrentNonDissip}
	\delta n_\x^a = -\frac{1}{2}n_\x^a g^{bc}\delta g_{bc} -\frac{1}{3!} \veps^{bcda}\mathcal{L}_{\xi_\x} n^\x_{bcd} \;.  
\end{equation}

As a result of \cref{eq:ConvectiveVariation4CurrentNonDissip} we can write the \textit{constrained} variation of the master function as
\begin{equation}\label{eq:ConstrainedVariation}
    \delta \big(\sqrt{-g}\Lambda\big) = \sqrt{-g} \bigg[\frac{1}{2}\Big( \Psi g^{ab} + \sum_\x n_\x^a\mu_\x^b\Big) \delta g_{ab} - \sum_\x \Big( f^\x_a + \Gamma_x \mu^\x_a\Big) \xi_\x^a\bigg] \;,
\end{equation}
where we introduced the ``generalized pressure'' $\Psi$ (not to be confused with the map introduced earlier)
\begin{equation}
	\Psi= \Lambda - \sum_\x n_\x^a \mu^\x_a\;,
\end{equation}
while the force densities and creation rates (for each species)
\begin{equation}\label{eq:VariationalForceDensities}
	f^\x_a = 2n_\x^b\nabla_{[b}\mu^\x_{a]} \;,\quad \Gamma_\x = \nabla_a n_\x^a \;.
\end{equation}
Since the particle-flux three-forms $n^\x_{abc}$ are pulled-back from the matter space, they are automatically closed---because $n^\x_{ABC}$ is a three-form on a three dimensional matter space and the pull-back operation commutes with the exterior derivative. As a result the constrained variation gives zero creation rate for each particle-flux\footnote{Here $dn$ represents the exterior derivative of the differential form $n$.}:
\begin{equation}\label{eq:conservativeGamma}
\Gamma_\x= \nabla_a n_\x^a = \frac{1}{3!}\veps^{bcda} \nabla_{[a}n^\x_{bcd]}  = \frac{1}{4!}\veps^{bcda} (dn)_{abcd} =  0 \;.
\end{equation}
Therefore, the term proportional to $\Gamma_\x$ actually drops out of \cref{eq:ConstrainedVariation}. Still, it is interesting to observe that such terms are formally present in the fluid equations. 

We conclude this subsection by observing that the constrained variational principle also gives the total fluid stress-energy-momentum tensor as 
\begin{equation}
    T^{ab} = \Psi g^{ab} + \sum_\x n_\x^a\mu_\x^b \;,
\end{equation}
and that it follows as an identity that 
\begin{equation}
	\nabla_aT^{ab} = \sum_\x f^\x_a =0 \;,
\end{equation}
where the last equivalence is ultimately a consequence of the second Bianchi identity satisfied by the Riemann tensor. 
As a final point, let us go back to discuss an issue we briefly hinted at in \cref{sec:PerfectFluids}.
Working with the multi-fluid framework, we can in fact appreciate the difference between equations of motion $(f^\x_a =0 ) $ and energy-momentum conservation laws. In particular, the procedure is built in such a way that we automatically obtain as many equations as needed, whatever the number of particle fluxes is. 

Before we move on to discuss extensions of the variational framework to the modelling of dissipative fluids, let us briefly comment on the case of massless particles, i.e. radiation.
Our discussion of the variational model relies heavily on the notion of matter space, which we have introduced by ``assigning a label'' to each fluid element worldline. 
Next, by associating the fluid elements' worldlines to those of the particles in the system, we have used the matter space construction to automatically impose particles' number conservation (for each species separately), as appropriate for a multi-fluid non-dissipative model. This logic obviously appears problematic if we consider, say, photons since their number is not conserved. 
At the same time, however, it is not clear whether modelling radiation as a fluid is the right thing to do in the first place.
An acceptable description of radiation should in fact be able to model both the ``trapped'' regime---where the photons, say, interact sufficiently often that their mean-free-path is small enough and we can meaningfully introduce a notion of fluid element (see \cref{subsec:FluxesBasic})---and the ``free-streaming'' regime---where the photons do not interact often and are able to escape freely---as well as the transition between the two. 
As such, it appears to us that to model radiation it is best to start from a more fundamental approach, namely relativistic kinetic theory (see, e.g. \cite{Cercignani,RezzollaZanotti}). In particular, one can start from the relativistic Boltzmann equation and derive an infinite hierarchy of equations for the moments of the one-particle distribution function associated with the radiation field \cite{ThorneMomentsRadiation}. 
In practice, this hierarchy needs to be truncated at some level as we cannot work with or simulate an infinite set of equations. 
In particular, it is quite common to stop at second order (see, e.g. \cite{Rezzolla1994,Weih+20-M1schemes,Radice+THCM1,MusolinoMoments}). This means that only the first two moments of the distribution function are evolved, and leads to the equations of radiation hydrodynamics (see, e.g. \cite{MihalasRadiationHydro}). 
Whilst the end-result resembles a ``fluid model'' for radiation, the underlying theory is much more detailed, and hence better suited to describe radiation in the first place. 

\subsection{Carter-like dissipative models}
We now describe two dissipative extensions of the variational model we just discussed. We start with a model for an heat-conducting medium and then move on to briefly discuss a variational model proposed by Carter.

\subsubsection*{Andersson-L\'{o}pez-Monsalvo model for relativistic heat conduction}\label{subsubsec:AnderssonMonsalvoModel}

Here we will briefly describe of a model for an heat-conducting medium developed in \citet{Lopez2011} and \citet{NilsHeat2011}. The model linchpin is to include the heat flux through an entropy current $s^a$ that can flow differently from matter $n^a$. As a result, the variational approach to multifluids is a natural starting point. This model is substantially a ``correction'' of a previous attempt by Carter (see \citet{CarterOffThePeg}). In his model Carter set to zero the entrainment between entropy and matter. This turns out to have a significant impact as the model without entrainment has been shown to violate causality (see \citet{OlsonHisckockOffThePeg}). The take home message is that entrainment between entropy and matter is a fundamental ingredient in the description of an heat-conducting fluid---through entrainment the entropy current gains an effective mass and this results in an inertial heat response.  

The model starting point is the non-dissipative variational principle for a two-fluid system where the particle fluxes are $n^a$---that represents matter particles---and an entropy current $s^a$---which can be thought of as a gas of thermal excitations. Using the results of the previous subsection we can write the force densities as 
\begin{subequations}
\begin{align}
	f^\n_a &= 2n^b\nabla_{[b}\mu_{a]} + \mu_a\nabla_bn^b \;,\\
    f^\s_a &= 2s^b\nabla_{[b}\theta_{a]} + \theta_a \nabla_bs^b \;,
\end{align}
\end{subequations}
where---following \cite{Lopez2011}---we named the matter particles and entropy 4-momenta (respectively) $\mu_a,\,\theta_a$.  As we already pointed out earlier, even though the constrained variation is built in such a way that the particle fluxes are automatically conserved, the action-based force densities contain a term proportional to $\Gamma_\s$ or $\Gamma_\n$. As a result we can build an hybrid model in the sense that the form of the force densities---and stress-energy-momentum-tensor---stems from the variational principle and, at the same time, we let the particle creation rates differ from zero. Because there is just one matter particle flux, we require $\Gamma_\n = \nabla_b n^b =0$ while $\Gamma_\s = \nabla_b s^b \ge 0$ to retain consistency with thermodynamics. It is then convenient to work in the Eckart frame, that is we introduce the matter 4-velocity $u^a$ such that $n^a = n u^a$ while for the entropy current we have 
\begin{equation}
	s^a = s^* (u^a + w^a) \;,
\end{equation}
where $s^*$ is the entropy density measured by the matter particles. We also introduce a similar decomposition for the entropy 4-momentum
\begin{equation}
	\theta_a = \big(\B^\s s^* + \A^{\n\s} n\big)u_a + \B^\s \s^* w_a \doteq \theta^* u_a + p_a \;,
\end{equation}
and then rewrite the heat-flux as 
\begin{equation}
	q_a = -\perp_{ab} u_c T^{bc} = s^* \theta^* w_a \doteq \theta^* \sigma_a \;.
\end{equation}
Next, the energy density measured by matter particles is $\veps = u_au_bT^{ab}= -\Lambda + p_a\sigma^a$. Therefore, when the system is out of equilibrium and the heat flows relative to matter the energy density depends also on the heat-flux (encoded in the variables $\sigma^a$ and $p_a$), i.e. we have an \textit{extended Gibbs relation}
\begin{equation}
	\d \veps = \mu\d n + \theta^* \d s^* + \sigma^a \d p_a \;.
\end{equation}
We stress that this result arises automatically in the model and is consistent with the EIT picture---the key difference is that it is derived, not postulated. 
Finally, it is possible to rewrite the equation of motion for the entropy current as an equation for the heat-flux as
\begin{equation}
	\tau\big(\dot q^a + q_\c\nabla^au^c\big) + q^a = \kappa \perp^{ab} \big(\nabla_b\theta^* + \theta^* a_b\big) \;,
\end{equation}
where the relaxation timescale $\tau$ can be rewritten in terms of the entrainment parameter $\A_{\n\s}$ and $\kappa$ is the heat conductivity. The last equation is clearly a general relativistic generalization of the Cattaneo law (see \cref{eq:CattaneoLaw}) and has been shown to be consistent with the Israel-Stewart model in the linear regime (see \cite{Lopez2011}). 

\subsubsection*{Carter variational principle for dissipative fluids}
Another important step forward, at least from the formal point of view, is the variational model proposed by Carter. 
We now sketch the main ideas and assumptions of the model, while referring to the original work for the details \cite{CarterDissipativeModel}. We do so mainly to highlight the differences with what comes next.

The key quantity is again the master function $\Lambda$ which now is assumed to depend also on a set of  additional rank-2 (symmetric) tensors $\tau^{ab}_\Sigma$. These new dynamical fields should be identified with viscous tensors and the label $\Sigma$ is introduced to allow for different sources of viscosity separately---such as bulk and shear viscosity. The variation of the master function $\Lambda(n_\x^a,\tau^{ab}_\Sigma,g_{ab})$ can then be written as 
\begin{equation}
	\delta \Lambda = \sum_\x \mu^\x_a \delta n_\x^a +\frac{1}{2}\sum_\Sigma \frac{\partial\Lambda}{\partial\tau^{ab}_\Sigma}\delta \tau^{ab}_\Sigma + \frac{\partial\Lambda}{\partial g_{ab}}\delta g_{ab} \;.
\end{equation}
In this approach, however, the action is only used to obtain the structure of the force terms, and the stress–energy–momentum conservation as a Noether identity.
The final equations of motion are not obtained by setting to zero the variation of the action, but enforcing consistency with the second law of thermodynamics. Moreover, the identification of the new dynamical fields $\tau^{ab}_\Sigma$ with viscous tensors is so far just formal. To complete the identification with the usual thermodynamic fluxes a specific expansion in deviations away  from thermal equilibrium had to be introduced, and the resulting model has been shown to belong to the same family as those of the MIS variety \cite{Priou1991}. 

\subsection{Andersson and Comer formalism}\label{subsec:ACformalism}
We now describe a recent action-based formalism for dissipative (general relativistic) multi-fluids proposed by \citet{2015CQAnderssonComer}, which is built by generalizing the non-dissipative model presented in \cref{subsec:VariationalNonDissipativeModels}. 
As this is the starting point for the original results described in \cref{ch:Linearizing}, we will spend more time going into the details. We also note two important aspects of the model that distinguish it from the ones presented in the previous subsections. First, the approach is ``fully variational'' in the sense that the final equations of motion are obtained as Euler–Lagrange equations starting from an action---while the models discussed in the earlier section took the variational equations as starting point, and then modified them appropriately.
This makes the model well suited for describing dissipative multi-fluids\footnote{As opposed to simple fluids with, possibly, heat flow, which the standard approaches are designed for.}. Second, the model does not introduce any new dynamical field,  focusing instead on the particle fluxes $n_\x^a$.

In order to not get lost in the algebra, it is useful to start with two simple observations. First, at the microscopic level dissipation is the product of interaction between particles, which, at the fluid-dynamical level, intuitively translates into the idea of \textit{interacting matter spaces}. 
Second, a central feature of dissipative fluids is having non-conserved fluxes, $\Gamma_\x \neq 0$. As can be seen going back to \cref{eq:conservativeGamma},  flux conservation is---in the constrained variation---a direct consequence of having an associated closed three-form $n^\x_{abc}$. As a result, if we want to keep working with the matter space construction and, at the same time, let the particle fluxes be non-conserved, we have to break the closure property of the three-forms. 

Let us implement these two ideas by reviewing the constrained variation procedure. Clearly, we still have that the particle labels do not change if we follow them (in the Lagrangian sense) $\Delta_\x X^A_\x =0$. Also, it is still true that
\begin{equation}
	\Delta_\x \Psi^A_{\x\,a } = \Delta_\x \Bigg(\frac{\partial X^A_\x}{\partial x^a}\Bigg) = \frac{\partial}{\partial x^a} \big(\Delta_\x X^A_\x\big) = 0 \;.
\end{equation}
We can now work out (again) the particle flux variation $\delta n_\x^a$ without assuming the three-form $n^\x_{ABC}$ to be closed to get 
\begin{equation}\label{eq:ConvectiveVariation4Current}
	\delta n_\x^a = -\frac{1}{2}n_\x^a g^{bc}\delta g_{bc} -\frac{1}{3!} \veps^{bcda}\bigg(\mathcal{L}_{\xi_\x} n^\x_{bcd} - \Psi^B_{\x\,[b}\Psi^C_{\x\,c}\Psi^D_{\x\,d]}\,\Delta_{\x} n^\x_{BCD} \bigg) \;.
\end{equation}
But there is a deeper point to be made here. Formally, we can take $n^\x_{ABC}$ to be a particle measure form on the matter space, which ``counts'' the total number of species $\x$ particles in the system. If it is a tensor on matter space then it must be  a function only of the matter space coordinates $X^A_\x$. The fact that $n^\x_{ABC} = n^\x_{ABC} (X^A_\x)$ implies $\Delta_\x n^\x_{ABC}= 0$, and the flux variation above reduces to the result for non-dissipative fluids. Therefore, to get the non-dissipative equations of motion one simply has to impose that the number of particles is conserved in the variation, or, equivalently, that the particle creation rates $\Gamma_\x = \nabla_a n^a_\x$ vanish. It then follows that a way to include dissipative processes (read: $\Gamma_\x \neq 0$) at the level of the action principle is to break the tensorial nature of the matter space particle measure form $n^\x_{ABC}$, and allow it to be a function of more than just the $X^A_{\x}$. In other words, we are breaking the closure property of the $n^\x_{abc}$. 

Before we move on to discuss the model more in detail, let us briefly comment on the differences between dissipative fluids and radiation. 
Using the variational approach to model radiation seems problematic as the photon number is not conserved---although we argued in \cref{subsec:VariationalNonDissipativeModels} that using a fluid-scheme to model radiation appears too restrictive in the first place. 
As the matter space is introduced by labelling the worldlines, and we naively tend to associate the worldlines to the particles, one may be led to think that using the matter space construction to model a dissipative system is similarly problematic. 
We now consider a multi-fluid system undergoing reactions to argue that this is not quite the case. 
First of all, let us begin by recalling that all fluid models inevitably involve some notion of averaging (see also \cref{sec:introAverageFluids}). 
As such, it is probably more correct to think of the worldlines as associated with fluid elements---and we are going to discuss more in detail how these are defined in \cref{subsec:FluxesBasic}---rather than the individual particles themselves.
Moreover, while the system is undergoing reactions, it is still subject to, say, the conservation of baryons, leptons and so on. 
In essence, one could imagine some notion of a ``global'' matter space that, in the non-dissipative limit factorizes as the direct sum of the individual matter spaces associated with each species. 
While the factorization in terms of matter spaces associated with the different species breaks down in the dissipative case, and one may be tempted to think of factorizing it in terms of baryons, leptons and so on, it still makes sense to consider fluid elements associated with the various species. 
The reason being that the dynamical behaviour of each of these is different---as they have, say, different charge and hence behave differently when immersed in a magnetic field. 
Admittedly, this discussion is somewhat hand-waving, and we are urged to make it more precise---which motivates the analysis and new results presented in \cref{ch:Linearizing}.
Before we get there though, let us see where this idea brings us by considering two explicit examples.

Following the discussion in the original paper, we assume that the x-particle three-form depends also on the matter space coordinates of the y-matter spaces $X^A_\y$. As a result we have 
\begin{equation}
\begin{split}
	\Delta_\x n^\x_{ABC} &= \sum_{\y\neq\x} \frac{\partial n^\x_{ABC}}{\partial X_\y^D}\Delta_\x X_\y^D = \sum_{\y\neq\x} \frac{\partial n^\x_{ABC}}{\partial X_\y^D} \Big(\delta X_\y^D - \Lie_{\xi_\x}X^D_\y\Big) \\
    &= \sum_{\y\neq\x} \frac{\partial n^\x_{ABC}}{\partial X_\y^D}\Big(\xi^a_\x - \xi^a_\y\Big)\partial_aX^D_\y \;.
\end{split}
\end{equation}
Using the last equation in \cref{eq:ConvectiveVariation4Current} and defining a ``resistivity coefficient'' as 
\begin{equation}\label{eq:PurelyReactive}
	\text{R}^{\x\y}_a = \frac{1}{3!} \mu_\x^{ABC} \frac{\partial n^\x_{ABC}}{\partial X_\y^D}\,\Psi^D_{\y\,a}  \;,
\end{equation}
it is possible to write the fluid-part in the Lagrangian variation as 
\begin{equation}\label{eq:example1variation}
	\mu^\x_a \delta n_\x^a = \text{``non-dissipative terms''} + \sum_{\y\neq\x} \text{R}^{\x\y}_a \Big(\xi^a_\y - \xi^a_\x\Big) \;.
\end{equation}
The additional piece in the variation then changes the equations of motion to
\begin{equation}
	f^\x_a + \Gamma_\x\mu^\x_a = \sum_{\y\neq\x}\big( \text{R}^{\y\x}_a -\text{R}^{\x\y}_a\big) \;,
\end{equation}
where $f^\x_a$ is the same as in \cref{subsec:VariationalNonDissipativeModels}. The take home message from this example is that by enlarging the set of quantities which the three-form can depend on we obtain additional terms that look like resistivity coefficients. 

The natural follow-up question then is: can we perform a similar calculation and obtain, again from an action principle, additional terms in the equations that look like viscous tensors? The answer---demonstrated by \citet{2015CQAnderssonComer}---is yes. Let us, in fact, consider a situation where the three form depends both on $X_\y^A$ and the projected metric
\begin{equation}
    g_\x^{AB} = \Psi^A_{\x\,a} \Psi^B_{\x\,b}\,g^{ab} \;.
\end{equation}
We then have 
\begin{equation}
\begin{split}
\Delta_\x n^\x_{ABC} &= \sum_{\y\neq\x} \frac{\partial n^\x_{ABC}}{\partial X_\y^D}\Delta_\x X_\y^D + \frac{\partial n^\x_{ABC}}{\partial g_\x^{DE}}\Delta_\x g_\x^{DE} \;.
\end{split}
\end{equation}
The novel terms arising from $\Delta_\x g_\x^{DE}$ in the fluid-part of the Lagrangian variation then read
\begin{equation}\label{eq:deltn_duegAB}
\begin{split}
	\mu^\x_a \delta n_\x^a &= \ldots + \frac{1}{3!} \mu_\x^{ABC} \frac{\partial n^\x_{ABC}}{\partial g_\x^{DE}}\,\Delta_\x g_\x^{DE} \\
    &=\ldots + \frac{1}{3!} \mu_\x^{ABC} \frac{\partial n^\x_{ABC}}{\partial g_\x^{DE}}\,\Psi^D_{\x\,a} \,\Psi^E_{\x\,b} \Big(\delta g^{ab} - 2\nabla^{(a}\xi_\x^{b)}\Big)  \\
    &=\ldots + \frac{1}{2} S_\x^{ab}\delta g_{ab} - S_{ab}^x \nabla^b\xi_\x^a \;,
\end{split}
\end{equation}
where we have defined the ``viscosity tensor''
\begin{equation}\label{eq:StdViscousTensor}
S^\x_{ab} = \frac{1}{3} \mu_\x^{ABC} \frac{\partial n^\x_{ABC}}{\partial g_\x^{DE}}\,\Psi^D_{\x\,a} \,\Psi^E_{\x\,b} \;,
\end{equation}
and the $\dots$ represent the non-dissipative terms plus the ones in \cref{eq:example1variation}.
From \cref{eq:deltn_duegAB} we the intuitively see that the additional dependence of the matter space three-form is going to affect both the force densities and the total stress-energy tensor of the system. 
Cranking through the algebra, the action variation can be written as
\begin{equation}
\begin{split}
	\delta \big(\sqrt{-g}\Lambda\big) = \sqrt{-g} \bigg[&\frac{1}{2}\Big( \Psi g^{ab} + \sum_\x n_\x^a\mu_\x^b + S^{ab}_\x\Big) \delta g_{ab}  \\
    &-\sum_\x \Big( f^\x_a + \Gamma_x \mu^\x_a + \nabla^bS^\x_{ba} - \sum_{\y\neq\x}\big( \text{R}^{\y\x}_a -\text{R}^{\x\y}_a\big) \Big) \xi_\x^a\bigg] \;.
\end{split}
\end{equation}
Therefore, the action-based model provides us with the total stress-energy-momentum tensor 
\begin{equation}
	T^{ab} = \Psi g^{ab} + \sum_\x n_\x^a\mu_\x^b + S^{ab}_\x \;,
\end{equation}
which contains the additional viscous stress-tensors $S^{ab}_\x$. Also the resulting fluid equations of motion in this second case are
\begin{equation}
	f^\x_a + \Gamma_\x\mu^\x_a + \nabla^bS^\x_{ba}  = \sum_{\y\neq\x}\big( \text{R}^{\y\x}_a -\text{R}^{\x\y}_a\big) \;,
\end{equation}
and thus contain both a resistivity term and the four divergence of the new viscous tensor. 

Having described these two examples, let us move on to the general formalism presented in \citet{2015CQAnderssonComer}. The authors consider the case where the particle three-forms depend on 
\begin{equation}
	n^\x_{ABC}= n^\x_{ABC}(X_\x^A,\,X_\y^A,\,g_\x^{AB},\,g_\y^{AB},\,g_{\x\y}^{AB}) \;,
\end{equation}
and the ``mixed projected metrics'' are defined as
\begin{equation}
	g_{\x\y}^{AB} = \Psi^A_{\x\,a} \Psi^B_{\y\,b}\,g^{ab} \;.
\end{equation}
Performing the variations as in the examples above they arrive at the following equations of motion
\begin{equation}\label{eq:DissipativeVariationalEoM}
	f^\x_a + \Gamma_\x\mu^\x_a + \nabla^bD^\x_{ba}  =  R^{\x}_a \;,
\end{equation}
where 
\begin{subequations}
\begin{align}
    D^\x_{ab} &= S^\x_{ab} + \sum_{\y\neq\x} s^{\y\x}_{ab} + \frac{1}{2}\Big(\mathcal{S}^{\x\y}_{ba} + \mathcal{S}^{\y\x}_{ab}\Big) \;, \\
    R^\x_a &= \sum_{\y\neq\x} \bigg(\text{R}^{\y\x}_a - \text{R}^{\x\y}_a\bigg) + \bigg(r^{\y\x}_a-r^{\x\y}_a\bigg) + \bigg(\mathcal{R}^{\y\x}_a -\mathcal{R}^{\x\y}_a \bigg) \;.
\end{align}
\end{subequations}
and
\begin{subequations}
\begin{align}
s^{\x\y}_{ab} &= \frac{1}{3} \mu_\x^{ABC} \frac{\partial n^\x_{ABC}}{\partial g_\y^{DE}}\,\Psi^D_{\y\,a} \,\Psi^E_{\y\,b} \;,\\
	\mathcal{S}^{\x\y}_{ab}& = \frac{1}{3} \mu_\x^{ABC} \frac{\partial n^\x_{ABC}}{\partial g_{\x\y}^{DE}} \,\Psi^D_{\x\,a} \,\Psi^E_{\y\,b} \;,\\
    r_a^{\x\y} &= \frac{1}{3!} \mu_\x^{ABC} \frac{\partial n^\x_{ABC}}{\partial g_\y^{DE}}\, \nabla_a \big(g^{bc} \Psi^D_{\y\,b} \Psi^E_{\y\,c}\big)\;, \\
    \mathcal{R}_a^{\x\y} &= \frac{1}{3!} \mu_\x^{ABC} \frac{\partial n^\x_{ABC}}{\partial g_{\x\y}^{DE}}\, g^{bc} \Psi^D_{\x\,b} \nabla_a \big( \Psi^E_{\y\,c}\big) \;,
\end{align}
\end{subequations}
while $\text{R}^{\x\y}_a,\,S^\x_{ab}$ are defined as in \cref{eq:PurelyReactive,eq:StdViscousTensor}. Projecting the field equation along $u^a_\x = n^a_\x/n_\x$, we see that
\begin{equation}\label{eq:gamx}
	\left(- u^a_\x \mu^\x_a\right) \Gamma_\x = u^a_\x \nabla^b D^\x_{b a} - u^a_\x R^\x_a \; , 
\end{equation}
while the stress-energy-momentum tensor is
\begin{equation}
	T^{ab} = \Psi g^{ab} + \sum_\x n_\x^a\mu_\x^b + D^{ab} \;,
\end{equation}
and $D^{ab}=\sum_\x D^{ab}_\x$ is the sum of each species' total viscous tensor. Let us observe that it follows, as an identity, that $\sum_\x R^\x_a = 0$, and because of this we have automatically $\nabla_a T^{a b} = 0$.  We also note that the ``resistive terms'' $r^{\x\y}_a,\,\mathcal{R}^{\x\y}_a$ as well as the viscous tensors $s^{\x\y}_{ab} ,\,\mathcal{S}^{\x\y}_{ab}$ arise because we assume that $n^\x_{ABC}$ depends on $g_\y^{AB}$ and $g^{AB}_{\x\y}$, respectively. Finally, it is easy to see that, in general, the x-species total viscous tensor $D^\x_{ab}$ is not necessarily symmetric because $\mathcal{S}^{\x\y}_{ab}$ is not. This property is, however, not inherited by the total viscous tensor of the system meaning $D_{ab}= \sum_\x D^\x_{ab} = D_{ba}$.

Let us conclude this section with some comments. First, the action-based dissipative formalism presented here is quite general and, as such usable---at least in principle---in a large number of astrophysical situations. For instance, it has already been used to build models beyond ideal magneto-hydrodynamics \cite{BeyondIdealMHD,BeyondMHD3+1,AnderssonCQG2017VariationalPlasmas} due to the fact that it is intuitively clear how to couple the model to electromagnetism. 
Also, it is important to note that the action and the field equations are fully non-linear. The ``variational'' aspect of the approach is in the context of the action principle, and there is nothing in the variational process that says the field equations themselves have to be linear in the fields\footnote{An obvious and familiar example illustrating this same feature comes to mind: the Einstein-Hilbert action yields the Einstein equations, which are notoriously non-linear in the metric.}. In fact, \cref{ch:Linearizing} deals with a linearization of the present model in terms of deviations from---a self-consistently defined notion of---thermodynamic equilibrium. 

\section{Hydrodynamics as an effective field theory}\label{sec:HydrodynamicFieldTheory}

After having discussed the ``traditional'' approaches and the recent variational efforts, we now turn our attention to some fairly recent work based on a field-theory take on the problem. As we will see, this work represent a significant change in perspective which is worth exploring. Moreover, we will adapt some of the ideas behind these strategies later on (see \cref{subsec:stability}), although in a different spirit.

Hydrodynamics can be viewed as the classical low energy limit of a more fundamental quantum (many body and thermal) field theory. Such a perspective is quite natural in the context of heavy-ion-collisions (see \cite{RomatschkeRFL}). We also note that, as these theories have been developed having in mind heavy-ion-collision applications, they are mostly discussed in a special relativistic setting. An extension to General Relativity is not necessarily straightforward, and may require some careful thinking---as discussed in the next chapter. 

We also mention that a similar view on hydrodynamics as a coarse-grained theory has been explored for non-dissipative fluid models as well. For instance, in \citet{BatthacharyaJHEP13} the authors proposed a variational non-dissipative hydrodynamic model based on a derivative expansion. Roughly speaking, the Lagrangian of the theory is built as the coarse-grained limit of some more fundamental one---the cutoff being the microscopic mean-free-path of the particles. Such an action is constructed summing all possible terms with a specific numbers of derivatives, each with its own coupling. In natural units, each derivative has the dimension of a mass.  Since all terms in the action must have the same dimension, those with higher derivatives correspond to lower (mass) dimension couplings. These lower dimension couplings will take smaller values in the low energy limit, so that they are suppressed at the hydrodynamic level. 

Moving on to  dissipative fluids, two important lessons can be drawn from the traditional approaches and suggests we should investigate the effective field theory point of view. First, most of the dissipative theories---with the notable exception of the variational model of \cite{2015CQAnderssonComer}---are intrinsically only valid in the ``linear regime'', i.e. close to some reference equilibrium state. Second, the relaxation effect we have described in \cref{subsec:divergence-type} suggests that while second-order theories may be required to overcome the problems of the Landau-Eckart models, their physical predictions are going to be practically indistinguishable form those of first-order ones.
Furthermore, when fluctuations about equilibrium are included in the modelling, new problems arise. In particular, it turns out that corrections to the correlation functions coming from second order terms are smaller than those due to interactions between fluctuations. This is the so called \textit{breakdown of second order hydrodynamics}. It is a relatively well known problem that is not specific to relativistic theories, and leads to stochastic hydrodynamics (see \citet{LandauFLuidMechanics} and the recent review by \citet{KovtunLN}). For all these reasons, it makes sense to ask whether one can fix the stability and causality issues directly at first order. 

In the effective field-theory framework, dissipative hydrodynamics equations for single fluids can be constructed as follows. The equations of motion are given by the (special relativistic) conservation laws 
\begin{subequations}
\begin{align}
     \partial_a n^a = 0\;, \\
    \partial_a T^{ab} = 0\;, 
\end{align}
\end{subequations}
where the conserved currents are decomposed as usual
\begin{subequations}
\begin{align}
    n^a &= n u^a + \nu^a \;, \\
    T^{ab} &= \veps u^a u^b + (p + \chi) \perp^{ab} + 2q^{(a}u^{b)} + \chi^{ab} \;.
\end{align}
\end{subequations}
To close the system, we need to provide explicit expressions for the dissipative fluxes via some constitutive equations. These are obtained through the most general gradient expansion in the (chosen) equilibrium hydrodynamic variables. Typically these are taken to be a four-velocity $u^a$, the chemical potential $\mu$ and the temperature $T$: 
\begin{subequations}\label{eq:KovtunGeneralExpansion}
\begin{align}
    \veps &= \veps_{\text{eq}} + \veps_1 \dot T/T + \veps_2 \partial_au^a + \veps_3 u^a\partial_a (\mu/T )\;, \\
    \chi &= \pi_1\dot T/T  + \pi_2\partial_au^a + \pi_3 u^a\partial_a (\mu/T) \;, \\
    q^a &= \theta_1 \dot u ^a + \frac{\theta_2}{T} \perp^{ab}\partial_b T + \theta_3 \perp^{ab}\partial_b (\mu/T ) \;, \\
    \chi^{ab}& = \eta \sigma^{ab} \;, \\
    n &= n_{\text{eq}} +  \nu_1 \dot T/T + \nu_2 \partial_au^a + \nu_3 u^a\partial_a (\mu/T \;, \\
    j^a &= \gamma_1 \dot u ^a + \frac{\gamma_2}{T} \perp^{ab}\partial_b T + \gamma_3 \perp^{ab}\partial_b (\mu/T )   \;,
\end{align}
\end{subequations}
where the ``dots'' represent a derivative along $u^a$, that is $\dot T = u^a\partial_a T$. The causality and stability properties of these theories have been studied in a number of recent papers, such as, for example \cite{Bemfica2018,KovtunStable,HoultKovtun2020}. The results depend on the equation of state, but it has been demonstrated---at least for some simple cases, like conformal fluids---that one can derive a set of constraints on the expansion parameters, and guarantee stability and causality. Even more recently, an extension of this approach to General Relativity was explored in \cite{Bemfica2020}. 

These results may come a bit of a surprise after the earlier discussion of first order theories, so let us briefly comment on why it is in fact possible to satisfy the stability and causality constraints at first order. 
A first and key ingredient for this is the larger number of free coefficients in the gradient expansion. This is ultimately motivated by the fact that quantities like the temperature or chemical potential are not uniquely defined out of equilibrium---different definitions are possible as long as they agree in equilibrium. This is essentially the reason why there are many more coefficients in 
\cref{eq:KovtunGeneralExpansion} when compared to the Landau or Eckart models. 
Moreover, we recall that, according to the effective field theory picture, hydrodynamics is a coarse grained theory whose validity is restricted to long wavelengths and low frequencies---and the gradient expansion makes sense as long as the gradients are in fact small. 
Given this, one is urged to respect the constraints coming from the second law of thermodynamics only in the regime of validity of the theory, namely ``on-shell''.
The trick consists of stabilising the unstable modes that appear in first-order theories by allowing for violation of the second law of thermodynamics (out of the ``hydrodynamic regime''). 
For more details on how this additional freedom can be used to ensure covariant stability, we point to \cref{subsec:stability} where a similar strategy is used in the context of turbulence modelling.

Even though the discussion in this section is at a broad-brush level, it is clear that these theories represent a radical change with respect to the MIS paradigm. 
This is mainly with regards to the way they solve the instability by ``killing'' the unstable modes, and also because of the change in perspective with respect to EIT. As such, they offer interesting prospects and we point to \cite{FrontiersGavassinoAntonelli} for a pedagogical discussion of the differences with the EIT paradigm. 

%% file: Parts/DissipativeHydro/Linearizing.tex
\chapter{Linearizing an action-based formalism for dissipative (multi-)fluids} \label{ch:Linearizing}

The main goal of this chapter is to compare the action-based formulation of \citet{2015CQAnderssonComer} with previous approaches---such as MIS (see \cref{subsec:MIS}). The key point is that the \citet{2015CQAnderssonComer} action principle does not reference any sort of chemical, dynamical, or thermal equilibrium, other than to start with the assumption that the physics can be modelled as fluid phenomena. Conversely, traditional strategies for dissipative fluids---and recent works based on an effective field-theory perspective (see \cref{sec:HydrodynamicFieldTheory})---use an expansion to create an approximate set of field equations to describe dissipative phenomena. Since the action-based model already provides a set of equations (at least in principle) valid in every regime, we can make the comparison using standard perturbation techniques. The dissipation terms are assumed to generate first-order deviations away from equilibria obtained using the non-dissipative limit of the field equations. Working this way we hope to also understand better the role of length- and time-scales of fluid elements on the large scale behavior of the system; in particular, how to link the micro-scale dynamics of the many particles in a fluid element with the macro-scale dynamics between the fluid elements themselves, and the role of the Equivalence Principle in setting these scales. The results presented in this chapter have been published in \citet{ActionDissip1st}.

\section{Main assumptions of the model}

Let us get started with an introductory section in which we expand on the central assumptions behind the fluid-modelling scheme and the nature of the non-closed three forms $n^\x_{ABC}$ introduced in \cref{subsec:ACformalism}---central to the dissipative model of \citet{2015CQAnderssonComer}.  

\subsection{Flux definition}\label{subsec:FluxesBasic}
The crux of the fluid modelling scheme is to assume that knowledge of the total mass-energy and momentum flux obtained by tracking the worldlines of individual particles can be replaced with tracking the worldlines of fluid elements. These are defined in the following way: Take a multi-particle system at some initial time having, say, total spatial size $V$, total number of particles $N$, total mass-energy $E$, and total entropy $S$. At the same time, fill-up side-to-side, top-to-bottom, and front-to-back the entire system with $I = 1 ... M$ local conceptual boxes---the fluid elements. Each element has its own volume $\delta V_I$, number of particles $\delta N_I$, mass-energy $\delta E_I$, and entropy $\delta S_I$. Roughly speaking, if there are characteristic values $\delta \, V_I \sim \delta \, V$, $\delta \, N_I \sim \delta \, N$, etcetera, representative of the fluid elements, then $V \sim M \, \delta V$, $N \sim M \, \delta N$, $E \sim M \, \delta E$, and $S \sim M \, \delta S$. Clearly, as the number $M$ is increased the ratios $\delta V/V$, $\delta N/N$, etcetera decrease, and the elements become ``ultra-local'', implying that the change in the spacetime metric across them is small. 

Now consider the $I^{\rm th}$-fluid element. It moves through spacetime and, if the element is small enough, the trajectory  can be accurately represented by a single unit four-velocity $u^a_I$. When taken together, and in the limit $M \to \infty$, all the $u^a_I$ form a vector field on spacetime and this field plays a role in the fluid system's degrees of freedom. If a local typical scattering length $\lambda_I$ between the particles exists, and the size of fluid elements is commensurate with that length ($\delta V_I \sim \lambda^3_I$), then the average four-velocity of the $\delta N_I$ particles will be $u^a_I$. 
In principle, we now have everything we need to define the actual fluid degrees of freedom, which are the particle fluxes $n^a_I = (\delta N_I/\delta V_I) u^a_I$. 

Even though this may be familiar, we went through the details of the formal process to define fluid elements to point out what are the fundamental assumptions behind a fluid description. 
We have introduced typical scattering lengths and average velocities as part of our fluid element definition, and therefore we must assume that fluid elements contain enough particles to warrant a statistical/thermodynamical treatment. In the formal procedure there is no requirement of being close to thermodynamic equilibrium. It can be shown (see, for instance \cite{Reichl}) that fluid dynamics can be obtained as a limit of kinetic theory (via a Chapman-Enskog type expansion), but the realm of hydrodynamics is potentially vaster.

\subsection{Matter space volume forms}

All dissipative terms that enter the action-based equations are obtained by assuming that the fundamental current three-forms $n^\x_{abc}$ depend on an additional set of quantities which breaks their closure ($\nabla_{[a} n^\x_{bcd]} \neq 0$). We now want to expand on how this can happen, but begin by introducing a bit of notation. 

We need to distinguish between the Levi-Civita symbol $\eta_{ABC}$ and a volume measure form $\varepsilon^\x_{ABC}$ on the matter space. The Levi-Civita symbol is defined as $\eta_{ABC} = [A\,B\,C]$ for every chosen set of coordinates (and thus is not a tensor but a tensor density) while the volume measure form $\veps^\x_{ABC}$ can be defined\footnote{This is tricky for a couple of reasons: It is well known from work on general relativistic elastic bodies \citep{Karlovini1} that this is not the only possible choice. Also, the projected metric $g^{AB}_\x$ is not ``fixed'' in the sense that the spacetime metric $g_{ab}$ changes, in a general curved spacetime, as a fluid element moves from point-to-point along its worldline.} by means of the push-forward of the metric: 
\begin{subequations}
\begin{align}
	&g_\x = \frac{1}{3!} \eta_{ABC}\eta_{DEF}\,g_\x^{AD}	g_\x^{BE}g_\x^{CF} =\text{det} (g_\x^{AB}) \;, \\
    &\veps^\x_{ABC} = \sqrt{g^\x}\eta_{ABC} = \sqrt{g^\x}[A\,B\,C]    \;,
\end{align}
\end{subequations}
where $g^\x = (g_\x)^{-1}$ is the determinant of the inverse matrix $g^\x_{AB}$; i.e.~$g^\x_{AC} g_\x^{CB} = \delta^B_A$. As a result, $\veps^\x_{ABC}$ is a three form and transforms as a tensor under coordinates transformations on the matter space. 

This volume measure form provides a way to measure the volume of ``matter elements'', infinitesimal volumes in the matter space manifold. We can relate these quantities to the current and momentum three-forms: 
\begin{subequations}
\begin{align}
	n^\x_{ABC} &= \N_\x\,\varepsilon^\x_{ABC} = \bar \N_\x \eta^\x_{ABC} \,,\\
    \mu_\x^{ABC} &= \M_\x\,\veps_\x^{ABC} = \bar \M_\x \,\eta_\x^{ABC}  \;.
\end{align}
\end{subequations}
The point we want to make here is that the barred quantities look more like scalar densities on the x-matter space, while the non-barred ones look more like scalars. The relation between the two normalizations is simply
\begin{subequations}\label{eq:normalizationsRel}
\begin{align}
	\N_\x &= \sqrt{g_\x}\,\bar \N_\x \; , \\
	\M_\x &= \sqrt{g^\x}\,\bar \M_\x \;.
\end{align}
\end{subequations}
We can use this to expedite our use of the variational principle by focusing the additional functional dependence of $n^\x_{ABC}$ into 
\begin{equation}
    \N_\x = \N_\x(X_\x^A,\,X_\y^A,\,g_\x^{AB},\,g_\y^{AB},\,g_{\x\y}^{AB})\ . \label{eq:mspnorm}
\end{equation}  

To make contact with proper quantities measured in spacetime---that is, with the rest frame density and rest frame momentum for each fluid component---it is useful to introduce an appropriate tetrad $e^{\ha}_a$ for each species; an orthonormal basis whose timelike unit vector ${\boldsymbol e}_{\hat 0} = {\boldsymbol u}_\x$, so that $u_\x^\ha =({\boldsymbol e}_{\hat 0})^\ha =\delta_{\hat 0}^\ha = (1,0,0,0)^\top$.  The components of the spacetime measure form in this tetrad basis are\footnote{Recall that, since $g_{ab}=e^\ha_a e^\hb_b \eta_{\ha\hb}$, the determinant of the tetrad $ e = \sqrt{|g|}$.}
\begin{equation}
	\veps^{\ha\hb\hc\hd} = \veps^{abcd}  e^\ha_a e^\hb_b  e^\hc_c e^\hd_d = \eta^{\ha\hb\hc\hd}
\end{equation} \;,
where $\eta^{\ha\hb\hc\hd} = - [\ha\, \hb\,\hc\,\hd ]$ and we have omitted the chemical index. Now, since push-forward (and pull-back) is a linear map between vector spaces (the tangent space), it transforms as a linear map under coordinate changes, and we can write
\begin{equation}
	A^A = \frac{\partial X_\x^A}{\partial x^a} A^a = \Psi^A_{\x\,\ha}A^\ha \;,
\end{equation}
where we have introduced the short-hand notation\footnote{Following \cite{GRCarroll} we denote the inverse matrix of the tetrad as $e^a_\ha$.}
\begin{equation}
	\Psi^A_{\x\,\ha}   \doteq \Psi^A_{\x\,a} \, e^a_\ha  = \frac{\partial X^A_\x}{\partial x^a}  \, e^a_\ha \;.
\end{equation}
Making use of the fact that $0 = u^\ha_\x \Psi^A_{\x\,\hat a} =\Psi^A_{\x\,\hat 0}$ we then get\footnote{The index $\hi$  runs over the $1,2,3$ components of the tetrad basis, and $\ha = \hat 0, \hi$.}
\begin{equation}
	g_\x^{AB} = \Psi^A_{\x\,\ha} \Psi^B_{\x\,\hb}\eta^{\ha\hb} \Longrightarrow g_\x = \text{det}\big(\Psi^A_{\x\,\hi}\big)^2 \;,
\end{equation}
which leads to\footnote{Note that, because of the standard convention we use $\eta^{\hat 0\hb\hc\hd} = -\varepsilon^{\hb\hc\hd}$ with $\hb,\hc,\hd = 1,2,3$.} 
\begin{equation}
\begin{split}
	\M_\x &= \frac{1}{3!}\, \mu_\x^{ABC}\,\veps_{ABC}^\x = \\
        &= \frac{1}{3!} \sqrt{g^\x}\eta_{ABC} \, \Psi^A_{\x\,\ha}\Psi^B_{\x\,\hb}\Psi^C_{\x\,\hc} \, \varepsilon^{\hat 0\ha\hb\hc } \mu^\x_{\hat 0} = \mu_\x \;,
\end{split}
\end{equation}
where we have used $ \mu_\x = -\mu^\x_a u^a_\x = -\mu_\x^{\hat 0}$. This fact is important because it makes clear that only the (rest-frame) energy content of the four-momentum co-vector $\mu^\x_a$ is stored in the normalization of the matter space momentum three-form $\mu_\x^{ABC}$. Similarly, one can show that $\N_\x = n_\x$
\begin{equation}
\begin{split}
	n_\x &= -\frac{1}{3!}u^\x_a\,\veps^{bcda}\,n^\x_{bcd} = -\frac{1}{3!} u^\x_\ha\,\veps^{\hb\hc\hd\ha}\,\Psi^B_{\x\,[\hb}\Psi^C_{\x\,\hc} \Psi^D_{\x\,\d]} \,\veps^\x_{BCD}\N_\x \\
    &=-\frac{1}{3!}u^\x_{\hat 0} \,\veps^{\hb\hc\hd\hat 0} \veps_{\hb\hc\hd}\,\N_\x = - u^\x_{\hat 0} \N_\x = \N_\x \;.
\end{split}
\end{equation}
These relations are not surprising. It is, in fact, quite intuitive that the  non-barred quantities are related to spacetime (rest-frame) densities given that the three-forms $\veps^\x_{ABC}$ measure the volume of the matter space elements. 

We can also use the tetrad formalism to prove another result that will be needed later on; the intimate connection between a non-zero particle creation rate and an extended functional dependence of the current three-form. In fact, we have (see \cref{eq:conservativeGamma})
\begin{equation}
\begin{split}
    \Gamma_\x = \nabla_a n_\x^a = \frac{1}{3!} \veps^{bcda} \,\Psi^B_{\x\,[b}\Psi^C_{\x\,c}\Psi^D_{\x\,d}\nabla_{a]} n^\x_{BCD} \;,
\end{split}
\end{equation}
where we used $\nabla_{[a}\Psi^B_{\x\,b}\Psi^C_{\x\,c}\Psi^D_{\x\,d]}=0$. Introducing (again) a tetrad comoving with the x-species, and multiplying by $\mu_\x$ we have 
\begin{equation}\label{eq:NonConservativeGamma}
	\mu_\x\Gamma_\x = \frac{1}{3!} \mu_\x^{ABC} \,u_\x^a\nabla_a n^\x_{ABC} \equiv \frac{1}{3!} \mu_\x^{ABC} \frac{d n^\x_{ABC}}{d \tau_\x} \;.
\end{equation}
As explained earlier, the right-hand-side of this equation vanishes identically if $n^\x_{ABC} = n^\x_{ABC} (X^A_\x)$, while it is in general non-zero if we assume the extended functional dependence given in \cref{eq:mspnorm}.

We can now use the introduced normalizations to slim the notation (with respect to that used in \cite{2015CQAnderssonComer}) for the various pieces of $R^\x_a$ and $D^\x_{ab}$. For instance, the ``purely reactive'' term from \cite{2015CQAnderssonComer} becomes
\begin{equation}\label{eq:PurelyReactiveSlim}
	\text{R}^{\x\y}_a = \frac{1}{3!} \mu_\x^{ABC} \frac{\partial n^\x_{ABC}}{\partial X_\y^D}\,\Psi^D_{\y\,a} = \M_\x\frac{\partial\N_\x}{\partial X_\y^D} \,\Psi^D_{\y\,a}\equiv \text R^{\x\y}_D\,\Psi^D_{\y\,a} \;. 
\end{equation}
Similarly we can write 
\begin{subequations}\label{eq:AdditionalViscousTensorsSlim}
\begin{align}
\begin{split}
		s^{\x\y}_{ab} &= \frac{1}{3} \mu_\x^{ABC} \frac{\partial n^\x_{ABC}}{\partial g_\y^{DE}}\,\Psi^D_{\y\,a} \,\Psi^E_{\y\,b} = 2\M_\x \frac{\partial \N_\x}{\partial g_\y^{DE}}\,\Psi^D_{\y\,a} \,\Psi^E_{\y\,b}   \\ 
	&\equiv s^{\x\y}_{DE}  \,\Psi^D_{\y\,a} \,\Psi^E_{\y\,b} \;, 
\end{split} \\
\begin{split}
	\mathcal{S}^{\x\y}_{ab}& = \frac{1}{3} \mu_\x^{ABC} \frac{\partial n^\x_{ABC}}{\partial g_{\x\y}^{DE}} \,\Psi^D_{\x\,a} \,\Psi^E_{\y\,b} = 2 \M_\x \frac{\partial\N_\x}{\partial g_{\x\y}^{DE}}\Psi^D_{\x\,a} \,\Psi^E_{\y\,b} \\
    &\equiv \mathcal{S}^{\x\y}_{DE}\,\Psi^D_{\x\,a} \,\Psi^E_{\y\,b} \;,
\end{split}
\end{align}
\end{subequations}
where we have used the fact that the partial derivatives are performed, say, with respect to the metric $g^{AB}_\y$ keeping fixed $g^{AB}_\x$ and $g^{AB}_{\x\y}$. We will consider the validity of this assumption later. The remaining viscous stress tensor, $S_{ab}^\x$, leads to a slightly more involved expression, because of the presence of $g^\x$ in \cref{eq:normalizationsRel}. We have
\begin{equation}\label{eq:StdViscousTensorSlim}
\begin{split}
	S^\x_{ab} &= \frac{1}{3} \mu_\x^{ABC} \frac{\partial n^\x_{ABC}}{\partial g_\x^{DE}}\,\Psi^D_{\x\,a} \,\Psi^E_{\x\,b} = 2\Bigg( \frac{\M_\x}{\sqrt{g^\x}} \frac{\partial \big(\N_\x\sqrt{g^\x}\big)}{\partial g_\x^{DE}} \Bigg)\,\Psi^D_{\x\,a} \,\Psi^E_{\x\,b} =\\
    &=  2\Bigg(\M_\x \frac{\partial \N_\x}{\partial g_\x^{DE}} - \frac{1}{2}\N_\x\M_\x\,g^\x_{DE}\Bigg)\Psi^D_{\x\,a} \,\Psi^E_{\x\,b} =\\
    &\equiv S^{\x}_{DE}  \,\Psi^D_{\x\,a} \,\Psi^E_{\x\,b}  \;.
\end{split}
\end{equation}
It is also obvious, by looking at the respective definitions, that the reactive terms that stem from the fact that $\N_\x$ can depend also on $g^{AB}_\y$ and $g^{AB}_{\x\y}$ can now be written  
\begin{subequations}
\begin{align}
	&r_a^{\x\y} = \frac{1}{2}s^{\x\y}_{DE}\, \nabla_a \big(g^{bc} \Psi^D_{\y\,b} \Psi^E_{\y\,c}\big) \;,\\
    & \mathcal{R}_a^{\x\y} = \frac{1}{2}\mathcal{S}^{\x\y}_{DE}\, g^{bc} \Psi^D_{\x\,b} \nabla_a \big( \Psi^E_{\y\,c}\big) \;.
\end{align}
\end{subequations}

Let us conclude this introductory section with a caveat on the slimmed notation introduced so far. In order to consider all the projected metrics to be independent, we need to make sure we are not actually adding degrees of freedom to the problem. Since the various projected metrics are ultimately combinations of $\Psi_{\x\,b}^A$ we need to make sure that the degrees of freedom associated with the metrics is less or equal to $12\,l$---where $l$ is the number of constituents. Because the number of mixed projected metrics is easily found to be $l(l-1)/2$ we have 
\begin{equation}
	6\frac{l(l+1)}{2} \leq 12 l \Longrightarrow l\leq 3 \;.
\end{equation}
Therefore, the slimmed notation introduced in this section applies to cases with less than (or equal to) three species moving independently. In what follows, the machinery is developed in a general setting, but we will focus to cases with less than three species for specific applications.

\section{The non-dissipative limit}\label{sec:thermoEq}

We now begin to develop the process for comparing standard relativistic models for dissipative fluids with that provided by the action principle. Standard approaches \cite{Muller:1967zza,ISRAEL1976,transientStewart77,IsraelStewart79,IsraelStewart79bis} assume a reference equilibrium state and then build in dissipation via deviations away from this state. The action principle formally does not require any sort of equilibrium, but provides a fully non-linear set of field equations. Obviously, our first task must be to extract from the non-linear equations a notion of equilibrium. This is not straightforward for various reasons, a key one being that an arbitrary spacetime in General Relativity does not have global temporal, spatial, and rotational invariance. As a first step, we will recall features of the typical laboratory set-ups within which the laws of chemistry, dynamics, and thermodynamics were first established. 

\subsection{Laboratory vs. general relativistic set-up}

A typical laboratory set-up is essentially local in the spacetime sense, implying there is---to a great deal of precision---temporal, spatial, and rotational invariance. Noether symmetries exist, which lead to energy, momentum, and angular momentum conservation. A clean separation between internal and external influences can be made, and these influences themselves can be manipulated. The effect of long-range, non-screenable forces on the system---for example, gravity---can be ignored. Well-defined (theoretical and experimental/observational) notions of total energy and entropy can be realized. Equilibrium can be defined in the broadest sense by saying the system evolves to a state where its total energy is minimized, or, equally, its total entropy is maximized.  

Internal interactions are due to, say, chemical reactions, whereas external interactions are those which distort the system's volume or allow particles and heat to enter or leave through the volume's surface. If a system is in chemical equilibrium internally, we can say that the reactions inside it are running forwards and backwards at such a rate that constituent particle number ratios remain fixed in time. If the given system is in chemical equilibrium with another system, then the chemical potentials of the two will be equal. A system in dynamical equilibrium just sits there, with no temporal evolution. Any pressure acting on the system's surface will be balanced by an internal pressure of the same value. Finally, we can say that two systems are in thermal equilibrium when there is no heat flow between them, the end result being equality of their respective temperatures. Now, let us return to the problem at hand---equilibrium when General Relativity cannot be neglected.


A general relativistic set-up is problematic from the get-go, because one is hard-pressed to find properties of equilibrium like those just discussed which are workable at all time- and length-scales. Broadly speaking, there seems to be no general relativistic rules on how the local thermodynamics of local (intensive) parameters---chemical potential $\mu$, pressure $p$, and temperature $T$---connect with some notion of global thermodynamics for global (extensive) parameters---such as the total energy $E$. An unambiguous extrapolation of the standard definitions of chemical,  dynamical, and thermal equilibrium given above to General Relativity is not possible, for reasons to be explained below. There is also the well-known difficulty of identifying the total energy of a region in an arbitrary spacetime, since the Equivalence Principle precludes an ultra-local definition of gravitational energy density.\footnote{Of course, for asymptotically flat spacetimes, one can define quantities like the Schwarzschild mass. Gravitational-wave energy can be defined but only after averaging over wavelengths, see \cite{maggioreGWs,Straumann}.}

The reason that the laboratory rules for chemical and thermal equilibrium are not viable in General Relativity was established long ago by Tolman and Ehrenfest \cite{TOLMAN1,TOLMAN2}: In General Relativity, all forms of energy react to gravity. Temperature and chemical potentials represent forms of energy and can undergo red-shift or blue-shift. There is no single temperature for an isolated system, and so saying ``system A is in thermal equilibrium with system B if their temperatures are the same'' becomes ambiguous; similarly for chemical equilibrium. As for dynamical equilibrium, a standard undergraduate physics calculation shows that pressure increases with depth in water which nevertheless remains at rest.\footnote{In this context, we can think of it as resulting from the breaking via gravity of spacelike Killing vectors which lead to space-translation invariance.}

Even the use of the word ``equilibrium'' is tricky because it tends to imply that a system in thermal and chemical equilibrium is independent of time, because the total entropy and total particle number do not evolve. In General Relativity, a system which is independent of time occurs only for special spacetimes which have a global timelike Killing vector field. Strictly speaking, this immediately puts the non-dissipative fluid models of Cosmology---the Friedman-Lemaitre-Robertson-Walker solutions---out of the discussion, as the universe is expanding, making it time-dependent. 

This points to another problem of the notion of total energy in General Relativity and arguments based on the standard understanding of energy conservation: In Special Relativity, the curvature is zero and there is a timelike Killing vector field leading directly to a Noether symmetry for the system and total energy conservation. (There are also Killing vector fields representing rotational and spatial invariance, which lead to Noether symmetries resulting in total angular and linear momentum conservation.) In an expanding universe this line of reasoning for energy conservation obviously breaks down. 

The main message is this: Important issues remain unsettled even after a century's worth of debate. We will not resolve these issues here; instead, what we will do is take the action-based formalism and see how its internal machinery can be manipulated to produce a self-consistent notion of the non-dissipative limit, without trying to resolve the deeper issues about the nature of equilibrium\footnote{We will still use the word ``equilibrium'' interchangeably with the non-dissipative limit.}. Our way forward is to take advantage of the fact that the action-based field equations are fully non-linear and complete. 

\subsection{Multiple equilibrium states}

The main mechanism for manipulating the machinery of the action-based field equations is to apply perturbation techniques similar to those used to determine, say, quasi-normal modes of neutron stars. The general idea for neutron stars is to analyze linear perturbations of configurations having particular symmetries generated by Killing vectors. Among the most studied neutron star ``ground-states'' are those having Killing vectors which generate staticity and spherical symmetry, and those with Killing vectors that generate axisymmetry and stationarity; basically, non-rotating and rotating backgrounds, respectively.

In an analogous way, we can expect different options for generating the non-dissipative limit of a multi-fluid system. For example, we can take the limit where the different dissipation coefficients (such as shear and bulk viscosities) are effectively zero. Another possibility is the limit where the dissipation coefficients are non-zero but the fluid motion itself is such that the dissipation mechanisms are not acting. The formalism developed by \citet{onsager31:_symmetry} (recall discussion in \cref{subsec:LIT}) is worthy of mention here, because the system of field equations it creates are more explicit in how the two limits can be implemented (see, for example, \cite{andersson05:_flux_con}). It is interesting also to note that the philosophy of the Onsager approach is not so much about how to expand away from an equilibrium, but rather how a non-equilibrium system gets driven back to the equilibrium state. Here, because the field equations are fully non-linear, they can, in principle, describe systems which are being driven toward or away from equilibrium. 

Next, we will explore some of the different options for equilibrium states. We will use a global analysis which assumes that the second law of thermodynamics applies and that a knowledge of the fluxes  throughout a region of spacetime is enough to determine whether or not dissipation is acting. A local analysis of the formalism will also be pursued, involving the field equations themselves. 
 
\subsection{Global analysis of the non-dissipative limit}

Recall that the fundamental dynamical variables are the particle fluxes $n^a_\x$ and the entropy $s^a=n_\s^a$.\footnote{Because we impose the second law of thermodynamics below, we are specifically separating out the entropy flux in this discussion.} The formalism's linchpin is the breaking of the closure of the particle-flux three-forms, $n^\x_{abc}$ and $s_{abc}$, which leads to non-zero creation rates $\Gamma_\x$ and $\Gamma_\s$. In turn, these non-zero creation rates lead to the resistive contribution $R^a_\x$ and the dissipation tensor $D^\x_{ab}$ terms in the equations of motion. The nice thing about fluxes, which we will exploit here, is that they can be integrated.

\begin{figure}
   \centering
   \includegraphics[width=0.8\linewidth]{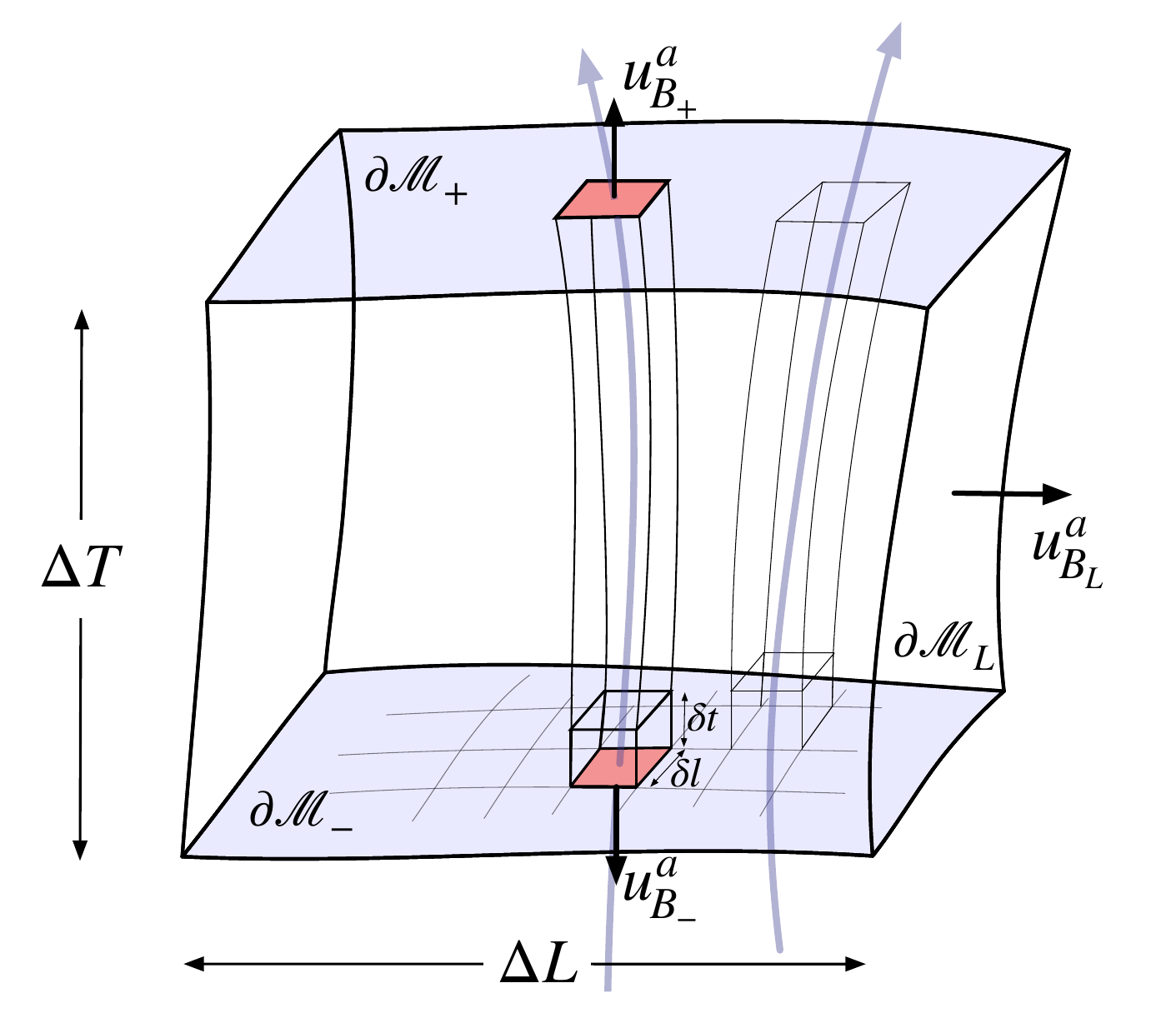}
   \caption{A depiction of the spacetime region ${\cal M}$, with one spatial axis suppressed. It has a characteristic spatial size $\Delta L$ and temporal size $\Delta T$. Inside ${\cal M}$ is a smaller region $\delta {\cal M}$ of characteristic spatial and temporal size $\delta l$ and $\delta t$, respectively. The boundary $\partial\cal M$ consists of the initial and final time-slices $\partial\cal M_-,\,\partial\cal M_+$ and the timelike hypersurface $\partial\mathcal{M}_L$. }
   \label{fig:FluidElem}
\end{figure}

When we use the Einstein equations and the field equations of a multi-fluid system, our goal is to get solutions for the metric and fluxes on a ``chunk'' of spacetime, for a given set of initial/boundary conditions. Suppose we pick an ad hoc region ${\cal M}$ of spacetime, as illustrated in \cref{fig:FluidElem}. The fact that it is a region implies there is a ``conceptual boundary'', meaning the whole spacetime is being divided up into smaller domains. Let $u^a_{\rm B}$ (collectively) denote the unit normal to the total boundary of the region, defined so that it always points ``out''. The boundary itself consists of two spacelike hypersurfaces $\partial {\cal M}_\pm$ (with unit normals $u^a_{\rm B_\pm}$, $u^a_{\rm B_\pm} u_a^{\rm B_\pm} = -1$), and a timelike hypersurface $\partial {\cal M}_L$ (with unit normal $u^a_{\rm B_L}$, $u^a_{\rm B_L} u_a^{\rm B_L} = + 1$); in essence, think of $\partial {\cal M}_-$ as a 3D region of characteristic volume $\Delta L^3$ on an initial time-slice of ${\cal M}$ and $\partial {\cal M}_+$ as the same volume on the final time-slice, and then $\partial {\cal M}_L$ will be similar to the union of the surface of the same volume on each leaf of some spacelike foliation of ${\cal M}$ between $\partial {\cal M}_-$ and $\partial {\cal M}_{+}$. The induced metric on $\partial {\cal M}_{\pm}$ is $h_\pm^{a b} = g^{ab} + u^a_{\rm B_\pm} u^b_{\rm B_\pm}$ and for $\partial {\cal M}_L$ it is $h_L^{a b} = g^{ab} - u^a_{\rm B_L} u^b_{\rm B_L}$.

There are three contributions to the total particle number change $\Delta N^\x$ and total entropy change $\Delta S$: (i) The total particle number $N_-^\x$ and entropy $S_-^\x$ which exist in $\partial {\cal M}_-$; (ii) The total particle number $N_+^\x$ and entropy $S_+^\x$ which exist in $\partial {\cal M}_+$; and, (iii) The number of particles $\Delta N^\x_L$ and amount of entropy $\Delta S_L$ which enter/leave $\partial {\cal M}_L$. Each contribution is obtainable from its associated flux: If $n^\x_\pm$ ($s_\pm$) are the particle number (entropy) densities as measured with respect to the volumes $\partial {\cal M}_\pm$, and $n^\x_L$ ($s_L$) is the number of particles (amount of entropy) per unit area per unit time entering/leaving $\partial {\cal M}_L$, then\footnote{We are denoting with $h_\pm$ the determinant of $h^\pm_{ab}$. We have also taken into account the fact that $u^a_{\rm B_-}$ points to the past, and that the signature of the induced metric $h_L^{ab}$ is $(-++)$ as $\partial \M_L$ is timelike.} 
\begin{subequations} \label{fluxdelnx}
\begin{align}
    N_+^\x &= \tintp n^\x_+ = \tintp \left(- u_a^{\rm B_+} n^a_\x\right) \ ,\\ 
    N_-^\x &= \tintm n^\x_- = \tintm \left(u_a^{\rm B_-} n^a_\x\right) \ , \\
    \Delta N_L^\x &= \tintL n^\x_L = \tintL \left(u_a^{\rm B_L} n^a_\x\right) \;,
\end{align}
\end{subequations}
and similarly for $\Delta S$. 
The changes in the total $\x$-particles $\Delta N_\x$ and entropy $\Delta S$ over the region ${\cal M}$ are therefore
\begin{subequations} \label{totdelNx}
\begin{align}
    \Delta N^\x = N_+^\x - N_-^\x + \Delta N_L^\x \ , \\
    \Delta S = S_+ - S_- + \Delta S_L \ .
\end{align}
\end{subequations}

If the length- and time-scales of spacetime region ${\cal M}$ are those typical of terrestrial labs (read: its curvature is zero throughout), then we have great confidence in asserting the second law of thermodynamics; namely, the net change of the total entropy must satisfy $\Delta S \geq 0$. We could even be confident that we could determine the total energy $E$ and volume $V$ of the system, and have a working first law of thermodynamics which connects $\Delta E$, $\Delta N^\x$, $\Delta V$, and $\Delta S$: 
\begin{equation}
    \Delta E = T \Delta S - p \Delta V + \sum_\x \mu_\x \Delta N^\x \ . \label{1stlaw}
\end{equation}
The temperature $T$, pressure $p$, and chemical potentials $\mu_\x$ would be well-defined and calculable. We could even use the standard notions of chemical, dynamical, and thermal equilibrium and say that system A of spacetime region ${\cal M}_A$ is in chemical, dynamical, and thermal equilibrium with system B of spacetime region ${\cal M}_B$ if, respectively, their chemical potentials are equal, their pressures are equal, and their temperatures are equal.  

Now, let us suppose we have a region large enough that spacetime curvature can no longer be ignored. Probably, it would be a safe bet to say that the second law still applies; i.e., $\Delta S \geq 0$. But, we are hard-pressed to employ the laboratory definitions of chemical, dynamical, and thermal equilibrium. Consequently, it is difficult to imagine a global first law of thermodynamics for general relativistic multifluid systems similar to that in \cref{1stlaw}; again, the reason being that intensive parameters are spacetime dependent, and an extensive parameter like total energy may not even be definable. Still, our task is to explore any possible link between parameters which require scales where spacetime curvature is necessary ($\Delta N^\x$ and $\Delta S$) to the local fluid variables ($n^a_\x$ and $s^a$) which enter the fluid field equations. Fortunately, the divergence theorem provides such a link.

Applying it to the divergence of both the particle and entropy fluxes gives
\begin{subequations}
\begin{align}
    \Delta N^\x &= \fint \nabla_a n^a_\x = \fint \Gamma_\x \ , \\
    \Delta S &= \fint \nabla_a s^a = \fint \Gamma_\s \ .
\end{align}
\end{subequations}
These are not new results, but they serve the purpose here of establishing a direct link between global and local variables, which we will use to formulate some aspects of the non-dissipative limit of our formalism.

Consider an idealized situation of a spacetime region ${\cal M}$ sub-divided into a region ${\cal M}_A$ for which $\Delta N^\x_A < 0$ and $\Delta S_A < 0$, and another region ${\cal M}_B$ for which $\Delta N^\x_B > 0$ and $\Delta S_B > 0$. The trick is that they are such that the total changes on ${\cal M}$ vanish:
\begin{equation}
       \Delta N^\x = \Delta N^\x_A + \Delta N^\x_B = 0
       \; , \;
       \Delta S = \Delta S_A + \Delta S_B = 0 \ .
\end{equation}
The point is that, even though $\Gamma_\x$ and $\Gamma_\s$ are not zero, this is an example of a global, fully general relativistic, non-dissipative system since there is no net total particle number or total entropy change. But is this realistic? Is this the kind of definition of the non-dissipative limit we are looking for? Probably not. What is more likely is that the non-dissipative limit is better understood by breaking up ${\cal M}$ into many small spacetime regions $\delta {\cal M}$, with characteristic temporal and volume scales $\delta t$ and $\left(\delta l\right)^3$, respectively, as illustrated in \cref{fig:FluidElem}. 

Once again, let us imagine that $\delta {\cal M}$  is subdivided into two regions $\delta {\cal M}_A$ and $\delta {\cal M}_B$. It is conceivable that on these scales statistical fluctuations could lead to positive creation rates in one region and negative in the other. If the regions are small enough, we can assume that $\Gamma_\x$ and $\Gamma_\s$ vary slowly across them so that we can approximate the integrals for $\delta N^\x_{\delta {\cal M}}$ and $\delta S_{\delta {\cal M}}$ as\footnote{We are also assuming that the $\delta \M$ to be small enough that we can transform away gravity by means of Riemann normal coordinates \cite{GravitationMTW}.}
\begin{equation}
       \delta N^\x_{\delta {\cal M}} \approx \Gamma_\x \delta t \left(\delta l\right)^3 
       \; , \;
       \delta S_{\delta {\cal M}} \approx \Gamma_\s \delta t \left(\delta l\right)^3 \ .
\end{equation}
However, the random nature of statistical fluctuations for a system purported to be in equilibrium implies that any non-zero creation rates inside $\delta {\cal M}_A$ and $\delta {\cal M}_B$ must balance on average so that
\begin{subequations}
\begin{align}
    \delta N^\x_{\delta {\cal M}} = \delta N^\x_{\delta {\cal M}_A} + \delta N^\x_{\delta {\cal M}_B} \approx \left(\Gamma^A_\x + \Gamma^B_\x\right) \delta t \left(\delta l\right)^3 = 0 \; \Longrightarrow \; \Gamma_\x = \Gamma^A_\x + \Gamma^B_\x = 0 \; , \\
    \delta S_{\delta {\cal M}} = \delta S_{\delta {\cal M}_A} + \delta S_{\delta {\cal M}_B} \approx \left(\Gamma^A_\s + \Gamma^B_\s\right) \delta t \left(\delta l\right)^3 = 0 \; \Longrightarrow \; \Gamma_\s = \Gamma^A_\s + \Gamma^B_\s = 0 \; \ .
\end{align}
\end{subequations}

One conclusion from this exercise is that the characteristic time and volume scales of $\delta {\cal M}$ must be large enough  that statistical fluctuations will, on average, balance out for a system in equilibrium. The second conclusion is that having $\delta N^\x_{\delta {\cal M}} = 0$ ($\delta S_{\delta {\cal M}} = 0$) on the one hand means $\Gamma_\x = 0$ ($\Gamma_\s = 0$) on the other, and vice versa. Putting both together we will assume that the equilibrium state for multi-fluid systems must be such that regions like $\delta {\cal M}$ set the scales for fluid elements and we have $\Gamma_\x = 0$ and $\Gamma_\s = 0$ everywhere in ${\cal M}$.

\subsection{Local analysis of the non-dissipative limit}

This next step begins where the previous one left off; that is, a necessary condition for a multi-fluid system to be in equilibrium is that the flux creation rates $\Gamma_\x$ (now including the entropy) vanish everywhere. We will use the field equations themselves to investigate three different ways for the action-based system to have zero particle creation rates: 1) The limit where the dissipation terms $R^\x_a$ and $D^\x_{ab}$ are zero, 2) the limit where the dissipation terms are non-zero but the fluid motion is such that the dissipative channels are dynamically suppressed, and 3) a combination of dynamical suppression with constraints between the dissipation terms that lead to Killing vector fields. 

But before moving on with the analysis, it is advantageous to consider the simplest non-dissipative fluid model which can be derived from the action above---the ordinary perfect fluid, where all particle species and entropy flow together and the total particle numbers and entropy are conserved individually. The calculation is straightforward \cite{comer93:_hamil_multi_con}. All the fluxes have the same four-velocity, say, $u^a$, and so $n^a_\x = n_\x u^a$. If each particle number flux is conserved individually, then
\begin{equation}
    \nabla_a n^a_\x = \nabla_a \left(n_\x u^a\right) 
      = u^a \nabla_a n_\x + n_\x \nabla _a u^a = 0 
      \; \Longrightarrow u^a \nabla_a \ln n_\x = - \nabla _a u^a \ .
\end{equation}
Obviously, the total particle flux $n^a = \sum_\x n^a_\x$ is also conserved and hence
\begin{equation}
     u^a \nabla_a \ln n = - \nabla _a u^a \; , \; n = \sum_\x n_\x \ .
\end{equation}
Therefore, we have
\begin{equation}
     u^a \nabla_a \ln n_\x - u^a \nabla _a \ln n = 0 
      \; \Longrightarrow u^a \nabla_a \left(\frac{n_\x}{n}\right) = 0 \ .
\end{equation}
The upshot is that each species fraction $n_\x/n$ must also be conserved along the flow, and this includes the entropy as well. This implies that only one matter space is required. In the action principle, this means that for each $\x$---including the entropy---we have $\xi^a_\x = \xi^a$, and there is only one Euler equation of the form
\begin{equation}
     \sum_\x f^\x_a = 0 \ , \label{singfl}
\end{equation}
where the $f^\x_a$ are exactly as in \cref{eq:VariationalForceDensities}.

With this example in mind we can now proceed with the description of three different non-dissipative limits consistent with the action-based model. Note that we will also impose another condition which defines the equilibrium. We will assume that all distinct fluids are comoving, so that we are not considering systems with superfluid/superconducting phases, or a perfect heat-conducting limit \citep{CarterNoto}. This means that there is a common four-velocity for all species, $u^a_\x = u^a$. However, it is important to point out a subtlety about this comoving limit: For a multi-fluid system each species has its own evolution equation. Even in the comoving limit there are still $\x$ fluid equations. Now consider the field equations for a multi-species, single fluid system---as we see from \cref{singfl}, it has only one fluid evolution equation. Therefore, the comoving limit of the multi-fluid system ($\x$ equations) is not equal to the single-fluid system (one equation). This is not an error, rather, it is a consequence of the fact that the number of independent field equations of the system is fixed by the number of independent fluids chosen before the action principle is applied. 
Note also that we can use the common four-velocity $u^a$ to introduce a spatial covariant derivative $D_a$---acting in directions perpendicular to $u^a$---and a time derivative $``\;\dot{}\;" = u^a \nabla_a$. For a scalar $A$ we have
\begin{equation}\label{eq:ProjectedDerivative}
	D_a A = \perp^b_a \nabla_b A = \left(\delta^b_a + u_a u^b\right) \nabla_b A = \nabla_a A + \dot A u_a \; ,
\end{equation}
and for a vector
\begin{equation}\label{eq:ProjectedDerivativevec}
	D_a A_b = \perp^c_a \perp^d_b \nabla_c A_d \; .
\end{equation}

\subsubsection*{Dynamical suppression of dissipation}
We start by considering the consequences of the non-dissipative limit  if the fluid flow is such that the dissipation mechanisms are not triggered. If we look at each species creation rate we have
\begin{equation}
	\mu_\x \Gamma_\x = - R^\x_a \,u^a - D^\x_{a b} \nabla^a u^b = 0 \;,
\end{equation}
so that, summing over all species
\begin{equation}\label{eq:EquilibriumRigidMotion}
	\sum_\x \mu_\x \Gamma_\x = - \left(\sum_\x R^\x_a\right) \,u^a - D_{a b} \nabla^a u^b = - D_{ab} D^{(a} u^{b)} = 0 \; ,
\end{equation}
where we have used the identities $u^b_\x D^\x_{ab} = 0$, $\sum_\x R^\x_a = 0$ and the fact that $D^{ab}$ is symmetric. Using the standard decomposition of the four velocity gradients it is easy to see that \cref{eq:EquilibriumRigidMotion} implies 
\begin{equation}
   D_{(a} u_{b)} = \perp^c_{(a} \perp^d_{b)} \nabla_c u_d = \nabla_{(a} u_{b)} + u_{(a} \dot{u}_{b)} = \sigma_{ab} + \frac{1}{3}\theta \perp_{ab} = 0 \; .
\end{equation}
In particular, this tells us that the (dynamically-suppressed) non-dissipative flow has zero expansion $\theta = 0$, and zero shear $\sigma_{a b} = 0$. What is left of the motion is captured by
\begin{equation}
    \nabla_a u_b = \omega_{ab} - \dot{u}_b u_a \; \ ,
\end{equation}
which is consistent with rigid rotation.

From the definition of creation rates, we can now write
\begin{equation}
	\Gamma_\x = \nabla_a n_\x^a = \dot n_\x + n_\x \theta = \dot n_\x = 0 \;.
\end{equation}
Assuming a thermodynamical relation in the standard way, namely that the energy functional of the system is\footnote{Note that we are here using a compact notation, so that the chemical index $\x$ in $\varepsilon(n_\x)$ runs over all the species/constituents in the sytem under consideration.} $\varepsilon = \varepsilon (n_\x)$, we see that the chemical potential of each species is $\mu_\x = \mu_\x(n_\x)$ and likewise for the pressure $p$. Therefore, we have $\dot \mu_\x = 0$ and $\dot p = 0$, as well. The proposed scenario is consistent with the minimum requirements for the system being non-dissipative---as explained above. This is not, however, the situation we will use as basis for the expansion.

\subsubsection*{The Euler limit}\label{vanRDeq}
Later (in \cref{sec:EnergyMinimized}) we will use thermodynamics arguments to show that the dissipative terms all vanish at equilibrium: $D_{ab}^{\x,\ \e} \text{ and } R_a^{\x,\ \e}=0$---where we have introduced the superscript ``e''  to stress that the dissipative terms are evaluated at equilibrium, consistently with the notation used later on. We now consider the non-dissipative limit with these additional constraints and show its compatibility with the Euler equations. Since the fluids are comoving at equilibrium we have for the fluxes $n_\x^a = n_\x u^a_\e$, and so the four-momenta become
\begin{equation}
	\mu_a^\x = \Big(\mathcal{B}_\x n_\x + \sum_{\y\neq\x}\mathcal{A}_{\x\y}n_\y \Big) u^\e_a = \mu_\x u^\e_a \;.
\end{equation}
and the equation of motion for the x-species is
\begin{equation}\label{eq:AlignedForce}
	f^\x_a =2n_\x^b \nabla_{[b}\mu^\x_{a]}= n_\x\mu^\x \dot u^\e_a + n_\x\Big(u^b_\e u^\e_a + \delta^b_a\Big)\nabla_b\mu^\x = n_\x \mu^\x \dot u^\e_a + n_\x D_a\mu^\x = 0 \;.
\end{equation}
The first term in $f^\x_a$ then looks like the mass/energy per volume times the acceleration while we can show that the second is a ``pressure-like'' term in the sense of being the gradient of a thermodynamic scalar. In fact, we have
\begin{equation} \label{muxcomov}
	\frac{\partial\Lambda}{\partial n_\x} = - \Bigg(\mathcal{B}_\x n_\x -\sum_{\y\neq\x} \mathcal{A}_{\x\y}n_\y^a u^\x_a \Bigg) =- \mu_\x \;,
\end{equation}
and the sum of these terms provides the derivative of the total pressure $\Psi$:
\begin{equation}
	\sum_\x n_\x D_a \mu^\x  = D_a \Big( \sum_\x n_\x \mu^\x +\Lambda\Big) = D_a \Psi \;.
\end{equation}
It is important to note that each individual term cannot (in general) be considered as the derivative of the x-species contribution to the total pressure. Partial pressures exist only when the various species do not interact.

Even though the comoving limit of the multi-fluid system is not the same as the single fluid, multi-species system, there is some overlap: Taking the sum over the chemical species of \cref{eq:AlignedForce} we find\footnote{We have used the standard Euler relation  $\sum_\x n_\x\mu^\x = p +\varepsilon$, where $p,\,\varepsilon$ are the equilibrium pressure and energy density, respectively.} the standard relativistic Euler equation we derived in \cref{sec:PerfectFluids}. One can also show that \cref{singfl} can be written in this form. This is an important self-consistency check, but because the multi-fluid comoving limit is not the same as the single fluid limit, we need to go back to the individual fluid equations of the multi-fluid system.

We can rewrite the individual equations of motion as
\begin{equation}\label{eq:MultiFluidEquilibriumEq}
	\dot u^\e_b = - D_b (\text{log }\mu_\x) \ ;
\end{equation}
thus, for each combination of $\x \neq \y$,
\begin{equation}
    D_a (\text{log }\mu_\x) = D_a (\text{log }\mu_\y) \; \Longrightarrow \; D_a \left(\text{log }\frac{\mu_\x}{\mu_\y}\right) = 0 \ .
\end{equation}
This self-consistency therefore requires the various chemical potentials $\mu_\x$ and $\mu_\y$ (as functions on spacetime) to be proportional to each other by some factor $C^\x_\y$, which is constant in the spatial directions; namely,
\begin{equation}
    \mu_\x = C^\x_\y \mu_\y \; , \; D_a C^\x_\y = 0 \ .
\end{equation}
This is to be contrasted with the single-fluid case, where there is no such restriction---in the sense of being forced by the evolution equations---between the chemical potentials. Usually, one must provide additional information. For example, for neutron stars one typically imposes that beta decay and inverse beta decay are in equilibrium. 
If we combine this with the ``dynamical suppression of dissipation'', the factor $C^\x_\y$ is in fact constant in all the space-time directions. 

\subsubsection*{Dynamical suppression and Killing vectors}

In a local region of spacetime, freely falling frames exist and the Killing equation will be satisfied approximately. In these local regions having an equilibrium will be consistent with the existence of Killing fields. However, local regions which are far removed from each other will not be (on the relevant dynamical timescale) in equilibrium with each other. This kind of ``quasi-local'' regression towards equilibrium has been discussed in the work of \citet{FukumaSakataniEntropic}, explicitly introducing two different spacetime scales to describe the  evolution of general relativistic dissipative systems. The hypothesis of Local Thermodynamic Equilibrium applies on the smaller scale---which is of the size of a fluid element---while the regression (in the sense of Onsager \cite{onsager31:_symmetry}) towards equilibrium takes place on the larger one, which can still be smaller than the body size.

A relation between the perfect fluid four-velocity and Killing vectors, for stationary axially symmetric rotating stars,\footnote{Note that \citet{RotatingStarsGourgoulhon} works with the enthalpy per particle instead of chemical potentials. However, this makes no difference for barotropic perfect fluids.} has been discussed by \citet{RotatingStarsGourgoulhon}. A similar discussion about thermodynamic equilibrium in General Relativity and the existence of Killing vectors was provided by \citet{Becattini16}. Specifically, he showed that there must be global Killing vector fields if the total entropy of the system is to be independent of the spacelike hypersurface over which the integration is performed. As for the work presented here, we will now show under what conditions the combination $\xi^a_\x = \mu_\x^{-1} u^a_\e$ can be turned into Killing vector fields. 

From \cref{eq:MultiFluidEquilibriumEq} it can be seen that 
\begin{equation}
\begin{split}
	\nabla_a \xi_b^\x + \nabla_b \xi_a^\x = \frac{1}{\mu_\x} \big(u^\e_{(a}u^\e_{b)} \dot{\text{log} \mu_\x} + D_{(a}u^\e_{b)}\big) \;,
\end{split} 
\end{equation}
so that, if dynamical suppression has worked, the right hand side becomes zero and the $\xi^a_\x$ will be a timelike Killing vector field, along which the local thermodynamical parameters $n_\x$, $\mu^\x$, $\veps$, and $p$ become constants of motion. Put in different words, if we want the system to be (at least quasi-locally) at equilibrium---i.e. stationary--- we also need to require rigid body motion.

\subsection{A final comment on equilibrium}

To conclude we will come full circle and consider again the change in total entropy given by \cref{totdelNx}. The result only references spacelike hypersurfaces as part of the (ad hoc) choice of the boundary of the spacetime region for which the entropy change is being determined. There are no restrictions placed on the spacetime geometry in this construct; in particular, no requirement of global Killing vectors. 
As a matter of practice, the change in entropy of a system is clearly dependent on the spatial size and the amount of time the system has had to evolve. Coupling this with the fact that a separation of space from time is always a choice---an arbitrary spacetime has no preferred directions, no natural ``moments-of-time"---we see that the ad hoc nature of the boundary in \cref{totdelNx} is not a drawback. This is precisely the freedom needed in order to incorporate a system's spatial extent and evolution time, and the fact that a separation of space from time in spacetime is always a choice.

The main reason why this is intriguing is that the second law of thermodynamics only refers to the change in total entropy, not the value of entropy itself at specific moments of time (i.e.~spacelike hypersurfaces). It may be that questions of equilibrium are not to be settled by the ``moment-to-moment'' behaviour of three-dimensional integrals, but rather by global statements of the type represented by \cref{totdelNx}. This is something we are currently investigating and hope to be able to give more detail on in a future work.

\section{Perturbations with respect to equilibrium}\label{sec:PerturbativeExpansion}

With the equations of motion obtained from an action principle, we can consider perturbations away from equilibrium configurations (of the kind described above) in a way that is closely related---at least from the formal perspective---to standard hydrodynamical perturbation theory. The general approach to Lagrangian perturbation theory is perhaps best described by \citet{FriedmannSchutz1975}. Roughly speaking, the evolution equations for the perturbed fields can be obtained by perturbing the equations that follow from the action. It is also clear---at least in principle---how to construct a Lagrangian whose variation gives the perturbed equations (see \textsection 2 of \cite{FriedmannSchutz1975}). However, since  we are not focussing on a stability analysis of fluid oscillations we will not consider this additional aspect here. 

To set the stage for the perturbative expansion, we consider the family of worldlines (not necessarily geodesics) that each constituent of a multifluid system traces out in spacetime. Our definition of equilibrium includes the assumption that all species are comoving. Therefore, our fiducial set of worldlines representing equilibrium are those the system would have followed if it were comoving throughout its history.
\begin{figure}
    \centering
    \includegraphics[width=0.8\textwidth]{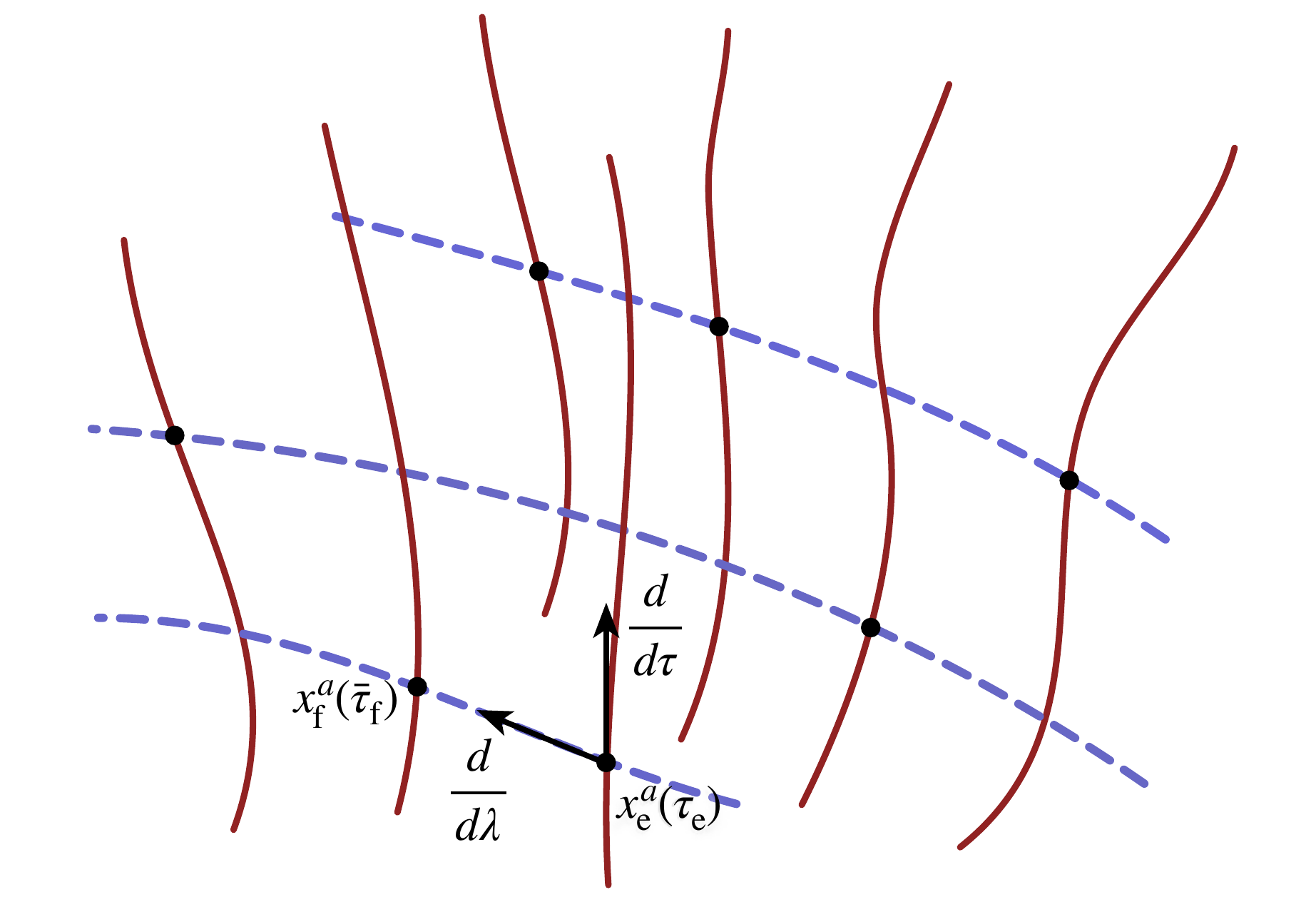}
    \caption{An illustration of worldlines associated with the fluid elements (solid vertical red lines, parameterized by $\tau,\,\bar\tau$) and “Lagrangian displacements” which connect fluid elements (dashed horizontal blue lines, parameterized by $\lambda$).}
    \label{fig:PerturbationWorldlines}
\end{figure}
This then allows us to view each of the ``final'' worldlines $x_\f^a(\bar \tau)$ as a curve in spacetime which is close to the equilibrium one $x_\e^a(\tau)$, with $\bar \tau$ and $\tau$ being the proper times of the respective curves. See \cref{fig:PerturbationWorldlines} for an illustration of the idea. The unit four-velocities associated with the two worldlines are
\begin{equation}
	u^a_\f = \frac{d x_\f^a}{d \bar \tau} \;, \qquad u^a_\e = \frac{d x_\e^a}{d \tau} \;.
\end{equation}
Obviously, $u^a_\e$ represents the comoving frame introduced earlier.

We assume another family of curves $x_{\e\f}^a(\lambda)$, where $\lambda$ is an affine parameter (say, the proper length), that connects the equilibrium worldline to the actual one. This means that for any point $x_\e^a(\tau_e)$ on the equilibrium worldline, there is a unique point $x_\f^a(\bar \tau_f)$ on the perturbed worldline, and a unique curve $x_{\e\f}^a(\lambda)$ between them having two points $x_{\e\f}^a\left(\lambda_\e\right)$ and $x_{\e\f}^a\left(\lambda_\f\right)$ such that
\begin{equation}
	x_\f^a\left(\bar \tau_\f\right) = x_{\e\f}^a\left(\lambda_\f\right) \;, \qquad x_\e^a \left(\tau_\e\right) = x_{\e\f}^a\left(\lambda_\e\right) \;.
\end{equation}
Taylor expanding the perturbed worldline about the equilibrium up to second order, we get
\begin{equation}
\begin{split}
	x_\f^a(\bar \tau_\f) &= x_\e^a(\tau_\e) + \frac{d x_{\e\f}^a}{d\lambda}\Big|_{\lambda_\e} (\lambda_\f - \lambda_\e) + \frac{1}{2} \frac{d^2 x_{\e\f}^a}{d^2\lambda}\Big|_{\lambda_\e} (\lambda_\f - \lambda_\e)^2 \\
 &= x_\e^a(\tau_\e) +\zeta^a \Delta\lambda + \frac{1}{2} \Big(\zeta^b\partial_b\zeta^a \Big)\Delta\lambda^2 \;,
\end{split}
\end{equation}
where we introduced the tangent vector 
\begin{equation}
	\frac{d}{d\lambda} = \frac{dx_{\e\f}^a}{d\lambda}\Big|_{\lambda_\e} \frac{\partial}{\partial x^a} = \zeta^a \partial_a \;.
\end{equation}

The first objects we want to perturb are the fluid element ``names''. That is, we attach a label $X^A$, where the index $A=1,2,3$, to each of the worldlines used to cover the region of spacetime occupied by the fluid. By definition of the Lagrangian variation \cite{FriedmannSchutz1975,Friedman1978a} we then have
\begin{equation}
	\Delta X^A = \Big(\phi_*  X^A(x_\f)\Big)(x_\e) - \bar X^A(x_\e) = X^A(x_\f) -\bar X^A(x_\e) = 0 \;, 
\end{equation}
where $\phi$ is the diffeomorphism that connects the perturbed and unperturbed worldlines, via the flow lines $x_{\e\f}^a$, and and $\phi_*$ denotes the pull-back from perturbed to equilibrium manifold.
The last equality then follows from the fact that the fluid label does not change as we follow it. As a result we have, to first order 
\begin{equation}
	\delta X^A = -\mathcal{L}_{\xi_\x} X^A = -\xi_\x^a \Psi^A_{\e \,a} = - \xi_\x^A\;, 
\end{equation}
where we introduced the Lagrangian displacement vector $\xi^a_\x= x_\f^a - x_\e^a$. 

It is important to note that these displacement vectors are different from the ones introduced when obtaining the equations of motion from the action principle (see \cref{eq:ConvectiveVariation4Current}), even though the mathematics appears the same. In the present case the displacement vector connects two configurations that are ``close'' in the space of physical solutions---in field-theory parlance they are both ``on-shell''. We also note that, to compute the second order variation we cannot rely on the simple relation that exists between Lagrangian and Eulerian variation (at first order). We need to perform such calculations explicitly. 

At this point, it is worth pausing to consider what is behind the perturbation scheme we are building. Since we assume the existence of a well defined equilibrium timelike congruence $x_\e^a$ with four velocity $u_\e^a$, we may imagine  riding along with the equilibrium fluid element  observing the evolution of the system (towards equilibrium) from this perspective. This means that the x-species four-velocity $u_\x^a$ can be decomposed (in the usual way) as 
\begin{equation}
	u^a_\x = \gamma_\x \Big( u_\e^a + w_\x^a\Big) \;, \text{   where }\quad w_\x^a u_a^\e = 0 \;, \quad\gamma_\x = \Big(1 - w^a_\x w_a^\x\Big)^{-1/2} \;.
\end{equation}
Moreover, since we are working up to first order we have  
\begin{equation}
	\gamma_\x = 1 +\frac{1}{2} w_\x^2 \approx 1 + \O(w_\x^2) \quad \Longrightarrow \quad u_\x^a = u_\e^a + w_\x^a \;.
\end{equation}
We note that this linear expansion in the relative velocities, although in a different spirit, has also been discussed in the context of extensions to magneto-hydrodynamics \cite{NilsPRD12,BeyondIdealMHD,BeyondMHD3+1}.

Also, it is interesting in itself (and necessary for perturbing the full set of fluid equations) to understand the relation between the spatial velocity $w_\x^a$ as measured by the equilibrium observer and the Lagrangian displacement $\xi^a_\x$. We consider the displacement to live in the local present of the equilibrium observer, i.e.,  to be such that $\xi_\x^a u^\e_a = \zeta_\x^au^\e_a =  0$.\footnote{This is essentially a gauge choice, see \cite{livrev} for discussion.} This implies that the vectors $\xi^a_\x$ and $\zeta^a_\x$ are spacelike non-null vector fields in spacetime. As a result, if we consider the proper time of the perturbed worldline, we have
\begin{equation}
	- d\bar \tau^2 = g_{ab}\,dx^a_\f\,dx^b_\f = g_{ab}\, dx_\e^a\,dx_\e^b + g_{ab} \Big( dx_\e^a\, \zeta^b \Delta \lambda + dx_\e^b\, \zeta^a \Delta \lambda\Big) = - d\tau^2 \;,
\end{equation}
where we used the fact that 
\begin{equation}
	x_\e^a = x_\e^a(\tau) \Longrightarrow dx_\e^a = u_\e^a d \tau \;.
\end{equation}
As a consequence, the proper time of the perturbed and equilibrium worldline is the same, so we have
\begin{equation}
	u_\x^a = \frac{dx_\f^a}{d \bar \tau} \approx \frac{dx_\e^a}{d\tau} + \frac{d}{d\tau} \xi^a_\x = u_\e^a + \dot\xi_\x^a \;,
\end{equation}
where (again) the dot represents the covariant directional derivative in the direction of the equilibrium four-velocity.\footnote{To be more precise, one should distinguish between $\frac{d}{d\tau}=u^b_\e \partial_b$ and $\frac{D}{D\tau}=u^b_\e \nabla_b$. Since we are introducing a decomposition of a vector as a sum of two, $\dot\xi^a_\x$ must be a vector as well so that the dot represents a covariant directional derivative. } We observe that from the construction we have $w_\x^a = \dot\xi^a_\x$ and it is clear that when pushing the expansion to second order their relation will become more involved---both because the difference between the proper times ($\bar \tau$ versus $\tau$) appears at second order and because the Taylor expansion gets more complicated. 

We now aim to understand how to construct the expansion directly in matter space.  We start by noting that, since we are considering each displacement $\xi_\x^a$ to be orthogonal to $u^a_\e$ there is no loss of information in projecting the Lagrangian displacements onto the equilibrium matter space and dealing with $\xi^A_\x$. The general picture is thus as follows: in the general non-linear theory each matter space can be considered as an independent but interacting manifold, but this changes when we consider a perturbative expansion.  In fact, the fundamental assumption of perturbation theory is that the two configurations (perturbed and unperturbed) are related by some diffeomorphism. This implies that the perturbed and unperturbed matter spaces\footnote{Recall that the matter space is obtained by taking the quotient of the spacetime over the corresponding worldline, i.e. identifying the worldline as a single point.} are diffeomorphic, that is they are \textit{the same} abstract manifold. Therefore we can use the same chart on the two manifolds $X^A$ (label the worldlines in the same way) and the difference will be only in that $X_\x^A(x^a) \ne X_\e^A(x^a)$. The difference between the two will be exactly what we found above, namely $-\xi_\x^A$. We also note that, by our definition of the unperturbed state, all the perturbed matter spaces are diffeomorphic to the same unperturbed one, and thus to each other. 

Given this, we can work out how a general matter space tensor transforms under diffeomorphisms \citep{GRCarroll}. For instance, if we consider the projected metric $g_\x^{AB}$ we have\footnote{For the Lie derivative we use the formula with partial derivatives in order to avoid the possible confusion arising from the choice of the connection used on the matter space.}
\begin{equation}
	\delta g_\x^{AB} = -\mathcal{L}_{-\xi_\x}g_\x^{AB} = \mathcal{L}_{\xi_\x} g_\x^{AB} = \xi_\x^C \partial_C g_\e^{AB} - g_\e^{CB}\partial_C\xi^A_\x - g_\e^{AC}\partial_C\xi_\x^B \;,
\end{equation}
where the partial derivatives are taken with respect to the equilibrium matter space coordinates. We now observe that, considering $\xi^A_\x$ as a scalar field in spacetime we can write
\begin{equation}
	-g_\e^{CB} \partial_C\xi^A = - g^{ab} \Psi^C_{\e\,a} \Psi^B_{\e\,b}\partial_C\xi_\x^A = - \Psi^B_{\e\,b} \nabla^b \xi^A_{\x} \;.
\end{equation}
We also note that, since\footnote{If this is not immediately convincing one can prove it by taking the explicit definition of a derivative on the coordinate functions $X^A(\bar X) = \delta^{A}_{\,C}\bar X^C = \bar X^A$ and using the linearity of the derivative.} $\partial_C \Psi^A_{\e\,a} = \partial_a \delta^A_C = 0,$ we have
\begin{equation}
	\partial_C g_\e^{AB} = 2\, g^{ab}\Bigg(\frac{\partial}{\partial X_\e^C} \Psi^A_{\e\,a}\Bigg) \Psi^B_{\e\,b} = 0 \;.
\end{equation}
As a result, the projected metrics transform as
\begin{equation}\label{eq:VariationGx}
	\delta g_\x^{AB} = - \Psi^B_{\e\,a} \nabla^a\xi^A_{\x} - \Psi^A_{\e\,a}  \nabla^a \xi^B_{\x} \;.
\end{equation}
This also tells us that building the variation of the metric tensor in this way, we are only comparing  the difference in the position of the particles, keeping fixed the spacetime metric.

We can now use the definition in \cref{eq:ProjectedDerivative} to decompose the displacement gradients as
\begin{equation}
	\nabla_a\xi_{\x}^A = -w_\x^A u^\e_a + D_a\xi_\x^A \;,
\end{equation}
and rewrite 
\begin{equation}\label{eq:DeltaMetricGradientsDisplacements}
\begin{split}
	\delta g^{AB}_\x &= \Psi^B_{\e\,a} (w_\x^A u^a_\e - D^a\xi_\x^A ) + \Psi^A_{\e\,a} (w_\x^B u^a_\e - D^a \xi_\x^B ) \\
    &= - D^B\xi_\x^A - D^A \xi_\x^B \;,
\end{split}
\end{equation}
where we introduced the short-hand notation $D^A = \Psi^A_{\e\,b}\,g^{ab}D_a$. It is  worth noting that \cref{eq:DeltaMetricGradientsDisplacements} is not a strain-rate tensor of the type usually introduced in fluid dynamics, because it involves gradients in the displacements instead of velocities. The usual strain rate tensor is in fact\footnote{To see this one has to use  $\mathcal{L}_{u_\x}\Psi^A_{\x\,a}=0$.}
\begin{equation}\label{eq:DotGAB}
\begin{split}
	\dot g_\x^{AB} &= -2\, \Psi^A_{\x\, (a}\Psi^B_{\x \,b)} \big[  -u_\x^b\dot u^a_\x + \varpi^{ab}_\x+ \sigma^{ab}_\x + \frac{1}{3}\theta_\x \perp^{ab}_\x\big] = \\ 
    & = - 2 \, \Psi^A_{\e \,(a}\Psi^B_{\e\, b)} \big(\sigma_\x^{ab} + \frac{1}{3}\theta_\x \perp^{ab}_\e\big)  + \mathcal{O}(2) = -2 \big(\sigma_\x^{AB} + \frac{1}{3}\theta_\x g^{AB}_\e\big)  \; ,
\end{split}
\end{equation}
We will comment on the implications of this difference later. 

Even if  it is not entirely obvious what kind of object the mixed projected metric $g^{AB}_{\x\y}$ is in the general non-linear case, in the context of a perturbative expansion there is no real difference between the various matter spaces (they are all diffeomorphic to the equilibrium one).  This means that we can use the same fundamental formula also for $g_{\x\y}$ to get
\begin{equation}\label{eq:VariationGxy}
\begin{split}
	\delta g_{\x\y}^{AB} &= g_{\x\y}^{AB} - g_\e^{AB} = g^{ab} \Big( \delta \Psi^A_{\x\,a} \Psi^B_{\e\,b} + \delta \Psi^B_{\y\,b} \Psi^A_{\e\,a}\Big) = \\
    & = - \Psi^B_{\e\,a} \nabla^a \xi^A_{\x} - \Psi^A_{\e\,a}\nabla^a \xi^B_{\y}  \;.
\end{split}
\end{equation}
It is interesting to note that since $\delta g^{ab} = 0$ we
have
\begin{equation}
	[\delta, \nabla_a] = [\delta , \partial_a ] = 0 \;.
\end{equation}
That is, the variation commutes with both partial and covariant derivatives. This will become relevant when we need to work out the variation of the resistive terms that stem from a dependence of the $\N_\x$ on $g_{\x\y}^{AB}$ and $g_\y^{AB}$. 

As there has been a number of recent efforts aimed at building first-order dissipative hydrodynamic models starting from a field-theory perspective (cf. \cref{sec:HydrodynamicFieldTheory}), it makes sense to point out the differences between the present expansion and the field-theory-based ones. 
In that context, the models are said to be of first order if the constitutive equations involve all permissible terms with just one derivative---as when the system is close to equilibrium one can expect the gradients in temperature, chemical potential etc$\dots$ to be small, so that terms with two or more derivatives are dominated by first-order ones. 
In contrast, we here assume the variables that define the physical state of the system to take values close to the equilibrium ones, and by ``first order'' we mean the deviations are expanded up to $\mathcal{O}(\xi_\x)$. It is therefore clear that the present approach differs from the field-theory-based (gradient) expansions. The ultimate reason is that the action-based model provides the exact equations, which we then approximate, while in the field-theory approach one is trying to build the full equations as successive expansions. 

\section{Energy density is stationary at equilibrium}\label{sec:EnergyMinimized}
As discussed in \cref{ch1subsec:EITCattaneo}, in order to describe out-of-equilibrium systems with the Extended Irreversible Thermodynamic (EIT) paradigm, one postulates the existence of a generalized entropy---a function of a larger set of Degrees of Freedom than the corresponding equilibrium ones---which is maximized at equilibrium. The starting point for the formalism used here is a generalized energy where the only degrees of freedom are the fluxes. The action-based model provides the total stress-energy-momentum tensor $T_{ab}$ of the system, so that we can easily extract the total energy density $\veps$ for some observer having four-velocity $u^a$ via the projection $\veps = u^a u^b T_{a b}$. We will now show that requiring the local energy density to be at a minimum in equilibrium means the viscous stress tensors have to be zero. 

When specific modeling is carried out, such as a numerical evolution, we would need to provide an equation of state and specify values for the microphysical input parameters. From the phenomenological point of view, this corresponds to assuming the existence of a function---in our case, energy density---defined on some ``thermodynamical manifold'' whose coordinates are the relevant degrees of freedom. Practically speaking, the formalism developed here suggests we may identify the thermodynamical manifold with the matter space used in the variational model. As the general discussion gets quite complex, we focus on the specific example of a two-component system, with the components representing matter and entropy (see \cite{Lopez2011,NilsHeat2011}). 

Let us first consider the non-dissipative limit.  Thermodynamics of a single fluid is described by some equilibrium energy $\varepsilon_\e(n,s)$ such that 
\begin{equation}
	d\varepsilon_\e = T ds + \mu dn = \sum_{\x=\n,\s}\mu^\x dn_\x \;.
\end{equation}
On the other hand, the conservative variational model is built using a master function $\Lambda(n_\n^2, n_\s^2, n_{\n\s}^2)$. Because of our assumption that all species are comoving in equilibrium there is no heat flux relative to the matter and therefore $n_{\n\s}^2 = - g_{ab} n_\n^a n_\s^b = + n_\n n_\s$, and the master function only depends on two variables, $\Lambda_\e = \Lambda_\e(n_\n,n_\s)$. It is indeed easy to see that the equilibrium energy density, as measured by the equilibrium observer, is
\begin{equation}
	\varepsilon_\e = T_{ab}^\e\,u_\e^a u_\e^b = \big[\Psi_\e g_{ab} + (\Psi_\e - \Lambda_\e)u^\e_a u^\e_b\big] u_\e^a u_\e^b = - \Lambda_\e\;.
\end{equation}
Since we have already identified the matter space normalizations of the three-forms with the rest frame densities $\N_\x = n_\x$, we can think of the thermodynamic energy as a function defined on the matter space, and write
\begin{equation}
	\varepsilon_\e = \varepsilon_\e(\N_\n,\N_\s) = - \Lambda_\e(\N_\n,\N_\s)\;.
\end{equation}
The equilibrium case suggests that we could try to extend this identification to the non-equilibrium setting, and ``build'' the thermodynamics on the matter space. This raises the (difficult) question of what the global matter space is in the full non-linear case. We  will not address that issue here. Instead, we focus on the near-equilibrium case, where we only have to deal with the equilibrium matter space. 

Because of the way we have built the expansion, it is natural to project tensor quantities ---fluxes, stress-energy-momentum tensor, etcetera---into the frame of the equilibrium observer, as defined by the equilibrium worldlines congruence $u^a_\e$. Quantities  measured in this frame will be indicated by a ``hat'' in the following. Objects without a hat are measured in fluid rest frames, which are defined by the $u^a_\x$. The equilibrium value of a quantity in the equilibrium frame will be indicated with a ``bar''. For instance, the particle density measured in the equilibrium frame is $\hat n_\x = - u^\e_a n^a_\x$; in the $\x$-fluid rest frame it is $n_\x = - u^\x_a n^a_\x$; and the equilibrium value in the equilibrium frame is $\bar n_\x = \hat n_\x\big|_\e$.

The ``out-of-equilibrium'' energy density $\hat\varepsilon_{\o.\e.}$ of the system as determined in the equilibrium rest frame is given by
\begin{equation}
	\hat\varepsilon_{\o.\e.} = \big( T^{ab}_{\n.\d.} + \sum_\x D^{ab}_\x \big) u^\e_a u^\e_b = \hat\varepsilon_{\o.\e.}^{\n.\d.} + D^{ab} u^\e_a u^\e_b \;,
\end{equation}
where we have separated the contribution from the viscous stress tensor $D_{ab}$ from those having the ``non-dissipative'' form
\begin{equation}
	T^{ab}_{\n.\d.} = \Big(\Lambda - \sum_\x n_\x^c \mu^\x_c\Big) g^{ab} + \sum_\x n_\x^a\mu_\x^b = \Psi\,g^{ab}+ \sum_\x n_\x^a\mu_\x^b \; .
\end{equation}
The expression for $\hat\varepsilon_{\o.\e.}$ can be made more explicit by means of \cref{muxcomov}, which leads to $\Psi = \Lambda + \sum_\x n_\x \mu_\x$ and
\begin{equation}
	\hat\varepsilon_{\o.\e.}^{\n.\d.} = u^\e_a u^\e_b T^{ab}_{\n.\d.} = - \Lambda - \sum_{\x= n,s} \big(n_\x \mu_\x - \hat n_\x \hat\mu_\x\big) \;.
\end{equation}

Because the flux is a vector, the two densities $\hat n_\x$ and $n_\x$ are easily shown to be related by
\begin{equation}\label{eq:restvseqdensity}
	\hat n_\x = - n_\x^a u_a =  -n_\x u_\x^a u_a = (1 - w_\x^aw^\x_a)^{-1/2} n_\x = \Big( 1 + \frac{1}{2}w_\x^2\Big)n_\x + \mathcal O(w_\x^3) \;.
\end{equation}
Meanwhile, the corresponding momentum relation is a bit more involved because of entrainment: 
\begin{equation}
\begin{split}
		\mu_\x &= -\mu^\x_b u_\x^b = -\gamma_\x (u^b + w_\x^b) \big(\mathcal B_\x n_\x u^\x_b + \sum_{\y \neq \x} \mathcal A_{\x\y} n_\y u_\y^b\big) \\
        &= \gamma_\x \Big( \hat\mu_\x - \mathcal B_\x n_\x \gamma_\x w_\x^2 -  \sum_{\y \neq \x} \mathcal A_{\x\y} n_\y \gamma_\y w_\x^aw^\y_a\Big) \;.
\end{split}
\end{equation}
We can rearrange this as 
\begin{equation}\label{eq:restvseqmomentum}
	\hat\mu_\x = \mu_\x - \frac{1}{2}\bar\mu_\x w_\x^2 + \bar {\mathcal B}_\x \bar n_\x w_\x^2 + \sum_{\y\neq \x} \mathcal{\bar A }_{\x\y}\bar n_\y w_\x^aw^\y_a \;,
\end{equation}
and, wrapping up, we get 
\begin{equation}\label{eq:nondissenergylambda}
\begin{split}
	\hat\varepsilon_{\o.\e.}^{n.d.} &= -\Lambda  + \bar {\mathcal B}_\n \bar n_\n^2 w_\n^2 +\bar {\mathcal B}_\s \bar n_\s^2 w_\s^2 + 2\bar \A_{\n\s}\bar n_\s\bar n_\n w_\n^aw^\s_a \\
    &=  -\Lambda +\bar\mu_\n\bar n_\n w_\n^2 +\bar\mu_\s\bar n_\s w_\s^2 - \A_{\n\s}\bar n_\n \bar n_\s w_{\n\s}^2\;,
\end{split}
\end{equation}
where 
\begin{equation}
 w_{\x\y}^2 = g_{a b} \big(w_\x^a - w_\y^a\big) \big(w_\x^b - w_\y^b\big)   \;. 
\end{equation}
It is now clear that, in order to proceed, we need an expansion for the master function, $\Lambda$. 

Note that the dissipative action model assumes $\Lambda$ depends on $(X_\n^A, X_\s^A, g_\n^{AB} , g_\s ^{AB}, g_{\n\s}^{AB})$ through the scalar product of the fluxes $n_\n^2, n_\s^2, n_{\n\s}^2$. Therefore, we can expand $\Lambda$ up to second order in the standard way (see \cite {livrev}). We thus have 
\begin{equation}\label{eq:Master2Ord}
\begin{split}
	\Lambda  = \Lambda_{\e} &-\frac{1}{2} \sum_{\x=n,s}  \B_\x\delta n_\x^2 - \A_{\n\s} \delta n_{\n\s}^2 - \frac{1}{4} \sum_{\x=n,s} \frac{\partial \B_\x}{\partial n_\x^2} (\delta n_\x^2)^2 -\frac{1}{2}\frac{\partial\A_{\n\s}}{\partial n_{\n\s}^2}  (\delta n_{\n\s}^2)^2 \\ 
    &-\frac{1}{2}\frac{\partial\B_\n}{\partial n_{\s}^2} (\delta n_\n^2)(\delta n_\s^2 )- \frac{\partial\A_{\n\s}}{\partial n_\n^2}(\delta n_\n^2)(\delta n_{\n\s}^2) - \frac{\partial\A_{\n\s}}{\partial n_\s^2} (\delta n_\s^2)(\delta n_{\n\s})^2 \;.
\end{split}
\end{equation}
To make contact with the previous expansion on the matter space we need  explicit expressions for $\delta n_\x^2$ and all  other similar terms that appear in this expression. 

For the four-current we have
\begin{equation}\label{eq:deltanexa}
\begin{split}
	\delta n_\x^a &= n_\x^a - \bar n_\x^a = (\bar n_\x + \delta n_\x )\Big[\big(1+\frac{1}{2}w_\x^2\big)u^a + w_\x^a\Big] - \bar n_\x u^a \\
    &= \frac{1}{2} \bar n_\x w_\x^2u^a + \bar n_\x w_\x^a +\delta n_\x u^a + \delta n_\x w_\x^a \;,
\end{split}
\end{equation}
and we see that it---quite intuitively---changes both as the density and the four-velocity change. By means of \cref{eq:deltanexa} we get
\begin{equation}\label{eq:deltanex2}
	\delta n_\x^2 = - \big(2\bar n_\x^a \delta n^\x_a+ \delta n_\x ^a \delta n^\x_a\big)  = 2 \bar n_\x \delta n_\x + (\delta n_\x)^2 \;.
\end{equation}
Similarly, we have
\begin{equation}\label{eq:deltanexway2}
\begin{split}
	\delta n_{\x\y}^2 &= - \Big( \bar n_\x^a \delta n_a^\y + \bar n_\y^a \delta n^\x_a + \delta n_\x^a \delta n^\y_a \Big)  \\
    &= \bar n_\x \delta n_\y+ \bar n_\y \delta n_\x + \delta n_\x \delta n_\y + \frac{1}{2} \bar n_\x \bar n_\y w_{\x\y}^2 \;.
\end{split}
\end{equation}

In order to complete the second order expansion of $\Lambda$ we also need the products (for every possible combination) of \cref{eq:deltanex2} and \cref{eq:deltanexway2}. These are found to be
\begin{subequations}
\begin{align}
	\big(\delta n_\x^2\big)^2  &= 4 \bar n_\x^2 (\delta n_\x)^2 \;,\\
      (\delta n_{\x\y}^2)^2 &= \bar n_\x^2 (\delta n_\y)^2 + \bar n_\y^2 (\delta n_\x)^2+ 2\bar n_\x \bar n_\y \delta n_\x \delta n_\y \;,\\
     (\delta n_\x^2) (\delta n_\y^2) &= 4 \bar n_\x \bar n_\y \delta n_\x \delta n_\y \;,\\
     (\delta n_{\x\y}^2)( \delta n_\x^2) &= 2\bar n_\x (\delta n_\x) \big(\bar n_\y \delta n_\x + \bar n_\x \delta n_\y\big)\;.
\end{align}
\end{subequations}
Plugging these expressions into \cref{eq:Master2Ord} we find (up to second order) 
\begin{equation}\label{eq:2OrdOffEqEnergy}
\begin{split}
	\hat\varepsilon^{n.d.}_{\o.\e.} &= \varepsilon_\e(\bar n_\n, \bar n_\s) + \bar \mu_\n \delta n_\n + \bar \mu_\s \delta n_\s  + \frac{1}{2} \big(\bar\B_\n \bar c_\n^2 -\bar\A^{\n\n}_{uu} \big)( \delta n_\n)^2 \\
    &+ \frac{1}{2} \big(\bar\B_\s \bar c_\s^2 -\bar\A^{\s\s}_{uu} \big)( \delta n_\s)^2 - \big(\bar\chi^{\s\n}_{uu} + \bar\A^{\n\s}_{uu} \big) (\delta n_\n)(\delta n_\s) + \bar\mu_\n\bar n_\n w_\n^2 \\
    &+ \bar\mu_\s\bar n_\s w_\s^2   - \frac{1}{2}\bar\A_{\n\s}\bar n_\n \bar n_\s w_{\n\s}^2  \;,
\end{split}
\end{equation}
where we have made use of \cref{eq:nondissenergylambda} and defined (see \citet{livrev})
\begin{subequations}\label{eq:SecondOrderMasterUsefulDef}
\begin{align}
	\bar c_\x^2 &= 1 + 2\frac{\bar n_\x^2}{\bar\B_\x}\frac{\partial\bar\B_\x}{\partial n_\x^2} \;,\\
  \bar  \A^{\x\x} _{ab} &= -\Big(\bar n_\y^2 \frac{\partial\bar\A_{\x\y}}{\partial n_{\x\y}^2} + 4 \bar n_\x \bar n_\y \frac{\partial\bar\A_{\x\y}}{\partial n_\x^2}\Big) u^\e_a u^\e_b \doteq \bar  \A^{\x\x} _{uu} u^\e_a u^\e_b \;,\\
  \begin{split}
  \bar  \A^{\n\s}_{ab} &= \bar\A^{ns}\perp_{ab} -\Big(\bar\A_{\n\s} + 2\bar n_\n^2 \frac{\partial\bar\A^{\n\s}}{\partial n_\n^2} + 2\bar n_\s^2 \frac{\partial\bar\A^{\n\s}}{\partial n_\s^2} + \bar n_\n\bar n_\s \frac{\partial\bar\A^{\n\s}}{\partial n_{\n\s}^2}\Big) u^\e_a u^\e_b \\
  &\doteq\bar\A^{\n\s}\perp_{ab} + \bar  \A^{\n\s}_{uu} u^\e_a u^\e_b
  \end{split} \;,\\
    \bar\chi^{\n\s}_{uu} &= -2\bar n_\n\bar n_\s \frac{\partial\bar\B^\n}{\partial n_\s^2} = -2\bar n_\n\bar n_\s \frac{\partial\bar\B^\s}{\partial n_\n^2}\;.
\end{align}
\end{subequations}
Noting that the quantity $\delta n_\x$ is the variation of the rest frame density, we can relate it to a variation of $\N_\x$ and ``close the loop''. Since the $\N_\x$ are functions on matter space of the variables $(X_\n,\;X_\s,\;g^{AB}_\n,\;g^{AB}_\s,\;g^{AB}_{\n\s})$ the expression for the energy is actually a second order expansion in terms of those variables. We note also that, because of the ``two-layer structure'', the $\delta n_\x$ above contain second-order terms.

A priori, the expression in \cref{eq:2OrdOffEqEnergy} does not provide the total out-of-equilibrium energy because we also need to account for the dissipative terms. However, we will now show that these actually do not contribute. To do this, we assume an expansion for all the viscous stress tensors of form 
\begin{equation}
	S_{AB} = S^{\e}_{AB} + S_{AB}^1 + S^2_{AB} + \mathcal{O}(\xi^3) \;,
\end{equation}
without providing (for now) the explicit expressions. Recalling $\Psi^A_{\e\,a} u^a_\e = 0$, we can write
\begin{equation}
	S_{ab} u^a_\e u^b_\e = S_{AB} (X_{\e}^A + \delta X^A)_{,a}(X_{\e}^B + \delta X^B)_{,b} u^a_\e u^b_\e = S_{AB}^{\e}\,\delta X^A_{\;,a}\,\delta X^B_{\;,b} u^a_\e u^b_\e \;,
\end{equation}
where the expansion is up to second order. It is clear that this argument is valid for each viscous stress tensor, and for $D_{ab} u^a_\e u^b_\e$ as well, so that the dissipative contributions to the off-equilibrium energy are, at least, of second order. Assuming that the energy is stationary, that is
\begin{equation}
	\hat\varepsilon^{n.d.}_{\o.\e.} - \varepsilon_\e(\bar n_\n, \bar n_\s) = \mathcal{O}(\xi^2) \;,
\end{equation}
we then have
\begin{equation}\label{eq:MinimumeEnergyCondition}
	\bar\mu_\n \delta n_\n + \bar\mu_\s\delta n_\s = \mathcal{O}(\xi^2) \;,
\end{equation}
which has a clear thermodynamical interpretation and is consistent with the EIT picture, since, up to first order, the generalized energy is a function of the $n_\x$ only. 

We want to translate the above result into conditions for the matter space functions $\N_\x$. We start by observing that in the conservative case, the three-form $n_{ABC}^\x$ is a function of the $X_\x^A$ coordinates only. Therefore, $\bar\N_\x$ is just a function of $X_\x^A$, while, because  $\N_\x = \bar\N_\x\sqrt{g^\x}$, the latter depends also on the projected metric 
\begin{equation}
	\frac{\partial \N_\x}{\partial g_\x^{AB}} = \frac{1}{2} \sqrt{g^\x}\bar\N_\x g^\x_{AB} = \frac{1}{2}\N_\x g^\x_{AB} \;.
\end{equation}
When considering the expansion of $n_\x$ (and hence $\N_\x$) we assume that we can write 
\begin{equation}
	\N_\x = \N_\x^\e + \N_\x^\d \;,
\end{equation}
where $\N_\x^\e$ is the same as in the non-dissipative limit while the dissipative contribution $\N_\x^\d$ is a function also of the additional variables that encode the dissipation. Given the separation of $\N_\x$ into two pieces it is natural to assume that $\N_\x^\d$, but not its derivatives, vanishes at equilibrium. 

Since the equilibrium evolves in a conservative fashion, we can write
\begin{equation}
\begin{split}
		\delta n_\x &\equiv \N_\x - \N_\x^\e = \N_\x^\d = \N_\x^\d -\N_\x^\d\Big|_{\e} \\
    &= \frac{\partial \N_\x^\d}{\partial X_\x^A} \delta X_\x^A +  \frac{\partial \N_\x^\d}{\partial X_\y^A} \delta X_\y^A +  \frac{\partial\N_\x^\d}{\partial g_\x^{AB}} \delta g_\x^{AB} + \frac{\partial\N_\x^\d}{\partial g_\y^{AB}} \delta g_\y^{AB} + \frac{\partial \N_\x^\d}{\partial g_{\x\y}^{AB}} \delta g_{\x\y}^{AB} +\mathcal{O}(2) \;,
\end{split}
\end{equation}
where here, and in similar expansions below, each quantity is to be evaluated at equilibrium. With this assumption it is easy to read off from \cref{eq:MinimumeEnergyCondition} the first order relation
\begin{equation}\label{eq:1stOrderEquilCondition}
    	\M_\n \delta\N_\n^\d+ \M_\s \delta\N_\s^\d =0 \ .
\end{equation}
This leads to
\begin{equation}
  	\M_\n \frac{\partial\N_\n^\d}{\partial X_\n^A} + \M_\s \frac{\partial\N_\s^\d}{\partial X_\n^A} =0 \ , 
\end{equation}
and analogous results for variations with respect to $X_\s^A$, $g_\n^{AB}$, $g_\s^{AB}$ and $g_{\n\s}^{AB}$ follow immediately.
In particular, this shows that the total viscous stress tensor, acting on each component $D^\x_{ab}$, vanishes when the energy is stationary. 

To see this explicitly we note that (see \cref{eq:StdViscousTensorSlim,eq:AdditionalViscousTensorsSlim})
\begin{equation}\label{eq:mixedviscousdie}
	\mathcal{S}^{\x\y,\,\e} _{AB} \equiv 2\M_\x\frac{\partial\N_\x}{\partial g_{\x\y}^{AB}}= - 2\M_\y\frac{\partial\N_\y}{\partial g_{\x\y}^{AB}}= -\mathcal{S}^{\y\x,\,\e}_{BA} \;,
\end{equation}
where we made use of the symmetry property of the mixed metric, namely $g_{\x\y}^{AB} = g_{\y\x}^{BA}$. 
Similarly,
\begin{equation}\label{eq:viscousdie}
\begin{split}
	S^{\x,\,\e}_{AB} & \equiv 2 \M_\x\Big(\frac{\partial \N_\x}{\partial g_\x^{AB}} -\frac{1}{2} \N_\x g^\x_{AB}\Big)  \\
    &=  2 \M_\x\Big[\frac{\partial \N_\x^\d}{\partial g_\x^{AB}} -\frac{1}{2} \big( \N_\x -\N_\x^\e \big)g^\x_{AB}\Big]  \\
    &=2 \M_\x \frac{\partial \N_\x^\d}{\partial g_\x^{AB}}= -  2 \M_\y\frac{\partial \N_\y^\d}{\partial g_\x^{AB}}= - s^{\y\x,\,\e}_{AB} \;.
\end{split}
\end{equation}
It is now clear that, by means of \cref{eq:mixedviscousdie} and \cref{eq:viscousdie}, the x-species viscous stress tensor vanishes:
\begin{equation}
	D^{\x,\,\e}_{AB} = S^{\x,\,\e}_{AB} + s^{\y\x,\,\e}_{AB} +\frac{1}{2}(\mathcal{S}^{\x\y,\,\e}_{AB} + \mathcal{S}^{\y\x,\,\e}_{BA}) =0\;.
\end{equation}

We have considered the fully general case with all the additional dependences in $\N_\x$ and all the viscous tensors $S^\x_{ab}$, $\mathcal{S}^{\x\y}_{ab}$ and $s^{\x\y}_{ab}$. The same result---that is, each $D^{\x,\,\e}_{ab}$ vanishes ---holds even in a less rich situation when the model is built from fewer viscous tensors. In that case we have to go back to \cref{eq:1stOrderEquilCondition} and modify the argument accordingly. It is important to stress that we have shown that the full stress-energy-momentum tensor at equilibrium is made out of just the non-dissipative part, and that the dissipative parts of the total stress-energy-momentum tensor do not contribute to the total energy density at second order. 

However, we note that the energy minimum conditions in \cref{eq:1stOrderEquilCondition} do not set the purely resistive terms to zero (\cref{eq:PurelyReactiveSlim}). In fact, it only leads to 
\begin{subequations}
\begin{align}
	&\M_\n\frac{\partial \N^\d_\n}{\partial X_\n^A} = -\text R^{\s\n,\,\e}_A \;,\\
    &\M_\s\frac{\partial \N^\d_\s}{\partial X_\s^A} = -\text R^{\n\s,\,\e}_A \;.
\end{align}
\end{subequations}
The reason for this is pretty clear as these terms do not enter the expression for the energy density. We nonetheless might want to consider the case where the equilibrium equations are exactly as the conservative ones. The motivation for this can be found in the derivation of the purely resistive terms. If the different species are comoving at the action level, there is no distinction between the different $X_\x^A$ and no resistive term of this form would appear. We can enforce consistency with this observation in two ways: either we assume that we use the complete dependence on $X_\x^A$ in the conservative part, in which case
\begin{equation}
	\frac{\partial  \N^\d_\x}{\partial X_\x^A}\Big|_{\e} =0  \Longrightarrow \text R^{\x\y,\,\e}_A = 0 \;,
\end{equation}
or, we just set the terms ${\text R}^\x_a$ to zero, so that
\begin{equation}
	\M_\n\frac{\partial \N^\d_\n}{\partial X_\n^A}\Big|_{\e} = \M_\s\frac{\partial \N^\d_\s}{\partial X_\s^A}\Big|_{\e} \;.
\end{equation}
The latter, less restrictive assumption reminds us of  the dynamical nature of chemical equilibrium in nature. Reactions happen also at equilibrium, although they do so in such a way that there is no net particle production.

Finally, it is  quite easy to see that if we choose a different observer, such as the ones associated with the Eckart or Landau frame, the differences in the energy density will be of second order. Crucially, the equilibrium conditions in \cref{eq:1stOrderEquilCondition} do not depend on the choice of frame.

\section{The last piece of the puzzle}\label{sec:LastPiece}

In order to work out the perturbative expressions we need to expand the various dissipative terms. It should now be clear that for the viscous stress tensors we can write\footnote{All the derivatives are intended to be evaluated at equilibrium.}
\begin{subequations}\label{eq:ExpansionViscousTensorsMatter}
\begin{align}
	\delta s^{\x\y}_{AB} &= 2\frac{\partial\N_\x^\d}{\partial g_\y^{AB}}\delta \M_\x 	+ 2\M_\x \delta \Bigg(\frac{\partial\N_\x^\d}{\partial g_\y^{AB}}\Bigg) \;, \\
    \delta \mathcal{S}^{\x\y}_{AB} &= 2\frac{\partial\N_\x^\d}{\partial g_{\x\y}^{AB}}\delta \M_\x  + 2\M_\x\delta \Bigg(\frac{\partial\N_\x^\d}{\partial g_{\x\y}^{AB}}\Bigg) \;,\\
    \delta S^{\x}_{AB} &= 2\frac{\partial\N_\x^\d}{\partial g_\x^{AB}}\delta \M_\x 	+ 2\M_\x\delta \Bigg(\frac{\partial\N_\x^\d}{\partial g_\x^{AB}}\Bigg) -\M_\x (\delta \N_\x^\d) g_{AB}^\e \;,
    \end{align}
\end{subequations}
where we recall that
\begin{subequations}
\begin{align}
    & s^{\x\y}_{AB} = 2\M_\x \frac{\partial \N_\x^\d}{\partial g_\y^{AB}} \;,\\
    &\mathcal{S}^{\x\y}_{AB} = 2 \M_\x \frac{\partial\N^\d_\x}{\partial g_{\x\y}^{AB}} \;,\\
    &S^\x_{AB} = 2\Bigg( \M_\x \frac{\partial \N_\x^\d}{\partial g_\x^{AB}} - \frac{1}{2}\M^\x\N_\x^\d\,g^\x_{AB} \Bigg) \;.
\end{align}
\end{subequations}
Similarly, for the ``purely resistive'' terms we have 
\begin{equation}\label{eq:ExpansionPurelyReactiveMatter}
	\delta \text R^{\x\y}_A =  \frac{\partial \N_\x^\d}{\partial X_\y^{A}}\delta \M_\x	+ \M_\x\delta \Bigg(\frac{\partial \N_\x^\d}{\partial X_\y^{A}}\Bigg) \;.
\end{equation}
Since $\N_\x^\d$ is a function of $(X_\x,\,X_\y,\,g^{AB}_\x,\,g^{AB}_\y,\,g^{AB}_{\x\y})$, its derivatives are as well, so that we have
\begin{equation}\label{eq:PertDerivBarN}
\begin{split}
	\delta \Bigg(\frac{\partial\N_\x^\d}{\partial X_\y^A}\Bigg) = &\frac{\partial ^2 \N_\x^\d}{ \partial X_\x^B \partial X_\y^A} \delta X^B_\x +  \frac{\partial ^2 \N_\x^\d}{ \partial X_\y^B \partial X_\y^A} \delta X^B_\y +\frac{\partial ^2 \N_\x^\d}{ \partial g_\x^{BC} \partial X_\y^A}\delta g_\x^{BC} \\
    &+\frac{\partial ^2 \N_\x^\d}{ \partial g_\y^{BC} \partial X_\y^A} \delta g_\y^{BC} +\frac{\partial ^2 \N_\x^\d}{ \partial g_{\x\y}^{BC} \partial X_\y^A} \delta g_{\x\y}^{BC} \;.
\end{split}
\end{equation}
Similar results hold for the other variations that were not explicitly written in \cref{eq:ExpansionViscousTensorsMatter} and \cref{eq:ExpansionPurelyReactiveMatter}. 

Concerning the purely resistive term we note that $u_\x^a\text R^{\y\x}_a=0$ by construction. Because we are doing an  expansion with undetermined coefficients, we need to impose this by hand at every order; specifically, at the linear level. This then leads to 
\begin{equation}
	\delta \Big(u_\x^a \text R^{\y\x}_a\Big) = \text R^{\y\x,\,\e}_A \Big(w_\x^A - \dot\xi^A_\x \Big) = 0\;,
\end{equation}
so that not only do we have $w_\x^a = \dot\xi^a_\x$ but also $w_\x^A = \dot \xi^A_\x$. This then means that we must have\footnote{Here the semicolon is, as usual, a short-hand notation for covariant derivative $A_{;a}= \nabla_aA$.} $u_\e^a\xi_\x^b X^D_{\e\,;ba} = 0$, which in turn implies  that the orthogonality conditions for the viscous stress tensors
\begin{equation}
	S^\x_{ab}u_\x^a = \mathcal{S}^{\x\y}_{ab} u_\x^a = s^{\x\y}_{ab}u_\y^a = 0 \;,
\end{equation}
are automatically satisfied at linear order. 

From \cref{eq:ExpansionViscousTensorsMatter} we can find the expansion for the spacetime viscous tensors through 
\begin{subequations} \label{eq:PertExpViscStressesSpacetime}
\begin{align}
	&\delta S^\x_{ab} = \delta S^\x_{DE} \Psi^D_{\e\,a} \Psi^E_{\e\,b} - S^\x_{DE} \Big( \xi^D_{\x\,,a}\Psi^E_{\e\,b} +\Psi^D_{\e\,a} \xi^E_{\x\,,b}\Big) \;,\\ 
	& \delta s^{\x\y}_{ab} = \delta s^{\x\y}_{DE} \Psi^D_{\e\,a} \Psi^E_{\e\,b} - s^{\x\y}_{DE}\Big( \xi^D_{\y\,,a}\Psi^E_{\e\,b} + \Psi^D_{\e\,a}\xi^E_{\y\,,b}\Big) \;,\\
	&\delta \mathcal{S}^{\x\y}_{ab} = \delta \mathcal{S}^{\x\y}_{DE} \Psi^D_{\e\,a} \Psi^E_{\e\,b} - \mathcal{S}^{\x\y}_{DE}\Big( \xi^D_{\x,a}\Psi^E_{\e\,b} +\Psi^D_{\e\,a} \xi^E_{\y\,,b}\Big) \;,
\end{align}
\end{subequations}
while for the resistive terms associated with $s^{\x\y}_{ab}$ and $\mathcal{S}^{\x\y}_{ab}$ we have 
\begin{subequations}\label{eq:PertReactiveViscousTensor}
\begin{align}
	\delta r^{\x\y}_a &= \frac{1}{2} \delta s^{\x\y}_{DE} \nabla_ag_\e^{DE} - \frac{1}{2} s^{\x\y}_{DE}  \partial_a \big[ g^{bc} (\xi^D_{\y\,,b} \Psi^E_{\e\,c} + \Psi^D_{\e\,b} \xi^E_{\y\,,c})\big] \;,\\
     \delta \mathcal{R}^{\x\y}_a &= \frac{1}{4}\delta \mathcal{S}^{\x\y}_{DE} \nabla_ag^{DE}_\e - \frac{1}{2}\mathcal{S}^{\x\y}_{DE}  g^{bc}\Big(\xi^D_{\x\,,b} \nabla_a \Psi^E_{\e\,c} + \Psi^D_{\e\,b} \nabla_a\xi^E_{\y\,,c}\Big) \;,
\end{align}
\end{subequations}
where we made use of the fact that $[\delta,\nabla_a]=0$ because of $\delta g^{ab}=0$ (see the discussion at the end of \cref{sec:PerturbativeExpansion}). 

Having ``understood'' how we may perturb the terms $R^\x_a$ and $D^\x_{ab}$, let us focus on the remaining pieces of the equation of motion. A quick look back at \cref{eq:DissipativeVariationalEoM} reveals that the only terms we still have to discuss are $\delta\Gamma_\x $ and $\delta \mu^\x_a$. For the particle creation rate we have (see \cref{eq:deltanexa})
\begin{equation}
	\delta \Gamma_\x = \nabla_a \delta n_\x^a = \dot{\delta n_\x} + \nabla_a(\bar n_\x w_\x^a) \;,
\end{equation}
while for the x-species momentum, we get 
\begin{equation}
	\delta \mu^\x_a = \delta (\mathcal{B}_\x  n_\x) u^\e_b + \bar{\mathcal{B}}_\x \bar n_\x w^\x_b + \sum_{\y\neq \x} \delta(\mathcal{A}_{\x\y} n_\y) u^\e_b + \bar{\mathcal{A}}_{\x\y}\bar n_\y w^\y_b \;.
\end{equation}
Using the fact that we identified $\M_\x$ with $\mu_\x$ we have 
\begin{equation}\label{eq:deltaBarMMx}
\begin{split}
	\delta\M_\x &=   \delta \big(-\mu_\x^a u^\x_a\big) =   -\big(\bar\mu^\x_a  w_\x^a + \delta \mu_\x^a u^\e_a  \big)\\
	& = \delta \Big(\mathcal{B}_\x n_\x + \sum_{\y\neq \x} \mathcal{A}_{\x\y}n_\y\Big) \;,
\end{split}    
\end{equation}
and since  $\B_{\x}$ and $\A_{\x\y}$ are ultimately functions of $n_\x^2$ and $n_{\x\y}^2$, we may use 
\begin{subequations}
\begin{align}
& \delta \mathcal{B}_\x = \Bigg( 2n_\x\frac{\partial \mathcal{B}_\x}{\partial n_\x^2}  + n_\y\frac{\partial \mathcal{B}_\x}{\partial n_{\x\y}^2} \Bigg) \delta n_\x +  \Bigg( 2n_\y\frac{\partial \mathcal{B}_\x}{\partial n_\y^2}  + n_\x\frac{\partial \mathcal{B}_\x}{\partial n_{\x\y}^2} \Bigg) \delta n_\y \;,\\
& \delta \mathcal{A}_{\x\y} = \Bigg( 2n_\x\frac{\partial \mathcal{A}_{\x\y}}{\partial n_\x^2}  + n_\y\frac{\partial \mathcal{A}_{\x\y}}{\partial n_{\x\y}^2} \Bigg) \delta n_\x +  \Bigg( 2n_\y\frac{\partial \mathcal{A}_{\x\y}}{\partial n_\y^2}  + n_\x\frac{\partial \mathcal{A}_{\x\y}}{\partial n_{\x\y}^2} \Bigg) \delta n_\y \;,
\end{align}
\end{subequations}
in \cref{eq:deltaBarMMx}. This way, making use of definitions in \cref{eq:SecondOrderMasterUsefulDef}, we arrive at
\begin{equation}\label{eq:PertBarMMxFico}
	\delta \M_\x = \big(\bar\B_\x c_\x^2 - \bar\A^{\x\x}_{uu}\big)\delta n_\x -\big(\bar\A^{\n\s}_{uu} +\bar\chi ^{\n\s}_{uu}\big)\delta n_\y \;,
\end{equation}
and we see that the parameters that enter the dissipative fluid equations are the entrainment coefficients (and  first derivatives; that is, second order derivatives of $\Lambda(n_\x^2, n_{\x\y}^2)$) and the (up to second order) derivatives of the function $\N_\x(X_\x,\,X_\y,\,g^{AB}_\x,\,g^{AB}_\y,\,g^{AB}_{\x\y})$.  

Having outlined the perturbative framework, it is natural to ask how many dissipative channels the (general) model contains. Or, to be more specific, how many ``dissipation coefficients'' would have to be determined from microphysics? According to the expansion scheme we have developed so far, the perturbative expressions for the dissipative terms will ultimately involve all second and first order derivatives of the $\N_\x^\d$ when considered as functions of $X_\x,\,X_\y,\,g^{AB}_\x,\,g^{AB}_\y,\,g^{AB}_{\x\y}$. Also, to make use of the model we need to specify the entrainment coefficients and their derivatives in the combinations from \cref{eq:SecondOrderMasterUsefulDef}. This means that the most general model one can think of contains a large number of coefficients. However, they should be, in general, known once a specific model is chosen; that is, once the explicit functional forms of $\Lambda$ and the $\N_\x^\d$ have been provided. For example, if nuclear physics calculations are used to determine these explicit forms, they must be done in such a way that the constraints which arise from requiring a meaningful equilibrium configuration are taken into account, and they must ensure that the second law of thermodynamics is obeyed. If Onsager-type reasoning is invoked to ensure $\Gamma_\s$ is positive (up to second order), then explicit use of
\begin{equation}\label{eq:GeneralEntropyProductionRate}
	T \Gamma_\s = - D_{ba}^\s \nabla^bu_\s^a - u_\s^a \text R^\s_a  \; ,
\end{equation}
where $T = - u^a_\s \mu^\s_a$ is the temperature, would have to be made.

\section{Model comparison}\label{sec:StdQuantities} 

As an intuitive application of the formalism we have developed, it is useful to make contact with existing models for general relativistic dissipative fluids, in particular, the classic work of Landau-Lifschitz and Eckart and the second-order M\"uller-Israel-Stewart model. Specifically, we want to understand how  standard quantities (like shear and bulk viscosity) enter the present formalism. Therefore, we need to see if the dissipative terms of the existing models can be matched with terms in the action-based description. This procedure is fairly straightforward. 

The action-based model provides the total fluid stress-energy-momentum tensor, so  we only have to decompose it in the usual way (cf. \cref{eq:Stress-energyDissipativeDecomposition}):
\begin{equation}
	T^{ab} = (\bar p+\chi)\perp^{ab} + \varepsilon u^a u^b + 2q^{(a}u^{b)} + \chi^{ab} \;,
\end{equation}
In this expression, the fluxes are defined with respect to some observer with four-velocity $u^a$. In order to be consistent with the perturbative expansion outlined above, we take this observer to be associated with the thermodynamical equilibrium, i.e.~$u^a = u^a_\e$. Finally, we have split the isotropic pressure into an equilibrium contribution (denoted as $\bar p$ as discussed above) and a non-equilibrium one. 

\subsection{Equating the flux currents}\label{sec:FluxCurrents}

Let us first consider  the heat. We can read off the heat flux from the total stress-energy-momentum tensor as 
\begin{equation}
	q^a = -\varepsilon u^a_\e - T^{ab} u^\e_b = - \perp ^a_b T^{bc}u^\e_c \;.
\end{equation}
First, let us note that there is no contribution at linear order coming from the dissipative part of the stress-energy-momentum tensor $D^{ab}$. In fact, making use of \cref{eq:mixedviscousdie,eq:viscousdie,eq:PertExpViscStressesSpacetime}, it is easy to show that $D^{ab}u^\e_a = (\delta D^{ab})u^\e_a = \mathcal{O}(2)$. Let us therefore consider the non-dissipative part of $T^{ab}$. For the generalized pressure we have to first order  
\begin{equation}
	\Psi = \Lambda + \sum_{x} n_\x\mu_\x = -\bar\varepsilon_\e + \bar\mu \bar n + \bar T \bar s + \sum_{x=n,s} \bar n_\x \delta \mu_\x = \bar p +  \sum_{x=n,s} \bar n_\x \delta \mu_\x \;,
\end{equation}
where we have used the minimum energy condition (\cref{eq:MinimumeEnergyCondition}) and the equilibrium Euler relation. Using 
\begin{equation}
	\sum_\x n_\x^a\hat\mu_\x = \sum_\x n_\x^a\mu_\x + \mathcal{O}(2) \approx \sum_\x \Big[\bar n_\x\bar\mu_\x u^a_\e +\bar n_\x\delta\mu_\x u^a_\e +\bar\mu_\x\big(\delta n_\x u^a_\e + \bar n_\x w_\x^a\big)\Big] \;,
\end{equation}
we then identify the heat flux as
\begin{equation}\label{eq:modelheatflux}
	q^a = \sum_\x \bar\mu_\x \delta n_\x^a = \bar\mu \bar n \,w_\n^a +\bar T\bar s \,w_\s^a \ . 
\end{equation}
Here, we have repeatedly used the Euler relation and the minimum energy condition \cref{eq:MinimumeEnergyCondition}. We note that this quantity is consistent with the definition used in the classic models, see \citep{livrev}. 

Let us now move on to the other fluxes and, as before, first focus on the non-dissipative contribution. It is easy to check that
\begin{equation}
\begin{split}
	T^{ab}_{\n.\d.} = &\bigg(\bar p + \sum_\x \bar n_\x \delta \mu_\x\bigg)g^{ab} + (\bar p+\bar\varepsilon_{\e})u^a_\e u^b_\e \\
    &+\sum_\x \Big[\bar\mu_\x\bar n_\x u^b_\e w_\x^a + \bar n_\x u^a_\e \big(\delta\mu_\x u^b_\e + \B_\x\bar n_\x w_\x^b + \sum_{\y\neq \x} \A_{\x\y} \bar n_\y w_\y^b\big)\Big]\;,
\end{split}
\end{equation}
so that, using the standard decomposition above one arrives at  
\begin{equation} 
(\bar p+\chi)\perp^{ab} + \chi^{ab} = \perp^a_c\perp^b_dT^{cd}=T^{ab} +T^{ad} u^\e_d u^b_\e + T^{cb} u^\e_c u^a_\e + \varepsilon u^a_\e u^b_\e \;.
\end{equation}
If we now use  the non-dissipative contribution $T^{ab}_{\n.\d.}$ in this  equation, we get
\begin{equation}
	(\bar p+\chi)\perp^{ab} + \chi^{ab} = (\bar p + \sum_\x \bar n_\x \, \delta \mu_\x)\perp^{ab} = (\bar p + \delta \Psi) \perp^{ab} \;.
\end{equation}
That is, there may be a first-order correction in the pressure coming from $T^{ab}_{\n.\d.}$. Next, let us consider the contribution due to the non-dissipative part. From \cref{eq:PertExpViscStressesSpacetime} we see that %
\begin{equation}
	\perp^a_c\perp^b_d D^{cd}= D^{ab}=\delta D^{ab} \;.
\end{equation}
Putting everything together, we have identified
\begin{subequations}
\begin{align}
	\hat\chi &= \delta \Psi + \frac{1}{3} g_{ab}\delta D^{ab} \;,\\
	\hat\chi^{ab} &= \delta D^{\langle ab \rangle} \;, \\
	\hat q^a &= \bar\mu\bar n w_\n^a + \bar T\bar s w_\s^a \;,
\end{align}
\end{subequations}
where we reintroduced the ``hat'' to stress that these fluxes are measured by the equilibrium observer while the angle brackets mean that we are taking the trace-free symmetric part of the tensor. 

\subsection{Example: A viscous single fluid}\label{sec:ViscousFluid}
We now consider the specific example of a two-component, single viscous fluid. The two species are matter, with non-equilibrium flux $n^a = n u^a_\f$, and entropy, with non-equilibrium flux $s^a = s u^a_\f$. In this simple case, we assume that the non-equilibrium fluxes remain parallel\footnote{We note that a ``real'' two-fluid model would involve two independent fluid degrees of freedom $n^a_\n,\,n^a_\s$. By forcing them to move together we are imposing quite strong constraints on the model. Basically, we are assuming that the timescale over which the entropy current relaxes to the particle flow is short enough that it may be neglected.}, meaning $w^a_\n = w^a_\s = w^a$ and therefore
\begin{subequations}
\begin{align}
    n_\n^a &= n u^a_\f = n(u^a_\e + w^a) \;,\\
    n_\s^a &= s u^a_\f = s(u^a_\e + w^a) \;,
\end{align}
\end{subequations}
where again $u^a_\e$ is the equilibrium flow. In this case we do not have resistive terms because the two fluids are locked together from the beginning. Dissipation enters by assuming both currents depend on the (single) projected metric
\begin{subequations}
\begin{align}
	\N_\n = \N_\n(X^A,\, g^{AB}) \;, \\
    \N_\s = \N_\s(X^A,\, g^{AB}) \;. 
\end{align}
\end{subequations}
In practice, this means that we will have additional terms due to $S^\s_{ab}$ and $S^\n_{ab}$ in the equations of motion.

Also, the creation rate $\Gamma_\n$ has to vanish\footnote{The matter particle flux $n_\n^a$ is conserved as it is identified with the baryon current.}; this implies
\begin{equation}
	\Gamma_\n = -\frac{1}{\mu_\n} S^\n_{ab} \nabla^a u^b_\f = 0 \Longrightarrow S^\n_{ab} = 0 \;,
\end{equation}
as, by construction, the viscous stress tensor $S^\n_{ab}$ must be orthogonal to $u_\f^a$ (see \cref{eq:StdViscousTensorSlim} and \cite{2015CQAnderssonComer} for further details).
As a result, the final form of the non-linear equation of motion is 
\begin{equation}\label{eq:ViscousOneFLuidEq}
	2n_\n^a\nabla_{[a}\mu^\n_{b]} + 2n_\s^a\nabla_{[a}\mu^\s_{b]} + \Gamma_\s \mu_b^\s = - \nabla^a S^\s_{ab} \;.
\end{equation}
Note that, when we linearize, the term involving $\Gamma_\s$ will not appear in the equations, because $\Gamma_\s$ has no linear contributions---entropy is expanded around a maximum, leaving only second-order terms.

Our next step is to use the expansion formalism developed in the previous sections to determine the explicit form of the viscous stress tensor $S^\s_{ab}$.
Let us start by considering the equilibrium (minimum energy) conditions. Clearly, we should have
\begin{equation}
	S^{\s,\,\e}_{AB} = 2\M_\s \frac{\partial \N_\s^\d}{\partial g^{AB}} = 0 \Longrightarrow \frac{\partial \N_\s^\d}{\partial g^{AB}} = 0 \;.
\end{equation}
It also makes sense to assume $\partial \N_\s^\d/\partial X^A = 0$. To see why, let us forget for the moment that the two species are locked together and consider:
\begin{equation}
\begin{split}
	R^{\s}_A &= \M_\n \frac{\partial \N_\n^\d}{\partial X^A_\s} - \M_\s \frac{\partial \N_\s^\d}{\partial X^A_\n} = -\M_\s \frac{\partial \N_\s^\d}{\partial X^A_\s} - \M_\s \frac{\partial \N_\s^\d}{\partial X^A_\n}  \\
    &= - 2\M_\s \frac{\partial \N_\s^\d}{\partial X^A_\s} =  - 2\M_\s \frac{\partial \N_\s^\d}{\partial X^A} = 0 \;,
\end{split}
\end{equation}
where we initially distinguished between the two constituents' matter-space coordinates, and used the equilibrium condition. The condition $\partial \N_\x^\d/\partial X^A = 0$ is thus motivated by the fact that the resistive term vanishes (because the two currents are effectively locked).

As a result of these constraints we have $\delta \N_\x^\d = \mathcal{O}(2)$, $\delta \Psi = \mathcal{O}(2)$ and the viscous stress tensor becomes (see \cref{eq:ExpansionViscousTensorsMatter,eq:PertExpViscStressesSpacetime}) 
\begin{equation}\label{eq:ViscousFluidSAB}
\begin{split}
	\delta S^\s_{ab} &= \Bigg[ 2\M_\s \delta \bigg(\frac{\partial \N_\s^\d}{\partial g^{AB}}\bigg) \Bigg]\Psi^A_{\e\,a}\Psi^B_{\e\,b} \\
&= 2\bar T \Bigg[ \frac{\partial \N_\s^\d}{\partial X^C \partial g^{AB}}\delta X^C + 2 \frac{\partial \N_\s^\d}{\partial g^{DE} \partial g^{AB}}\delta g^{DE}  \Bigg]\Psi^A_{\e\,a}\Psi^B_{\e\,b} \;,
\end{split}
\end{equation}
Let us now rewrite the terms within the square brackets as 
\begin{subequations}
\begin{align}
	A_{CAB} &=  2\frac{\partial \N_\s^\d}{\partial X^C \partial g^{AB}} \;,\\
    \Sigma_{DEAB} &= 4 \frac{\partial \N_\s^\d}{\partial g^{DE} \partial g^{AB}} \;.
\end{align}
\end{subequations}
We further assume that $A_{CAB}$ is zero as this involves degrees of freedom we do not need to recover Navier-Stokes equations.

Before moving on with the model specification, let us introduce a slight generalization to the original model from \cite{2015CQAnderssonComer}. The idea is to take the normalizations $\N_\x^\d$ as functionals---instead of functions---of the additional variables. This does not constitute a major difference as the equations of motion, and particle production rate formulae, remain unchanged. Still, the step can be taken subject to the following caveat: The  (functional) integration should extend (at most) to the spacetime region that is causally connected with each point. Here, we will assume that the analysis is done locally in space but not necessarily in time, i.e. on the world-tube formed by the spatial part of the region $\delta {\cal M}$ in \cref{fig:FluidElem}.  
In the present example this would mean 
\begin{equation}
\begin{split}
	\N_\x^\d[g^{AB}] = &\N_\x^\d[g^{AB}_\e] + \int  \frac{\delta \N_\x^\d}{\delta g^{AB}(x)}\delta g^{AB}(x) \text{d}^4x+ \\
    &+\frac{1}{2}\int \frac{\delta ^2 \N_\x^\d}{\delta g^{AB}(x)\delta g^{CD}(y)}  \delta g^{AB}(x)\delta g^{CD}(y)\text{d}^4x\text{d}^4y 
\end{split}
\end{equation}
where the first two terms vanish because (i) $\N^d_\x$ vanishes at equilibrium, and (ii) the minimum energy condition. 
The key step then is to replace the ordinary partial derivatives with functional derivatives in the various expressions we have discussed, so that the viscous stress tensor will be
\begin{equation}
	S^\s_{AB}(x) =  2 \bar T \delta \Bigg( \frac{\delta \N_\s^\d}{\delta g^{AB}} \Bigg)(x) = 2 \bar T \int  \frac{\delta^2 \N_\s^\d}{\delta g^{AB}(x)\delta g^{CD}(y)}\delta g^{CD}(y) \text{d}^4y \;.
\end{equation}
We can now formally introduce a set of spatial coordinates $\bar x$ comoving with the equilibrium observer and attached to the world-tube, and take the time coordinate to be the equilibrium worldline's proper time $\tau$. Also, to enforce locality in space we let
\begin{equation}
	 \frac{\delta^2 \N_\s^\d}{\delta g^{AB}(x)\delta g^{CD}(y)} = \frac{1}{4}\Sigma_{ABCD}(\bar x, \tau_x-\tau_y) \, \delta^3(\bar x - \bar y)  \; .
\end{equation}
where the causality condition $\tau_x - \tau_y \ge 0$ is assumed to be encoded within $\Sigma_{ABCD}$.  

Let us first of all, as a consistency check, show that the formula for the particle production rate remains unaltered by these modifications. We have (see \cref{eq:NonConservativeGamma})
\begin{equation}
\begin{split}
	\mu_\x\Gamma_\x &= \frac{1}{3!} \mu_\x^{ABC}\frac{d n^\x_{ABC}}{d\tau_\x} = \bar\M_\x \frac{d}{d\tau_\x}\big(\bar\N_\x^\e + \bar\N_\x^\d\big) \\
    &= \M_\x \Bigg( \frac{d\N_\x^\d}{d\tau_\x} + \frac{1}{2}\N_\x^\d g^\x_{AB}\frac{dg_\x^{AB}}{d\tau_\x}\Bigg) \;.
\end{split}
\end{equation}
where, for a single viscous fluid the result simplifies to 
\begin{equation}
	\Gamma_\x =  \frac{d\N_\x^\d}{d\tau} +  \mathcal{O}(3) \;.
\end{equation}
If in particular we consider the entropy production rate $\Gamma_\s$ we have 
\begin{equation}
\begin{split}
	\N_\s^\d &= \frac{1}{8} \int  \Sigma_{ABCD}(\bar x,\tau-\tau') \delta g^{AB}(\bar x,\tau) \delta g^{CD}(\bar x,\tau') \text{d}^3\bar x\, \text{d}\tau \,\text{d}\tau' \;,
\end{split}
\end{equation}
To compute the entropy creation rate we have to use the chain rule (generalized to functionals) on $\N_\s^\d[g^{AB}(x)]$
\begin{equation}
	\frac{d\N_\s^\d}{d\tau} = \int  \frac{\delta N_\s^d}{\delta g^{AB}(y)} \frac{\delta g^{AB}(y)}{\delta \tau(x)} \text{d}^4y\;.
\end{equation}
But, because $g^{AB}$ is a ``normal'' function of the spacetime coordinates 
\begin{equation}
	\frac{\delta g^{AB}(y)}{\delta \tau(x)} = -2\delta^4(x-y) D^{(A}w^{B)} (x)\;,
\end{equation}
so that we are left with 
\begin{equation}
\begin{split}
	\Gamma_\s &= -\frac{1}{\bar T} S_{AB}(\bar x,\tau) D^{(A}w^{B)} (\bar x,\tau) \;. 
\end{split}
\end{equation}

We can now make use of this ``functional generalization'' to recover the Navier-Stokes model for a bulk- and shear viscous fluid. To focus on the key point, let us consider first the purely bulk-viscous case
\begin{equation}\label{eq:PurelyBulkExample}
	S(\tau) = \int K(\tau-\tau') A(\tau') \text{d}\tau' \;,
\end{equation}
where $S$ represents the trace of the viscous-stress tensor while A stands for the trace $\text{tr }\delta g^{AB}$. Because of the difference between $\delta g^{AB}$ and $\dot g^{AB}$---the former involves gradients in the displacement while the latter depends on the velocity---in order to recover a Navier-Stokes model we need to take 
\begin{equation}
	K(\tau-\tau') = -T\zeta\partial_{\tau'}\delta (\tau-\tau') \;,
\end{equation}
since this would give
\begin{equation}
	S(\tau) = T\zeta\int [-\partial_{\tau'}\delta (\tau-\tau')] A(\tau')\text{d}\tau' =T\zeta \int\delta (\tau-\tau') \partial_{\tau'}A(\tau')\text{d}\tau' = T\zeta\frac{dA}{\text{d}\tau} \;,
\end{equation}
as desired. We now implement this for the bulk- and shear model. We can do this using the standard decomposition of the bulk and shear response as
\begin{equation}\label{eq:SigmaDecomposition}
	\Sigma_{ABCD} = \Sigma_{ABCD}^\b + \Sigma_{ABCD}^\s \;,
\end{equation}
with 
\begin{subequations}\label{eq:functionalSigmas}
\begin{align}
    \Sigma_{ABCD}^\b &=  \frac{\zeta(x)}{\bar T}\, g^\e_{AB}g^\e_{CD} \delta^3(\bar x - \bar y) \, q_\b(\tau_x-\tau_y)\;,\\
    \Sigma_{ABCD}^\s &= 2\frac{\eta(x)}{\bar T}\, \bigg( g^\e_{A(C}g^\e_{B)D} - \frac{2}{3}g^\e_{AB}g^\e_{CD}  \bigg )\delta^3(\bar x - \bar y) \, q_\s(\tau_x-\tau_y) \;,
\end{align}
\end{subequations}
where the two kernels would be\footnote{Note that we have chosen to separate the bulk- and shear channels as usual, even though the present construction allows for anisotropic response in the velocity gradients to viscosity relation. We have also introduced two independent kernels to allow for different response to bulk and shear strain rates.} the same  $q_\b = q_\s = - \partial_{\tau_y}\delta(\tau_x-\tau_y)$. It then follows that the only viscosity tensor of the model is
\begin{equation}
	 S^\s_{ab} = \chi_{ab} + \chi \perp_{ab} = \frac{1}{3}\zeta \, \theta \perp_{ab} + \eta\,\sigma_{ab} \;.
\end{equation}
It is also easy to check that we can enforce compatibility with the second law by fixing the sign of the bulk-and shear viscosity coefficients. In fact,
\begin{equation}
	\Gamma_s = \frac{\zeta}{\bar T}\, \theta^2 + \frac{\eta}{\bar T}\, \sigma^{ab} \sigma_{ab} \ge 0 \;.
\end{equation}%
where $\zeta,\eta \ge 0$. With these relations we have recovered the usual relativistic Navier-Stokes equations (the Landau-Lifschitz-Eckart model for a viscous fluid).

Let us conclude by pointing out that, to write down the full set of equations one should also expand the ``Euler part'' of the equation of motion, i.e. the left-hand-side of \cref{eq:ViscousOneFLuidEq}. We have provided all the ingredients necessary for the explicit calculation, but leave it out here---and point to \cite{livrev} for further details---as the result is not new and not key to the present discussion.

\section{Cattaneo-type equations}
\label{sec:Cattaneo}

As a practical example of the first-order expansion we outlined the model for a single (bulk and shear) viscous fluid, and showed how this leads to the expected form of the relativistic Navier-Stokes equations.
The derivation shows that the action-based formalism encodes the previous models.
It is also clear that the formalism allows us to consider much more complicated settings, should we need to do so. 
However, the discussion of the first-order results is clearly not complete, because the final set of equations is widely known to suffer from causality/stability issues (see discussion in \cref{sec:TraditionalStrategies}) and it is then natural to wonder how we may fix this.
When it comes to heat-conducting systems, the way forward has been discussed in \cite{Lopez2011,NilsHeat2011}, where it is demonstrated that one can resolve the stability/causality issues at first order by properly accounting for the entrainment between matter and entropy currents---retaining the compatibility with the second law. 
However, for the single viscous fluid under consideration, the problem must be addressed in a different way, as the key ingredient used to solve the heat-flux case accounting for the entropy inertia, will not work for the present case as our model setting does not involve relative flows.
We now show how we can make progress in the single viscous fluid case by using a functional form that is different from \cref{eq:functionalSigmas}. 
Notably, the argument stresses the importance of the ``principle of memory or heredity'' (see \cite{EIT}). 

Let us first focus on the bulk viscosity case, and then extend the results to the bulk- and shear viscous model. Recalling \cref{eq:PurelyBulkExample}, the first step would again be to assume
\begin{equation}
	K(\tau-\tau') = -\partial_{\tau'}g (\tau-\tau') \;,
\end{equation}
so that 
\begin{equation}
	S(\tau)= \int K(\tau-\tau') A(\tau')\text{d}\tau' = \int g(\tau-\tau') \frac{dA(\tau') }{\text{d}\tau'}\text{d}\tau'  \;.
\end{equation}
We can then look for the convolution kernel $g$ such that the bulk-viscous scalar $S$ satisfies an equation of the Cattaneo type
\begin{equation}\label{eq:CattaneoS}
	t_\b \dot S = -S -\zeta \frac{dA}{d\tau} \;,
\end{equation}
where $t_\b$ is the relaxation time-scale of the bulk-viscosity response. In terms of the convolution kernel $g$ this would mean
\begin{equation}
	g(\tau-\tau') = -\frac{\zeta}{t_\b} e^{-(\tau-\tau')/t_\b}\theta(\tau-\tau') \;,
\end{equation}
and one can check by direct computation this leads to  \cref{eq:CattaneoS} as
\begin{equation}
	t_\b\partial_\tau g(\tau-\tau') + g(\tau-\tau') = -\zeta\delta(\tau-\tau')\;.
\end{equation}
By inspecting the last expression, we also see that in the ``fast relaxation limit'' $t_\b \to 0$ we recover a Navier-Stokes-type response as we would intuitively expect. 
We also point to \cref{ch:BVinSIM} where this fast relaxation limit is discussed in light of the (perhaps inevitable) resolution limitations faced in numerical implementations. 

We are now ready to go back to the full bulk- and shear viscous model.  We will retain the structure and symmetries from before (see \cref{eq:functionalSigmas,eq:SigmaDecomposition}), but introduce two different convolutions $q_\b$ and $q_\s$ to account for retarded response to bulk and shear strain rates. In essence, we have shown how we can implement a retarded response of the Cattaneo type in the action-based model by assuming that $S_{AB}$ (and therefore $D_{ab}$ as well) is an integral function of $g^{AB}$. The question then is, does this mean that the final fluid equations are integro-differential equations? Fortunately the answer is no. In fact, we have shown that, by a suitable choice of the response function $q(\tau-\tau')$, the fluxes satisfy an equation of the Cattaneo-type. Therefore, instead of solving an integro-differential equation, one should treat $S^\s_{ab}= \chi \perp_{ab} + \chi_{ab} $ as an unknown in \cref{eq:ViscousOneFLuidEq}, and add two equations to the system
\begin{subequations}\label{eq:Cattaneo-typeEquations}
\begin{align}
	\chi + t_\b\,\dot \chi &= - \zeta \,\theta\;, \\
    \chi_{ab} + t_\s\,\dot\chi_{ab} &= -\eta\,\sigma_{ab} \;.		
\end{align}
\end{subequations}

This means that, at the end of the day, to actually solve a set of differential dissipative equations at first order, we have to treat the fluxes as additional unknowns, for which one has to provide equations that are not given by the stress-energy-momentum conservation law $\nabla_aT^{ab}=0$. This is reminiscent of the EIT paradigm, where one postulates from the beginning an entropy function that depends on an additional set of quantities---the thermodynamical fluxes. The difference is that the microphysical origin of the equation for the fluxes is now more clear. It is also worth noting that equations of the Cattaneo-type for the fluxes cannot be obtained in the field-theory-based models, as the constitutive equations are given in terms of the usual equilibrium variables (like $\mu,\,T$) and their derivatives---so that terms with derivatives of the fluxes (like $\dot\chi$) do not appear. 

Equations \ref{eq:Cattaneo-typeEquations} are (formally) the same as in the linearized version of M\"ueller-Israel-Stewart model, which has been shown to be stable and causal. In theory, nothing prevents us from choosing a different form for the retarded response $q$ which could  lead to acausal/unstable behaviour. However, the form of $q$ suggested above has a clear physical interpretation and is microphysically motivated. If one wants to come up with an alternative, this would need to be motivated by microphysical arguments, as well. 

Let us now consider the implications of the Cattaneo laws for the entropy production rate. As in the Navier-Stokes model sketched above we have 
\begin{equation}\label{eq:BulkShearEntropy}
\begin{split}
	\Gamma_\s &= \int  \Sigma_{ABCD}(x,\tau-\tau') D^{(C}w^{D)}(\bar x,\tau')D^{(A}w^{B)}(\bar x,\tau) \text{d}\tau'= \\
    &= \frac{1}{\bar T} S_{AB}(\bar x,\tau) D^{(A}w^{B)} (\bar x,\tau) \;.
\end{split}
\end{equation}
Again, let us use the bulk-viscous case to highlight the relevant features. We then have 
\begin{equation}\label{eq:CITentropy}
	\Gamma_\s = \frac{\zeta}{t_\b} \int_{-\infty}^\tau  e^{-(\tau-\tau')/t_\b}\theta(\tau)\theta(\tau') \,\text{d}\tau' \;,
\end{equation}
where $\theta$ is the expansion rate. It is clear that, because the expansion rate is evaluated at different times, we cannot guarantee the positivity of the entropy production rate just by fixing the sign of the bulk-viscosity coefficient as in the previous Navier-Stokes case. However, we will argue this result is not as dramatic as it may appear at first sight---at least not in the regimes relevant for physical predictions. Because of the exponential in the integral, we can assume that the values for the expansion at times $\tau'$ can be neglected for $\tau'$ that is a few $t_\b$ away from $\tau$. As a result we can expand $\theta(\tau')$ as 
\begin{equation}
	\theta(\tau') = \theta(\tau) + \frac{d\theta(\tau)}{\text{d}\tau} (\tau'-\tau) = \theta(\tau) + \dot\theta(\tau)(\tau'-\tau) \;,
\end{equation}
so that we obtain
\begin{equation}\label{eq:CITentropySimplified}
	\Gamma_\s = \zeta \big[\theta^2(\tau) -t_\b \theta(\tau)\dot\theta(\tau)\big] \;.
\end{equation}
Again, because of the product of the expansion rate with its time derivative, the entropy production rate cannot be made generally positive simply by fixing the sign of the bulk-viscosity coefficient $\zeta$. However, as discussed in \cite{Geroch1995Relaxation,Lindblom1996Relaxation}, physical fluid states relax---on a timescale characteristic of the microscopic particle interactions---to ones that are essentially indistinguishable from the simple relativistic Navier-Stokes description. Translated to the present context this would mean 
\begin{equation}
	t_\b \dot\theta(\tau) \approx t_\b \frac{\Delta \theta}{\tau_{\text{hydro}}} \to 0 \;.
\end{equation}
because the ratio between the relaxation timescale $t_\b$ and the hydrodynamical timescale $\tau_{\text{hydro}}$ is effectively negligible. As a result, for actual physical applications we can neglect the second term in the entropy production rate and compatibility with thermodynamic second law is restored. In a way, this would mean that for bulk- and shear viscous fluids, we can introduce Cattaneo laws for the fluxes---to fix the causality and stability issues of the Navier-Stokes model---while the physical content will be precisely that of Navier-Stokes. 

\section{Summary and outlook}\label{sec:LinearizingSummary}

We have considered the close-to-equilibrium regime of the action-based model of \citet{2015CQAnderssonComer} for dissipative multi-fluid systems. In particular, we have shown that, starting from a set of fully non-linear dynamical equations with only the fluxes as the degrees of freedom, an expansion with respect to (a self-consistently defined) equilibrium can be introduced in a clear fashion, with the line of reasoning being similar to that of usual hydrodynamical perturbation theory. 

After discussing the aspects of equilibrium which can be inferred from the action-based model itself, we established how to construct an expansion in deviations away from equilibrium in a general setting, so that the framework is of wider relevance. In the process we demonstrated the importance of the frame-of-reference of the equilibrium observer. We also noted that the construction promotes the role of the matter space: Instead of it being a mathematical ``trick'' to facilitate a constrained variation, it might well be the arena where the microphysical details are encoded. This is a novel perspective that needs further discussion and consideration.

We then focused on a particular first-order viscous fluid model, with shear- and bulk-viscosity, paying particular attention to the key causality issues. We showed that causal behaviour can be linked to a retarded response function that keeps track of a system's history. The specific form of the response function can be modelled in a phenomenological way---as we did---but should ideally be provided by specific microphysical calculations, for instance by means of the fluctuation-dissipation theorem (see \cite{Reichl} for a general discussion and \cite{EIT} for comments on its role from the EIT perspective).\footnote{We also note that there have been recent efforts to make explicit use of the fluctuation-dissipation theorem to compute response coefficients through Green-Kubo-like formulae, see \cite{MontenegroTorrierCausal,MontenegroTorrierKubo}---although in a special relativistic setting. Nevertheless, there are similarities with the way we deal with causality issues here.} In a sense, the action-based model provides the ``context'', determining the geometric structure and form of the equations of motion, while the detailed microphysics is encoded in the specific response function. 

Building the first-order expansion we made this connection clear, and showed how and where the microphysics enters the discussion. 
An interesting outcome of this analysis is that we showed that there is no need to go to second order in deviations from equilibrium to implement a causal response in the model. This has already been demonstrated for the heat-flux problem (see \citep{NilsHeat2011,Lopez2011}), where the  Cattaneo-type equation for the heat flux is ultimately related to the multi-fluid nature of the problem. The entrainment effect (through which the entropy current gains an effective mass \citep{andersson10:_caus_heat}) results in an inertial heat response.  The case of a single viscous fluid is  different since its retarded response cannot be associated with the multifluid nature of the problem. 

As the variational model is designed for dealing with multi-fluids, the route to further extensions is---at least at the formal level---quite clear.
A natural next step would be the modelling of a viscous fluid allowing for the heat to flow differently from the matter. This application should be fairly straightforward since the two main issues of the problem have now been studied separately. A more challenging  step will be the inclusion of superfluidity. The presence of currents that persist for very long times changes drastically the non-dissipative limit. The model would require the use of more than one equilibrium worldline congruence \cite{2015CQAnderssonComer,ThermoGavassino}, one for each ``superfluid condensate'' and one for all the remaining constituents.

%% file: Parts/TurbulenceAndLES/part2macro.tex
\input{Parts/TurbulenceAndLES/LEShydro}

\input{Parts/TurbulenceAndLES/LESmhd}

%% file: Parts/TurbulenceAndLES/LEShydro.tex
\chapter{Filtering relativistic hydrodynamics}\label{ch:LES}

In the first part of this thesis we focused on modelling dissipative fluids in relativity, presenting the different ideas/schemes currently on the market, and then focusing on the only scheme that naturally lends itself to multi-fluid extensions. 
However, any discussion of dissipative relativistic fluids, would not be complete unless it touches upon (at the very least) the issues that arise when modelling turbulent flows. 
Turbulent flows are not only ubiquitous in the real-world---as turbulence is a manifestation of highly non-linear behaviour intrinsic to hydrodynamic equations---but also known to transport quantities like energy and momentum at a much faster rate than we would expect from microphysical transport mechanisms.

We start with a brief introduction to hydrodynamic turbulence in \cref{sec:IntroTurbulence}, whilst the rest of this chapter is devoted to address some of the issues that arise when modelling turbulent flows in relativity. 
We focus on the formal aspects associated with averaging/filtering the fluid dynamics which enter most of the recent developments/discussions. 
To make progress, we develop a new covariant framework for filtering/averaging based on the fibration of spacetime associated with fluid elements and the use of Fermi coordinates to facilitate a meaningful local analysis.
We demonstrate how ``effective'' dissipative terms arise because of the coarse-graining, paying particular attention to the thermodynamical interpretation of the resolved quantities.
In particular, as the smoothing of the fluid dynamics inevitably leads to a closure problem, we discuss a new closure scheme inspired by the recent progress in modelling dissipative relativistic fluids that we briefly touched upon in \cref{sec:HydrodynamicFieldTheory}. 
The results presented in this chapter have been published in \citet{fibrLES}.

We continue in \cref{ch:LESMHD} by discussing the first steps towards extending the framework to charged multi-fluids mixtures. 
In particular, we argue it is somewhat natural to begin with a discussion of magneto-hydrodynamics (MHD). 
We will do so after having introduced and discussed the main differences between hydrodynamics and MHD turbulence. 
There we will also derive the relativistic MHD equations, given that electromagnetic aspects have not been discussed in the earlier parts of this thesis. 

\section{A brief introduction to hydrodynamic turbulence}\label{sec:IntroTurbulence}
As a warm up, we begin with a very brief introduction to hydrodynamic turbulence, while referring to monographs such as \cite{Lesieur} for an exhaustive discussion. 
Everyday life gives an intuitive understanding of hydrodynamic turbulence: the flow of a river downstream of an obstacle or atmospheric/oceanic currents are just two examples. Turbulence typically involves an abrupt change in, say, the velocity field (or better, changes over very small scales) and is often driven by the development of fluid instabilities. A turbulent flow is also often associated with chaotic behaviour and a lack of predictability. While ``chaos'' can be precisely defined for simple mechanical systems---and turbulent flows are not necessarily chaotic in this sense---it is true nonetheless that uncertainties in the observation of the initial state forbid a precise prediction at later times. This is, in fact the reason why turbulent flows appear chaotic. On the other hand, there is ample evidence that Navier-Stokes equations describe well turbulent flows. An important quantity in the characterization of turbulence is the Reynolds number
\begin{equation}\label{eq:Renumber}
    \text{Re} = \frac{\rho\,L^2/\eta}{L/V}=\frac{L\,V\,\rho}{\eta} \;,
\end{equation}
where $L$ and $V$ are the characteristic length and velocity of the flow, while $\rho,\,\eta$ are the mass density and (dynamic) shear-viscosity coefficient---which has units (in cgs) of $\text{g cm}^{-1} \text{s}^{-1}$. The Reynolds number is formed as the ratio of the timescales over which flow properties are transferred by molecular diffusion as compared to macroscopic convection. It then quantifies the importance of inertial over viscous effects\footnote{Similarly one can define the Peclet number $\text{Pe} = L\,V/ \kappa$ for turbulent heat diffusion, where $\kappa$ is the heat conductivity.}. As viscosity tends to make neighbouring fluid elements move together, it is intuitively clear that for high Reynolds numbers we expect the fluid flow to change over very small scales. 

\begin{figure}
   \centering
   \includegraphics[width=0.8\linewidth]{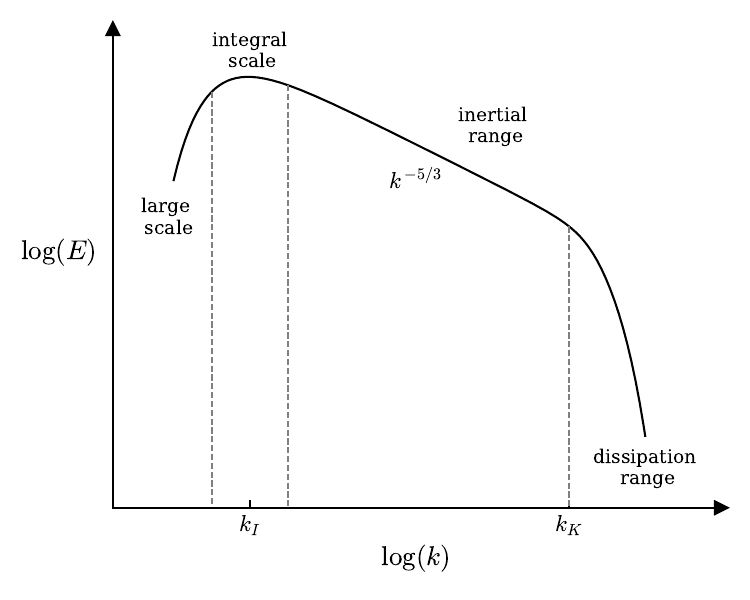}
   \caption{Cartoon of the scales of the turbulent energy spectrum. Figure adapted from \citet{mcdonough}.}
   \label{fig:EnergyCascade}
\end{figure}

From these considerations, we can conclude that fluid turbulence is (or can be described as) a deterministic phenomenon, although the evolution in time is very much complicated by the non-linearities in the fluid equations. Having said that, while a precise definition of turbulence does not exist, we will try to define it anyway pointing out some of the common properties of turbulent flows. First, a turbulent flow is disordered in (space and time) and often presents well-organized structures such as vortices. Second, turbulent flows are able to mix transported quantities---like energy and momentum---much faster than if only molecular diffusion processes were involved. Thirdly, it involves a wide range of spatial wavelengths. The latter are typically divided into three broad ranges (see \cref{fig:EnergyCascade}) : 
\begin{itemize}
    \item the \textit{large scale} is defined by the problem domain geometry.
    \item the \textit{integral scale} is a fraction of the large scale, often associated with a single wavelength $k_I$. It is defined as the maximum of the energy spectrum, and often associated with the scale at which energy is put into the system to sustain turbulence. 
    \item the \textit{inertial range}, which covers a wider range of length scales, is characterized by the fact that viscous effects are negligible.  
    \item the \textit{dissipation range}, typically associated with the Kolmogorov scale $k_K$, where viscosity effects are dominant over inertial ones. 
\end{itemize}

Because of the complexity of turbulent flows, a deterministic analytical description is practically excluded. One option, not further developed in the present work, is to resort to statistical analysis. A very important result in this direction was obtained by Kolmogorov in 1941 (see \cite{Kolmogorov1,Kolmogorov2,Kolmogorov3} and the more recent discussion by \citet{Frisch}). Assuming statistical isotropy and homogeneity, Kolmogorov derived a simple set of scaling laws that are in very good agreement with observations---the most famous being the ``$5/3$ law''. Using simple dimensional arguments he showed that
\begin{equation}\label{eq:Kolm53}
    E(k) = C_K \veps^{2/3} k^{-5/3} \;,
\end{equation}
where $E(k)$ is the energy spectrum, $\veps$ is the rate with which energy is pumped into the system at large scales, and $C_K$ is a constant---the Kolmogorov constant. 
For stationary, homogeneous and isotropic turbulence $\veps$ is also the constant energy flux from scale to scale, and hence the rate at which energy is dissipated once we hit the Kolmogorov scale.

An alternative option, which is often preferred, especially with the increase of computing power, is to rely on numerical simulations. For relatively simple flows one can try to evolve directly the Navier-Stokes equations. However, it has been shown that the number of grid points needed to fully resolve the flow scale as\footnote{This is for the three dimensional case, the two dimensional scaling is $\text{Re}^{2}$ \cite{lesbook}.} $\text{Re}^{9/4}$. This means that most of the interesting turbulent flows, characterized by very high Reynolds numbers of the order of $10^4$ or bigger, are in fact out of reach for direct numerical simulations---now and in the not so near future.
Then, what do we do? The strategy, which motivates the analysis in the rest of this chapter, is to use mean flow models or large-eddy-simulations. This boils down to averaging/smoothing the fluid dynamics, and resolving scales only up to some value in the inertial sub-range, while taking into account the smaller scales through sub-grid models. 

\section{Averaging turbulent flows}\label{sec:introAverageFluids}
Fluid models inevitably involve aspects of averaging---we have to average over a large number of particles in order to describe a fluid system in terms of a small number of macroscopic quantities (in the thermodynamic sense). However, in the simplest settings we do not have to worry (too much) about the actual process of averaging. For example, the transition from particle kinetic theory to a fluid model follows intuitively when the momentum distribution develops a well-defined peak. Similarly, the notion of a fluid element enters naturally on scales much larger than the individual particle mean-free paths (cf. discussion in \cref{subsec:FluxesBasic}). However, the story changes when we turn to dynamical simulations and problems involving, for example, turbulence. When we consider the problem from a simulation point of view, we have to consider the scale associated with the numerical resolution. This numerical scale tends to be vast compared to (say) the size of a fluid element. For example, in the high-density core of a neutron star we would typically deal with mean-free paths of a fraction of a millimeter, while the best current large-scale numerical simulations of neutron-star mergers involve a resolution of order 10 meters (see \cite{Kiuchi_2018}). This scale discrepancy has ``uncomfortable'' implications. In a highly dynamical situation we may not be able to resolve the full range of scales involved. Quite a lot of action can be hidden inside each computational cell. As we discussed in \cref{sec:IntroTurbulence}, this is a well-known fact that motivates the (considerable) effort going into developing ``large-eddy'' simulations schemes in computational fluid dynamics (the subject of numerous textbooks, see for example \cite{lesbook,mcdonough,Lesieur}). 

The astrophysical significance of the problem is obvious given that many relevant situations  involve/require the modelling of hydro(-magnetic) turbulence. Again, the dynamics of neutron stars  comes to mind. A topical example is the turbulent flow caused by the Kelvin-Helmholtz instability, which in turn drives the amplification of the magnetic field (through the dynamo effect) in binary neutron star mergers (\cite{PriceRosswog,PRD.92.124034,PRL.100.191101,Zrake}). Traditionally---in the context of Newtonian theory---turbulent flows have been studied in terms of Reynolds averaged equations, which, roughly speaking, are obtained via time-averaging the fluid dynamics. This smoothing requires the modelling of  features that are not captured by the resolved flow, bringing in the need to introduce suitable ``closure conditions'' (necessary as the averaged scheme introduces more degrees of freedom than there are equations of motion). Most recent work replaces the averaging with  spatial filtering, leading to what is known as Large Eddy Simulations (LES).
This strategy is often preferred because it involves less modelling \cite{mcdonough}, although Reynolds-type averaging is still widely used in the context of magnetic dynamos, see the reviews by \citet{Brandeburg2005} and \citet{Rincon2019}.

Because of the relevance for, in particular, binary neutron-star merger simulations, there have been several recent efforts to extend the ``familiar logic'' from the Newtonian setting to relativity. These range from the more formal discussion in \cite{eyink} to the actual simulations in \cite{radice1,radice2}. Also worth noting is the recent work in \cite{duez} in which the results from \cite{radice1} are contrasted with those obtained modelling the turbulent flow as effectively viscous on a larger scale (see e.g. \cite{shibata}). The general relativistic magneto-hydrodynamics (MHD) merger simulations of \cite{Giacomazzo} are another relevant example. Most of these results are based on spatial averaging, with subgrid models tailored to account for  small-scale dynamo action. Recently, a more refined gradient subgrid-scale model for general relativistic simulations was developed in \cite{viganoNR,carrasco,viganoGR} and  applied to binary neutron-star mergers with impressive results \cite{aguil,PalenzuelaBampinBNS,Aguilera-Miret-Universality}.

In short, while there has been notable effort to carry out large-eddy simulations in relativity, the formal underpinnings for this effort are not as firmly established as one might like---the exception being the discussion in \cite{eyink}. This is the gap  we are trying to bridge here. Starting from the beginning, we bring to the fore the fundamental issues associated with any effort to ``average'' or ``filter'' in a curved spacetime.

We want to consider the problem from a covariant spacetime point of view, a key point being that the underlying principles---for both  time-averaging and space filtering--- should be the same (or at least ``similar''). Both strategies combine ``smoothing''  with suitable closure relations to determine contributions that may not be ``directly'' calculable (i.e. represented on the resolved scale).  The issues we are interested in can be approached at the level of an ``effective theory'' based on fairly simple rules, avoiding a detailed discussion of the underlying averaging/filtering process. This is a useful strategy as it leads to a relatively straightforward derivation of the dynamical equations. At the same time, one has to pay attention to the details as a number of issues come into play when we consider the problem from the covariant perspective of General Relativity. In essence, we want to establish what a consistent spacetime scheme for averaging or filtering should look like, and highlight issues relating to the formulation of such a scheme. 
As the final aim is to develop a consistent  set-up for simulations, there are important numerical issues to be discussed; e.g. the implicit filtering associated with numerical discretisation. In order not to confuse these with the foundational issues, we leave them out of the present discussion.

As we need to keep track of the relevant scales and quantities associated with different ``observers'', the notation easily gets somewhat messy. This may be inevitable, but let us try to ease the pain by explaining the notation used in this chapter from the outset.  First of all, we need to distinguish between fine-scale and coarse-scale quantities. To do so, we  use bars and angle brackets---i.e. $\overline{A}$ and $\la A\ra$---for averaged and filtered quantities (respectively), obtained from the fine-scale one, $A$. However, as we will see, these quantities are not necessarily the most natural to evolve. Thus, we use  tildes, e.g. $\tilde A$, to identify  the evolved/resolved quantities. Finally, while discussing the linear stability of the proposed closure scheme  (\cref{subsec:stability}), we  drop the tilde notation---as all the quantities  are then assumed to be evolved and there is no need to make the distinction---and use instead sub/superscripts to represent quantities evaluated on the background, like $A_0$. This subscript should not be confused with the spacetime indices, which are represented by  latin letters $a,b,c \ldots = 0,1,2,3$ throughout.

\section{Averaging vs filtering}
\label{sec:AverVsFilter}

In order to provide the appropriate context and establish the general strategy, it is useful to briefly summarise the standard approach for  (typically incompressible) fluid dynamics in Newtonian gravity. 
Traditionally, small scale fluctuations are considered in terms of averaging, following the pioneering work of Reynolds and others (see \cite{lesbook}). In effect, this means that we have $A=\overline A+ \delta A$, with the fluctuations represented by $\delta A$ at each spacetime point. Introducing this formal split has the advantage of providing a straightforward derivation of the dynamical equations and a relatively clear interpretation of the involved quantities. One may also resort to an expansion for ``small'' $\delta A$ (see \cite{duez} for a relevant example of this). Typically, progress is made by assuming that the average of the linear fluctuations vanishes, which may not be a faithful representation of the physics the model aims to describe (see \cite{mcdonough} for a more extensive discussion). Noting this, the typical strategy for spatial filtering---forming the basis for modern large-eddy simulations---is different. In particular, the filtered fluctuations are not taken to vanish, nor does the argument involve expanding in the fluctuations. Instead, one typically proceeds by introducing a new set of variables to represent the filtered dynamics. From the conceptual point of view, each of the two strategies has attractive features and---as we are interested in the formal aspects of the problem---we will consider both of them in the following.

Let us first consider the standard averaging problem.
The standard strategy is to derive the averaged equations by applying a simple set of rules. In essence, one would \emph{assume} that (using a bar over quantities to represent averaging) 
\begin{subequations}\label{eq:OperationalRules}
\begin{align}
    \overline c &= c \ , && \mbox{for\ constants} \ , \\
    \overline{ A + B} &= \overline A  + \overline B \ , && \mbox{linearity of the procedure} \ ,   \\
    \overline{ \partial_a A}  &= \partial_a \overline A \ , && \mbox{averaging commutes with derivatives} \ .
\end{align}
\label{assum0}
\end{subequations}
It is immediately clear that, while the last of these relations is intuitive for a time average in Newtonian physics, we need to tread carefully when we turn to the relativistic setting. First of all, we need to face the fact that we do not have an observer-independent space-time split. Secondly, the derivatives we need to consider will be covariant, and hence we must consider the spacetime curvature. Simply noting these reservations for the moment (they will be discussed in \cref{sec:FCcoords}), the stated rules imply that
\begin{equation}
    \overline{ cA}  = c\, \overline A \ .
\end{equation} 
Moreover---and this is where the main distinction from large-eddy models comes in---it is common to further assume that the average of the fluctuations vanishes so we have
\begin{equation}\label{eq:AverFluct}
    \overline{ \delta A} = 0 \;.
\end{equation}
It then follows that
\begin{equation}\label{eq:DoubleAverage}
    \overline{\overline{A}} = \overline{A} \;,
\end{equation}
which means that the field $\overline{A}$ remains unchanged after the averaging.
In effect, this additional rule leads to 
\begin{equation}
    \overline{ \overline A B} = \overline A\, \overline B  \;.
\end{equation}
This simplifies the discussion considerably as we can ignore all linear fluctuation terms in the averaged equations. 

Time averaging is the (conceptually) simplest approach to the problem, but (strictly speaking) it removes dynamical features associated with the fluctuations\footnote{The averaged quantity is strictly speaking not time dependent, even though the time derivative term is usually retained in the evolution of Reynolds-Averaged Navier-Stokes equations (see \cite{mcdonough}).}, which is unlikely to be realistic. A faithful representation of the physics may require a different prescription. One option would be to not introduce the assumption from \eqref{eq:AverFluct}. The typical  description then involves (spatial) filtering, using some specified kernel to define the separation of scales (see \cite{lesbook}). This (effectively) leads to the same kind of rules as before---with the exception of \cref{eq:AverFluct,eq:DoubleAverage}---although we now have (indicating filtering by angle brackets)
\begin{equation}
    \langle \langle A\rangle B \rangle \neq \langle A\rangle \langle B \rangle \ ,
\end{equation}
which means that filtered fluctuations do not have to vanish. That is, in general we have $\langle \delta A \rangle \neq 0$. 
\section{The spacetime view: Fermi Coordinates}
\label{sec:FCcoords}

As a first---and essential---step towards a relativistic model for averaging/filtering, we have to consider the spacetime aspects of the problem. In particular, we need to introduce an unambiguous space-time decomposition---otherwise we cannot meaningfully consider ``time'' averages or ``space'' filtering. This is more than semantics \cite{eyink}. An interesting discussion of the problem (mainly from the cosmology perspective) has been provided by Ellis \cite{EllisInhomCosmo}, and it is evident that the issue is conceptually problematic since the notions of time and space are observer dependent. The problem is particularly vexing for a foliation based approach to spacetime  (as assumed in numerical relativity \cite{duez}, where the spacetime foliation is manifestly gauge dependent). However, for a fluid there does exist a natural fibration of spacetime \cite{livrev}. If we take the associated fluid frame as our starting point, we can introduce a meaningful ``local analysis'' which allows us to make progress. Moreover, it is natural to use the fluid frame to make the all-important connection with the  microphysics and the equation of state \cite{livrev}. The strategy also allows us to consider thermodynamical aspects of the averaging/filtering scheme. 

Let us explore the steps involved in an averaging/filtering procedure based on a spacetime fibration. In particular, we want to establish under which conditions we may assume that the covariant derivative commutes with the averaging/filtering procedure. Intuitively, we need to assume from the outset that there is a separation of scales between the metric fluctuations and the fluid fluctuations. The natural approach to the problem then involves Fermi-type coordinates (cf. \cref{app:FermiCoord}). In order to establish the logic, consider the following situation: The fluid four-velocity (and other physical properties) varies over a resolved spacetime region (this can be thought of as a numerical cell, even though such numerical cells would typically be defined in terms of a foliation). However, we assume that it is still possible to identify a family of observers associated with  a four-velocity vector field $U^a$ which can be taken to be constant over the resolved region and which  is ``close enough'' to the actual fluid four-velocity. (The latter assumption is not strictly required for the definition of an averaging procedure, nor for the spacetime decomposition, but it helps develop the logic). Then we can use the worldlines with tangent $U^a$ to construct  Fermi-type coordinates and explore the details of a given averaging/filtering procedure. 

Fermi coordinates were first introduced by Fermi in $1922$ \cite{fermi1,fermi2} and then developed by, in particular, Manasse and Misner \cite{ManasseMisner} (see also, for example, \cite{synge,Rakhmanov_2014}). We will not dwell on the construction itself here as this is not a new result (we point to \cref{app:FermiCoord} for more details).
Instead, we focus on the properties and  region of validity of the associated coordinate system. The set of coordinates is essentially built from a spacetime tetrad transported along a central worldline (naturally taken to be timelike in our case). This is convenient because the metric and the Christoffel symbols take a very simple form along the central curve. 

Let us introduce Fermi coordinates $x^\ha = \{x^\hZ,\,x^\hU,\,x^\hD,\,x^\hT\}$ (distinguished by hats on the indices) such that, on the central worldline $G$, the metric reduces to the Minkowski form $g_{\ha\hb}= \eta_{\ha\hb}$ while its first derivatives can be obtained from the  Christoffel symbols (see \cite{GravitationMTW}) 
\begin{subequations}
\begin{align}
	g_{\ha\hb,\hZ} &=   0 \;, \\
    g_{\hZ\hZ,\hj} &=  -2 a_\hj \;,\\
    g_{\hZ\hj,\hk} &= 0\;,\\
    g_{\hj\hk,\hm} &=  0 \;,
\end{align}	
\end{subequations}
where the commas represent partial derivatives.
We have introduced the non-vanishing piece of the four acceleration, $a_\hj$, of the worldline\footnote{That is, the four acceleration is $a_b = U^a\nabla_aU_b$ here.} and chosen to construct the tetrad in such a way that the associated observer  is non-rotating (which seems natural). With this construction we can formulate an expansion of the metric in the neighbourhood of the worldline. This leads to 
\begin{subequations}\label{eq:MetricFermiExp}
\begin{align}
	g_{\hZ\hZ} &= g_{\hZ\hZ}\big|_G + g_{\hZ\hZ,\ha} x^\ha = -(1 + 2a_\hj x^\hj) + \mathcal{O}(x^\hj)^2 \;,\\
    g_{\hZ\hj} &= g_{\hZ\hj}\big|_G + g_{\hZ\hj,\ha} x^\ha =  \mathcal{O}(x^\hj)^2 \;,\\
    g_{\hi\hj} &=   g_{\hi\hj}\big|_G + g_{\hi\hj,\ha} x^\ha = \eta_{\hi\hj}  + \mathcal{O}(x^\hj)^2 \;,
\end{align}	
\end{subequations}
where $|_G$ indicates that the quantity is evaluated on the worldline.
This is just a Taylor expansion for the metric where the ``small parameter'', $s$ (say), is taken to be the proper distance from the central curve. That is, we have $s^2 = (x^\hU )^2 + (x^\hD )^2 + (x^\hT )^2$. We see that, if the worldline is a geodesic then $a^\hj = 0$ and there are no corrections up to second order in the metric. However,  there will always be corrections at second order due to the spacetime curvature. These corrections can be expressed in terms of the Riemann tensor (again evaluated on the worldline $G$), but we will not need the explicit results here. Because we are assuming that the metric fluctuations happen on a larger scale (with respect to the fluid variations), we can make use of these expansions in the following.

Next, we can use the coordinates we have introduced to define a formal averaging or filtering procedure. 
Focusing on time-averaging first, we may use the spacetime split associated with the coordinates and define the procedure as 
\begin{equation}
	\overline A (\hat x) = \lim_{T\to\infty}\frac{1}{T}\int_{0}^{ T}  d\hat \tau A(\hat x,\hat \tau)  \;.
\end{equation}
That is,  given a point on the fluid element trajectory  we average in the proper time associated with the worldline. In terms of the Fermi coordinates, the time coordinate is exactly the proper time of the central curve. Note that there is no problem in taking the limit $T \to \infty$ since the Fermi coordinates are formally defined over the entire worldline. The region of validity is only limited in the spatial directions orthogonal to the central curve. From the definition, it is clear that time-averaged quantities must be time-independent, and it immediately follows that $\overline{\overline A}= \overline{A}$ (making contact with \eqref{eq:DoubleAverage}). 
We stress that this property follows if and only if we take the limit $T\to\infty$ in the definition, exactly as in the Newtonian case. 
As such, this implies that we should (strictly) neglect time derivatives in the averaged equations. The upshot is that the time-averaging strategy is (formally) valid for stationary flows only. The same is true in the Newtonian context---even though time derivatives are typically retained in the equations\footnote{See \cite{mcdonough} for a more extensive discussion on this.}. In fact, this is one of the main motivations in favour of spatial filtering and large-eddy models. In the following, we follow the ``tradition'' and retain terms involving time derivatives, as the main point of our discussion of the time-averaging case is pedagogical. 

As an application, let us consider the averaged metric. From the metric expansion above, we immediately see that the ``time-time'' component gains a correction due to the acceleration, which does depend on the proper time. However, we only need to integrate over points on the central worldline, where we have the Minkowski metric (in Cartesian coordinates) by construction. The situation is similar for all the remaining components. 

As a result, each component of the averaged metric takes the non-averaged value from the central worldline. That is, the metric is constant (in the sense described in \cref{sec:AverVsFilter}), and we have

\begin{equation}
\overline g_{\ha\hb} = g_{\ha\hb} \;.
\end{equation}
Similarly, $U^a$ is constant under averaging. To see this it is sufficient to note that (in terms of the Fermi coordinates) we have $U^\ha = (1,0,0,0)^\top$ so that $\overline U^a = U^a$. 

Analogously, we can use the spacetime split to define a space-filtering. We first have to assume that the width $L$ of the region over which we are filtering---the ``resolved box''---is small enough (in terms of the distance $\lambda$)  that the Fermi coordinates are well defined on it. For instance, one such condition is $L < 1/a$ where $a$ is the magnitude of $a^\hj$ (see \cite{Nesterov} for a detailed discussion). In this case, the  filtering procedure may be defined through 
\begin{equation}
	\la A\ra (\hat x, \hat t) = \int dV \,A(\hat x+\hat y,\hat t)f(\hat y)   \ , 
\end{equation}
where we introduced the filter $f$ (normalised over the resolved box). The expression simplifies further if we note that, by construction,  the spacetime point $(\hat x,\hat t)$ lies on the central worldline $(x^\hZ =\hat t = \tau,\, x^\hi =0)$ where $\tau$ is the relevant proper time. Also, in terms of the Fermi coordinates, the volume element is
\begin{equation}
	dV = U^\hZ \sqrt{-g} dy^\hU  dy^\hD  dy^\hT = (1 + 2a_\hj y^\hj)^{1/2} dy^\hU  dy^\hD  dy^\hT \approx (1 + a_\hj y^\hj) d^3\hat y \;.
\end{equation}
Note that we have not specified the exact filter to be used. This is not required at this stage, but we will assume the filter to be an even function (noting that this is the case for the three most common filters used in large-eddy models, see \cite{lesbook} for instance) and normalized over the resolved box such that we have\footnote{If the filter has a sharp boundary---i.e. vanishes at the boundary of the resolved box---the argument does not involve extending the spatial integral to infinity. However, one can think of filters with no sharp boundary, like a Gaussian filter (see \cite{mcdonough}), which may give rise to formal issues. In practice though, the exponential tail of the Gaussian should suppress anything beyond the Fermi coordinate boundary.}
\begin{equation}
	\int d^3\hat y \,f(\hat y) = 1 \quad\,,\quad\int d^3\hat y \,y^\hj f(\hat y) = 0 \ .
\end{equation}

Again, let us first apply this  to the filtered metric. Since there are no first-order corrections in the  expansion in \cref{eq:MetricFermiExp}, $g_{\hZ\hi}$ and $g_{\hi\hj}$ are constant over the box, and we have 
\begin{subequations}
\begin{align}
\la g_{\hZ\hi} \ra & = 0 =  g_{\hZ\hi} \big|_G \ , \\
\la g_{\hi\hj}\ra &= \int d^3\hat y(1 + a_\hj y^\hj)  f(\hat y)  \eta_{\hi\hj} = \eta_{\hi\hj} =g_{\hi\hj} \big|_G \ .
\end{align}
\end{subequations}
We also have
\begin{equation}
\begin{split}
	\la g_{\hZ\hZ} \ra &= -\int d^3\hat y(1 + 2a_\hi y^\hi)^{3/2}f(\hat y)  \\
    &= -1 - 3a_\hi \int d^3\hat y\,y^\hi f(\hat y)  = -1 = g_{\hZ\hZ}\big|_G \; \ ,
\end{split}
\end{equation}
where the last integral vanishes because of the assumed symmetry of the kernel. Once again, each component of the filtered metric takes the  non-averaged value from the central worldline throughout the region under consideration. That is, the metric is constant
\begin{equation}
	\la g_{\ha\hb}\ra =  g_{\ha\hb} .
\end{equation}
We also note that, by construction $U^a$ is constant over the box so we have $\la U^a\ra = U^a$.

Finally, since we have shown that the metric can (effectively) be taken to be constant under both averaging and filtering, it is easy to show that partial derivatives commute with each procedure. That is (obviously connecting with \eqref{assum0})
\begin{equation}\label{eq:partialscommute}
	\partial_a \la A\ra = \la\partial_a A \ra \quad \text{and} \quad \partial_a \overline A = \overline {\partial_a A} .
\end{equation}
We now show that these relations hold given the definitions of the average/filter above.
Let us start with the time-averaging case. When we consider the partial derivative with respect to the spatial coordinates the argument is straightforward, as the partial derivative (in the spatial direction) can be brought inside the integral. For the time derivative, we have 
\begin{equation}
	\overline{\partial_{\hat t} A}(\hat x,\hat t) = \lim_{T\to\infty} \int_0^T d\hat\tau \partial_{\hat\tau} A(\hat x,\hat\tau) =  \lim_{T\to\infty} \frac{A(T) - A(0)}{T} = 0 \;.
\end{equation}
On the other hand, if we take the time derivative of the averaged quantity, this trivially vanishes as it is time-independent. 

The argument for the time derivative in the filtering case is similarly straightforward. For spatial derivatives, we have on the one hand
\begin{equation}
	\la \partial_{\hat i} A\ra (\hat x,\hat t) = \int dV \left(\partial_{\hat i} A\right)(\hat x+\hat y,\hat t)f(\hat y) \;,
\end{equation}
while, on the other hand,
\begin{equation}
	\frac{\partial}{\partial x^{\hat i}} \la A\ra (\hat x,\hat t) \big|_{\hat x_0}= \int dV  \frac{\partial}{\partial x^{\hat i}} A(\hat x+\hat y,\hat t)\big|_{\hat x_0} f(\hat y) \;.
\end{equation}
Using the  chain-rule in the last equation, we see that
\begin{equation}
	\frac{\partial}{\partial x^{\hat i}}A(\hat x+\hat y)\big|_{\hat x_0} = \frac{\partial A(\hat z)}{\partial z^{\hat i}}\big|_{\hat z = \hat x_0+ \hat y} \ , 
\end{equation}
and it is clear that the two relations lead to the same result.

We  now have all the ingredients we need to prove that covariant derivatives commute with the averaging/filtering procedure. In fact, given that the metric is constant (in the sense of \cref{eq:OperationalRules}), we have 
\begin{subequations}\label{eq:gabprimeconst}
\begin{align}
		\la \partial_c g_{ab}\ra &= 	\partial_c  \la g_{ab}\ra = \partial_c g_{ab} \;, \\
        \overline{\partial_c g_{ab}} & = \partial_c \overline{g_{ab}} = \partial_c g_{ab} \;.
\end{align}
\end{subequations}
As a result, the Christoffel symbols---which are obtained from combinations of first derivatives of the metric---are (locally) constant under the procedure, as well. Therefore, we have 
\begin{equation}
	\la\nabla_a A^b \ra = \partial_a \la A^b \ra + \Gamma^b_{ac} \la A^c\ra = \nabla_a \la A^b\ra \,,
\end{equation}
with an analogous result for the time-averaging case. At the end of the day, the argument is quite intuitive.

\subsection{On covariance and the Einstein equations}\label{subsec:LEScovariance}

It makes sense now, before we move on, to spell out the covariance of the proposed averaging/filtering procedure, and comment on its compatibility with the field equations of General Relativity. It is, in fact, clear that the integrals we used  to define the averaging/filtering procedure are not of the usual type. We have to define each procedure in such a way that the   integration preserves the tensorial nature of the input\footnote{Recall that the usual integral on a (sub-)manifold of dimension $p$ takes a $p$-form as input and outputs a scalar \cite{GourghoulonSR}.}. 
Given this, we define the procedure for scalar quantities and then apply it to each component\footnote{Intuitively, because the procedures are based on the Fermi-coordinates construction in terms of a non-rotating tetrad with respect to $U^a$, we can effectively think of the components as scalars.} of, say, the metric tensor. We then require the averaged/filtered quantity to transform as a tensor on the resolved scale. The proposed definition reduces to the one of \cite{eyink} in the special relativity context and it also leads back to the Newtonian ones  \cite{lesbook,mcdonough}. The main difference is that, in special relativity such integrals are chart independent---as long as the integration is performed on each component using a fixed basis (see \cite{GourghoulonSR})---while this is not the case in General Relativity. 
However, even though the special relativistic integrals
are chart independent, the results are not, because the notions of length- and time scales are observer dependent. We also note that, as shown in \cite{eyink}, the observer-dependence cannot be resolved by the introduction of some kind of ``spacetime'' filter. However,  we are setting up the averaging/filtering using Fermi coordinates defined from the fibration. As this is
naturally associated with the fluid motion, the ``gauge'' dependence of the procedure is more physical. We execute the smoothing in ``some'' local frame $U^a$ which we can choose to ``associate'' to the ``micro-scale'' fluid motion.

Let us now discuss the averaging/filtering of the Einstein equations, focusing on the geometry. First of all, consider the Einstein tensor $G^{ab}$. From \cref{eq:partialscommute,eq:gabprimeconst} it follows that
\begin{subequations}
\begin{align}
		\la  g_{ab,cd}\ra &=  g_{ab,cd} \;, \\
        \overline{ g_{ab,cd}} &=  g_{ab,cd} \;.
\end{align}
\end{subequations}
Since the Einstein tensor is ultimately a combination of the metric and its (up-to-second order) derivatives $\boldsymbol{G}=\boldsymbol{G}(g,\,\partial g,\, \partial^2 g)$, this implies that we must have
\begin{equation}
    \overline{\boldsymbol{G}(g,\,\partial g,\, \partial^2 g)}  =\boldsymbol{G}(\overline g,\,\partial \overline g,\, \partial^2 \overline g) = \boldsymbol{G}(g,\,\partial g,\, \partial^2 g) \;,
\end{equation}
and analogously for the filtering case. The net result is that the coarse-grained theory remains consistent with General Relativity. In particular, the Einstein equations become
\begin{subequations}
\begin{align}
    G^{ab} &= 8\pi \overline{T^{ab}} \;, \\
    G^{ab} &= 8\pi \la T^{ab}\ra \;.
\end{align}
\end{subequations}
These results follow from the Fermi-coordinate construction, and the assumed separation of scales in the metric fluctuations with respect to the fluid variables. This should be a safe assumption for binary neutron star merger applications, but not necessarily for problems relating to the  very early universe (where quantum fluctuations in the gravitational field may play an important role). In this sense the Fermi-coordinate construction should be regarded as a pragmatic argument rather than a mathematical proof.
Having said that, we can now focus the discussion on the matter side, i.e. the stress-energy-momentum tensor $T^{ab}$. 

\section{Averaging in the fluid frame}
\label{sec:average}

Backed up by the Fermi-coordinate argument, let us first explore the problem of spacetime averaging. We start by introducing a fine-grained congruence  of wordlines with tangent vector field $u^a$, e.g. associated with individual fluid elements. Then,  working on a slightly larger (coarse grained) scale, we introduce another vector field $U^a$---the one used to define Fermi coordinates---such that small scale features are smoothed. In effect, we can then use the decomposition
\begin{equation}
u^a = \gamma (U^a + \delta v^a) \ , 
\end{equation}
with
\begin{equation}
    U_a \delta v^a = 0  \ ,
  \label{con0}
\end{equation}
and
\begin{equation}
\gamma = \left( 1 - \delta v^2 \right)^{-1/2} \approx 1 + \frac{1}{2} g_{ab} \delta v^a \delta v^b \ .
\end{equation}
At this point we take the view that it is natural to assume $\delta v\ll 1$ (the speed of the fluctuations is well below that of light) as this should be a safe assumption for the problems we are interested in \cite{duez}. This allows us to develop the logic more explicitly, even though we will drop this assumption later. One may  view the linear assumption as an additional  constraint (alongside the assumptions of the Fermi frame) on the size of the region we average over, although we will not try to make this statement precise. Finally, let us assume that it makes sense to work with an ordered expansion in the fluctuations. Working to second order---throughout the discussion of averaging but not in the filtering case that follows, where the expressions are not expanded in this sense---we then have
 \begin{equation}
u^a \approx  \left( 1 + \frac{1}{2} g_{bc} \delta v^b \delta v^c\right) U^a + \delta v^a \;.
\label{bigU}
\end{equation}

Let us now consider the average of this four velocity. Given the set-up we have  $\overline U^a = U^a$ and one might expect to have $\overline u^a = U^a$, as well. However, the problem turns out to be a little bit more intricate than that. First of all, from the discussion of time-averaging in \cref{sec:AverVsFilter} we assume 
\begin{equation}\label{eq:bardeltava}
    \overline{ \delta v^a } = 0  \ .
\end{equation}
It is also worth noting that the averaging procedure preserves directionality; that is 
\begin{equation}
    U_a \delta v^a = 0 \ \Longrightarrow U_a  \overline{\delta v^a} = 0 \;,
\end{equation}
This holds as long as we satisfy the conditions laid out in \cref{sec:FCcoords}---not because of \cref{eq:bardeltava}. 
We also get (to second order)
\begin{equation}
    \overline \gamma = 1 + \frac{1}{2} g_{ab} \overline{ \delta v^a \delta v^b } \ ,
\end{equation}
and it follows that
\begin{equation}
\overline {u}^a = \left( 1 +  \frac{1}{2} g_{bc} \overline { \delta v^b \delta v^c} \right) U^a
= \overline \gamma U^a \ .
\end{equation}

At this point we reach an impasse. It is clear that $\overline u^a$ is not (automatically) normalised and therefore can not serve as a four velocity. We would have to re-calibrate co-moving clocks to depend on the averaged fluctuations. This is problematic as we need  a projection to effect the space-time split and one would expect this to involve the fluid four velocity. There seems to be two ways to proceed. First, we could (perhaps pragmatically) opt to work with $U^a$ as the variable representing the flow. Alternatively, we may constrain the fluctuations to ensure that the averaging procedure returns $\overline u^a=U^a$. This would follow if we were to assume  the fluctuations to be such that 
\begin{equation}
   g_{ab} \overline { \delta v^a \delta v^b} = 0 \ \Longrightarrow \ \overline \gamma =1 \ .
   \label{con1}
\end{equation}
This allows us to move on, working with $\overline u^a$ to represent the flow, which  might seem the natural generalisation of the  Newtonian logic. However, considering the expected nature of small-scale turbulence, condition \eqref{con1} seems too restrictive. By assuming that the variance of the velocity fluctuations vanishes, we effectively remove the small scale kinetic energy that links to large-scale features in standard eddy-based models for turbulence. The kinetic energy (per particle) of the fluctuations is defined as
\begin{equation}
k = \frac{1}{2} g_{ab} \overline { \delta v^a \delta v^b}  \ ,
\end{equation}
which clearly vanishes if we impose \eqref{con1}. The second of the suggested approaches thus seems unattractive and we will not pursue it further. A third---indeed, likely preferred---possibility will become apparent when we consider the equation for the conserved matter flux.

\subsection{Baryon number conservation}\label{subsec:BaryonConserv_aver}

Having discussed the issues associated with averaging the four-velocity, the natural next step is to consider baryon number conservation. Letting $n = \overline n + \delta n$ represent the baryon number, the matter flux takes the form
\begin{equation}
    n^a = n u^a \approx \left( \overline n + \delta n \right)  ( \gamma  U^a + \delta v^a) \ , 
\end{equation}
such that (since $\overline{ \delta n} =0$ from the averaging) 
\begin{equation}
    \overline n^a = \overline n\, \overline \gamma \, U^a + \overline{ \delta n \delta v^a} \ .
\end{equation}
The number density measured by an observer moving along with $U^a$ is then given by
\begin{equation}
     n_0 = - U_a \overline n^a = \overline n\, \overline \gamma \ , 
\end{equation}
and we can write the averaged flux as
\begin{equation}
    \overline n^a = n_0 U^a + \overline{ \delta n \delta v^a} \ .
\end{equation}

The continuity equation then becomes
\begin{equation}
 \overline{\nabla_a n^a} = \nabla_a \overline n^a = \nabla_a (  n_0  U^a) + \nabla_ a \left( \overline{ \delta n \delta v^a }\right) = 0 \ ,
\end{equation}
or 
\begin{equation}
U^a\nabla_a n_0 + n_0 \nabla_a  U^a = - \nabla_ a \overline{ \delta n \delta v^a } \;.
 \label{bary1}
\end{equation}
This last equation shows that there is particle diffusion at second order (relative to $U^a$). Fluctuations lead to drift from large scale elements to their neighbours.

While nothing prevents us from taking this as given and moving on to the stress-energy-momentum tensor, the other equations of motion and the equation of state, it is useful to consider a density-weighted velocity, now starting from the flux $n^a$. The advantage of this is that we can arrange things in such a way that the new four velocity is normalised, while the non-linear fluctuations are hidden in its definition. In essence, we use the weighting with the number density to adjust the co-moving clocks in the desired way. Suppose we define 
\begin{equation}\label{eq:tilde_na}
    \overline n^a = \tilde n \tilde u^a \ , 
\end{equation}
while insisting that $\tilde u_a \tilde u^a = -1$. This immediately leads to\footnote{From \cref{eq:tilde_na,eq:tilde_n} we see that this density-weighted average corresponds to the Favre-type averaging often used in the Newtonian context (see, for instance, \cite{Lesieur,SchmidtLES,dartevelle2005comprehensive}).}
\begin{equation}\label{eq:tilde_n}
    \tilde n = \overline n \,  \overline \gamma \ , 
\end{equation}
and
\begin{equation}
    \tilde u^a = U^a + \frac{1}{\tilde n} \overline {\delta n \delta v^a}
    \label{FavreU} \ .
\end{equation}
It is easy to see that this retains the required normalisation as long as we ignore terms beyond second order. Crucially, this would remain true (by construction) if we did not expand in small fluctuations. We can now meaningfully introduce the projection with respect to the ``Favre''-filtered observer $\tu^a$, namely 
\begin{equation}
    \tilde \perp^a_b = \delta^a_b + \tilde u^a \tilde u_b \ .
\end{equation}
As anticipated, it also follows that 
\begin{equation}
    \tilde n = - \tilde u_a \overline n^a \ , 
\end{equation}
and the continuity equation takes the form
\begin{equation}\label{bary2}
    \nabla_a ( \tilde n \tilde u^a ) = \dot{\tilde n}  + \tilde n \nabla_a \tilde u^a = 0 \ ,
\end{equation} 
where the ``dot'' represent the covariant derivative with respect to $\tu^a$, i.e. $\dot{\tilde n} = \tu^a\nabla_a\tn$. This is attractive because---as the fluctuations are ``hidden''---we are left with a conservation law of the pre-averaged form. Note that we can remove the need of a closure in this equation, but this forces us to use the $\tu^a$ observer. One can imagine making a different choice, which would lead to drift terms entering  the continuity equation, as in \eqref{bary1}.

\subsection{Averaged matter dynamics}\label{subsec:AverEoM}

Next, consider the perfect fluid stress-energy-momentum tensor (derived in \cref{sec:PerfectFluids}). 
Starting from 
\begin{equation}
T^{ab} = (p+\varepsilon) u^a u^b + p g^{ab} \ , 
\end{equation}
we make use of \eqref{bigU} (noting that, to second order in the fluctuations, we have $\overline{\gamma^2} = \overline \gamma^2$), then after averaging we introduce $\tilde u^a$ according to  \eqref{FavreU}. This leads to
\begin{equation}
\overline T^{ab} = 
(\overline p+ \overline \varepsilon )\overline \gamma^2 \tilde u^a \tilde u^b  + \overline p g^{ab} + 2 \tilde u^{(a} q^{b)} + s^{ab} \ , 
\end{equation}
with 
\begin{equation}\label{eq:average_q}
    q^a = -\frac{\overline p + \overline \varepsilon}{\overline n} \overline {\delta n \delta v^a} + \overline{ (\delta p + \delta\varepsilon) \delta v^a} \;,
\end{equation}
and
\begin{equation}
    s^{ab} = (\overline p + \overline \varepsilon) \overline{\delta v^a \delta v^b} \;.
\end{equation}
It is convenient to work with the energy density measured by an observer moving along with $\tilde u^a$. This follows from 
\begin{equation}
    \tilde \varepsilon = \tilde u_a \tilde u_b \overline T^{ab} =  \overline \gamma^2\overline \varepsilon + \left( \overline \gamma^2 - 1\right) \overline p \;.
\end{equation}
As a result, we have
\begin{equation}
    \overline p + \tilde \varepsilon = ( \overline p + \overline \varepsilon) \overline \gamma^2 \;,
    \label{newp}
\end{equation}
which means that we can rewrite the stress-energy-momentum tensor as
\begin{equation}
\overline T^{ab} = 
(\overline p+ \tilde \varepsilon )\tilde u^a \tilde u^b  + \overline p g^{ab} + 2 \tilde u^{(a} q^{b)} + s^{ab} \;.
\label{tabav}
\end{equation}
The equations of motion 
\begin{equation}
\nabla_a \overline T^{ab} = 0 \;,
\end{equation}
then lead to the energy equation;
\begin{equation}
\dot\teps + (\overline p + \tilde \varepsilon) \nabla_a \tilde u^a = \tilde u_b \tilde u^a \nabla_a q^b - \nabla_a q^a + \tilde u_b \nabla_a s^{ab} \;,
\label{enav}
\end{equation}
and the momentum equation;
\begin{equation}
    (\overline p + \tilde \varepsilon)\,\ta^b + \tilde \perp^{ab} \nabla_a \overline p = -  \tilde \perp^b_{\ c} \tilde u^a \nabla_a q^c - q^b \nabla_a \tilde u^a - q^a \nabla_a \tilde u^b - \tilde \perp^b_{\ c} \nabla_a s^{ac}\;,
\label{momav}
\end{equation}
where $\ta^b = \tu^a\nabla_a\tu^b$.
It is worth remarking that with this choice of resolved variables, the final equations of motion resemble those of a general dissipative fluid with viscosity and heat-flux (cf. discussion in \cref{ch:DissipationLiteratureReview}).

\subsection{The equation of state}\label{subsec:BarotropicEoS}

A key step of any fluid model involves the connection to the microphysics as represented by the equation of state. As a first stab at this, let us outline the logic for the simple barotropic case (and then return to the issue in \cref{sec:2parEoS}, with a more realistic model in mind). In a barotropic model the starting point is a one-parameter energy density $\varepsilon = \varepsilon(n)$, which leads to the thermodynamical (Gibbs) relation
\begin{equation}\label{eq:1parGibbs}
    p + \varepsilon = n\mu \;,
\end{equation}
where the chemical potential is defined as
\begin{equation}
    \mu = \frac{d \varepsilon}{dn} \;.
\end{equation}
Introducing fluctuations in all the scalars, as before (eg. $p=\overline p + \delta p$), we have
\begin{equation}
    \overline p + \delta p + \overline \varepsilon + \delta \varepsilon = (\overline n + \delta n ) (\overline \mu + \delta \mu) 
= \overline n\ \overline \mu + \overline n \delta \mu + \overline \mu \delta n + \delta n \delta \mu \;.
\end{equation}
Since the average of linear fluctuations are all taken to vanish in the present case, this leads to 
\begin{equation}\label{pgibbs1}
 \overline p  + \overline \varepsilon  
= \overline n\ \overline \mu + \overline{ \delta n \delta \mu }  \;, 
\end{equation}
which shows that the number density fluctuations impact on the equation of state inversion required to evolve the system. In order to proceed, we need to provide a closure relation for $\overline{ \delta n \delta \mu}$.

Noting that the energy was assumed to depend only on the number density, the  fluctuations in \eqref{pgibbs1} should not be independent. In order to take a closer look at this, we may Taylor expand for small fluctuations $\delta n$ (recalling that we have already considered expanding the Lorentz factor in this way). This then leads to 
\begin{equation}
\mu \approx \mu(\overline n) + \mu'(\overline n) \delta n + \frac{1}{2} \mu''(\overline n ) \delta n^2 \;.
\end{equation}
That is, we identify
\begin{equation}
    \overline \mu =  \mu(\overline n)+  \frac{1}{2} \mu''(\overline n ) \overline {\delta n^2} \;,
\end{equation}
and
\begin{equation}
\delta \mu = \mu'(\overline n) \delta n \;. 
\end{equation}
The averaged Gibbs relation then becomes
\begin{equation}
    \overline p + \overline \varepsilon = \overline n \mu (\overline n) + \left[ \frac{\overline n} {2} \mu''(\overline n) + \mu'(\overline n)  \right] \overline { \delta n^2} \;, 
    \label{gibbs}
\end{equation}
showing that---in addition to the derivatives of the chemical potential--- we need to provide a (closure) relation for $\overline {\delta n^2}$. The other fluctuations are similarly slaved to $\delta n$. We get 
\begin{equation}
    \varepsilon(n) \approx \varepsilon(\overline n) + \varepsilon'(\overline n) \delta n + \frac{1}{2}\varepsilon''(\overline n) \delta n^2 \;,
\end{equation}
which leads to
\begin{equation}
    \overline \varepsilon = \varepsilon (\overline n) + \frac{1}{2}\varepsilon''(\overline n) \overline{ \delta n^2 }\;,
\end{equation}
and 
\begin{equation}
    \delta \varepsilon = \varepsilon'(\overline n) \delta n = \mu(\overline n) \delta n\;.
\end{equation}
 
In summary, we  have  equations describing the (proper time) evolution of  $\tilde n$, $\tilde \varepsilon$ and $\tilde u^a$, represented by equations \eqref{bary2}, \eqref{enav} and \eqref{momav}. In order to be able to solve these equations, we  need to provide closure relations for the fluctuations involved in $\overline \gamma$, $q^a$ and $s^{ab}$, that is $\overline{\delta n \delta v^a}$ and $\overline{\delta v^a \delta v^b}$. Once these, and $\overline{\delta n^2}$, are provided, we can work out $\overline{n}$ from $\tn$ and---assuming that we have access to the derivatives of the chemical potential---then we have $\overline \varepsilon$, as well. Using either \eqref{gibbs} or \eqref{newp} we can then rewrite $\overline p$ in terms of resolved variables and the closure terms, which completes the set of quantities we need to close the system and carry on the evolution. In principle, the fibration based averaging model is complete.

\section{Fluid element filtering}
\label{sec:filter}

Having established a workable procedure for averaging in the spacetime setting, let us turn to the issue of filtering. This is important because most actual numerical simulations assume  filtering rather than averaging. Formally, we expect the two problems to be similar, but we know from \cref{sec:AverVsFilter} that the filtering problem involves a slightly different logic. In particular, it is less natural---and also not desirable since the filtered fluctuations are unlikely to vanish---to consider an expansion in terms of the fluctuations. This is an important difference, so we need to carefully consider the steps in the analysis.

The natural way to proceed is to use a weighted average, in the spirit of \eqref{FavreU}. We would then start from\footnote{It is important to note that quantities like $\tilde n$ are not the same in the averaging and filtering cases. Still, we are using the same notation because they play the same role in each evolution scheme. Note also that our definition does not mean that $\la \tilde{u}^a \ra = \tilde{u}^a$. }
\begin{equation}
\langle n^a \rangle \equiv \tilde n \tilde u^a \ , 
\end{equation}
with (by construction) $\tilde u_a \tilde u^a=-1$,
leading to 
\begin{equation}
    \tilde n = - \tilde u_a \langle n^a \rangle \ .
\end{equation}
As in the averaging case, it is easy to see that the continuity equation then becomes
\begin{equation}
\la \nabla_a n^a\ra = \nabla_a \langle n^a \rangle = 0 \quad \Longrightarrow \quad \dot\tn+ \tilde n \nabla_a \tilde u^a = 0 \ .
\label{bary3}
\end{equation}

The main lesson is that, formally, the equation we arrive at takes the same form as in the averaging case; equation \eqref{bary2}. Still, there are differences relating to i) the  nonlinear quantities that have to be provided by a closure model and ii) the interpretation of the resolved/evolved variables. 

Deferring the discussion of more general aspects of  the equation of state for a moment (these will be discussed in \cref{sec:2parEoS}), let us move on to write down the equations of motion consistent with \eqref{bary3}. The filtered version of the perfect fluid stress-energy-momentum tensor can be written as
\begin{equation}
\langle T^{ab} \rangle 
= (\langle p \rangle + \langle \varepsilon \rangle ) \tilde u^a \tilde u^b + \langle  p \rangle g^{ab} + \tau^{ab} \ , 
\label{filtab}
\end{equation}
where  
\begin{equation}
\tau^{ab} =  \langle ( p + \varepsilon)  u^a  u^b \rangle - ( \langle p \rangle +  \langle \varepsilon \rangle ) \tilde u^a \tilde u^b \;,
\end{equation}
requires a closure relation.

Introducing, as before, the energy density measured by $\tilde u ^a$:
\begin{equation}
    \tilde \varepsilon = \tilde u_a \tilde u_b \langle T^{ab} \rangle = \langle \varepsilon\rangle + \tilde u_a \tilde u_b \tau^{ab} \ , 
\end{equation}
we can rewrite the filtered stress-energy-momentum tensor as
\begin{equation}
\label{eq:filtered_tab}
\langle T^{ab} \rangle 
= (\langle p \rangle + \tilde \varepsilon  ) \tilde u^a \tilde u^b + \langle  p \rangle g^{ab} + 2 \tilde u^{(a} q^{b)} + s^{ab} \ , 
\end{equation}
with
\begin{equation}\label{eq:filtered_tab_q}
    q^a = - \tilde \perp^a_b \tilde u_c \tau^{cb} = - \tilde \perp^a_b \tilde u_c \,\langle ( p + \varepsilon)  u^a  u^b \rangle\ , 
\end{equation}
and
\begin{equation}\label{eq:filtered_tab_sab}
    s^{ab} = \tilde \perp^a_c \tilde \perp^b_d \tau^{cd} = \tilde \perp^a_c \tilde \perp^b_d \, \langle ( p + \varepsilon)  u^a  u^b \rangle \ .
\end{equation}
It is easy to see that the energy and momentum equations  take exactly the same form (once we change $\langle p\rangle \to \overline p$) as in the averaged case (cf. equations \eqref{enav} and \eqref{momav}). As in  that case, the equations of motion provide all the information we need to solve the system once the relevant closure relations are provided. The key to a workable model is to provide appropriate closure relations, so it is important to understand what this involves. 

\section{Filtered Thermodynamics}
\label{sec:2parEoS}

So far we have only outlined the argument for the simple case of a barotropic fluid. As the equation of state is a central issue for any realistic model---and, obviously, any numerical simulation---let us rethink this.
A quick look back at \cref{pgibbs1} shows that, even if we start from a barotropic Gibbs relation at the fine scale, the averaged/filtered result is effectively ``non-barotropic''. In fact, the  $\overline{\delta n \delta\mu}$ closure term could be interpreted as an 
``entropy-like'' contribution associated with the fluctuations. This suggests that it makes sense to start straight away from a non-barotropic model, i.e. with a Gibbs relation of the form 
\begin{equation}\label{eq:2parGibbs}
	p + \veps = \mu n + T s \;,
\end{equation}
where $s$ is the entropy density and $T$ is the associated temperature.

The barotropic example also shows that the ``effective'' term that stems from the averaging or filtering procedure does not relate to the actual entropy, in the sense that it is not associated with some dissipative process and/or entropy production rate. This is evident from the fact that we have freedom in the choice of the averaging/filtering observer, and therefore in the variables to be evolved, and one might choose to frame the model in such a way that the fluctuation terms are reabsorbed in the definition of the variables themselves, just like we did for the weighted four-velocity $\tilde u^a$ in \cref{FavreU}.

Let us try to make these points more concrete, by considering a model that is non-barotropic from the get-go. Because there is no formal difference in the resulting equations, we will do this without distinguishing between the averaging and filtering cases (although, as we are now familiar with the logic, using the slightly more abstract notation from the latter). 

We start by discussing some subtleties of the barotropic example that we previously left aside. The energy density of a barotropic fluid is a function of the matter density only, which means there is no need to evolve both quantities separately. On the fine-scale, the energy equation contains the same information as the continuity equation:
\begin{equation}
\begin{split}
	\frac{d\veps}{d\tau} + (p + \veps)\nabla_au^a = \mu \frac{d n}{d\tau} + \mu n\nabla_au^a = \mu (\nabla_an^a) = 0 \;,
\end{split}
\end{equation}
where $\tau$ is the proper time associated with the ``actual'' fluid worldlines with tangent $u^a$. Note that the Gibbs relation \cref{eq:1parGibbs} is crucial for this argument. However, the situation changes on the coarse-grained scale. The link between the micro-scale equation of state and the resolved energy is not trivial---it may have to be established by a set of high resolution simulations, even though this may not be practical/feasible. In effect, the equation of state we are working with here is not the one you get from nuclear physics.  This is, in fact, true also in the simpler case where we set to zero the contribution coming from the $\tau^{ab}$ residual as the evolved density is not obtained by averaging the fine-scale one $\tn \ne \la n \ra$. The net result is that we have to treat the resolved energy $\teps$ and the resolved density $\tn$ as independent variables, and evolve both of them. 

\subsection{The effective entropy}

This subtle difference in the counting of independent variables between the fine- and coarse scale models obviously no longer exists for a non-barotropic fluid. For a two-parameter equation of state, the energy and particle density can be taken as independent variables already at the fine scale. This is, indeed, standard practice in numerical relativity simulations.
Let us also recall that, if the fluid is ideal there is no additional information gained from evolving the entropy current. 
In fact, we have seen in \cref{sec:PerfectFluids} that the entropy current is automatically advected as a consequence of the perfect fluid equations, provided this is a function of the energy and particle number densities, $s = s(n,\veps)$.

To complete the model set-up, however, we still have to clarify how the filtered pressure relates to the evolved variables. The barotropic model has been discussed in \cref{subsec:BarotropicEoS} for the averaging case, and the filtering case would work analogously. We now look at the non-barotropic fluid case. We can work with a resolved entropy defined as the usual thermodynamic potential $\ts \doteq s(\teps,\,\tn)$. The resolved temperature and chemical potential then follow from the standard definitions
\begin{subequations}
\begin{align}
	\frac{1}{\tilde T} &\doteq \Big(\frac{\partial \ts}{\partial \tilde \veps}\Big)_{\tilde n}(\teps,\,\tn) \;,\\
    -\frac{\tmu}{\tilde T} &\doteq \Big(\frac{\partial \ts}{\partial \tilde n}\Big)_{\tilde \veps}(\teps,\,\tn) \;.
\end{align}
\end{subequations}
We stress that we chose to use as thermodynamic potential the entropy, as it is a function of the chosen independent variables in the equations of motion, $\tilde n$ and $\tilde \veps$. Note also that, $\ts$ does not represent the true entropy and $\tT$ is not  the actual temperature, either. We are simply assuming that the usual thermodynamical definitions ``make sense'' at the filtering scale. Then, following the same logic we used for the barotropic case (see \cref{subsec:BarotropicEoS}) we filter \cref{eq:2parGibbs} and rewrite it as
\begin{equation}
	\la p \ra= -\teps + \tmu\tn + \tilde T\ts  + M \;,
\end{equation}
with 
\begin{subequations} \label{eq:M_Tau_residuals}
\begin{align}
    M &= \Big(\la Ts\ra - \tilde T\ts\Big) +\Big ( \la\mu n\ra  - \tmu\tn\Big) -\Big(\la\veps \ra - \teps\Big)\;.
\end{align}
\end{subequations}
The argument is now complete. We have explained how to express the averaged/filtered pressure that enters the equations of motion in terms of the resolved variables $\teps,\,\tn$ and the (new) residual $M$.

Having considered the resolved thermodynamics, we can turn to the ``entropy production'' associated with the averaging/filtering procedure. The final equations (i.e. \cref{enav,momav}) clearly remind us of the result for a dissipative fluid (see \cite{livrev}) so we are motivated to consider possible constraints stemming from the second law of thermodynamics. To do this we can work through steps analogous to \cref{eq:EntropyAdvection}  to establish the impact of the averaging/filtering procedure. Because the resolved entropy $\ts$ is taken to be a function of the resolved energy $\teps$ and the number density $\tn$ we have 
\begin{equation}
    \tT \nabla_a (\ts\tu^a) = \tT \ts \nabla_a \tu^a + \tT \dot \ts = \tT \ts \nabla_a \tu^a + \dot\teps - \tmu\dot\tn \;.
\end{equation}
Now, by means of \cref{bary2,enav} we obtain 
\begin{equation}\label{eq:effective_s_production}
\begin{split}
     \tT \nabla_a (\ts\tu^a) &= \big( \tT\ts +\tmu\tn - \la p \ra - \teps \big)\nabla_a\tu^a - q^a\tilde a_a - \nabla_aq^a - s^{ab}\nabla_a\tu_b \\
     &= - M \nabla_a\tu^a - q^a\tilde a_a - \nabla_aq^a - s^{ab}\nabla_a\tu_b \;.
\end{split}
\end{equation}
This shows that the entropy is no longer advected at the coarse scale, as a result of the averaging/filtering procedure. However, the fine scale (``exact'') theory is ideal, so the actual entropy is advected. 
For this reason, the model is not constrained by the second law at the coarse-grained scale. This is a very important point as it impacts on the closure relations (see below), which (evidently) can be discussed without considering the thermodynamical restrictions for ``real'' dissipative fluids. Effectively, the heat-flux term in \cref{eq:average_q} or \cref{eq:filtered_tab_q} is associated with energy transfer from  large eddies to small ones (or vice versa) rather than being a faithful heat transfer. We also note that this would not change even if we started from a fluid that is dissipative already at the fine scale.  Restrictions stemming from the second law of thermodynamics apply only at the fine-scale level, not at the coarse one.

\subsection{Energy cascade argument}\label{sec:energy_cascade}

Having discussed the thermodynamical interpretation of the quantities that enter the equations of motion, it makes sense to consider the involved energy cascade. This is relevant because an analogous argument is used in standard work on turbulence (see, for instance, \cite{Lilly,Leonard}) to motivate the closure of the fluid equations. It is useful to spell out the relativistic analogue of the classical argument. 

The starting point is the energy equation (see \cref{enav}) rewritten as 
\begin{equation}
    \underbrace{\dot\teps + (\la p\ra + \teps)\nabla_a \tu^a}_{\text{macro}} = \overbrace{- q^b \tilde a_b + \nabla_a q^a + \tu_b\nabla_a s^{ab}}^{\text{mixed}} \;.
\end{equation}
Here, we have highlighted that the terms on the left-hand side can be considered as macroscopic, in the sense that they involve only resolved quantities and describe an ideal evolution---intended to correctly capture the large-scale dynamics. In contrast, the terms on the right-hand side are ``mixed'' as they involve unresolved quantities---the residuals---and couple macro- and micro-scale terms. In effect, they can be thought of as transferring energy from one scale to another. 

To see this, we may, for a moment, assume a steady state evolution. As a consequence of the matter continuity equation, we then have $\nabla_a\tu^a = 0$ and therefore rewrite the energy equation as (setting to zero terms involving time derivatives with respect to $\tu^a$) 
\begin{equation}\label{eq:ClassicaCascade}
    \nabla_a q^a = s^{ab} \nabla_a\tu_b \;.
\end{equation}
In this relation, the term on the left-hand side should represent the energy sink (source) due to the (inverse) energy cascade---subtracting energy from the macro-scale into the micro one (or vice versa). 
In analogy with Newton's law of viscosity, Boussinesq suggested that one should relate the turbulent stress to the mean shear flow (see, e.g., \cite{mcdonough}). In our case, this leads  to
\begin{equation}\label{eq:sab_propto_sigma}
    s^{ab} \propto \tsig^{ab} \;,
\end{equation}
where the shear rate $\tsig^{ab}$ is defined in the usual way but in terms of the filtered four velocity, namely
\begin{equation}
    \tsig_{ab} = \left[  \tilde{\perp}^c_{(a} \tilde{\perp}^d_{b)}  - \frac{1}{3} \tilde{\perp}^{cd} \tilde{\perp}_{ab} \right] \nabla_c \tilde{u}_d.
\end{equation}
Motivated by this argument we move on to develop a closure scheme to complete the ``fibration framework'' we are proposing.

\section{An explicit closure model}\label{sec:closure}

Our ultimate aim is to develop a consistent scheme for large-eddy simulations in relativity. Even though this involves numerical aspects which we will not touch upon here, we need to provide a strategy for closing the system of equations already at the fibration level. This is the problem we focus on now.

As we have seen, in order to carry out an evolution we need to provide some prescription for the residual terms, that is $\tau^{ab},\,M$ in the filtering case or, equivalently,  $\overline{\delta n\delta n},\,\overline{\delta n\delta v^a},\,\overline{\delta v^a\delta v^b}$ in the averaging one. In classical computational fluid dynamics, one of the earliest closures proposed---still widely used---is due to Smagorinsky \cite{Smagorinksy}. This model effectively boils down to retaining only the $s^{ab}$ term and modelling it as a traceless tensor proportional to the (resolved) shear-flow. Such a closure is motivated by arguments of the kind we provided in the previous section. However, this model may be too simplistic to capture all relevant features of a turbulent flow\footnote{We do not want to comment on the validity of the Boussinesq hypothesis here, so simply refer to \cite{mcdonough} where it is discussed, albeit in a non-relativistic setting.}. In view of this, we aim to set up a scheme that can  be used to describe turbulent flows for which the Smagorinsky model gives unsatisfactory results. Finally, it is important to note that the Smagorinsky model is typically implemented---both in recent relativistic numerical work as well as in the Newtonian context---in the Eulerian frame associated with a foliation. The simple fact that the translation between fibration and foliation leads to a ``mixing'' of the different terms in the stress-energy-momentum tensor,  suggests that we need to consider a more general closure model.

In effect, we propose to model the residuals in terms of a general expansion in derivatives of the resolved variables, $\tn,\,\tu^a$ and $\teps$. For  practical reasons we halt the derivative expansion at first order, and decompose the gradients of the resolved quantities as
\begin{subequations}
\begin{align}
	\nabla_a \tn &= \tilde{\perp} ^b_a \nabla_b\tn - \tu_a \dot{\tn} \;, \\
    \nabla_a \teps &= \tilde{\perp} ^b_a \nabla_b\teps - \tu_a \dot{\teps} \;,\\
    \nabla_a\tu_b &= - \ta_b\tu_a + \tvort_{ab} + \tsig_{ab} + \frac{1}{3}\ttheta\tilde{\perp}_{ab} \;,
\end{align}
\end{subequations}
where, as before, $\dot{\tilde n} = \tilde u^a\nabla_a \tilde n$ (similarly for $\dot{\tilde \veps}$) and the filtered four velocity gradients $\nabla_a\tu_b$ are decomposed as usual. 
This closure scheme is analogous, although in a different spirit, to the most general constitutive relations discussed for dissipative hydrodynamics (at the linear level), see \cref{sec:HydrodynamicFieldTheory}.

Because there is no formal difference in the modelling of the sub-filter scale terms between the averaging and filtering cases, let us set up the closure scheme for the filtering case. We also immediately consider the case of a two-parameter equation of state, as the barotropic limit can be easily recovered from the more general results. We then have to model the residuals $q^{a} \text{ and }s^{ab}$. Recalling the definitions \cref{eq:filtered_tab_q,eq:filtered_tab_sab} we express these as
\begin{subequations}\label{eq:KovtunClosure_qs}
\begin{align}
    s^{ab} &= -\eta\tsig^{ab} + (\pi_1 \ttheta + \pi_2 \dot \tn + \pi_3 \dot \teps) \tilde{\perp}^{ab}\;,\\
    q^a &= \theta_1 \ta^a + \theta_2 \tilde{\perp}^a_b\nabla^b\tn + \theta_3\tilde{\perp}^a_b\nabla^b\teps\;.
\end{align}
\end{subequations}
In order to evolve the system, we also need to express $\la p\ra$ in terms of the resolved variables. To do so, we have to provide $M$. As this is a scalar, we model it as (see \cref{eq:M_Tau_residuals})
\begin{subequations}\label{eq:KovtunClosure_Mtau}
\begin{align}
	M &= \chi_1 \ttheta + \chi_2 \dot \tn + \chi_3 \dot \teps \;,
\end{align}
\end{subequations}
and  the filtered pressure then takes the form
\begin{equation}\label{eq:FilteredPressure}
	\la p \ra = - \teps + \tilde{T}\tilde{s}+ \tn\mu + M \;.
\end{equation}

We have now introduced a total of $10$ parameters to be used in the actual large-eddy model. These parameters---potentially validated/calibrated through high-resolution simulations---can be considered as functions of the resolved energy, density etcetera. Therefore, when we focus on a small region of the fluid they can be treated as simple constants.
 
\subsection{Stability Analysis}\label{subsec:stability}

Let us turn to the issue of linear stability, as this is a necessary condition for the system of equations to be (numerically) solved. 
Moreover, the fact that the ``effective'' theory we arrive at the resolved scale resembles that of a dissipative fluid further motivates this analysis.
After all,  it is well known that the standard/textbook relativistic viscous hydrodynamics equations are unstable (cf. \cref{ch:DissipationLiteratureReview}). 

The linear stability of the effective theory obviously depends on the closure used, so let us focus on the specific relations proposed above. However, the aim is not to discuss the stability of the closure model in full generality, only to provide a ``proof of principle'' argument. 


As the averaging/filtering residuals have been expressed in terms of gradients of the evolved variables and we are considering a local region, it makes sense to assume that the background configuration---the stability of which we want to study---is that of a homogeneous fluid at rest. We will also consider, as usual, a flat background spacetime and ignore metric perturbations. This is justified---even in the general relativistic context---since the stability analysis is intended to be local, so that the effects of gravity can be transformed away (using a local inertial frame argument, in the spirit of the Fermi frame logic). Finally, we simplify the notation in order not to clutter up the equations. We drop the ``tildes'' used to identify the resolved variables, as we no longer need to make the distinction. Instead, we identify background quantities with a subscript ($0$).  For instance, we write the background four-velocity as $u_0^a$ and the chemical potential in the background configuration as $\mu_0$.

Let us start by expanding the perturbed fields (indicated by a $\delta$) in Fourier modes
\begin{subequations}
\begin{align}
    \delta u^a &= \B^a \pw \;,\\
    \delta n &= \A \pw\;,\\
    \delta \veps &= \E \pw \;.
\end{align}
\end{subequations}
where $\B^a$ is orthogonal to $u_0^a$ because $u^a_0\delta u_a = 0$ as a result of the four-velocity normalization. We also decompose the wave-vector $k^a$ as
\begin{equation}
     k^a = \omega u_0^a + k\hat k^a \;,
\end{equation}
where $\omega$ is the frequency, $k$ is the wavelength and $\hat k^a$ is a unit four-vector orthogonal to $u_0^a$ which describes each mode's direction. Because of the metric signature convention (+2), the system will be linearly stable (in time) if all solutions to the dispersion relation---written as $\omega = \omega(k)$---have a negative (or vanishing) imaginary part. We will also use $\hk^a\hk^b$ and\footnote{Recall the definition of orthogonal projection with respect to a time-like vector. The sign difference stems from $\hat k^a$ being a unit space-like vector.} $\delta^{ab} -\hk^a\hk^b$ to decompose the momentum equation as well as $\B^a= (0,\,B_L,\,\B_{T1},\,\B_{T2})^\top$ into its longitudinal and transverse part (with respect to wave direction).

In order to write the linearized equations in terms of the perturbed fields, we have to clarify how to perturb the pressure. As can be seen from \cref{eq:FilteredPressure}, its explicit expression depends on $M$. Let us first focus on the non-residual contribution and come back to $M$ later:
\begin{equation}
    \delta p =  (\C \A +\D\E)\pw + M\;,
\end{equation}
where we have defined 
\begin{subequations}\label{eq:CDdefinitions}
\begin{align}
    \C &= \Big(\frac{\partial p}{\partial n}\Big)_\veps (n_0,\,\veps_0) \;,\\
    \D &= \Big(\frac{\partial p}{\partial \veps}\Big)_n(n_0,\,\veps_0) \;,
\end{align}
\end{subequations}
to simplify the expressions that follow.

Let us start by linearizing first the non-residual part of the equations of motion. The result is 
\begin{subequations}
\begin{align}
    &-i\omega \A + i k n_0 \B_L = 0\;,\\
    &-i\omega \E + i h_0 k \B_L = 0\;, \\
    &-i h_0 \omega\B_L + ik(\C \A +\D\E) = 0\;, \\
    &-i h_0 \omega\B_{T1} = 0\;,\\
     &-i h_0 \omega\B_{T2}  = 0\;,
\end{align}
\end{subequations}
where we have introduced the usual enthalpy density, $h_0 = p_0 + \veps_0$. To work out the full linearized system of equations, let us consider each residual at a time. We start by looking at the trace-free part of $s^{ab}$, as this would correspond to the (fibration version of the) model  proposed by Smagorinsky in the Newtonian context. A straightforward calculation leads to
\begin{equation}
\begin{split}
    \delta\sigma^{ab} &= \delta\big(\perp^{ac}_0\perp^{bd}_0\partial^{(c}u^{d)} - \frac{1}{3} (\partial_cu^c)\perp_0^{ab}\big) \\
    &= ik\big(\hat{k}^{(a}\B^{b)} - \frac{1}{3}\B_L \perp_0^{ab}\big)\pw \;,
\end{split}
\end{equation}
where $\perp_0^{ab} = \eta^{ab}+ u_0^a u_0^b$ is the projection orthogonal to the background velocity and $\eta^{ab}$ is the Minkowski metric. As for the trace part of $s^{ab}$ we have  
\begin{equation}
\begin{split}
	&\delta\Big[\big(\pi_1 \partial_cu^c + \pi_2 u^c\partial_c n + \pi_3 u^c\partial_c\veps\big) \perp_0^{ab}\Big] = \\
    &= \big(i \pi_1 k \B_L - i\pi_2 \omega \A - i\pi_3 \omega\E\big) \perp_0^{ab} \;,
\end{split}
\end{equation}
and it is easy to see that these additional terms only affect the longitudinal projection of the momentum equation. Next we have the heat-flux $q^a$. It is fairly easy to see that only two (out of five) terms will contribute to the linearized equations. These terms lead to 
\begin{equation}
	\delta (\partial_aq^a) = \theta_1 \omega k \B_L - \theta_2 k^2 \A - \theta_3 k^2\E \;,
\end{equation}
which enters the energy equation, while 
\begin{equation}
	\delta (\perp^b_c u^a\partial_aq^c) = -\theta_1 \omega^2 B^b + \theta_2 \omega k\hk^b\A + \theta_3 \omega k\hk^b\E \;,
\end{equation}
affects the momentum equation. We note that the last two terms in the expression above affect only the longitudinal projection, while the first term modifies both the longitudinal and transverse components. Last but not least, we consider the residuals that arise from the Gibbs relation:
\begin{equation}
	M = \chi_1\theta + \chi_2 \dot n + \chi_3 \dot\veps \;.
\end{equation}
It is easy to see that this residual will not affect the (linearized) energy equation, while it contributes to the longitudinal momentum equation as
\begin{equation}
	\delta \big(\perp^{ab}\partial_a M\big) = \big[-\chi_1k^2\B_L + \chi_2 k\omega \A + \chi_3 k\omega \E\big] \hk^b\;.
\end{equation}
Collecting everything together, we can write the linearized equations\footnote{Note that the coefficient matrix depends only on $\omega = -u^ak_a$ and $k^2 = k^ak_a + \omega^2$, in accordance with Lorentz invariance.} as
\begin{equation}\label{eq:GeneralMatrix}
    \begin{pmatrix}
    \pmb{L} & \pmb{0} \\
    \pmb{0} & \pmb{T} 
    \end{pmatrix}
    \cdot  \begin{pmatrix} 
    \A & \E &\B_L & \B_{T1} & \B_{T2} 
    \end{pmatrix}^\top = 0 \;,
\end{equation}
where
\begin{equation}\label{eq:GenLongMatrix}
    \pmb{L}  = \begin{pmatrix}
    -i\omega  & 0 & i n_0 k  \\
    -\theta_2 k^2  & -i\omega -\theta_3k^2 & i h_0 k +\theta_1 k\omega \\
     ik\C + (\zeta_2+ \theta_2) k\omega & i \D k + (\zeta_3 + \theta_3) k\omega& -\big(i h_0\omega -\frac{2}{3}\eta k^2 + \zeta_1 k^2 + \theta_1 \omega^2\big)   \\
    \end{pmatrix} \;,
\end{equation}
and 
\begin{equation}\label{eq:GenTransverseMatrix}
    \pmb{T}  = \begin{pmatrix}
    -\big( i h_0\omega - \frac{\eta}{2} k^2 + \theta_1\omega^2 \big) & 0 \\
    0 &-\big( i h_0 \omega - \frac{\eta}{2} k^2 + \theta_1\omega^2 \big) 
    \end{pmatrix} \;,
\end{equation}
and we have introduced $\zeta_i =  \chi_i  + \pi_i$ with $i = 1,2,3$. A similar analysis is carried out, for instance, in \cite{KovtunStable}. However, the different gradient expansions (cf. \cref{eq:KovtunGeneralExpansion} and \cref{eq:KovtunClosure_qs,eq:KovtunClosure_Mtau}) lead to slightly different results.

The stability analysis for the general case is perhaps best considered numerically. We also stress (again) that, because the entropy $\ts$ does not represent the true one, we are allowed to violate the second law of thermodynamics. This will give us more freedom (with respect to faithful dissipative fluids) to control the stability of the closure. 

\subsection{Smagorinsky model}

As a first step, let us consider the simple case where the only non-vanishing parameter in \cref{eq:KovtunClosure_qs,eq:KovtunClosure_Mtau} is $\eta$. This would correspond to the (fibration version of the) model proposed by Smagorinsky in the Newtonian context. Starting from  \cref{eq:GeneralMatrix} we easily obtain the linearized equations for this case
\begin{equation}
    \begin{pmatrix}
    -i\omega & 0 & i n_0 k & 0 & 0 \\
    0 & -i\omega & i h_0 k &0 & 0   \\
    ik\C & i\D k& -i h_0 \omega + \frac{2}{3}\eta k^2 & 0 & 0 \\
    0 & 0 & 0 & -i h_0 \omega + \frac{\eta}{2} k^2 & 0  \\
   0 & 0 & 0 & 0  & -i h_0 \omega + \frac{\eta}{2} k^2   
    \end{pmatrix} \cdot 
    \begin{pmatrix}
    \A \\ \E \\ \B_L \\ \B_{T1} \\ \B_{T2} 
    \end{pmatrix} = 0\;.
\end{equation}
The required dispersion relations are obtained by setting to zero the determinant of the coefficient matrix. Working this out, we find that the transverse modes decouple, and the corresponding dispersion relation is 
\begin{equation}\label{eq:SmagorinskyTransverse}
    \omega = - i \frac{\eta}{2 h_0} k^2 \;.
\end{equation}
The other non-trivial modes are longitudinal, with dispersion relation 
\begin{equation}
 h_0 \omega^2 + i\frac{2}{3} \eta k^2\omega  - (h_0 \D + \C n_0)k^2 = 0\;.
\end{equation}
Solving this equation we obtain 
\begin{equation}\label{eq:SmagorinskyLongitudinal}
	\omega = \pm c_s k - i \frac{\eta}{3 h_0}k^2 + \mathcal{O}(k^3) \;,
\end{equation}
where
\begin{equation}\label{eq:Lmode_speed}
    c_s^2 = (\D + \C n_0/h_0) \;,
\end{equation}
is the usual sound speed. The longitudinal modes represent sound waves, while the transverse modes are not propagating. Both sets of modes are stable for $\eta>0$.  As a simple consistency check, it is easy to see that in the ideal limit, when $\eta = 0 $, one obtains a single non-trivial solution representing an undamped sound wave. 

The result demonstrates that the simple Smagorinsky model is stable according to the fibration observer. This is a key conclusion for cosmological applications, as these tend to involve a cosmological time associated with a co-moving observer; in essence, a fibration. The situation is different for numerical relativity simulations, which tend to be based on a spacetime foliation. The matter description  (formally) involves a fibration associated with fluid element worldlines, but the evolution is carried out in a different frame. Our stability demonstration does not (yet) cover this case. In order to complete the argument,  we have to consider the stability issue in a different frame. We therefore introduce (as usual) the Eulerian observer $N^a$ as 
\begin{equation}
    u^a = W\left(N^a + v^a\right) \;, \quad \mbox{with} \quad
    W = (1 - v^2)^{-1/2} \;,
\end{equation} 
and note that  the two frames are related by a  Lorentz boost. 
This turns out to cause trouble. The simple Smagorinsky model, while  stable in the fibration frame, becomes unstable in the boosted frame. 

In order to demonstrate this result, we start by noting that (using primes to indicate boosted quantities)
\begin{subequations}
\begin{align}
    \partial'_a\veps' &= \Lambda^b_a\partial_b\veps = 0 \;,\\
    \partial'_a n' &= \Lambda^b_a\partial_b n = 0 \;,\\
    \partial'_a u'_b &= \Lambda^c_a\Lambda^d_b\partial_cu_d = 0 \;,
\end{align}
\end{subequations}
where $\Lambda$ is the Lorentz boost matrix. As we are linearizing with respect to a homogeneous (in spacetime) background this confirms that the gradient-based closure scheme we are proposing  still makes sense. We have to work out the dispersion relations in a non-comoving frame, but because these are expressed in terms of $\omega = -k^au_a$ and $k^2 = k^a k_a + \om^2$, we just have to boost these quantities. We can then take (without loss of generality) $v^a$ to be in the $x$-direction, while $\hat k$ lies in the $x-y$ plane. Then we introduce the angle $\phi$ between the wave-vector and $v^a$ as $\hat k^a v_a = v\,\cos{\phi}$ and write the Lorentz boost as:
\begin{subequations}\label{eq:LorentzBoost}
\begin{align}
    \om &= W(\om' - vk'\cos{\phi}) \;, \\
    k_x &= W(k'\cos{\phi} - v\om') \;,\\
    k_y &= k'\sin{\phi} = k'_y \;,\\
    k_z &= 0 = k'_z\;.
\end{align}
\end{subequations}

Applying this to the transverse dispersion relation in \cref{eq:SmagorinskyTransverse} we obtain (dropping the primes for clarity)
\begin{multline}
     \left(\eta W^2 v^2\right)\om^2 -2 \left(i h_0W - \eta W^2 v k\cos{\phi}  \right)\om - \\ -\left( \eta W^2 k^2\cos^2{\phi}+ \eta k^2 \sin^2{\phi}- 2ih_0\cos{\phi} Wvk\right) = 0 \;.
\end{multline}
We note that \cref{eq:SmagorinskyTransverse} was a first order polynomial, while the boost made it second order, thus generating an additional solution. For long wavelengths, the two solutions are
\begin{subequations}
\begin{align}
    \om &= vk\cos{\phi} - i \frac{\eta}{2h_0 W^3} \left(\cos^2\phi + W^{-2}\sin^2\phi\right) k^2 + \O (k^3)\\
    \om &= i\frac{2h_0}{\eta W v^2} + \O(k)\;.
\end{align}
\end{subequations}
The first solution is the boosted version of the mode we obtained in the fibration. It is stable for $\eta> 0$ (as in the comoving frame), propagating with phase velocity $v\cos{\phi}$ and the decay rate reduces to the original value as $v\to 0,\,W\to 1$. There is, however, an additional solution which is non-vanishing for $k=0$ (in \cite{KovtunStable,HoultKovtun2020} these are referred to as ``gapped'' modes). This second mode is evidently unstable for $\eta > 0$. 

This result demonstrates that the simple Smagorinsky model is unstable when ``observed'' from a non-comoving frame. This is a well-known problem of Eckart-Landau models for dissipative fluids (cf. \cref{ch:DissipationLiteratureReview}). 
As we are not dealing with a dissipative model that describes linear deviation from a thermodynamical equilibrium state, this is not intrinsically problematic.
For real dissipative systems, stability of equilibrium is not only required for numerical implementations, but also guided by intrinsic consistency. 
A system slightly out of equilibrium
must evolve, “by definition,” toward thermodynamical
equilibrium, no matter if the fluid in equilibrium is at rest or not.
Our case is different. 
However, because we are setting up the filtering scheme in the fibration---in order to retain consistency with the covariance of General Relativity, as discussed in \cref{subsec:LEScovariance}---while the simulations will be carried out in the foliation, we  have to ensure that the model is ``covariantly'' stable. 
The LES model of \cite{radice1} gets away with a simple Smagorinsky closure because it is directly implemented in the foliation frame, where the simulation is then performed.

\subsection{Fixing the Smagorinsky instability}

The aim now is to show how we can fix the instability problem (in the boosted frame) by introducing more parameters in the closure. Focusing first on the transverse modes, we see from \cref{eq:GenTransverseMatrix} that the only way to fix the problem is by considering a non-zero $\theta_1$. The co-moving transverse dispersion relation then becomes 
\begin{equation}
    2ih_0 \om - \eta k^2 + \theta_1 \om ^2 = 0 \;,
\end{equation}
with solutions for long wavelengths (i.e. small $k$):
\begin{subequations}
\begin{align}
    \om_+ &= -i \frac{\eta}{2h_0}k^2 + \O(k^4) \;,\\
    \om_- &= -2 i \frac{h_0}{\theta_1} + \O(k^2)\;.
\end{align}
\end{subequations}
Because the dispersion relation is now quadratic we obtain two solutions: the ``un-gapped'' mode from  the Smagorinsky model, and an additional gapped mode that appears already in the co-moving frame. The (long wavelength) stability in the un-boosted frame  is guaranteed by taking $\eta>0 $ (as before) alongside $\theta_1>0$. We can further check the stability in the co-moving frame at all wavelengths by means of the Routh-Hurwitz criterion (cf. \cref{app:RH} and \cite{korn2013mathematical}). In order to do so we introduce $\Delta = - i \om$ (to deal with real algebraic equations) and rewrite the dispersion relation as 
\begin{equation}
   \theta_1 \Delta^2 + 2 h_0 \Delta + \eta k^2 = 0 \;.
\end{equation}
Stability requires the solutions to have negative real part $\text{Re} \Delta< 0$. The Routh-Hurwitz criterion then guarantees the stability (at all wavelengths) as long as $\theta_1 > 0$ and $\eta > 0$. These conditions are identical to the ones obtained at long wavelengths. 

In order to check the stability in the boosted frame, we boost the transverse modes dispersion relation (as before) to get
\begin{multline}\label{eq:transverseBoosted}
     \left(\theta_1-\eta v^2  \right)W^2\om^2 + 2 \left(i h_0W + (\eta + \theta_1) W^2 v k\cos{\phi}  \right)\om - \\
    - \left(\eta W^2 k^2\cos^2{\phi}  + \eta k^2\sin^2{\phi} + 2 i h_0 W v k - \theta_1 W^2 v^2 k^2\cos^2\phi\right) = 0 \;.
\end{multline}
To work out the long wavelength stability conditions, we may solve this equation perturbatively. This means that we introduce
\begin{subequations}\label{eq:IterativeExpansion}
\begin{align}
    \om &= \om_0  + \om_1 k+ \om_2 k^2+ \om_3 k^3 \;,\\
    \om^2 &= \om_0^2+ 2\om_0\om_1 k + (\om_1^2 + 2 \om_0\om_2 )k^2 + (2\om_1\om_2 + 2 \om_0\om_3) k^3  \;,\\
    \om^3 &= \om_0^3+3\om_0^2\om_1k + (3\om_0\om_1^2 + 3 \om_0^2\om_2) k^2 + (\om_1^3 + 6\om_0\om_1\om_2 + 3 \om_0^2\om_3)k^3\;,
\end{align}
\end{subequations}
and solve order by order. Solving \cref{eq:transverseBoosted} to lowest order we find two solutions: the first is  the un-gapped mode ($\om_0$ = 0) while the second is given by
\begin{equation}
    \om_0 = - i \frac{2h_0 W^{-1}}{\theta_1 - \eta v^2 }  \,.
\end{equation}
We may focus on the un-gapped mode as we already have the imaginary part (to lowest order) for the gapped mode. Working to first order we obtain the phase velocity, while at second order we get the damping rate. Collecting the results, the small $k$ solutions to the boosted dispersion relation can be written
\begin{subequations}
\begin{align}
    \om &= - i \frac{2h_0 W^{-1}}{\theta_1 -\eta v^2} +  \O(k) \;,\\
    \om &= v k \cos\phi + i \frac{\eta}{2h_0W^3}\left(\cos^2\phi + W^{-2}\sin^2\phi\right) + \O(k^3) \;.
\end{align}
\end{subequations}
We  see that stability in the boosted frame requires $\eta>0$ and $\theta_1> \eta v^2$. 
To the best of our knowledge, stability at all wavelengths cannot be studied analytically in the boosted case. This is because the Routh-Hurwitz criterion applies to real polynomials only. The stability is then perhaps  best studied numerically, once a specific equation of state model has been chosen. However, our demonstration shows that the LES model passes the key stability tests.


Next, we  turn our attention to the longitudinal modes, assuming again that the only non-vanishing parameters in \cref{eq:GenLongMatrix} are $\eta$ and $\theta_1$. The (comoving) longitudinal modes dispersion relation is then 
\begin{equation}\label{eq:ComovingLDisp}
    \om \left[\theta_1 \om^3 + i h_0 \om^2 - \left(\frac{2\eta}{3} +\theta_1 \D\right)k^2 \om - i h_0 c_s^2 k^2 \right] = 0\;.
\end{equation}
As before, we can work out the non-trivial longitudinal modes  using \cref{eq:IterativeExpansion}. Again, working to lowest order we find that one mode is ``gapped'' and two are not. These long wavelength modes are 
\begin{subequations}
\begin{align}
    \om &= -i \frac{h_0}{\theta_1} + \O(k^2) \;, \\
    \om &= \pm c_s k - \frac{i}{3h_0} \left(\eta - \frac{3}{2}(c_s^2 - \D)\theta_1 \right)k^2 + \O(k^3) \;.
\end{align}
\end{subequations}
In order to make sure the longitudinal modes are also stable we have to take $\theta_1 >0,\, \eta > 0$ and 
\begin{equation}\label{eq:UnboostedStabL+T}
    \frac{3}{2} (c_s^2 - \D) < \frac{\eta}{\theta_1} < \frac{1}{v^2} \;.
\end{equation}
We can then study the stability condition at all wavelengths using the Routh-Hurwitz criterion. To do so, we again introduce $\Delta = -i\om$ in \cref{eq:ComovingLDisp} to make it a real algebraic equation: 
\begin{equation}
    \theta_1 \Delta^3 + h_0 \Delta^2 + A \Delta k^2 + h_0 c_s^2 k^2 = 0 \;,
\end{equation}
where $A = 2/3\eta + \theta_1 \D$. From this it is easy to see that the Routh-Hurwitz criterion guarantees stability for \cref{eq:UnboostedStabL+T}. Again, the general $k$ case does not change the stability requirements. 

\begin{figure}\centering
\includegraphics[width=.95\linewidth]{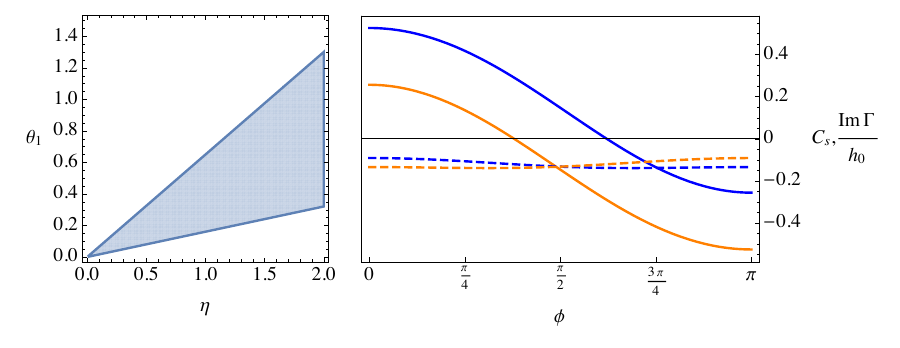}
 \caption{Left: shaded region resulting from combining  the stability constraints obtained for the  i) transverse modes (gapped, un-gapped, boosted, un-boosted), ii) longitudinal un-boosted modes (gapped and un-gapped), and iii) longitudinal boosted gapped modes. 
Right: sound speeds (solid) and damping rates (dashed) for the un-gapped longitudinal modes in the boosted frame. Colours match sound speeds with the corresponding damping rates. 
For illustrative purposes we used: $c_s = 0.16,\, v= 0.4,\,\eta = 1.2,\,\theta_1 = 0.5$ and a barotropic equation of state (both figures). }
  \label{fig:StabilityLES}
\end{figure}

As in the case of transverse modes, the story does not end here. We still have to establish the stability in the boosted frame. To do so, we ``boost'' \cref{eq:ComovingLDisp} using \cref{eq:LorentzBoost} to obtain 
\begin{equation}
    a\,\om^3 + b\, \om^2 + c\,\om + d = 0 \;,
\end{equation}
where
\begin{subequations}
\begin{align}
    a &= W^3 \left(\theta_1 - A\,v^2\right)  \;,\\
    b &= - W^2 \left[(3\theta_1 - 2A) Wvk\cos\phi - ih_0 - WAv^3k\cos\phi + ih_0c_s^2v^2\right] \;,\\
    c &=\begin{multlined}[t] 
    W \Big[(3\theta_1 - 2A)W^2v^2k^2\cos^2\phi - Ak^2\left(1 + W^2v^2 \cos^2\phi\right) \\- 2ih_0 Wv k (1 - c_s^2)\cos\phi\Big] \;,
    \end{multlined} \\
    d &=\begin{multlined}[t]
        - \Big[\theta_1 v^3 W^3\cos^3\phi -W A v k^3 \cos^3\phi - W^3 v^3 A k^3 \cos^3\phi \\
    -i h_0k^2 \left(W^2(v^2 - c_s^2)\cos^2\phi - c_s^2(1 - \cos^2\phi)\right)\Big] \;.
    \end{multlined}
\end{align}
\end{subequations}
Again we solve the problem using the expansion in \cref{eq:IterativeExpansion}. At lowest order we find two ``un-gapped'' modes and one ``gapped'' solution. The latter is given by 
\begin{equation}
    \om = - i \frac{h_0(1-c_s^2v^2)}{(\theta_1 - Av^2)W} + \O(k) \;,
\end{equation}
which is stable for 
\begin{equation}\label{eq:BoostedGappedLStab}
    \eta < \frac{3}{2}\theta_1 \frac{1- \D v^2}{v^2} \;.
\end{equation}
As in the case of the ``un-gapped'' modes, working to $\O(k^2)$ (as the $\O(k)$ problem is trivial) we obtain the boosted sound speed $\om_1 = \C_s$. This is found by solving 
\begin{equation}
    W^2 (1- v^2c_s^2)\C_s^2 - 2W^2 v \cos\phi(1- c_s^2) \C_s + W^2( v^2 - c_s^2)\cos^2\phi  - c_s^2 \sin^2\phi = 0\;.
\end{equation}
To see that this result actually makes sense, we provide the solution for the two cases where $v^a$ is parallel/orthogonal to $\hat k^a$ (respectively)
\begin{subequations}
\begin{align}
    \C_s &= \frac{v\pm c_s}{1\pm v\,c_s}  \;,\\
    \C_s &= \pm\frac{\sqrt{1- v^2}}{\sqrt{1- v^2c_s^2}}c_s \;.
\end{align}
\end{subequations}
The solution for different values of $\phi$  is best understood by considering specific examples, see \cref{fig:StabilityLES}, noting that it only depends on the thermodynamic speed of sound $c_s$ and the relative velocity $v$. 
 
In order to work out the longitudinal mode damping we work at $\O(k^3)$. This  leads to a purely imaginary $\om_2 = \Gamma(\phi)$. The solution involves the boosted sound speed $\C_s$, and it is not particularly illuminating, so  it is also best understood by specific examples, see \cref{fig:StabilityLES}. 

As we  see from the illustrations in \cref{fig:StabilityLES} and \cref{fig:StabilityLES} there are regions of the ($\eta,\theta$) parameter space  where all  modes are stable. As we are only aiming at a  proof of principle (not a comprehensive stability analysis) this concludes the argument. The stability conditions depend on the equation of state (which enters through $c_s,\,\D,\,\C$) and the relative velocity $v$. Therefore, an exhaustive study of the stability in the  general case is  best done once a specific equation of state has been chosen, following the logic outlined here. 

\section{Summary}\label{sec:LESremarks}
Averaging and filtering are the standard strategies for dealing with the problem of simulating (computationally demanding) turbulent flows. Both approaches are complicated by the covariance of General Relativity, where the split between space and time is an observer-dependent notion. As the problem is beginning to be considered from the point of view of numerical relativity (relevant examples are \cite{Giacomazzo,radice1,radice2,aguil,viganoGR, duez}) , it is important to understand the underpinning theory. Hence, we decided to start from the beginning and  considered how the different strategies should be implemented in the curved spacetime setting  relevant for, say, binary neutron-star mergers. 

After clarifying the sense in which consistency with the principles of General Relativity poses interesting foundational questions, we argued that it is natural to set up the analysis in the ``fibration'' associated with individual fluid elements.
This then allowed us to introduce a meaningful local analysis via the use of Fermi coordinates (\cite{fermi1,fermi2}), which defines the covariant averaging/filtering procedure. Building on this, we worked out the coarse-grained fluid dynamics, and considered  the impact  averaging/filtering has on the (thermodynamical) interpretation of the resolved variables.
Finally, because smoothing the fluid dynamics inevitably introduces a closure issue, we proposed a closure scheme and discussed its linear stability. This  completed the formal development of the fibration-based model. 
In order for this work to have practical relevance, however, we need to make contact with actual simulations.
Whilst this goes beyond the scope of this thesis, we will go back to (briefly) comment on this at the end of \cref{ch:LESMHD}.

%% file: Parts/TurbulenceAndLES/LESmhd.tex
\chapter{Filtering relativistic magneto-hydrodynamics}\label{ch:LESMHD}

In this chapter we take the first steps towards extending the covariant filtering scheme discussed in \cref{ch:LES} to charged multi-fluids. 
The relevance of this effort is clear in the context of neutron star astrophysics, as the discussion in the previous chapter left out aspects that are important for (if not crucial to) neutron stars modelling---from electromagnetism to superfluidity/superconductivity. 
Keeping in mind the bigger picture and final aim, it is natural to start by focusing on magneto-hydrodynamics (MHD), which can be derived as the single-fluid limit of an underlying two-fluid plasma model. 

We will start in \cref{sec:IntroMHDTurbulence} with a brief introduction to magneto-hydrodynamics turbulence, mainly focusing on the differences with hydrodynamics. 
The discussion is based on the recent reviews by \citet{SchekochihinRev} and \citet{Beresnyak2019Rev}, which we refer to for more details (and references to the relevant original papers). 
We then introduce the relativistic magneto-hydrodynamics equations in \cref{sec:MHDfibration}. As anticipated, these can be derived starting from a two-fluid plasma model, but we here take a shortcut, namely we start from the Euler-Maxwell system instead. 
We conclude this chapter (and this part) by discussing in \cref{sec:LESMHD} the first steps towards a covariant filtering scheme for magneto-hydrodynamics based, as the approach discussed in \cref{ch:LES}, on the fibration associated with individual fluid elements. 

\section{A brief introduction to magneto-hydrodynamic turbulence}\label{sec:IntroMHDTurbulence}

In a similar fashion as for hydrodynamic turbulence---discussed in \cref{sec:IntroTurbulence}---useful quantities in the characterization of MHD turbulence are the magnetic Reynolds number $\text{Re}_m $ and the Lundquist number $S$. 
These are defined as
\begin{equation}
    \text{Re}_m = \frac{\rho V L }{\beta} \;, \quad S = \frac{L B}{\sqrt{\mu_0 \rho}\beta} = \frac{L v_A}{\beta} \;,
\end{equation}
where $\beta$ is the magnetic diffusivity (or resistivity, proportional to the inverse of the conductivity) and $v_A$ is the Alfvén velocity, whilst $\rho,\,V,\,L$ are characteristic values for the density, velocity and lengthscale of the flow (as in \cref{sec:IntroTurbulence}). 
The magnetic Reynolds number measures the importance of fluid convection over resistive diffusion, while the Lundquist number has a similar interpretation, but compares the Alfvén wave crossing time to the timescale of resistive diffusion.
When these numbers\footnote{One can also define the magnetic Prandtl number $\text{Pm}= \text{Re}_m / \text{Re} = \eta /\beta$, quantifying the importance of fluid viscosity over magnetic diffusivity.} (together with the Reynolds number defined in \cref{eq:Renumber}) are very large, we expect the MHD flow to be turbulent. 

MHD turbulence, however, is expected to be substantially different from hydrodynamics. 
This is intuitive, as we can eliminate a mean flow by a proper choice of frame, while a mean magnetic field cannot be removed by such a transformation. 
In essence, we may say that the magnetic field is the only large-scale feature that does not ``go away'' at small scales.
This simple fact is what makes MHD turbulence \emph{a priori} different from hydrodynamics: the large-scale mean magnetic field makes the system anisotropic.  
This simple fact means we should expect MHD turbulence to be ``more complicated'', as it would not be possible to, for example, justify scaling laws of the form in \cref{eq:Kolm53}---which assume (statistical) homogeneity and isotropy. 

Dynamically, we expect transport along the magnetic field lines (on a scale $l_{{\parallelsum}}$) to be associated with Alfvén wave, while transport across the field lines (on a scale $\lambda$) to be associated with non-linear interactions. 
This leads us to a key feature of MHD turbulence, the so-called \emph{critical balance}. Originally conjectured by \citet{GoldreichSridhar95,GoldreichSridhar97}, critical balance boils down to postulating a balance between parallel and perpendicular transport in the inertial range. 
To see where this brings us, we introduce the Alfvén time and the non-linear time 
\begin{equation}
    \tau_A = \frac{l_{{\parallelsum}}}{v_A} \;, \qquad \tau_{nl} = \frac{\lambda}{ \delta u_\lambda} \;,
\end{equation}
where $\delta u_\lambda$ is the typical velocity increment between points separated by $\lambda$, and assume the two timescales are of the same order $\tau_A \sim \tau_{nl}\sim \tau_c$---where $\tau_c$ is the cascade time, that is the typical time it takes to transfer energy from one (perpendicular) scale to the next. 
Considering perpendicular transport first, and noting that on dimensional grounds we expect the energy spectrum $E \sim \delta  u_\lambda^2 \lambda$, we have on the one hand, 
\begin{equation}
    \veps \sim \frac{\delta u_\lambda^2}{\tau_{c}} \sim \frac{\delta u_\lambda^2}{\tau_{nl}} \Longrightarrow \delta u_\lambda \sim (\veps \lambda)^{1/3} \Longrightarrow E(k_\perp) \sim \veps^{2/3} k_\perp^{-5/3} \;.
\end{equation}
that is, the Kolmogorov spectrum. 
If we focus instead on transport along the magnetic field, a similar logic leads to 
\begin{equation}
    \veps \sim \frac{\delta u_{l_{\parallelsum}}^2}{\tau_c} \sim \frac{\delta u_{l_{\parallelsum}}^2}{\tau_A} \Longrightarrow \delta u_{l_{\parallelsum}} \sim (\veps \tau_A)^{1/2} \Longrightarrow E(k_{\parallelsum}) \sim \frac{\veps}{v_A}k_{\parallelsum}^{-2} \;.
\end{equation}
In essence, we find that critical balance leads to an anisotropic spectrum, with a Kolmogorov type spectrum in the perpendicular directions. 

Critical balance can be justified as follows. 
On the one hand we have the observation (based on theoretical arguments and numerical evidence) that weak MHD turbulence---where perturbation amplitudes are small enough that non-linear interactions are negligible ($\tau_{nl} \gg \tau_A$) and the perpendicular spectrum presents the same scaling as the parallel one---leads naturally to the strong turbulence regime where $\tau_{nl} \sim\tau_A$. 
On the other hand, the opposite regime with $\tau_A \gg \tau_{nl}$ is unsustainable as information in MHD propagates predominantly along the magnetic field lines with velocity $v_A$ (the Alfvén waves) and hence no structure with $l_{\parallelsum}$ larger than $v_A \tau_{nl}$ can be kept coherent and will break up. 
In fact, critical balance appears to be a robust property of MHD turbulence. 

\begin{figure}
    \centering
    \includegraphics[width=0.95\textwidth]{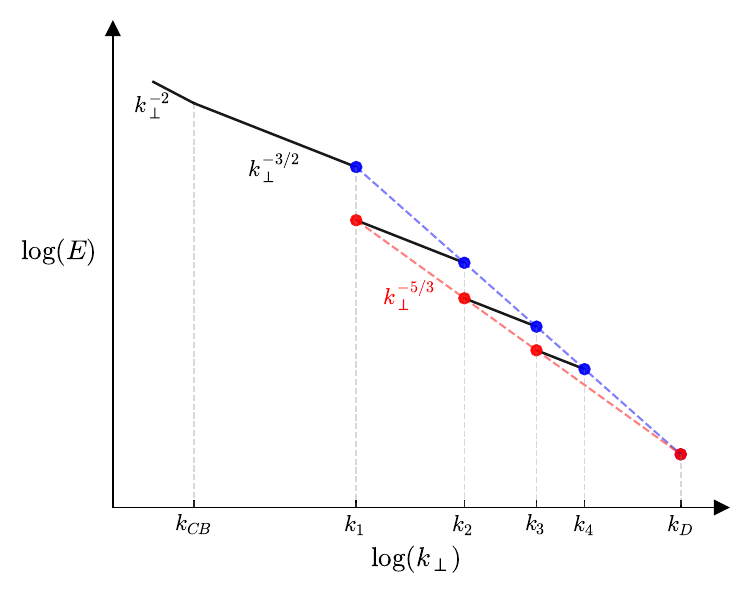}
    \caption{Cartoon of the perpendicular MHD turbulence spectrum. At scales larger than critical balance (i.e. $k < k_{CB}$) is shown the same scaling as in weak turbulence (i.e. $\propto k_\perp^{-2}$). At scales smaller than critical balance is shown the aligned cascade (i.e. $\propto k_\perp^{-3/2}$) periodically interrupted at $k_1,k_2$ and so on, while $k_D$ represents the scale at which dissipative/resistive effects begin to prevail over inertial ones. Figure adapted from \citet{SchekochihinRev}. }
    \label{fig:MHDturb_cascade}
\end{figure}

Whilst the parallel spectrum, with scaling $k_{\parallelsum}^{-2}$, appears to be robust, there is currently no definitive consensus in the community with regards to the perpendicular spectrum. 
Initially, in fact, solar wind observations favoured the $-5/3$ Kolmogorov scaling, but then sets of (direct) numerical simulations started to suggest a different scaling $E(k_\perp) \sim k_\perp^{-3/2}$---although issues with such results have been raised and some authors claim to see a better convergence with the Kolmogorov scaling (e.g. \citet{Beresnyak20115/3}). 
This leads us to another important feature of MHD turbulence, \emph{dynamic alignment}, according to which the velocity and magnetic field tend to shear each other into alignment (in the plane perpendicular to the mangetic field) and hence form sheet-like structures as we approach smaller scales. 
Phenomenological models based on dynamic alignment (and critical balance) can, in fact, reproduce both the parallel spectrum we found above and the $-3/2$ scaling in the perpendicular spectrum. 
The key point, however, is that such sheet-like configurations are not sustainable asymptotically at ever smaller scales due to different known instabilities---including, for example, a magnetized version of the Kelvin-Helmoltz instability (cf. \cref{ch:MRI})---and other non-ideal processes like magnetic reconnection, thus eventually leading to the break up of these sheets into islands. 
When enough of these islands form the flow becomes isotropic, and the cascade starts up again in Kolmogorov form. 
This isotropic-to-sheet-to-islands transition is expected to repeat, giving a periodically interrupted, or intermittent, turbulent cascade (cf. \cref{fig:MHDturb_cascade}). 

In essence, it appears that a complete theory of MHD turbulence should contain (to some degree) a theory of
reconnection, thus making the story even more complicated\footnote{Not to mention the fact that we have here discussed the so-called balanced MHD turbulence regime---where the averaged/total cross helicity  $\vec v \cdot \vec B$, where $\vec v$ is the velocity and $\vec B$ is the magnetic field, vanishes---and did not touch upon the link between turbulence and dynamo processes leading to the amplification of the magnetic field \cite{Brandeburg2005,Rincon2019}.}. 
As for what matters for the present discussion, it is fair to say that sub-grid models of MHD turbulence constitute a very challenging problem because of the local anisotropy and complicated dissipative processes like reconnection \cite{SchmidtLES}.

\section{Magneto-hydrodynamics in the fibration: a shortcut}\label{sec:MHDfibration}

We argued that, from a theory perspective, it would be natural to start the discussion at the level of a charged multi-fluid model, and derive the magneto-hydrodynamics equations as the single-fluid limit of a two-fluid plasma model. 
The underlying multi-fluid nature is, in fact, also important for the development of consistent models beyond ideal. 
This motivates several discussions in the literature, both connecting to the variational (dissipative) multi-fluid framework discussed in \cref{sec:VariationalModels} (see \cite{NilsPRD12,AnderssonCQG2017VariationalPlasmas,BeyondIdealMHD}) as well as linking to ``3+1 formulations'' geared towards numerical implementations (see \cite{BeyondMHD3+1,dMHD31}). 
Our aim here is, however, less ambitious. 
We aim to provide a streamlined derivation of the magneto-hydrodynamics equations to set the stage for discussing extensions of the covariant filtering scheme developed in \cref{ch:LES}.
With this aim in mind, we will start from the Euler-Maxwell system as a proxy for magneto-hydrodynamics. 
We will mainly focus on the electromagnetic degrees of freedom, while the fluid equations are obtained from the stress-energy-momentum conservation law (as before).

The total stress-energy-momentum tensor is written as the sum of a perfect fluid part
\begin{equation}
    T^{ab}_{\text{p.f.}} = \veps u^a u^b + p (g^{ab} + u^a u^b) \;,
\end{equation}
where $u^a$ is the fluid four-velocity, $p$ is the pressure and $\veps$ is the energy density (not to be confused with the energy flux from scale to scale in the previous section), augmented with the Maxwell stress-energy-momentum tensor
\begin{equation}
    T^{ab}_{EM} = \frac{1}{\mu_0}  \left[ F^{ad}F^b_{\phantom{b}d} - \frac{1}{4} g^{ab}F^{cd}F_{cd}\right]\;,
\end{equation}
where $F^{ab}$ is the Faraday tensor. 
The fluid equations are then obtained from\footnote{The last step here involves using the Maxwell equations, \cref{eq:MaxwellCovariant}.}
\begin{equation}\label{eq:MHDproxyFluid}
	\nabla_a T^{ab}_{\text{p.f.}} = - \nabla_aT^{ab}_{EM} = -j_a F^{ab}\;,
\end{equation}
where the term on the right-hand side is the Lorentz four-force and $j^a$ is the charge four-current.
We consider the pressure to be a function of two thermodynamic variables---recall the discussion in \cref{sec:2parEoS} to see why barotropic models are of lesser interest for the present discussion---which we conveniently take as the energy $\veps$ and the baryon density $n$. 
As such, we also need an equation for $n$, given by the continuity equation
\begin{equation}\label{eq:MHDproxyContinuity}
	u^a\nabla_a n + n \nabla_a u^a = 0.
\end{equation}
Before we move on to focus on the electromagnetic degrees of freedom, let us stress that there is a formal inconsistency in the fluid equations we wrote down. 
Deriving the equations of magneto-hydrodynamic as the single fluid limit of a two-fluid plasma model, one can show that a perfect fluid form for the matter stress-energy-momentum tensor holds in the ``centre of momentum'' frame---the analogue of Landau frame in this context---but this would lead to a continuity equation for the baryon current with drift terms\footnote{To get rid of drift terms we would have to work in the (analogue of) Eckart frame \cite{NilsPRD12,BeyondIdealMHD}, but this in turn would lead to momentum flux terms in the fluid stress-energy-momentum tensor.}.
As discussed in detail in \cite{dMHD31}, working with \cref{eq:MHDproxyFluid,eq:MHDproxyContinuity} involves additional steps such as assuming that the electron mass is much smaller than the baryon mass. 
This assumption may or may not apply depending on the physical system under consideration. It is certainly well motivated for non-relativistic systems, but less obvious for a neutron star core model where the electron effective mass may be up to ten per cent of baryon rest mass \cite{comer03:_rel_ent}.
Let us further note that an evolution equation (or a constraint) for the charge four-current is missing from the fluid equations obtained---although the charge current is intuitively associated with a drift velocity between the two species constituting the plasma. 
Simply noting these reservations here, we will work with \cref{eq:MHDproxyFluid,eq:MHDproxyContinuity} as the fluid equations, and move on to focus on the electromagnetic degrees of freedom. 

Assuming we want to work with the electric and magnetic fields, and because the filtering scheme of \cref{ch:LES} is tied to the fluid, it is natural to write down the Maxwell equations in the ``fluid frame'' \cite{livrev}---leading to a (fibration) formulation commonly used in cosmology \cite{Barrow2007}. 
We start from Maxwell equations in covariant form\footnote{
Here we are working using a ``mixture'' of SI units and geometric units.
In the SI system, the electric constant (or vacuum permittivity) is given by $\veps_0 = 1/\mu_0c^2$---where $c$ is the speed of light and $\mu_0$ is the magnetic constant (or vacuum permeability)---and the electric and magnetic field have different units.
This would appear to be in contrast with \cref{eq:FaradayEBDecomp} as $e^a$ and $b^a$ seem to have the same dimension. 
However, this is due to the fact that we are also using (at the same time) geometric units, where the speed of light $c=1$ and suppressed, so that $\veps_0 = 1/\mu_0$.
This detail is however irrelevant for the present discussion.}
\begin{equation}\label{eq:MaxwellCovariant}
    \nabla_aF^{ba} = \mu_0 j^b \;, \qquad \nabla_{[a}F_{bc]} = 0 \;,
\end{equation}
and introduce a four-velocity $U^a$ associated with a generic observer. 
We decompose the Faraday tensor and the charge current as 
\begin{equation}\label{eq:FaradayEBDecomp}
    F_{ab} = 2U_{[a}e_{b]} + \veps_{abc}b^c  \;, \qquad j^a = \sigma U^a + J^a \;.
\end{equation}
where the electric and magnetic field (as measured by $U^a$) are defined as 
\begin{equation}
    e_a =  F_{ab}U^b \;, \;\; b_a =  \frac{1}{2} \veps_{abc}F^{bc} \;, \text{ and } \veps_{abc} = \veps_{dabc}U^d \;.
\end{equation}
With these definitions we can work out the parallel and orthogonal projections---with respect to $U^a$---of  \cref{eq:MaxwellCovariant} and rewrite the Maxwell equations as \cite{EllisCargese,NilsPRD12,livrev}
\begin{subequations}\label{eq:Maxwell_fibration}
\begin{align}
    \perp^{a}_{b}\nabla_ae^b - \mu_0 \sigma &=  2 W^a b_a \;, \\
    \perp^{a}_{b}\nabla_ab^b  &=  -2W^ae_a \;, \\
    \perp_{ab} \dot e^b - \veps_{abc}\nabla^bb^c + \mu_0J_a &= e^b\left(\sigma_{ba} + \omega_{ba} - \frac{2}{3}\theta\perp_{ba}\right) + \veps_{abc}a^bb^c \;, \\
     \perp_{ab}\dot b^b + \veps_{abc}\nabla^be^c &= b^b \left(\sigma_{ba} + \omega_{ba} - \frac{2}{3}\theta\perp_{ba} \right) - \veps_{abc}a^be^c\;,
\end{align}
\end{subequations}
where dots---which stand for co-moving time derivatives---and the projection operator refer to the observer\footnote{ The same is true for the observer four velocity gradients, and we also note for clarity that $W^a$ is the vorticity vector, defined as $W^a =\frac{1}{2}\veps^{abc}\om_{bc}$.} $U^a$.
In \cref{eq:Maxwell_fibration}, the terms on the left-hand side should be familiar, while those on the right-hand side are associated with gradients of the observer four-velocity. 
As such, they vanish identically for an inertial observer, and hence do not appear in most textbook discussions.
We also note that the system of \cref{eq:MHDproxyFluid,eq:MHDproxyContinuity,eq:Maxwell_fibration} is not closed. 
We need an additional equation linking the charge four-current to the other quantities.
This can be derived starting from a two-fluid model \cite{NilsPRD12,BeyondIdealMHD,dMHD31}.
For ideal models, however, this additional equation is often given by a phenomenological Ohm's law based on the standard argument that in a perfect conductor---where charges easily flow---one would expect the electric field to ``short out'' as the matter becomes locally charge neutral \cite{BekensteinOron}. 
Although this is sufficient for getting to a workable (i.e. closed) set of equations \cite{WhiskyMHD,Spritz}, we here want to tread a bit more carefully. 
We do so for two reasons: First, our discussion will allow us to better appreciate the assumptions involved as it connects directly to the Newtonian argument for dropping the displacement current. 
Second, we will derive the induction equation according to a general (i.e. non inertial) observer, which will be used later in \cref{ch:MRI}.

While both the electric and magnetic fields are dynamical degrees of freedom at the level of Maxwell equations, in textbook magneto-hydrodynamics the electric field is demoted to a lesser role and the magnetic field evolution is described by the induction equation.
The Newtonian argument (see, for example, \cite{bellan}) for deriving the induction equation from Maxwell involves assuming that the dynamics is associated with  characteristic length- and timescales, $L$ and $T$,  leading to an associated velocity $V\sim L/T$ which is much smaller than the speed of light. 
With this in mind, we can ``massage'' the Faraday equation in the form
\begin{equation}
    \underbrace{\perp_{ab}\dot b^b - b^b \left(\sigma_{ba} + \omega_{ba} - \frac{2}{3}\theta\perp_{ba} \right)}_{\sim b/T} + \underbrace{\veps_{abc}\left( \nabla^be^c  + a^be^c\right)}_{\sim e/L}  = 0 \;,
\end{equation}
to see that $e \sim L b / T \sim V b$. 
As long as the electric and magnetic fields are slowly evolving, a similar dimensional analysis then leads us to neglecting terms involving the electric field (i.e. the displacement current) in the Ampère law, so that
\begin{equation}\label{eq:noninertial_Ampere_pre}
    J^a = \frac{1}{\mu_0} \left(\veps_{abc}\nabla^bb^c + \veps_{abc}a^bb^c\right) \;.
\end{equation}
We then immediately see where this is going to take us. 
By effectively working with the pre-Maxwell  form of Ampère law (leaving out the displacement current) the charge current is slaved to the magnetic field.
At the same time, we no longer have an evolution equation for the electric field $e^a$, so we need an additional relation between the electric and magnetic fields.
This is where the issue of (infinite) conductivity enters the discussion.
Connecting the local fluid four-velocity $u^a$ to the observer $U^a$ as\footnote{It would seem natural, in writing down the Maxwell equations in the fibration formulation, to identify the fibration observer as the local fluid four-velocity \citep{Barrow2007,livrev}. However, as we want here to make contact with the discussion in \cref{ch:LES}, it is natural to distinguish the two (cf. discussion in \cref{sec:average}).}
\begin{equation}
	u^a = \gamma(U^a + v^a) \;, \quad U^a v_a = 0 \;, \gamma = \left(1 - v^a v_a\right)^{-1/2} \;,
\end{equation}
where $v^a$ is the spatial fluid velocity as measured by the observer $U^a$, we see that the electric field as measured by the fluid is linked to $e^a,\,b^a$ (measured by $U^a$) as 
\begin{equation}
    F_{ab} u^b = \gamma \left[e_a + \veps_{abc} v^b b^c + U_a (v^b e_b)\right]\;.
\end{equation}
Assuming this vanishes yields
\begin{equation}\label{eq:ideal_Ohm}
    e_a + \veps_{abc}v^bb^c = 0 \;.
\end{equation}
With this constraint, we can derive the induction equation from Faraday's law.
To do so, note that 
\begin{multline}\label{eq:3LCcontraction_fibr}
    \veps_{abc}\veps^{cde} = U^fU_g\veps_{fabc}\veps^{gdec} = -3!U^fU_g\delta^{[g}_f\delta^{d}_a\delta^{e]}_b \\ =
    \left(\delta^d_a\delta^e_b - \delta^e_a\delta^b_d\right)- \left({\parallelsum}^d_a\delta^e_b - {\parallelsum}^e_a\delta^d_b\right)-\left(\delta^d_a{\parallelsum}^e_b-\delta^e_a{\parallelsum}^b_d\right) \;,
\end{multline}
where we introduced the parallel projection ${\parallelsum} ^a_b = -U^a U_b$. 
When this is contracted with a spatial tensor (with respect to $U^a$) the last two terms in \cref{eq:3LCcontraction_fibr} can be dropped. 
It follows that 
\begin{equation}
    \veps_{abc}\veps^{cde}\left(a^b v_db_e\right) =  \left(a^b b_b\right)v_a - \left(a^b v_b\right) b_a  \;.
\end{equation}
We also need to take care of the curl of $e^a$ term in the Faraday equation. 
This can be written
\begin{equation}
    -\veps_{abc} \nabla^b\left(\veps^{cde}v_db_e\right) = \left[ -\veps_{abc} \left(\nabla^b\veps^{cde}\right)v_db_e\right] - \left[ \veps_{abc} \veps^{cde}\nabla^b \left( v_db_e\right)\right] \;,
\end{equation}
where it is convenient to consider the two terms separately. We start from the second term and, even if $U^a$ is not necessarily surface forming (i.e. has non-vanishing vorticity), we introduce a ``spatial'' covariant derivative $D$ in the usual way (projecting each index in the sub-space orthogonal to $U^a$). 
Then, it is easy to see that 
\begin{equation}
    \veps_{abc}\veps^{cde}D^b\left(v_db_e\right) \doteq \veps_{abc}\veps^{cde} \left(\perp^b_f\perp^g_d\perp^h_e\right)\nabla^f \left(v_gb_h\right) = \veps_{abc}\veps^{cde} \nabla^b \left(v_db_e\right) \;,
\end{equation}
and hence
\begin{equation}
    -\veps_{abc}\veps^{cde} \nabla^b \left(v_db_e\right) = D^b\left(v_bb_a\right) -D^b\left(v_ab_b\right) \;.
\end{equation}
As for the other term, writing it it as
\begin{multline}
    -\veps_{abc} \left(\nabla^b\veps^{cde}\right)v_db_e =-U^g\veps_{gabc}\veps^{fcde}\left(\nabla^b U_f\right) v_db_e \\
    =  -U^g\delta^{[f}_g \delta^d_a\delta^{e]}_b\,g_{fh}\left(-U^ba^h + \om^{bh} + \sigma^{bh} + \frac{1}{3}\theta\perp^{bh}\right)v_db_e \;,
\end{multline}
we see that---given the anti-symmetrization---it vanishes identically.
In summary, the induction equation according to a generic observer\footnote{The worldlines of the generic observer $U^a$ constitute a fibration of the spacetime, hence we may call this the ideal induction equation in the fibration framework, as opposed to the corresponding ``3+1 form'' derived by, for example, \citet{dMHD31}.} can be written as 
\begin{equation}\label{eq:RelIndNI}
    \perp^{ab}\dot b_b  + D_b(v^b b^a) - D_b (v^a b^b)=  \left(\sigma^{ab} - \omega^{ab} - \frac{2}{3}\theta\perp^{ab}\right)b_b + v^a (a_b b^b) - b^a (a_b v^b) \;,
\end{equation}
where the terms on the left should be familiar, while those on the right vanish for an inertial observer. 

Let us now pause for a second, and ponder the implications of the argument we put forward---which is the intuitive extension to the curved spacetime setting of the Newtonian one. 
Clearly, the argument is non-controversial at the Newtonian/non-relativistic level. 
At the relativistic level, however, dropping the displacement current might not be fully justified, at least not in general.
Whilst this may be seen as a flaw in the logic, it is actually the reason why we derived the induction equation this way. 
We could have, in fact, argued for a relation of the form in \cref{eq:ideal_Ohm} from the beginning, thus arriving at the same induction equation but ``sweeping under the rug'' the controversial issue of dropping the displacement current.
The derivation provided suggests that, in a sense, magneto-hydrodynamics is intrinsically---with a slight abuse of nomenclature---a ``post Newtonian'' theory as it necessarily involves a low-frequency/low-velocity approximation (the timescale over which the local electric field shorts out is tiny but not zero). 
In essence, we have to apply the
magneto-hydrodynamics approximation with some level of caution. 
We also point to the discussion in \cite{dMHD31} where the same argument is detailed in the so-called ``3+1 formulation''---thus showing in which sense the approximation may still be used (on a case by case basis) and how gauge issues (in the choice of lapse and shift) play a key role for the validity of the approximation.

We conclude this section by showing how the derived equation further simplifies in the Newtonian limit. 
On dimensional grounds, we observe that the last two terms on the right hand side of \cref{eq:RelIndNI} contain an extra factor of $1/c^2$ (with respect to the rest, where $c$ is the speed of light), and will as a result be negligible in the non-relativistic limit ($c^2\to \infty$). 
Similarly, let us consider the absence of monopoles constraint. 
From \cref{eq:Maxwell_fibration} and \cref{eq:ideal_Ohm} we immediately obtain
\begin{equation}
    \perp^a_b \nabla_ab^b = 2 W^a \veps_{abc} v^b b^c \;,
\end{equation}
and we observe that the term on the left hand side is $\sim b/L$ while that on the right is $\sim b L/T^2$.
Dimensional consistency implies the term on the right-hand side contains an extra factor of $1/c^2$ and should be neglected in the Newtonian limit.
In essence, non-inertial effects do not affect the absence of monopoles constraint at the Newtonian level. 
When it comes to the Lorentz force, we expect it not to change at the Newtonian level, but let us nonetheless check this for consistency. 
The Lorentz four-force can be written
\begin{multline}
     -j_bF^{ba} = - \left(\sigma U_b + J_b \right)\left(U^b e^a + U^a e^b + \veps^{bacd} U_c b_d\right) \\
     = -U^a \left(J_b\veps^{bcd}v_cb_d\right) + \veps^{abc}\left( J_b-\sigma v_b \right)b_c \;,
\end{multline}
where we used the ideal magneto-hydrodynamics relation \eqref{eq:ideal_Ohm} in the last step.
The Lorentz three force corresponds to the second term, where the charge density is measured by the observer, hence does not (in general) vanish.
However, if we insist on the local charge density to be zero (consistently with \eqref{eq:ideal_Ohm}), then we have 
\begin{equation}
    -u^a j_a = W(\sigma - v_a J^a) = 0 \Longrightarrow J_b - \sigma v_b = \left(g^a_b - v^a v_b\right)J_a \;.
\end{equation}
Re-inserting the factor of $1/c^2$ we see that the second term is negligible with respect to the first.
As also the second term in  \cref{eq:noninertial_Ampere_pre} is negligible in the Newtonian limit, we see that the Lorentz force in the Euler equation is unchanged (as expected).

\section{MHD covariant filtering: first steps}\label{sec:LESMHD}

We now want to make contact with the filtering strategy developed in \cref{ch:LES}.
Whilst this is very much unfinished business, we discuss some of the issues that arise along the way, and what seems to be the right strategy to make progress.

First of all, we note that it does not seem to make much sense to take the MHD-approximation---that is, neglecting the displacement current---first and then applying the filtering procedure. 
The main reason for this has to do with the choice of filtering observer---recall the discussion of the normalization issue in \cref{sec:average}---and the simple fact that electric and magnetic fields are observer dependent quantities.
That such a strategy would be problematic can also be appreciated by considering the filtered charge current, noting that the issue of local charge neutrality and vanishing electric field are intrinsically linked \cite{dMHD31}.
Using the filtering notation introduced in \cref{ch:LES}, we can consider the filtered charge four-current and decompose it as 
\begin{equation}
    \la j^a \ra = \tsig \tu^a + \tJ^a\;, \quad \tsig = \tu_a \la j^a \ra \text{ and } \tJ^a = \tperp^a_b \la j^b \ra \;,
\end{equation}
where $\tu^a$ is the filtered four-velocity (say, the Favre-filtered velocity).
Using an analogous decomposition of the fine-scale charge-current with respect to the fine-scale fluid velocity we get 
\begin{equation}
    \tsig = \tu_a \left( \la\sigma u^a\ra + \la J^a\ra\right) \;.
\end{equation}
The upshot is that even if we assume the system to be charge neutral at the fine-scale (i.e. $\sigma = 0$), we are left with 
\begin{equation}
    \tsig = \tu_a  \la J^a\ra \;,
\end{equation}
which clearly does not have to vanish.
Conversely, if we impose charge neutrality on the filtered flow then this condition may not hold on the small scale (which might be an issue for equation of state inversion in numerical relativity simulations). 

The upshot is that the correct strategy seems to involve a filtering at the level of the Maxwell equations. 
We can then use the fact that the filtering procedure defined in \cref{ch:LES} commutes with partial derivatives to arrive at 
\begin{equation}
    \nabla_a \la F^{ba} \ra = \mu_0 \la j^b\ra \;. 
\end{equation}
These equations can then be written in terms of the coarse-grained electromagnetic fields---defined decomposing $\la F^{ab}\ra$ as in \cref{eq:FaradayEBDecomp}, now in terms of the filtered velocity $\tu^a$. 
The net result is that the filtered Maxwell equations retain the usual form (as the Maxwell system is linear). 
In particular, the filtered electric and magnetic field are given in terms of the fine scale ones as
\begin{equation}
    \tE^a = \tu^b\la u_ae_b \ra -\tu^b\la u_be_a\ra -\teps_{abc} \la u^b b^c \ra \;,
\end{equation}
and
\begin{equation}
    \tB^a = \tu^b\la u_ab_b \ra -\tu^b\la u_bb_a\ra + \teps_{abc} \la u^b e^c \ra\;.
\end{equation}
From this, we immediately see that even if we take the fine-scale electric field to vanish (due to the assumption on local charge neutrality) we would still have a non-vanishing coarse grained electric field:
\begin{equation}
    \tE^a = - \teps_{abc}\la u^b b^c \ra \;, \qquad \tB^a = \tu^b \left(  \la u_ab_b \ra - \la u_bb_a \ra\right) \;.
\end{equation}
Nonetheless, we may, pragmatically, move on and take the MHD-approximation at the filtered level (i.e. neglecting the displacement current in the filtered equations) and provide a closure relation for the filtered electric field. 
For example, we could model the filtered electric field using an algebraic decomposition in terms of the filtered variables as 
\begin{equation}
    \tE^a = \gamma_1 \tB^a + \gamma_2 \tJ^a + \gamma_3 \teps^{abc}\tB_b \tJ_c \;.
\end{equation}
Whilst this is a very simple closure scheme, it is attractive as it allows for an immediate physical interpretation of the expansion coefficients. 
It is, in fact natural to interpret the second term as an effective resistivity and the third one as an effective Hall term.
The first term instead reminds us of the ``classic'' alpha-dynamo term, which enters many mean-field-dynamo models  \cite{BucciantiniDelZanna2013,Brandeburg2005,Rincon2019}. 

We conclude this section by commenting on the filtering of the fluid equations. 
When it comes to the baryon continuity equation, it is natural to work with the Favre-filtered four-velocity as in \cref{sec:filter} so that it retains the pre-filter form. 
As for the (fluid part of the) stress-energy-momentum tensor, we know already from the hydrodynamic analysis in \cref{ch:LES} that filtering will introduce additional terms in the equations that can be interpreted as effective dissipative terms.
In addition to these, we also have terms coming from filtering the Lorentz four-force
\begin{equation}
    \F^a = \la j_b F^{ba}\ra - \la j_b \ra \la F^{ba}\ra \;.
\end{equation}
This contribution also requires modelling. 

\section{Outlook}\label{sec:MHDLESsummary}

As anticipated at the end of the previous chapter, an important point missing from the current discussion is the link with actual numerical relativity simulations. 
These tend to involve a foliation of spacetime and the associated “3+1” spacetime split \cite{Gourgoulhon3+1,RezzollaZanotti}, thus adding a new observer (the ``Eulerian'' observer) to the game.
In principle, one might make progress simply by ``translating'' our results to the foliation picture, but it is by no means clear that this is the most sensible way to proceed. 
Moreover, any discussion of numerical simulations should consider a number of additional issues, such as the role of numerical discretization errors.
Models developed directly in the foliation---such as the gradient sub-grid model developed in \cite{viganoNR,carrasco}---appear to be more suited for dealing with issues like the lack of numerical resolution. 
At the same time, the fibration scheme we developed provides a clear link with the underlying thermodynamics and the equation of state information we would like to extract from numerical simulations. 
Moreover, the covariance of our framework, as well as the fact that the filtering observer acquires a clear physical meaning,  allow us to ``lift'' the results of box simulations to any space-time (at least in principle). 

In essence, the scheme we have discussed suggests that the large-eddy strategy can be enhanced to a tool for linking models valid at ``mesoscopic'' and ``macroscopic'' scales. 
This is relevant as it appears that large-eddy simulations based on gradient sub-grid models need to minimally resolve the relevant physical effects in order to describe turbulence at even smaller scales---as demonstrated by \citet{Miravet-Tenes:2022ebf} for the specific case of the magneto-rotational instability (cf. \cref{ch:MRI}). 
The (although minimal and incomplete) discussion provided here is a first step in this direction.

%% file: Parts/BNSapplications/part3macro.tex
\input{Parts/BNSapplications/BVinSim}
\input{Parts/BNSapplications/MRI}

%% file: Parts/BNSapplications/BVinSim.tex
\chapter{Formulating bulk-viscosity for neutron star simulations}\label{ch:BVinSIM}

In the previous parts of this thesis, we focused on modelling dissipation and turbulence in general relativistic fluids.
This part is instead dedicated to neutron star merger applications. 
In particular, we will consider in this chapter the role of bulk viscosity associated with nuclear reactions---which may, or may not, leave an observable imprint on (say) the gravitational-wave signal \cite{Most2021BV,PeteGWreactions,Most2022BV,Radice+THCM1,Zappa+BVM1test}---and ask how this mechanism can be implemented in nonlinear simulations. 
We highlight the formal aspects of the problem and establish how the inevitable ``limitations'' of a numerical simulation (in terms of resolution) enter the discussion. 
The aim is to establish to what extent simulations based on an effective bulk viscosity are viable and (perhaps more importantly) when they are not. 
This understanding will be crucial for future numerical implementations.  
The results discussed in this chapter have been published in \cite{BV_in_sim}. 

In the following chapter we discuss the magneto-rotational instability (MRI) using a (Newtonian) local analysis. 
We do so as the MRI is thought to play a key role in the development of (MHD) turbulence in the outer envelope of a merger remnant. 
In particular, the discussion we provide is suited for highly-dynamical systems such as mergers, highlighting the importance of global properties for the standard results (and criteria) to be valid.

\section{Simplifications must be made}
The underlying physical model of a neutron star merger is expected to be a system of multiple interacting ``fluids'' of different charged particle species, coupled to an electromagnetic field and radiation through, for example, neutrinos, all evolving on a dynamical relativistic spacetime \cite{livrev}. The complexity of this model makes it impractical for use in either theoretical calculations or numerical simulations. Instead, simplifications must be made, with heuristic arguments needed to justify each assumption that is introduced.

To illustrate the key argument with a simple toy model, consider the problem of heat propagation. When the underlying model is required to be \emph{causal}, the starting point is often taken to be the Cattaneo equation (see \cref{ch1subsec:EITCattaneo})
%
%
%
\begin{equation}\label{eq:cattaneo}
	\tau \pdv{q}{t} + q = - \kappa \pdv{T}{x} \;.
\end{equation}
In the ``fast relaxation limit'', when the relaxation time $\tau \to 0$, we recover Fourier's law,
relating the heat flux to temperature gradients, and leading to the familiar heat equation. While the underlying model is hyperbolic, the fast relaxation limit is parabolic and hence not causal.
Specifically, working out the characteristic velocities in the problem\footnote{Computed, for example, taking the large wavenumber limit of the (real part of the) phase speed.} one finds that the Cattaneo equation~\eqref{eq:cattaneo} is causal with finite propagation speeds bounded by $\pm (\kappa / \tau)^{1/2}$. At the same time, there is a critical wavenumber $\gtrsim (\kappa \tau)^{-1/2}$ below which the behaviour is purely parabolic and the solution is diffused away. An illustration of the transition from second sound to diffusion can be found in Figure 16 of~\cite{livrev}.

From a theory point of view it would be natural to argue that we should base our models on the Cattaneo formulation, but from a numerical perspective this may be problematic. We would need to resolve the (presumably fast) relaxation towards equilibrium and this may not be possible/practical. In this sense, the parabolic prescription may be preferable. 

A heuristic argument for using the parabolic heat equation within a relativistic model---for which causality would be a prerequisite---would be as follows. We assume that on the length scales $L$ relevant for our model we have $\tau \sim L / c$, where $c$ is the speed of light. By causality there can be no (propagating!) scales of physical relevance faster than $c$, hence with timescales smaller than $\tau$ or (equivalently) frequency scales larger than $\tau^{-1}$. Therefore the only relevant behaviour for heat propagation is the purely parabolic case where heat fluctuations are rapidly damped, which is well modelled by the standard heat equation.
It is possible to use the internal consistency of the underlying model to check when this heuristic argument is valid. For example, to be consistent with causality the dispersion relation at low frequencies requires that $\kappa \le \tau c^2 \sim c L$.

The key point here is the existence of a single scale (length, time, or frequency) at which some physical effect acts or changes type. This issue is often considered for turbulence (for example, are the length scales probed sufficient to trigger the magneto-rotational instability \cite{Duez+MRI,SiegelMRI,KiuchiMRI,GuiletMRI}), for reactions (are physical regions of parameter space probed so that out-of-equilibrium physics, such as a  bulk viscosity has to be accounted for \cite{PRLAflordBulkMergers}), and for radiation (are neutrinos propagating or trapped, see e.g.\ \cite{PeteThermal}). When this key scale is outside that which can be included in the model then the physical effect is either ignored (by using a purely ideal model, or by assuming instantaneous relaxation to an equilibrium) or modelled (by approximating the additional physics through a closure term, such as an effective equation of state, or a large-eddy approach \cite{PeteThermal,Giacomazzo,radice1,carrasco}).

The fundamental issue regards how the physics, and hence the model, should behave when key scales cross or overlap at different parts of the required parameter space. This is particularly relevant for nonlinear numerical simulations, where the discretization length introduces a scale (or multiple scales with uneven grids or mesh refinement), and nonlinearity can lead to the relevant physical scales varying over many orders of magnitude. The interaction between the different scales makes  heuristic simplifications dubious.

With these issues in mind we will study, analytically and numerically,  issues relating to bulk viscosity in reactive fluids. Here the underlying model involves nuclear reactions, specifically the direct and modified Urca reactions (although the analytical calculations presented apply more generally). These microphysical reactions can, in some regimes, be approximated as a (resonant) bulk viscosity \cite{Andreas_WB_chapter}. However, the timescales on which the reactions take place strongly depend on (for example) the temperature and may as a result be close to those that can be captured in numerical calculations. The timescales may also interact with, for example, large eddy closure terms required for turbulent regions (cf. \cref{sec:LES}).

\section{The reactive system}\label{sec:hydro-thermodynamics}

We want to consider the hydrodynamics of an isotropic reactive system consisting of comoving neutrons, electrons and protons, the simplest meaningful matter composition for a neutron star core.\footnote{This may seem somewhat reductionist, given that the high density region is likely to bring other matter constituents into play, but our main interest is to establish the principles involved. If richer matter models are required then the extension of our discussion  will be conceptually straightforward.} Further, because we assume charge neutrality, there are only two independent number densities (or, equivalently, one density and one species fraction). These are conveniently taken as $n$, representing the baryon number density, and $Y_\e = {n_\e}/{n}$ the electron (lepton) fraction. The proton fraction then follows as $Y_\p = n_\p/n = Y_\e$ while the neutron fraction is given by $Y_\n = n_\n/n = 1 - Y_\e $. Because we assume isotropy, the stress-energy-momentum tensor takes the perfect fluid form (cf. \cref{sec:PerfectFluids})
\begin{equation}\label{eq:SEMtensor}
	 T^{ab} = \veps  u^a u^b +  p (g^{ab} + u^a u^b) \;,
\end{equation}
where the four-velocity $u^a$ is uniquely defined as the one associated with the flow of all particle species. The pressure and energy are identified with the corresponding thermodynamical quantities. As we will not impose that the system is in chemical equilibrium (we allow reactions) we assume the equation of state to involve three parameters. That is, the pressure follows from $p = p (n,\veps, Y_\e)$. This relation is assumed to be provided in tabulated form (suitable for a numerical simulation, see \cref{app:compOSE} and \cite{compOSE}). 
The energy-momentum conservation laws and baryon continuity are the same as in \cref{sec:PerfectFluids}. 
Nonetheless, we repeat them here for convenience:
\begin{subequations}\label{eqs:Euler}
\begin{align}
    u^b\nabla_b \veps &= - (p + \veps) \theta \;,\\
    (p+ \veps) a_b &= - \perp_b^c \nabla_c p \;,
\end{align}
\end{subequations}
and
\begin{equation}\label{eq:baryonCont}
	u^a \nabla_a n + n \theta = 0 \;.
\end{equation}
As we are presently working with a three parameter equation of state, we need an evolution equation for the electron fraction. 
We write this as 
\begin{equation}\label{eq:ElectronFractionEq}
	u^a \nabla_a Y_\e = \frac{\Gamma_\e}{n}\;,
\end{equation}
where the rate $\Gamma_\e$ is generally non-vanishing as we are considering a reactive system. Once the reaction rate is provided by the microphysics (and tabulated as a function of the other variables), these equations constitute a closed system. 

From the perspective of  thermodynamics, it is natural to introduce the affinity 
\begin{equation}
    \beta = \mu_\n - \mu_\p - \mu_\e\;,
\end{equation}
as it quantifies how far the system is out of cold beta equilibrium. This has the advantage that we can work within the so-called Fermi surface approximation \cite{AlfordHarris18} and express relevant quantities, like the reaction rates, as expansions for small values of $\beta$ (with the coefficients in the expansion evaluated at equilibrium, $\beta=0$). In the following, and notably in the illustrations we provide, we assume that this strategy is appropriate. Pragmatically, this makes sense as we are only aiming to establish a proof of principle and these assumptions allow us to work out all required parameters for the model from a standard tabulated equation of state. However, it is important to keep in mind that the assumptions will not be appropriate for much of the parameter space (temperature and density) explored in the binary neutron star merger/post-merger phase, and they completely exclude any neutrino effects. At finite temperatures, the true notion of beta equilibrium is more complex, and may require the addition of an isospin chemical potential in the definition of $\beta$ (see \cite{AlfordHarris18,AlfordHarrisDamping18,Alford_Zhang21,AlfordHarutyunuanSedrakian19,AlfordHarutyunuanSedrakian21,PeteThermal}). A complete model should account for the correct notion of equilibrium, but  this will require the equation of state table to be extended to include all necessary information. As such data is not yet available for simulations, we are (pragmatically) doing the best we can given the information at hand.

As we will see in \cref{subsec:ThermoReactive}, the affinity is (thermodynamically) conjugate to $Y_\e$, meaning that either of the two variables can be used ``equivalently'' in the discussion. This is important because state-of-the-art simulations tend to involve $Y_\e$ while the theory is somewhat more transparent when expressed in terms of $\beta$. In the following we will develop the model both in terms of $\beta$ and $Y_\e$, showing the expected consistency, and emphasising the different perspectives the two complementary approaches bring on the problem. The evolution equation for $\beta$ is easily obtained by considering it as a function of $(\veps,n,Y_\e)$------which follows from $\beta$ and $Y_\e$ being thermodynamically conjugated. We arrive at 
\begin{equation}\label{eq:BetaEqGeneral}
	u^a \nabla_a \beta = \left(\frac{\partial \beta}{\partial Y_\e}\right)_{n,\veps}\frac{\Gamma_\e}{n} - n\B\theta \;,
\end{equation}
with 
\begin{equation}\label{eq:DefB}
    \B = \left(\frac{\partial \beta}{\partial n}\right)_{\veps,Y_\e} + \frac{p+\veps}{n}\left(\frac{\partial \beta}{\partial \veps}\right)_{n,Y_\e} .
\end{equation}
We again see that the system of equations is closed once the reaction rate (as well as the relevant thermodynamical coefficients) is provided. 

A useful simplification occurs when the system is sub-thermal, when\footnote{Let us note for clarity that we use units where the Boltzmann constant $k_B =1$.} ${\beta}/{T}\ll 1$. Then we can expand the rate with respect to chemical equilibrium $\beta = 0$ to write it as\footnote{There is a sign convention here, and we are following \cite{AlfordBulk10}. The logic is, if $\beta > 0 $ then $\mu_\n >\mu_\e + \mu_\p$ and neutron decay is favoured (over electron-capture) as this will release energy. Therefore, we want the electron rate to be positive when $\beta$ is positive and vice versa. The sign of $\gamma$ should then be negative.} $\Gamma_\e = -  \gamma \beta$. The evolution equation for the affinity $\beta$ then simplifies to
\begin{equation}\label{eq:BetaEquationSubthermal}
	u^a \nabla_a \beta = - \A \beta - n\B \theta\;,
\end{equation}
where we introduced the new coefficient (with units of inverse time)
\begin{equation}\label{eq:Adef}
    \A = \frac{\gamma}{n}\left(\frac{\partial \beta}{\partial Y_\e}\right)_{n,\veps} \;.
\end{equation}
The information encoded in the reaction rate $\Gamma_\e$ is now ``stored'' in $\gamma$. We can then make progress and compute $\gamma$ from the equation of state tables provided in the compOSE database \cite{compOSE} \emph{as long as} we ignore finite temperature effects \cite{AlfordHarris18,AlfordHarrisDamping18,Alford_Zhang21,AlfordHarutyunuanSedrakian19,AlfordHarutyunuanSedrakian21} (see \cref{app:compOSE} and \cite{PetePHD} for more details).
While the coefficient $\B$ can be introduced without reference to an expansion around equilibrium, this is not the case for $\A$. In the sub-thermal limit we  retain only terms linear in $\beta$, so that $\A$ must be evaluated at $\beta = 0$. 

For completeness, let us also comment on the entropy density, viewed as a function of $(\veps, n, Y_\e)$. Using the equations of motion for these quantities (as well as the generalized Gibbs relation provided in \cref{sec:hydro-thermodynamics} below) we arrive at 
\begin{equation}
  T \nabla_a \left( su^a \right) =  \beta\Gamma_\e\;.
\end{equation}
As long as $\beta$ has the same sign as $\Gamma_\e$ the entropy  increases. This is guaranteed to be the case in the sub-thermal limit if $\gamma < 0$. Note that this assumes that a negligible amount of energy is deposited in neutrinos by the reactions, which will be a poor approximation at high temperatures. This important caveat will quantitatively affect our results without changing the qualitative conclusions.

\subsection{Thermodynamics of a reactive system}\label{subsec:ThermoReactive}

Having discussed the hydrodynamics, let us turn to the associated thermodynamics. Because the underlying model is required to be causal, and hence based on Cattaneo-type laws, it is natural to set the discussion within the Extended Irreversible Thermodynamics (EIT) framework (cf. \cref{ch1subsec:EITCattaneo}). 
We here provide a streamlined discussion, and refer to the monograph \cite{EIT} and references therein (see also \cite{GavassinoBulk} for a recent analysis focused specifically on bulk viscosity).
The first step is to \underline{assume} that the Gibbs relation takes the usual form---noting that the various quantities may not be in thermodynamical equilibrium
\begin{equation}\label{eq:OutEqGibbs}
    p+\veps = \sum_{\x = \n,\p,\s,\e} n_\x \mu_\x = n\mu_\n - n_\e \beta + T s \;,
\end{equation}
and
\begin{equation}\label{eq:pressuredifferential}
    dp =\sum_{\x = \n,\p,\s,\e} n_\x d\mu_\x  = n d\mu_\n - n_\e d\beta + s dT .
\end{equation}
As we will be working at the fluid level with either $(n,\veps,Y_\e)$ or $(n,\veps,\beta)$, we can use the entropy as a thermodynamical potential. This is convenient because if we also assume that the system is close to thermodynamical equilibrium, we can  expand the entropy as 
\begin{equation}
    s = s^\mathrm{eq}(n,\veps) + \frac{1}{2}s_2(n,\veps) \beta^2 \;, \quad \text{where } s_2 = \left(\frac{\partial^2s}{\partial\beta^2}\Big|_{\beta = 0}\right)_{n,\veps} .
\end{equation}
From this we can compute the out-of-equilibrium expansion of the thermodynamical variables. Linearizing in deviations from equilibrium, and assuming that the equation of state is expressed in terms of $Y_\e$ rather than $\beta$, we obtain
\begin{subequations}\label{eq:expansionsWithBeta}
\begin{align}
    T &= T^\mathrm{eq} \left[ 1 -n \left(\frac{\partial \beta}{\partial \veps}\right)_{n,Y_\e} \left(\frac{\partial \beta}{\partial Y_\e}\right)_{n,\veps}^{-1} \beta\right]  + \O(\beta^2) \;,\\
    \mu_\n &= \mu^\mathrm{eq} - \left[ n \mu^\mathrm{eq} \left(\frac{\partial \beta}{\partial \veps}\right)_{n,Y_\e} + n  \left(\frac{\partial \beta}{\partial n}\right)_{\veps,Y_\e}- Y_\e \right]\left(\frac{\partial \beta}{\partial Y_\e}\right)_{n,\veps}^{-1} \beta + \O(\beta^2) \;, \\
    s&= s^\mathrm{eq} + \frac{1}{2} \frac{n}{T^\mathrm{eq}}\left(\frac{\partial \beta}{\partial Y_\e}\right)_{n,\veps}^{-1}\beta^2 \;.
\end{align}
\end{subequations}
Note that the thermodynamical requirement on the entropy reaching a maximum at equilibrium, namely $s_2 < 0$, implies that ${\partial \beta}/{\partial Y_\e}$ must be negative. Recalling \cref{eq:BetaEquationSubthermal,eq:Adef}, and the fact that the restoring term $\gamma<0$, we see that $\A >0$ and therefore plays the role of an (inverse) relaxation rate.

Now that we have  expansions (in $\beta$) of the thermodynamical variables, we can use the Gibbs relation to work out the pressure. To linear order in the deviation from equilibrium we then have 
\begin{equation}
    p = p(n,\veps,\beta) = p^\mathrm{eq} + p_1 \beta \;,\quad \text{where } p^\mathrm{eq} = p(n,\veps,\beta = 0) \;,\; \text{and } p_1 = \left(\frac{\partial p}{\partial \beta}\Big|_{\beta=0}\right)_{n,\veps} \;,
\end{equation}
or, explicitly, 
\begin{align}
    p^\mathrm{eq} = -\veps + n\mu^\mathrm{eq} + T^\mathrm{eq} s^\mathrm{eq} \;,
\end{align}
and 
\begin{equation}\label{eq:p1WithBeta}
    p_1 = - n^2 \left(\frac{\partial \beta}{\partial Y_\e}\right)_{n,\veps}^{-1} \left[\frac{p^\mathrm{eq} + \veps}{n}\left(\frac{\partial \beta}{\partial \veps}\right)_{n,Y_\e} + \left(\frac{\partial \beta}{\partial n}\right)_{\veps,Y_\e} \right] = \frac{\gamma}{n}\frac{\B}{\A}\;.
\end{equation}
In essence, the thermodynamical expansion provides us with an expression for the out-of-equilibrium contribution to the pressure, which would naturally be interpreted as a bulk viscosity. We identify
\begin{equation}\label{eq:Pi_t}
    \chi_t = \left(\frac{\partial p}{\partial \beta}\Big|_{\beta=0}\right)_{n,\veps} \beta = p_1 \beta \;.
\end{equation}

Note that, even though we have outlined the derivation in the simplest case (where the system is subthermal and close to equilibrium), the argument applies more generally. A broader discussion would rely on a detailed description of the out-of-equilibrium physics, which is not included in equation of state tables currently used for numerical simulations. 
Specifically, as described in \cref{app:compOSE}, starting from a three-parameter equation of state from the compOSE database, the derivatives we need can be worked out before carrying out a simulation. We are relying on this in the specific example discussed later.

\subsection{Thermodynamics working with the equilibrium electron fraction}\label{subsec:ThermoYe}

As  mentioned earlier, we may equivalently work with the electron fraction $Y_\e$. Given this, we want to revisit the  path that we just followed, now  working with the electron fraction. As we will see, this also results in a correction term to the pressure, which will be naturally expressed in terms of derivatives involving a notion of equilibrium electron fraction.

In order to define the equilibrium electron fraction we consider the following thermodynamical potential 
\begin{equation}
    g = s - n_\e \frac{\beta}{T} \;,
\end{equation}
such that 
\begin{equation}
    dg = \frac{1}{T}d\veps - \frac{\mu_\n}{T}dn - n_\e d\left(\frac{\beta}{T}\right) \;.
\end{equation}
Exactly as we did for the entropy, we can then expand $g$ around equilibrium. There will now be a first order term in $\beta$, which provides the ``formal definition'' of the equilibrium electron number density, $n_\e^\mathrm{eq}$. We get
\begin{equation}
    g(n,\veps,{\beta}/{T}) = s^\mathrm{eq} (n,\veps) - n_\e^\mathrm{eq}\frac{\beta}{T} + \frac{1}{2} g_2 \left(\frac{\beta}{T}\right)^2 \;,
\end{equation}
where
\begin{equation}
    n_\e^\mathrm{eq} = \left(\frac{\partial g}{\partial ({\beta}/{T})}\Big|_{\beta = 0} \right)_{n,\veps}\;,\quad\text { and }\quad g_2 = \left(\frac{\partial ^2 g}{\partial \left( {\beta}/{T}\right)^2} \bigg|_{\beta = 0}\right)_{n,\veps} \;.
\end{equation}

We can then work out the expansion (to first order in $\beta$) for $n_\e$ as 
\begin{equation}
    n_\e (n,\veps,\beta) = n_\e^\mathrm{eq} (n,\veps) - g_2 \frac{\beta}{T}     \;,
\end{equation}
and use it in the definition of $g$ to arrive at 
\begin{equation}
    g = s^\mathrm{eq} + \frac{1}{2}s_2\beta^2 - n_\e^\mathrm{eq}\frac{\beta}{T}  + g_2 \left(\frac{\beta}{T} \right)^2 \;.
\end{equation}
Comparing this with the expansion above we identify
\begin{equation}
    g_2(n,\veps) = - T_\mathrm{eq}^2 s_2 = -  T_\mathrm{eq}\left(\frac{\partial \beta}{\partial n_\e}\right)_{n,\veps}^{-1} \;.
\end{equation}
Using this result and working out the expansion for the thermodynamical variables from $g$, we obtain (after linearizing in $\beta$ and  introducing $Y_\e^\mathrm{eq} = n_\e^\mathrm{eq} / n$) 
\begin{subequations}\label{eq:ExpansionsWithxp}
\begin{align}
    T &= T^\mathrm{eq} + n \left(\frac{\partial Y_\e^\mathrm{eq} }{\partial \veps}\right)_n T^\mathrm{eq} \beta \;, \\
    \mu_\n &= \mu^\mathrm{eq} + \left[\mu^\mathrm{eq}n \left(\frac{\partial Y_\e^\mathrm{eq} }{\partial \veps}\right)_n + n\left(\frac{\partial Y_\e^\mathrm{eq} }{\partial n}\right)_{\veps} - Y_\e^\mathrm{eq} \right]\beta \;,\\
    Y_\e &= Y_\e^\mathrm{eq} - \left(\frac{\partial \beta }{\partial Y_\e}\right)^{-1}_{n,\veps}\beta \;.
\end{align}
\end{subequations}
Combining the result with the Gibbs relation we can work out the pressure (and the ``thermodynamical'' bulk viscosity)
\begin{equation}
    p(n,\veps,\beta) = p ^\mathrm{eq}(n,\veps) + p_1 \beta = p^\mathrm{eq} + \chi_t \;,
\end{equation}
with 
\begin{equation}\label{eq:p1Withxp}
    p_1 =  n \left[(p^\mathrm{eq} + \veps) \left(\frac{\partial Y_\e^\mathrm{eq} }{\partial \veps}\right)_n + n \left(\frac{\partial Y_\e^\mathrm{eq} }{\partial n}\right)_\veps\right] \;.
\end{equation}
The take home message is that we have  two equivalent expressions for the bulk-viscosity purely from thermodynamical arguments: this is always written as $\chi_t = p_1 \beta$, where $p_1$ can be either written as in \cref{eq:expansionsWithBeta} or  \eqref{eq:p1Withxp}. These results are thermodynamically correct as long as the linearization in $\beta$ is valid (i.e., the system is close to chemical equilibrium), independently of the modelling of the relaxation towards equilibrium. 

\section{Approximating the reactive system}\label{sec:ApproxReactive}

At this point we have formulated the relaxation problem for the reactive system. The equations we have written down are, in principle, all we need to evolve the system. 
However, it may not be numerically practical to solve the full nonlinear system, for example when the physical reactions are fast compared to numerically resolvable timescales. Given this, it is natural to consider approximations.

To set up the discussion let us introduce the (proper) time derivative $d/dt = u^a\nabla_a$. We can then write the hydrodynamical equations in non-dimensional form (assuming for a moment that we work with the lepton fraction instead of $\beta$, an assumption that makes no practical difference here)%
\begin{subequations}\label{eqs:nonDimReactiveYe}
\begin{align}
    \frac{d\veps}{dt} &= - \frac{1}{\eps_{St}} (\veps + c_r^2 p )\theta \;, \\
    a_b &= - \frac{1}{\eps_{St}}\frac{1}{\eps^2_{Ma}} \frac{1}{\veps + c_r^2 p}\perp^{c}_b\nabla_c p \;,\\ 
    \frac{dn}{dt} &= -\frac{1}{\eps_{St}}n\theta \;,\\
    \frac{d Y_\e}{dt} &= -\frac{1}{\eps_A} (Y_\e - Y_\e^\mathrm{eq}) \;, 
\end{align}
\end{subequations}
where we have defined the dimensionless parameters 
\begin{equation}
    \eps_{St} = \frac{l_r}{u_r t_r} \;, \quad 
    \eps_{Ma} = \frac{u_r}{c_r} \;, \quad 
    \eps_{A} = \frac{1}{\A t_r} \;,
\end{equation}
and introduced a reference sound speed $c_r$ as well as $l_r, t_r, u_r$ as reference lengthscale, timescale and fluid velocity---so that $a_b$ has dimensions $u_r/t_r$. 
To write the equations in non-dimensional form as above, we have first taken the sub-thermal limit $\Gamma_\e = - \gamma \beta$ and then expanded around equilibrium\footnote{Let us note that, as is clear from the last of \cref{eq:ExpansionsWithxp}, $Y_\e = Y_\e^\mathrm{eq}$ if and only if $\beta = 0$, and that in the sub-thermal limit this means $\Gamma_\e (Y_\e = Y_\e^\mathrm{eq})= 0$.}.
From this we see that $n,\veps, u^a$ evolve on similar timescales (in terms of the proper time associated with $u^a$), while the electron fraction evolution timescale is given by $\eps_A$. Assuming---as expected for large regions in neutron star merger simulations---that reactions occur on a fast timescale (so that $\eps_A \ll 1$), we may consider three different regimes: i) The expansion rate\footnote{As can be seen from e.g. \cref{eq:BetaEqGeneral}, the expansion rate is a ``source-term'' in the affinity equation.} $\theta$ varies only (or primarily) on slow timescales---which makes sense for numerical simulations where the spatial dynamics are resolvable but the reaction timescale is not; ii) The expansion $\theta$ varies on the fast timescale and hence this must be resolved in a simulation; iii) The flow is turbulent and therefore all scales are coupled.
In the last two cases, we cannot analytically simplify the problem much---expensive direct numerical simulations are required, although the large-eddy strategy \cite{mcdonough,lesbook,carrasco,fibrLES} (see also \cref{sec:LES}) may provide a useful alternative. As our main interest here is to consider the regime where progress can be made through approximations, we focus on the first case, where we can use standard multi-scale methods (see, e.g.~\cite{Weinan}) to ``integrate out'' the fast behaviour.

\subsection{Multi-scale arguments and the reactive system}\label{subsec:MultiScaleReactive}

Bringing the multiscale argument from \cite{StuartPavliotis} to bear on the problem (see \cref{app:MultiScale} for more details on the results used throughout the rest of this chapter) and assuming that we continue to work with $\beta$, we have to compute the late-time behaviour for the affinity by integrating the $\beta$ equation considering the other variables as parameters, and then taking the limit $t\to\infty$. The approximated equations for the remaining variables are then unchanged to lowest order, but we have to evaluate every function of $\beta$ using the late-time result. 
This is intuitively motivated by the underlying assumption that the affinity evolves on a faster timescale, so that the remaining degrees of freedom are approximately frozen on short timescales. 
The inclusion of the first order corrections then guarantees the approximated equations to be correct to $\O(\eps_A^2)$ up to times $\O(1)$. 

The equations of motion using the affinity can be written
\begin{equation}
    \frac{d}{dt} \begin{pmatrix} n \\ \veps \\ \beta \end{pmatrix} = \begin{pmatrix} -n \theta \\ -(p+\veps) \theta \\ - \A \beta - n \B \theta \end{pmatrix}.
\end{equation}
The non-dimensional form in \cref{eqs:nonDimReactiveYe} indicates that the reaction timescale $\epsilon_A$ enters only through $\A$---and it can be easily seen that $\B$ relates to how quickly the equilibrium electron fraction adjusts to a change in number and energy density.  Assuming only $\A$ to be fast, and using the results from \cref{sec:appendix_multiscale_construct_fast} we split the affinity into fast and slow pieces, writing
\begin{subequations}
    \begin{align}
        \frac{d}{dt} \begin{pmatrix} n \\ \veps \\ \beta^{\text{slow}} \end{pmatrix} &= \begin{pmatrix} -n \theta \\ -(p+\veps) \theta \\ - n \B \theta \end{pmatrix}, \\
        \frac{d}{dt} \begin{pmatrix} \beta^{\text{fast}} \end{pmatrix} &= \begin{pmatrix} - \epsilon_A^{-1} \A \beta \end{pmatrix},
    \end{align}
\end{subequations}
where $\beta^{\text{fast}} + \beta^{\text{slow}} = \beta$. By construction, $\A$ is independent of $\beta$---as it is defined at $\beta=0$---so the fast dynamics are now linear and the results of~\cref{sec:multiscale_linear} apply. We see that the invariant manifold---describing the ``late-time behaviour''---is given by $\beta^{\text{fast}} = -\beta^{\text{slow}}$, which is equivalent to saying $\beta = 0$ on the invariant manifold. That is, the invariant manifold corresponds to beta equilibrium, as expected. 

The reduced system is
\begin{equation}
    \frac{d}{dt} \begin{pmatrix} n \\ \veps \\ \beta^{\text{slow}} \end{pmatrix} = \begin{pmatrix} -n \theta \\ -(p+\veps) \theta \\ - n \B \theta \end{pmatrix} - \epsilon_A \frac{n \B \theta}{\A} \begin{pmatrix} 0 \\ \frac{\partial p}{\partial \beta} \theta \\ 0 \end{pmatrix}\ ,
\end{equation}
where all terms (pressure and its derivatives, and $\B$) have to be evaluated at beta equilibrium. The final equation decouples as all quantities in the first two equations depend on the total $\beta$ evaluated at equilibrium, which is $\beta=0$.
We see from the energy equation that the pressure correction appears as
\begin{equation}\label{eq:Pi_dBeta}
    \chi_d = -\epsilon_A \left( \frac{\partial p}{\partial \beta} \right)_{n, \epsilon} \frac{n \B \theta}{\A} = - \left( \frac{\partial p}{\partial \beta} \right)_{n, \epsilon} \frac{n \B \theta}{\A},
\end{equation}
where in the second equality we have re-absorbed the scaling parameter into the reaction rate $\eps_A^{-1}\A \to \A$.

If we instead make the assumption that $\B$ is also a fast parameter, then it is easy to see that the affinity would be an entirely fast variable, that is $\beta^{\rm{slow}} = 0$ automatically. Repeating the analysis, the invariant manifold would now be $\varphi(X) = - n \B \theta / \A$, and as a consequence the pressure in the energy (as well as Euler) equation 
\begin{equation}\label{eq:expPbeta}
    p = p \left(n,\veps, - \frac{n \B\theta}{\A}\right) = p^\mathrm{eq} + \chi_d \;.
\end{equation}
This shows that, in the case when  $\B$ is fast the bulk-viscous correction enters already at lowest order. The  corrections to this are second order in $\beta$, beyond the  regime of validity of the theory (linear in $\beta$), and hence cannot be trusted. In essence, whatever assumption we make for  $\B$ (being fast or slow), the result is the same.

The ``interpretation'' of \cref{eq:Pi_dBeta} as the Navier-Stokes bulk-viscosity is supported by the analysis in \cref{sec:hydro-thermodynamics} (cf. \cref{eq:Pi_t})---recall also the argument around \cref{eq:cattaneo}, now specified to \cref{eq:BetaEquationSubthermal}. This result, while  in line with ``expectations'', is non-trivial. In fact, common arguments in favour of a representation of the net effect of under-resolved reactions via a bulk-viscous pressure are typically  perturbative in nature \cite{AlfordBulk10,AHSparticles20}.

\subsection{Invariant manifold method with the electron fraction}\label{subsec:InvMnfld_Ye}

Let us now run through the invariant manifold argument working with the electron fraction.
%
%
Consistently with the dimensional analysis of \cref{eqs:nonDimReactiveYe}, we assume that the electron fraction evolves on faster timescales than the other variables. 
This means we are effectively considering the system of equations
\begin{subequations}
\begin{align}
    \frac{d}{dt} \begin{pmatrix} n \\ \veps \end{pmatrix} &= - \begin{pmatrix} n\theta  \\ (p+\veps)\theta \end{pmatrix} \;, \\
    \frac{d}{dt} \begin{pmatrix} Y_\e \end{pmatrix} & = - \begin{pmatrix} \eps^{-1} \A(Y_\e - Y_\e^\mathrm{eq}) \end{pmatrix} \;.
\end{align}
\end{subequations}
where we have made explicit that the fast behaviour arise from $\A$ as explained in \cref{sec:appendix_multiscale_construct_fast}.
Again, the linear fast case of~\cref{sec:multiscale_linear} is relevant, giving that the invariant manifold is the equilibrium surface.
This immediately tells us that, to lowest order, the approximated equations describe a reactive fluid for which chemical equilibrium is restored immediately on the dynamical timescale. Including the first order corrections to the approximated equations---simply drawing on the results from \cite{StuartPavliotis} and \cref{sec:multiscale_linear}---we obtain
\begin{subequations}
\begin{align}
    \frac{d}{dt} \begin{pmatrix} n \\ \veps \end{pmatrix} &= - \begin{pmatrix} n\theta \\
    (p^\mathrm{eq} + \veps + \chi_d) \theta \end{pmatrix} \;,
\end{align} 
\end{subequations}
where 
\begin{equation}\label{eq:BulkPressure_xeq}
    \chi_d = \frac{n}{\gamma} \left(\frac{\partial \beta}{\partial Y_\e}\right)_{n,\veps} ^{-1} \left(\frac{\partial p}{\partial Y_\e}\right)_{n,\veps} \left[(p^\mathrm{eq} + \veps) \left(\frac{\partial Y_\e^\mathrm{eq} }{\partial \veps}\right)_n + n \left(\frac{\partial Y_\e^\mathrm{eq} }{\partial n}\right)_\veps\right]\theta \;.
\end{equation}
Now, using the fact that 
\begin{equation}
    \left(\frac{\partial p}{\partial \beta}\right)_{n,\veps} = \left(\frac{\partial p}{\partial Y_\e}\right)_{n,\veps}\left(\frac{\partial \beta}{\partial Y_\e}\right)_{n,\veps}^{-1}  \;,
\end{equation}
along with the two alternative formulae for $p_1$ \cref{eq:p1Withxp,eq:p1WithBeta} we observe the (pleasing) consistency with the result in \cref{eq:Pi_dBeta}. Exactly as when working with the affinity, by integrating out the fast variable we pick up an additional contribution to the pressure that corresponds to a bulk-viscous response. 

\subsection{Partially resolved reactions and double counting}\label{subsec:doublecounting}

Let us now consider the situation where some part of the reactions are slow enough that we can capture them, and the rest is not. We consider the case where we can split the reactions in two families,  fast ones (representative of, say, direct Urca processes) that cannot be captured by the numerics, and  slow ones (representative of, say, modified Urca processes) that can be resolved. 
Even though this might be somewhat artificial, the key point is that  we assume a clear scale-separation between  two types of reactions. 
We do so as we are only interested in a proof of principle argument here. 
We want to show the problems in constructing a consistent bulk viscous approximation in this case.

It would seem reasonable, after the discussion we just had, to model the impact of unresolved reactions as a bulk-viscosity. Whilst this may be a valid strategy, one has to be careful because the multi-scale methods suggest that the resolvable reaction rates should pick up a correction term as well.
This also provides a proof of principle argument for the ``double counting'' issue raised by \citet{PeteThermal}. The discussion in \cite{PeteThermal} points out that the effects of under-resolved reactions are already accounted for in schemes aimed at modelling neutrinos, and hence adding a bulk-viscous pressure on top of that may lead to a double-counting. 

Let us start from the equation for the electron fraction, written in terms of the reaction rate. We define the fast/slow part of the total creation rate as discussed in \cref{sec:appendix_multiscale_construct_fast}
\begin{equation}
    \Gamma_f = \lim_{\eps\to 0} ( \eps\Gamma_\e) \;, \qquad  \Gamma_s = \Gamma_\e - \frac{1}{\eps}\Gamma_f \;,
\end{equation}
and hence split the electron fraction into its fast and slow contributions $Y_\e = Y_s + Y_f$ according to
\begin{equation}
    \frac{d}{dt} Y_s = \frac{\Gamma_s}{n} \;,\quad \text{and}\quad \frac{d}{dt} Y_f = \frac{1}{\eps} \frac{\Gamma_f}{n} \; .
\end{equation}
We can think of $\Gamma_f$ as a function of $(n,\veps, Y_s+Y_f)$ and define a fast equilibrium fraction $Y_f^\mathrm{eq}$ such that $\Gamma_f(n,\veps,Y_s +Y_f^\mathrm{eq}) = 0$. It is then possible to expand the equation for the fast variable as
\begin{equation}
    \frac{d}{dt} Y_f = \frac{1}{\eps} \frac{\partial \Gamma_f}{\partial Y_\e}\Big|_{Y_\e =Y_s + Y_f^\mathrm{eq}} (Y_f - Y_f ^\mathrm{eq}) \; .
\end{equation}
Note that, because the two fractions must add up to the total electron fraction, we have $Y_f^\mathrm{eq}  = Y_\e -  Y_s$, namely the equilibrium fast fraction $Y_f^\mathrm{eq} = Y_f^\mathrm{eq} (n,\veps,Y_s)$. As for the slow reaction rate, we do not need to expand it around equilibrium because this is assumed to be resolved in the simulation. 
Applying the results of \cref{app:MultiScale} (including the first order corrections) we obtain
\begin{equation}
    \label{eq:DoubleCountingResult}
    \frac{d}{dt} \begin{pmatrix} n \\ \veps \\ Y_s \end{pmatrix} =  \begin{pmatrix} - n\theta  \\
    - (p + \veps + \chi_d) \theta \\
    \Gamma_s -  \frac{1}{n} \left(\frac{\partial \Gamma_s}{\partial Y_\e}\right)_{n,\veps}\left(\frac{\partial p}{\partial Y_\e}\right)_{n,\veps}^{-1}\chi_d  \end{pmatrix},
\end{equation}
with 
\begin{equation}
    \chi_d =\left\{ \left[(p + \veps) \left(\frac{\partial Y_f^\mathrm{eq} }{\partial \veps}\right)_{n,Y_s} + n \left(\frac{\partial Y_f^\mathrm{eq} }{\partial n}\right)_{\veps,Y_s}\right]\theta  - \left(\frac{\partial Y_f^\mathrm{eq}}{\partial Y_s}\right)_{n,\veps} \frac{\Gamma_s }{n} \right\} \left(\frac{\partial \Gamma_f }{\partial Y_\e}\right)_{n,\veps}^{-1} \left(\frac{\partial p }{\partial Y_e}\right)_{n,\veps} \;,
\end{equation}
and everything evaluated at $Y_f = Y_f^{\rm{eq}}$.
This argument then shows that, if there are both fast and slow reactions in the system, and we are trying to capture the effect of the fast/unresolved ones via a bulk-viscosity like contribution, we  need to tread carefully, as the introduction of the bulk viscosity also impacts on the resolved reaction rates, and the rates impact on the definition of equilibrium.

\section{Making contact with simulations}

Having discussed the approximate equations we obtain from the multi-scales approach, it makes sense to ``step back'' and ask to what extent we expect this approximation to make sense for numerical simulations. To set the stage for the discussion, let us rewrite \cref{eq:BetaEquationSubthermal} as
\begin{equation}
    \frac{d\beta}{dt}= - \A \beta + \B \frac{d n}{dt} \; .
\end{equation}
The parabolic limit---which corresponds to neglecting the time derivative of the affinity $ d\beta / dt$---in the Fourier domain is
\begin{equation}
    \beta_{NS} (\omega)= \frac{\B}{\A} n\omega \;,
\end{equation}
while the extended irreversible thermodynamics (EIT) result takes the form
\begin{equation}
    \beta_{EIT} (\omega)= n\B\A \frac{\omega}{\omega^2 + \A^2} = \beta_{NS}(\omega) \frac{\A^2}{\A^2 + \omega^2} \; .
\end{equation}
\begin{figure}
\centering
\includegraphics[width=0.9\textwidth]{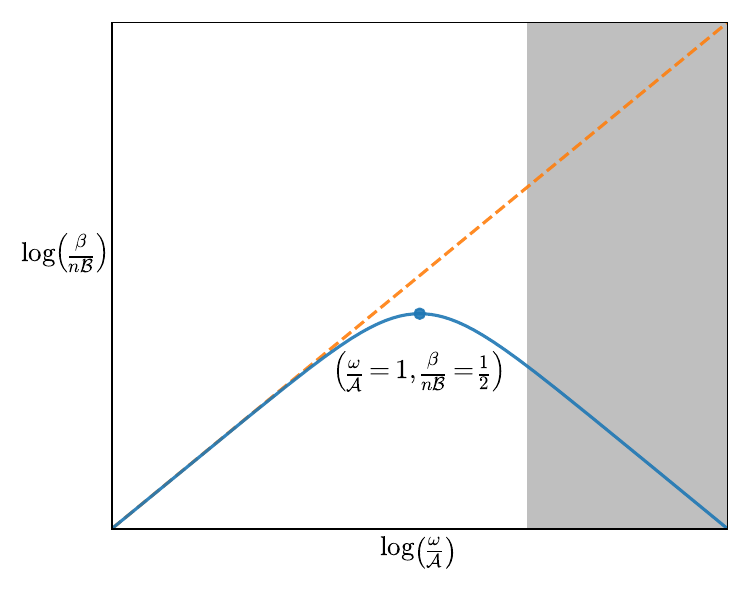}
\caption{Illustrating the behaviour of  $\beta(\omega)$ in two cases: the solution to the Cattaneo equation (i.e. the full extended irreversible thermodynamics (EIT) relaxation towards the Navier-Stokes limit, blue solid curve) and the  parabolic case (the Navier-Stokes limit, orange dashed line). The shaded region indicates  frequencies we assume we may not have ``access'' to numerically (this region  moves towards higher frequencies when the numerical resolution is increased).
}\label{fig:AffinityPlot}
\end{figure}
The two results are compared in \cref{fig:AffinityPlot}, from which we see that the EIT behaviour shows the ``expected'' resonance feature \cite{Sawyer89,AlfordBulk10,AlfordHarrisDamping18,AHSparticles20,AlfordHarutyunuanSedrakian19,AlfordHarutyunuanSedrakian21}. We note, however, that this is not the usual figure (see, for example, figure 2 in \cite{AlfordHarrisDamping18}), as we are plotting the affinity $\beta$ and not the bulk-viscosity coefficient. To link the results, we need to recall that $\chi = p_1 \beta $ and then define $\zeta$ via  $ \chi = - \zeta \theta$ . Then we can use \cref{eq:baryonCont} to write  $\theta = - {\dot n}/{n}$ to obtain
\begin{equation}
    \zeta_{NS} = p_1 \frac{\B}{\A} n  \;, \quad \text{and}\quad \zeta_{EIT} = \zeta_{NS} \frac{\A^2}{\omega^2 + \A^2} \; .
\end{equation}
The difference is subtle as both  $\beta_{EIT}$ and $\zeta_{EIT}$ present resonant features. However, $\zeta_{EIT}$ does so when the frequency $\om$ is kept fixed and $\A$ is varied, while $\beta_{EIT}$ exhibits the resonance even if we fix $\A$ and vary the frequency.

The difference may not seem particularly relevant, but the illustration in \cref{fig:AffinityPlot} allows us to draw useful conclusions. The figure shows that the parabolic limit (i.e. the limit we represent with the multi-scale argument) is a ``good'' approximation at low frequencies (we provide a quantitative argument in the next \cref{subsec:BoundTheError}), but becomes less accurate at higher frequencies\footnote{The illustration also provides an intuitive demonstration of why the high-frequency behaviour is non-causal and associated with a linear instability~\cite{HiscockInsta}.}. Keeping in mind that we consider the problem in the context of numerical simulations, we may also note that there will inevitably be frequencies that we do not have access to. This high-frequency cut-off is (schematically) represented by the shaded region in~\cref{fig:AffinityPlot}. Specifically, the resolution is limited by $\omega \sim \Delta t ^{-1}$ where $\Delta t $ is the numerical timestep. Now, because the resonance frequency is given by $\omega = \A$, we are left with two options: either the numerical timestep is large enough  ($\Delta t ^{-1} < \A$) that the peak is not resolved, or the simulation is precise enough that the region to the right of the resonant peak is (at least partially) resolved. In the first case, our simulation cannot resolve the (fast) relaxation towards the Navier-Stokes limit (and it is also likely that the expected instability \cite{HiscockInsta} associated with the Navier-Stokes behaviour will be suppressed). In the second case, the relaxation towards Navier-Stokes can be resolved by the numerics, so working with the Navier-Stokes approximation is just wrong.

\subsection{What bulk viscous pressure approximation is suitable?}\label{subsec:BoundTheError}

In the regime where the reactions need to be approximated as a bulk viscous pressure there are many possible ways in which such an approximation can be constructed. The standard first order form could be used, where the bulk pressure depends on a (data dependent) coefficient multiplied by the expansion associated with the fluid motion \cite{Most2021BV}. Alternatively, in frequency space the bulk pressure can be written as a term depending on the thermodynamics multiplied by a function of frequency. In this case the ``true'' (Cattaneo) result differs from the first order (``Navier Stokes'') result only in the form of the frequency term.

Intuitively we have argued that these two different bulk viscous approximations should be close to each other, as long as we are considering frequencies below the resonant peak. Here we make that argument more quantitative.

We are interested in the low frequency part of the pressure that can be captured in a numerical simulation. We will therefore assume that we are only interested in frequencies $\wh = \om / \A < 1$ (``to the left of the peak''), and that there is a hard cut-off at $\whd \sim 2 \pi / (\A \, \Delta t)$ (the numerical scheme is spectral like and captures all frequencies available on the grid). Therefore
\begin{equation}
    \label{eq:bulk_pi_freq}
    \chi = \int_{-\infty}^{\infty} \dd{\wh} H \left( 1 - \left| \frac{\wh}{\whd} \right| \right) p_1 \beta = 2 \int_{0}^{\whd} \dd{\wh} p_1 \beta\;.
\end{equation}

We want the relative difference between the two bulk viscous pressure approximations, which is
\begin{equation}
    \E = \frac{\left| \chi_{\text{Cattaneo}} - \chi_{\text{NS}} \right|}{\left| \chi_{\text{NS}} \right|} \;.
\end{equation}
To compute this we \emph{assume} that we can bound the correction terms as powers of the frequency, as
\begin{equation}
    \label{eq:bound_coeffs}
    C_{-} \wh^a < n \B p_1 < C_{+} \wh^b \;,
\end{equation}
where $0 > a > b > -2$ is needed for the results to converge. This seems reasonable for Kolmogorov turbulence, but the range of these coefficients ($C_{\pm}, a, b$) has an impact on the result.

From this assumption we have
\begin{equation}
    2 \int_{0}^{\whd} \dd{\wh} C_{-} \wh^{1+a} < \left| \chi_{\text{NS}} \right| < 2 \int_{0}^{\whd} \dd{\wh} C_{+} \wh^{1+b}\;,
\end{equation}
giving
\begin{equation}
    \label{eq:bulk_NS_bound_both}
    \frac{2 C_{-}}{2 + a} \whd^{2+a} < \left| \chi_{\text{NS}} \right| < \frac{2 C_{+}}{2 + b} \whd^{2 + b}\;.
\end{equation}

To bound the pressure difference, first write
\begin{equation}
\begin{split}
    \left| \chi_{\text{Cattaneo}} - \chi_{\text{NS}} \right| &= \left| n \B p_1 \wh \left( \frac{\wh^2}{1 + \wh^2} \right) \right| \\
        &= \left| n \B p_1 \wh^3 \right| + {\cal O}(\wh^5)\;.
\end{split}
\end{equation}
From this we get
\begin{equation}
    \label{eq:bulk_NS_bound_diff}
    \frac{2 C_{-}}{4 + a} \whd^{4+a} < \left| \chi_{\text{Cattaneo}} - \chi_{\text{NS}} \right| < \frac{2 C_{+}}{4 + b} \whd^{4 + b}\;.
\end{equation}

Using the appropriate bound for numerator and denominator in $\E$, it follows that
\begin{equation}
    \label{eq:bulk_error_bound}
    \E < \frac{C_{+}}{C_{-}} \frac{2 + a}{4 + b} \whd^{2 + b - a}\;.
\end{equation}

As $a > b$ and $a-b < 2$ we see the difference between the two approximations is, in the frequency range ($\whd < 1$) of interest, of order $\E < \mathcal{O}(\whd^c)$ where $c \in (0, 2)$. If we expect $a \simeq b$ then $\E < \mathcal{O}(\whd^2)$. Therefore the difference between the two approximations will be small until we are close to $\whd = 1$, which is the resonant frequency peak. This supports the arguments made in connection with the results in~\cref{fig:AffinityPlot}.

\subsection{How relevant is bulk viscosity in mergers?}
\label{sec:relevance}

In the previous sections we have described how a reactive system near equilibrium can be approximated as a single fluid model with a bulk viscous pressure. One question to tackle is whether this approximation is of any relevance for, for instance, a neutron star merger simulation. In order for the argument to be of interest we need multiple criteria to be met. First, the timescales that we can resolve in the simulation must not include the key timescales for the reactions. If we can resolve the reactions then we should, as the bulk viscosity approximation (as well as the Cattaneo-type law \cref{eq:BetaEquationSubthermal}) only includes the linear effects (in deviations from thermodynamic equilibrium), whilst the reactions account for the full nonlinear behaviour. Besides, the illustration in fig.~\ref{fig:AffinityPlot} shows that the approximation would simply break down in a resolved simulation. Second,  we would require that the correction to the total pressure from the bulk viscous approximation should not be negligible. If the bulk viscous term due to the reactions is tiny in comparison to the standard fluid pressure then it is pointless to include the reactions in the numerical simulation. Instead we should impose chemical equilibrium directly, and solve without reactions or a bulk viscosity approximation.

\begin{figure}
    \centering
    \includegraphics[width=\textwidth]{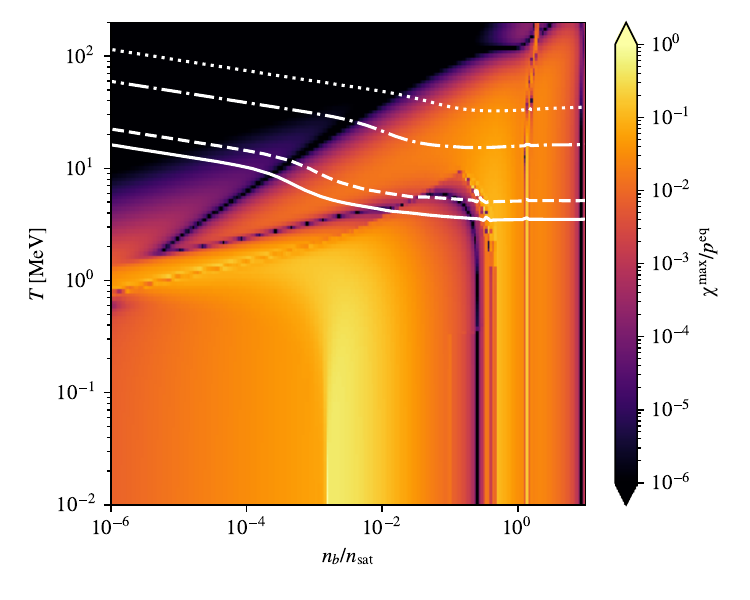}
    \caption{The maximum (for each point we assume that $\omega=\A$) potential relative contribution of the bulk viscous approximation $\chi^\mathrm{max}/p^\mathrm{eq}$ at each point in phase space using the APR \cite{compOSE,APReos} equation of state.
    We see that the bulk viscous pressure contribution can be large for most temperatures when $n \gtrsim 10^{-3} n_{\text{sat}}$. However, the bulk viscous approximation should only be used where the reaction rate cannot be resolved by the numerical simulation, which is where the grid frequency is greater than $\A$. Also shown are contours at $\A=\{10^3,10^4,10^7,10^9\}~\mathrm{s^{-1}}$ (solid, dashed, dot-dash, dot). For current simulations, frequencies of $\sim 10^6~\mathrm{s^{-1}}$ are resolvable. This shows that the bulk viscous approximation should only be used for $T \gtrsim 10$MeV, and as the grid resolution improves becomes less necessary.}
    \label{fig:P_test_new}
\end{figure}

Figure~\ref{fig:P_test_new} combines both these criteria in a single plot (based on data for the APR equation of state \cite{APReos,compOSE} used in the simulations discussed in \cite{PeteThermal}). In order for the bulk viscous approximation to be useful the timescale from the simulation must be slower, or at a lower angular frequency, than that given by the peak frequency of the resonance. This peak is defined by $\A$. Therefore every point in the $(T, n)$ plane \emph{below} a contour of fixed $\A$ has reactions acting on time or frequency scales that can (and hence should) be resolved by the numerical simulation. The contours given show that as the numerical grid resolution is improved, more of the $(T, n)$ plane should be modelled by solving directly for the reactions. The bottom two contours in the figure bracket frequencies relevant for bulk gravitational wave generation ($1-10$kHz), and so must be captured by any numerical simulation. The top contour lines are indicative of the grid frequencies achievable by current simulations (an angular frequency of $10^7$s${}^{-1}$ roughly corresponds to a grid spacing of $200$m) and the state of the art in maybe ten years (an angular frequency of $10^9$s${}^{-1}$ roughly corresponds to a grid spacing of $2$m). However, there remain points at densities above $10^{-2} n_{\text{sat}}$ and temperatures above $10$MeV where reactions can only be modelled using the bulk viscous approximation, even with the best resolution available in the near future. Current simulations such as~\cite{PeteThermal} do see substantial amounts of matter in the post-merger remnant within this part of phase space, meaning that the bulk viscous approximation will remain necessary for the foreseeable future.

In addition, \cref{fig:P_test_new} shows the maximum relative magnitude of the bulk viscosity pressure, again in the $(T, n)$ plane, with contours of fixed $\A$ overlaid. In this calculation the bulk viscous approximation requires a range of additional approximations, including assuming the Fermi surface approximation is valid in order to compute the rates. These approximations are laid out in detail in \cref{app:compOSE}. However, for the purposes of our argument, the key is to note that we are interested in regions of phase space where the bulk viscosity approximation may be valid (above the contours), and where the bulk viscosity makes a significant contribution to the pressure. We see that this holds again in the region with densities above $10^{-2} n_{\text{sat}}$ and temperatures above $10$MeV.

Figure \ref{fig:P_test_new} also shows a number of additional features. First, on the right hand side we see a vertical line. 
As this corresponds to a density just above saturation, it is natural to associate it with the APR phase transition. 
To the left of this there are two features that start vertically at similar densities.
The first one goes up until about $10$MeV before turning left and going down to just below $1$MeV. This is expected to be an artefact due to the way the equation of state is constructed since different techniques are used in different parts of the parameter space and then cobbled together. The second feature also turns left, but then goes down to about $2$MeV before turning right again and going to the top right of the figure. Even though we have not explored this feature in detail, we believe this has to do with points in parameter space where $\partial \beta / \partial Y_\e$ goes to zero.
We also stress that we see similar features when analogous figures are produced using different (but similarly constructed) equations of state.

As a final important point we note that there is no sharp transition between regions where the effects of the bulk viscous approximation are sizeable and where they are not, when considering contours of fixed $\A$ (which can be linked to the numerical resolution). Unless the numerical resolution can be increased by many orders of magnitude beyond the current state of the art, there will always be regions in spacetime where the bulk viscosity is significant but the approximation itself is debatable. Therefore we have to consider a model that is able to model reactions by directly evolving, for example, the species fraction in some parts of spacetime, makes the bulk viscous pressure approximation in other parts of the spacetime, and transitions between the two appropriately. This is difficult to do correctly.

\subsection{The impact of large-eddy filtering}\label{sec:LES}

We have demonstrated how the Navier-Stokes limit for bulk viscosity due to reactions can be obtained through a multi-scale argument, effectively integrating out the fast timescales of the problem. The argument is quite intuitive. However, the discussion is not yet complete. In the context of numerical simulations we also have to  consider other ``filtering'' aspects. In particular, we need to explore the link to (or conflict with) the large-eddy strategy. This problem is not straightforward. In a sense, the large-eddy approach is ``complementary'' to the approximate scheme as it aims to represent the regimes where multi-scale arguments do not  apply---namely turbulent flows and when all dynamics take place on fast timescales. 
{One may be tempted to view the multi-scale method and large-eddy space-filtering as ``orthogonal''. A space filter cuts off short length-scales, while  the invariant manifold method integrates out the fast dynamics (removing short timescales). 
However, the two issues are linked.} 

An actual numerical implementation introduces an implicit filtering associated with the discretized numerical grid. 
On a grid with fixed spacing $\Delta x$, the implicit filtering requires that any spatial feature on shorter lengthscales cannot be captured and must be modelled by some closure relation, as discussed in \cite{radice1,radice2,carrasco,viganoGR,duez}. Equally, the CFL bound (linked to causality) imposes that the timestep $\Delta t \propto \Delta x$, and so any physical feature happening on shorter timescales cannot be captured. Thus, increasing the accuracy by modifying the grid spacing automatically means the $\Delta t$ in \cref{fig:AffinityPlot} decreases and therefore the amount of the fast reactions that cannot be captured, shown by the grey area, decreases. There is a direct link between the amount of physics that must be modelled via a spatial filtering and via a time filtering or multi-scale argument.

In order to examine how the filtering associated with the large-eddy strategy impacts on the discussion of bulk viscosity, let us frame the argument using the ``fibration'' framework developed in \cref{ch:LES}.
Let us first consider  what happens when filtering is applied to the continuity equations for baryons and electrons, \cref{eq:baryonCont,eq:ElectronFractionEq}. As discussed in \cref{subsec:BaryonConserv_aver}, the equation for the baryons remains unchanged if (and only if!) we choose to work with the density-weighted four-velocity:
\begin{equation}
    \nabla_a n^a = 0 \Longrightarrow \tu^a \nabla_a \tn + \tn\nabla_a \tu^a = 0 \;,
\end{equation}
where $\la n^a\ra = \tn \tu^a$ defines the filtered baryon number density current (indicated with a tilde). In the case of  the electrons, we need to introduce the ``electron fraction residual''
\begin{equation}
    \tau^a_{Y_\e} = \la Y_\e n^a \ra  - \la Y_\e\ra \la n^a\ra \;,
\end{equation}
and work with a coarse-grained electron fraction,  defined by $\tn_\e = - \tu_a \la n_\e u^a\ra$ and leading to
\begin{equation}
    \tY_\e = \frac{\tn_\e}{\tn} = - \frac{\tu_a}{\tn} \la n_\e u^a\ra  \;.
\end{equation}
The filtered equation for the  electron fraction then becomes\footnote{Note that we could choose a different coarse-grained observer in such a way that the effective creation rate is re-absorbed in the macroscopic fluid four-velocity. However, this would come at a cost since the equation for the filtered baryon current would then have an additional diffusion term on the right-hand-side.}
\begin{equation}
    \tn \tu^a\nabla_a \tY_\e = \la\Gamma_\e\ra - \nabla_a \tnu_\e^a \;,
\end{equation}
where
\begin{align}
    \tY_\e = \la Y_\e \ra - \frac{1}{\tn} \tu_b\tau_{Y_\e}^b \;, \qquad \text{and } \tnu_\e ^a = \tperp^a_b \la n_\e u^b \ra =  \tperp^a_b\tau_{Y_\e}^b \;.
\end{align}
To avoid confusion, let us explicitly state that the orthogonal projection is here defined with respect to the filtered velocity $\tu^a$.
The take home message is simple. The large-eddy filtering introduces an effective creation rate in addition to the faithful microphysical one. As the effect of the reactions can be modelled as a bulk-viscosity this may clearly have an impact on the analysis (cf. \cref{eq:BetaEqGeneral}). 

Now turn to the remaining equations of motion, e.g. \cref{eqs:Euler}, which follow from the conservation of the stress-energy-momentum tensor. As this retains exactly the same form as in the perfect fluid case, we can simply draw on the results from \cref{subsec:AverEoM,sec:filter}. The only difference is in the pressure and the Gibbs relation, as we are now considering a reactive system: 
\begin{equation}
    \la p \ra = \la - \veps + Ts + \mu_\n n - n_\e \beta \ra\;.
\end{equation}
The  $n_\e\beta$ term was not considered in \cref{sec:2parEoS} as the fine-scale model considered there did not account for reactions. Nonetheless, we may simply adapt the same strategy: introduce an effective three-parameter equation of state at the resolved scale, and use it to define the macroscopic thermodynamic variables
\begin{subequations}
\begin{align}
    \frac{1}{\tilde T} &\doteq \left(\frac{\partial \ts}{\partial \tilde \veps}\right)_{\tn,\tY_\e}\;,\\
    - \frac{\tilde \mu_\n}{\tilde T} &\doteq  \left(\frac{\partial \ts}{\partial  \tn}\right)_{\teps,\tY_\e} - \frac{\tY_\e}{\tn} \left(\frac{\partial \ts}{\partial  \tY_\e}\right)_{\tn,\,\teps}\;,\\
    \frac{\tilde \beta}{\tilde T} &\doteq \frac{1}{\tn}\left(\frac{\partial \ts}{\partial \tY_\e}\right)_{\tn,\teps} \;.
\end{align}
\end{subequations}
With these definitions in hand, we can  rewrite the filtered Gibbs relation as
\begin{equation}
    \la p \ra = -\teps + \tn \tmu_\n + \tilde T \ts - \tn\tY_\e\tbeta + M \;,
\end{equation}
with the enhanced closure term
\begin{equation}\label{eq:FilteringOutEqPressure}
    M = \left[\left(\la n \mu_\n \ra  -  \tn\tmu_\n\right) + \left(\la Ts \ra - \tilde T\ts\right) - \left(\la \veps \ra - \teps \right) - \left(\la n_\e \beta\ra - \tn\tY_\e\tbeta \right)\right] \;.
\end{equation}
It makes sense to introduce the effective pressure $\tp = \la p \ra - M$ as this will satisfy a Gibbs relation of the pre-filtered form (but now in terms of the coarse-grained equation of state and the associated variables). Then, because the filtered energy and Euler equations will contain $\la p \ra$, the $M$ term will enter the final equations as a correction to the pressure---effectively providing a bulk-viscous-like contribution. Let us stress that while we are only providing a minimal discussion of a large-eddy model, this is everything we need here. To make real progress we would need to introduce an explicit closure scheme and perform numerical experiments, both of which go beyond the scope of this thesis. 

The essence of the argument is that the large-eddy filtering introduces an ``effective bulk-viscosity'' contribution to the coarse-grained equations. 
This happens in (at least) two ways: i) through the residual term $M$ stemming from filtering the Gibbs relation ii) by adding an effective creation rate, and the two effects are not (necessarily) linked as they depend on the introduced closure relations. 
In particular, the effective creation rate (which follows from the four-divergence of $\nu_\e^a$) affects the effective restoring term $\gamma$, and in turn $\A$---as is evident from inspection of \cref{eq:Adef}. 
As the resonance frequency in \cref{fig:AffinityPlot} is given by $\omega = \A$, and this essentially dictates whether or not the Navier-Stokes approximation is applicable, it makes sense to consider applying the filtering \underline{and} the multi-scale approach at the same time. 

We can intuitively see (and check explicitly) that the  coarse-grained equations will have a bulk-viscous contribution stemming from having integrated out the electron fraction degrees of freedom, and one from the filtering. The analytic expressions of these two terms depend on the order with which we take the steps: either we apply the multi-scale methods first and then filter, or the other way around. The results are unlikely to be the same. To see this, simply note that the multi-scale/invariant manifold method essentially boils down to an approximation of the equations around the equilibrium surface $Y_\e = Y_\e^\mathrm{eq}$. If we take the filtering step first, the notion of equilibrium changes---both because the (potentially different) equation of state is evaluated in terms of the coarse grained variables, and because of the effective rate. 
This highlights the importance of including all the relevant physics when constructing the closure terms in a large eddy model. This is problematic as the best closure terms require direct fine scale numerical simulations, and the analysis in this paper shows these expensive simulations need repeating each time additional physics is added. It is bound to be an expensive endeavour.

\section{Summary and Outlook} \label{sec:BVsummary}

Binary neutron star mergers offer a unique opportunity to explore several extremes of physics, but we need to improve our numerical simulation technology if we want to realize the discovery potential of future gravitational-wave instruments (required to catch the high-frequency dynamics of the merger events). 
In particular, we need to make sure that efforts to infer the detailed physics are not stumped by systematic errors associated with the numerical implementation.
An important step towards realism involves dealing with nuclear reactions. 
Motivated by this, we have considered the issue of reactions from the perspective of numerical simulations (and the associated limited resolution), aiming to provide ``new insights on an old problem''.
Specifically, we have discussed to what extent it makes sense to capture the net effect of reactions via a bulk viscosity prescription, taking explicitly into account the issue of resolution limitations. 
In essence, we assessed the impact of the reaction timescales on the way we should  frame the modelling, and represent the net effect of reactions. 

Our key messages link the reaction timescales to the numerical (grid) timescales. 
When the reactions are slow the ground truth result is found by evolving the reactions directly. 
When the timescales are comparable, particularly when the physical timescales are (slightly) faster, this is not numerically practical. Instead the evolution system must be approximated. 
To leading order the reactions relax the system to equilibrium instantly. 
However, the error incurred is proportional to the ratio in the timescales. 
The first order (in the ratio of scales) approximation introduces correction terms that act as a bulk viscosity. 
This demonstrates how  a multi-component reactive system can be approximated as a dissipative single fluid. 
This bulk viscous structure emerges regardless of whether we formulate the problem in terms of $\beta$ (which would be natural from a thermodynamics perspective) or $Y_\e$ (to connect more directly with simulations).

In a neutron star merger simulation the ratio of scales covers all ranges, with the scales being comparable (and hence a bulk viscosity approximation necessary) particularly in the core shortly after merger (see  \cite{AlfordPRL2018,AlfordHarrisDamping18} and cf. \cref{fig:P_test_new}). We also showed that the prescription for the bulk viscosity---either algebraically in a ``Navier-Stokes'' like fashion, or by providing an equation of motion in a ``Cattaneo'' or ``Israel-Stewart'' like fashion---is irrelevant, as they agree (to the order of the approximation) in the regime where the approximation is useful (at low frequencies where the ratio of scales is comparable). Finally, we demonstrated explicitly how the equations of motion are modified when some, but not all, of the scales are comparable. This flags up how the introduction of a bulk viscosity (either directly through approximating reactions, or indirectly through using large-eddy simulation techniques) can conflict with the definition of equilibrium. Consistently accounting for these issues to avoid ``double-counting'' is possible, {as outlined in \cref{subsec:doublecounting}}, but difficult. 

Our main conclusions are intuitive but this is the first time that they have been spelled out in detail. Contrasting theoretical work against the simulations in \cite{PeteThermal}, we find that different regions of the parameter space relevant for mergers would require different prescriptions. In particular, there are regions of the density-temperature phase space where reactions are slow enough that they can (and hence, should) be captured directly, and other regions where they are not---even with the best resolution available, now and in the foreseeable future---with no sharp boundaries between the two regimes. This indicates that, to properly account for reactions, we  need to develop numerical codes  capable of handling both regimes (reactions fast/slow compared to the resolved timescales), and the transition between them on the fly.

The issue of bulk viscosity is also closely linked to the role of large-eddy filtering---which enters the discussion implicitly or explicitly. Noting this, we provided general filtering arguments that help set the stage for further work, {and highlighted the coupling between bulk viscosity and large-eddy modelling. A} ``definitive'' prescription for reactive systems will require explicit numerical experiments and the introduction of an appropriate closure model. 

As a final comment, let us note that there have been recent papers supporting the idea that bulk viscosity effects can be important for gravitational waves \cite{PeteGWreactions,Most2022BV}, and also papers suggesting it has no impact at all \cite{Radice+THCM1,Zappa+BVM1test}. 
In particular, different approaches are used to tackle the stiffness issue, where reactions happen faster than the numerical scheme can stably capture them. 
Even though we would have to know the fineprints of the different numerical implementations to fully understand the origin of such discrepancies, we can add a final comment that may be part of the story.
By checking numerically the formal accuracy of the (bulk viscous) approximation detailed above, we found that the expected accuracy results are recovered when the system is started close to the equilibrium manifold. 
However, for the accuracy result to hold, additional boundary layer effects need to be accounted for if the system is kicked far out of equilibrium---as suggested, for example, by the simulations of \cite{PeteThermal}.

%% file: Parts/BNSapplications/MRI.tex
\chapter{Magneto-rotational instability in mergers: a local Newtonian analysis}\label{ch:MRI}

In the previous chapter we focused on modelling bulk viscosity from the perspective of numerical simulations. 
We also touched upon the links to turbulence and filtered models. 
In this chapter, instead, we will focus on the magneto-rotational instability or, as will become clear as we proceed, on more general magneto-shear instabilities. 
We do so as we often see this kind of instabilities in action---the Kelvin-Helmholtz instability is, for example, responsible for the formation of the billow clouds---and because these provide one of the most important mechanisms for developing and sustaining turbulence. 
Binary neutron star mergers are no exception. 
Before we march on, let us point out that the analysis in this chapter is \emph{Newtonian}, in marked contrast with the rest of this work.
The reason, quite naturally, is that the nature of the magneto-rotational instability is typically discussed in a Newtonian setting, with almost no exception to the best of our knowledge.

The magneto-rotational instability was discovered by Balbus and Hawley in the early 1990s~\citep{BalbusHawley1,BalbusHawley2,BalbusHawleyRev} (linking to earlier ideas from, for example, \citet{Chandrasekar1960} and \citet{velikhov1959}). Due to the fast instability  growth rate, this mechanism is considered the most promising candidate for developing/sustaining magneto-hydrodynamic turbulence in accretion disks as well as  explaining enhanced angular momentum transfer \citep{BalbusHawleyRev,Shakura+1973}. 
The instability is due to an interplay between the magnetic field and a sheared background flow. With few exceptions (see for example \cite{MahajanKrishan2008} and \cite{Shakura2022}) and due to its ``local'' nature, the magneto-rotational instability is  discussed in the so-called ``shearing sheet approximation'' \cite{GoldreichLynden-Bell65,Hill1878}. That is, the instability is established in  a frame that corotates with a fiducial point in the mid-plane of the undisturbed disk (see also \cite{GoodmanXu94}). 
This is convenient both for analytical studies as well as numerical analysis since local simulations can reach much higher resolution than global ones (see \cite{1995ApJ...440..742H,ZierVolker22} and references therein).

Although originally discussed in the context of accretion disks, the magneto-rotational instability is expected to play a role also in neutron-star mergers \cite{Duez+MRI,SiegelMRI,PalenzuelaBampinBNS,Margalit2022,Hayashi22_1s,Kiuchi2022_1s}, especially for sustaining a magneto-turbulent state in the outer envelope of the remnant, where the Kelvin-Helmholtz instability is less significant or, indeed, not active \citep{Kiuchi_2018}. 
To assess whether or not the magneto-rotational instability is  active and resolved in merger simulations, criteria discussed/established in the context of accretion disks \citep{Hawley:2011tq,Hawley:2013lga,shibataNR} are often used. 
However, because binary neutron star mergers are highly dynamical environments, framing a discussion of the magneto-rotational instability using criteria that exploits restrictive symmetry conditions might be misleading.
Motivated by this, we aim to discuss the impact of relaxing common assumptions---well-motivated in the accretion disks scenario, like an axi-symmetric and circular background flow---on the magneto-shear instability.
The results presented in this chapter will be published in \cite{MRIpaper}.

\section{Background gradients and plane-wave expansion}\label{sec:WKBlocalbox}

Let us begin by observing that the magneto-rotational instability is, in some sense, a ``global instability analyzed with local tools''. 
The local nature is  evident since the instability is established by means of a dispersion relation (hence involves a plane-wave expansion and, by assumption, a short-wavelength approximation). 
At the same time, one may appreciate the ``global nature'' of the instability by recalling the key aspects of the instability: The addition of a weak magnetic field turns  axisymmetric modes (which would otherwise be hydrodynamically stable) unstable.
The global axisymmetry of the background, then, plays a crucial role as the relevant hydrodynamic stability criterion---the Rayleigh criterion \citep{Rayleigh1917}---applies to axisymmetric modes only.
Although the standard derivation of the instability does not highlight this subtlety, this  aspect becomes apparent if we formulate the problem using a co-rotating local frame (cf. the discussion in \cref{app:MRIlocal}).

With these points in mind, let us  spell out how we intend to discuss the magneto-shear instability without referring to a given axisymmetric and circular background. Consistently with the shearing box idea \citep{GoldreichLynden-Bell65,Hill1878}, the strategy is to zoom in on a small region of fluid---small enough for the analysis to be local but large enough to allow for a meaningful hydrodynamic description. 
We then set up a local Cartesian frame comoving with the background flow---so that the background velocity vanishes at the origin of the local box. 
As this frame moves around with the flow---and hence cannot be expected to be inertial---we need to consider the (at this point Newtonian) ideal magneto-hydrodynamics equations in a non-inertial frame.
The non-inertial equations will then be perturbed---retaining gradients in the background quantities as explained below---and a local WKB-type dispersion relation will be derived and studied.  
This way we can account for the effects of a sheared background and its interplay with the magnetic field in a general setting. 

Strictly speaking, the plane-wave expansion only makes sense for a homogeneous background ---that is, the plane-wave amplitude is assumed to vary over the same scales as the background. 
At the same time, we know that a sheared background is key to the magneto-rotational instability. 
Therefore, given any quantity/field $a$, we first write it as a sum of background plus perturbations
\begin{equation}
    a = A + \delta A \;,
\end{equation}
and then introduce a WKB-type expansion of the form \citep{ThorneBlandford,Anile}
\begin{equation}\label{eq:WKBexp}
    \delta A = \bar\delta \left(\sum_{q=0} \epsilon^q \bar A_q \right)e^{i\theta/\eps}\;,
\end{equation}
with book-keeping parameters $\bar\delta$ and $\eps$ (see also \cite{Palapanidis}). 
The former ($\bar \delta$) is introduced to measure the relative magnitude of background vs. perturbations, while the latter ($\epsilon$) is given by $\eps \approx \lambda / L$ where $\lambda$ is the typical wavelength of the waves and $L$ is the typical lengthscale over which the wave amplitude, polarization and wavelength vary.
Having split the perturbations into amplitude and phase, we follow the standard convention \citep{GravitationMTW} and stick all ``post-geometric optics'' corrections into the amplitude $\bar A_q$. 
With this Ansatz, the background equations are obtained by collecting all terms of order $\O(\bar\delta^0, \eps^0)$, while the perturbation equations are obtained collecting terms of order $\O(\bar\delta^1, \eps^0)$. 
Terms of higher order in $\eps$ correspond to post-geometric optics, while those of higher order in $\bar \delta$ represent non-linear perturbations. 

Along with this WKB-type Ansatz, we need to introduce the concept of fast and slowly varying quantities. Given a specific choice of coordinates, a quantity is slow in the variable $x$ if $A = A(X) $ where $X = \eps x$ while it is fast if $A = A(x)$. 
Deciding which quantities are fast or slow corresponds to specifying (in a qualitative manner) the background configuration. 
As an illustration, consider the simple toy problem 
\begin{equation}
    a (\partial_x b + \partial_x c) =0 \;,
\end{equation}
together with the Ansatz from \cref{eq:WKBexp}. 
Let us first assume that both $B$ and $C$ are fast, so that $\partial_x B \approx \O(\bar\delta^0,\eps^0)$ and similarly for $C$. The background equation is then
\begin{equation}
    A\left(\partial_x B + \partial_x C\right) = 0 \;.
\end{equation}
If we  instead assume that, say, $B$ is fast while $C$ is slow, then $\partial_x B \approx \O(\bar\delta^0, \eps^0)$ while $\partial_x C \approx \O(\bar\delta^0,\eps)$ and the background equation becomes
\begin{equation}
    A \partial_xB =0\;.
\end{equation}
Clearly, the two problems are different already at the background level.
Let us now turn to the linear perturbations. 
Because we have explicitly introduced the book-keeping parameter $\eps$ in \cref{eq:WKBexp}, we take all amplitude terms as well as the phase to be slowly varying. 
Then, to order $\O(\bar\delta,\eps^0)$ we have
\begin{equation}
    \left(A + \bar\delta \bar A_0 e^{i\theta/\eps}\right) \partial_x  \left[B + \bar\delta \bar B_0 e^{i\theta/\eps} + C + \bar\delta \bar C_0 e^{i\theta/\eps}\right] =0 \;.
\end{equation}
Assuming again that the background quantity $B$ is fast, while $C$ is slow, the perturbation equation becomes 
\begin{equation}
    \bar A_0 \left(\partial_x B\right) e^{i\theta/\eps} + A\left(\bar B_0 \partial_x e^{i\theta/\eps} + \bar C_0 \partial_x e^{i\theta/\eps}\right) =0 \;.
\end{equation}
Next, Taylor expanding the phase---which is slowly varying---we get
\begin{equation}
    \frac{\theta(x)}{\eps} \approx \frac{\theta(0)}{\eps} + \pdv{\theta}{X}\Big|_{X=0}x + \dots = \theta(0)/\eps + k_x x + \O(\eps) \;,
\end{equation}
where we define the wave-vector $k_x = \partial\theta / \partial X$ from the first order term in the expansion,  while the overall constant can be neglected.
Then, introducing an analogous expansion for the fast background gradients 
$\partial_x B (x)= \partial_x B(0) + \O(\eps)$ we end up with
\begin{equation}
    \bar A_0 (\partial_x B) + A\left(ik_x \bar B_0 + i k_x \bar C_0\right) =0 \;,
\end{equation}
where both $\partial_x B$ and $A$ are evaluated at a point (conveniently chosen as the origin of the coordinate system).
Therefore, if all background quantities are ``slow'', we get back the dispersion relation we would have obtained ignoring all background gradients. This is quite intuitive. 
However, the strategy also allows us to account for the impact that ``fast'' background gradients have on the dispersion relation.
In short, as long as these terms are treated as constants, we may retain them and work out a dispersion relation in the usual way. 

\section{The slowly evolving background}\label{sec:SlowlyEvolving}

The starting point for any hydrodynamic perturbation analysis is the choice/identification of a stationary background flow configuration, which is then perturbed in order to establish stability (or not). 
Here, we want to frame the analysis of the magneto-shear instability without considering a specific background configuration (with constraining symmetries) stated from the outset.  Nonetheless, we need to clarify how we can refer to a suitable ``background'' in highly dynamical environments like binary neutron star mergers. Given real numerical simulation data, this discussion will inevitably involve some kind of filtering operation.  Anticipating that this can be done in a meaningful way, we consider perturbations evolving rapidly  with respect to the evolution time-scale of an unspecified ``background'' flow. 

To make this statement more precise, let us consider the inertial ideal MHD equations and introduce reference values for each quantity (indicated with an ``$r$'' subscript) such as $\rho = \rho_r \tilde \rho$. 
We introduce the (dimensionless)
Strouhal, Mach, Froude and magnetic interaction numbers as
\begin{equation}
    \veps_\text{St} = \frac{l_r}{t_rv_r} \;,  \quad \veps_\text{Ma} = \frac{v_r}{c_r}\;, \quad \veps_\text{Fr} = \frac{v_r}{\sqrt{\Phi_r}} \;, \quad \veps_{B} = \frac{B_r^2}{\mu_0\rho_r v_r^2} \;,
\end{equation}
where $l_r,\,t_r,\,v_r$ are characteristic lengthscale, timescale and velocity (respectively) while $B_r,\,\phi_r,\,\rho_r$ are reference values for the magnetic field, gravitational potential and density and $c_r$ is the (adiabatic) speed of sound. 
This way, the non-dimensional inertial ideal MHD equations read (now dropping the ``tilde''s for notational clarity)
\begin{subequations}
\begin{align}
    \veps_\text{St} \,\pd{t}{\rho} &= - \rho \nabla_i v^i - v^i \nabla_i \rho \;, \\
    \veps_\text{St} \,\pd{t}{B^i} &= - v^j \nabla_j B^i + B^j\nabla_j v^i - B^i \nabla_jv^j \;,\\
    \veps_\text{St}\, \pd{t}{v^i} &= - v^j \nabla_j v^i -\frac{1}{\eps_{\text{Ma}}^2} \frac{1}{\rho}\nabla^i\rho -\frac{1}{\eps_{\text{Fr}}^2} \nabla^i \Phi - \eps_{B}\frac{1}{\rho} \left[  B^j  \nabla_j B^i -\nabla^i \left(\frac{B^2}{2}\right) \right] \;.
\end{align}
\end{subequations}
From this we see that a generic flow configuration can be considered slowly evolving (in time) as long as the corresponding Strouhal number is small. 
In practice, given a characteristic lengthscale $l_r$ and velocity $v_r$ of a generic flow, we consider disturbances evolving on timescales $t_r$ such that $\eps_{\text{St}}\ll 1$---over which the background can be effectively taken as stationary.
In turn, this  determines the time-scales over which we expect the following results to be reliable.

\subsection{Velocity gradient decomposition}\label{subsec:vel_gradients}
In the following we will consider the impact that gradients in the background flow velocity have on the time evolution of perturbations.
It is then convenient to introduce the standard decomposition of the velocity gradient into expansion, shear and vorticity.
Even though this has been used many times before in this thesis, it makes sense to write it down explicitly here as we are working at a Newtonian setting. 
The three vector velocity gradient decomposition in the Newtonian case reads 
\begin{equation}
     \nabla_i v_j =  \frac{1}{3}\theta g_{ij} + \sigma_{ij} + \om_{ij} \;,
\end{equation}
where
\begin{subequations}\label{eq:velGradDec}
\begin{equation}
    \theta = \nabla_i v^i \;, \quad \sigma_{ij} = \nabla_{(i}v_{j)} - \frac{1}{3}\theta g_{ij}= \frac{1}{2}\left(\nabla_{i}v_{j}+ \nabla_j v_i\right) - \frac{1}{3}\theta g_{ij}\;,
\end{equation}
and
\begin{equation}
    \omega_{ij}=  \nabla_{[i}v_{j]} = \frac{1}{2}\left(\nabla_{i}v_{j}- \nabla_j v_i\right)  \;.
\end{equation}
\end{subequations}
In order to bring out the magneto-shear nature of the instability, we will consider the impact of having a background with non-negligible shear and vorticity separately. 
We will, however, not consider the impact of a background expansion rate as exact non-linear results are sufficient to predict this.  In fact, 
due to the Alfvén theorem, we  know that the magnetic intensity must grow in a (ideal magneto-)fluid undergoing compression as the field lines are  squeezed together.  Similarly, the field will get weaker in an expanding fluid. In essence, 
we expect---and have verified explicitly---this non-linear prediction to emerge in the  analysis as a generic ``instability''. The background magnetic field cannot grow in time as it is assumed to be slowly evolving by construction, so the required growth must be represented by perturbations. 

In what follows, we will first analyse the problem analytically, and come back to discuss the link to/relevance for numerical simulations in the concluding remarks in \cref{sec:MRIsummary}.
Before we move on though, it is useful to take a brief detour and consider a realization of a flow with only non-negligible shear.
Because we are interested in flows that are ``slowly evolving'' we can start by assuming that $v^i = v^i(\eps t,x,y,z)$ and suppress the time dependence in the following.
We then take the velocity vector as mainly two-dimensional, specifically in the $x-y$ plane of a set of local Cartesian coordinates
\begin{equation}
    \vec v = v^x( x,y,z) \hat x+ v^y(x, y, z) \hat y + \O(\eps) \;,
\end{equation}
where, in order to make sure the expansion is small, we take $\pd{x}{v^x} = - \pd{y}{v^y} + \O(\eps)$.
The shear matrix  is then given by
\begin{equation}
    \gvec{\sigma} = \begin{pmatrix}
    \pd{x}{v^x} & \frac{1}{2}\left(\pd{x}{v^y} + \pd{y}{v^x}\right) & \frac{1}{2}\pd{z}{v^x} \\
    \frac{1}{2}\left(\pd{x}{v^y} + \pd{y}{v^x}\right) & - \pd{x}{v^x}  & \frac{1}{2}\pd{z}{v^y} \\
    \frac{1}{2}\pd{z}{v^x} & \frac{1}{2}\pd{z}{v^y} & 0
    \end{pmatrix} + \O(\eps) \;,
\end{equation}
while the curl of $\vec v$ becomes
\begin{equation}
    \nabla \times \vec v = -\left(\pd{z}{v^y}\right) \hat x + \left(\pd{z}{v^x} \right)\hat y  + \left( \pd{x}{v^y} - \pd{y}{v^x}\right) \hat z \;.
\end{equation}
This has to be of $\O(\eps)$ for background flows with only non-negligibile shear, in which case
\begin{equation}
    \pd{z}{v^x} = \pd{z}{v^y} =  \O(\eps) \;, \qquad \pd{y}{v^x} = \pd{x}{v^y} + \O(\eps), 
\end{equation}
and as result the determinant of the shear matrix vanishes (more precisely, is of order $\O(\eps)$).
This is equivalent to saying that two eigenvalues of the shear matrix are opposite and the third is zero (to order $\O(\eps)$). 
In essence, a mainly two-dimensional flow with negligible expansion and vorticity is characterized by a shear matrix with one zero eigenvalue, and hence a vanishing determinant.
This will turn out to be a useful observation later on. 
We also note that having negligible expansion (although relevant for the present analysis) is not strictly necessary to the argument. 

\section{Non-inertial equations and the local frame}\label{sec:NonInertialEq}

The ideal magneto-hydrodynamic equations above hold in an inertial frame. 
As an observer locally comoving with the fluid cannot be expected to be inertial (in general) we need to consider the equations according to a non-inertial observer. 
The non-inertial MHD equations have been derived earlier, so that we will here build on the results of \cref{sec:MHDfibration}.
However, as we are interested in a local analysis, we now take a step further and make contact with the concept of local frame associated with an observer (see \cite{GourghoulonSR,GravitationMTW}). 
In doing so we, for a moment, go back to use concepts and notations that are common in relativity. 
This is required as we need to make contact with the non-inertial induction equation derived in \cref{sec:MHDfibration} using a relativistic language. 
Given an observer worldline with tangent $U^a$, the local frame is constructed by considering three spatial unit vectors that complete $U^a$ to an orthonormal basis on the tangent space at a point (see \cref{app:FermiCoord} for more details). 
These three spatial vectors are then transported along the worldline as 
\begin{equation}\label{eq:frameTransport}
    \nabla_U e_{\hat a} = \Om^{\hat b}_{\;\hat a} e_{\hat b} \;,\quad \Om_{\hat a \hat b} = U_{\hat a}a_{\hat b}- a_{\hat a}U_{\hat b} -\eps_{\hat a\hat b\hat c\hat d}U^{\hat c} W^{\hat d} \;,
\end{equation}
where $a^{\hat a}$ is the four-acceleration of $U^a$ (an intrinsic property of the worldline) and $W^{\hat d}$ is the \emph{arbitrary} four-rotation of the local frame.
Let us then look at the non-inertial induction equation, which we report here for convenience\footnote{We drop the last two terms in equation \cref{eq:RelIndNI}, cf. discussion at the end of \cref{sec:MHDfibration}.}
\begin{equation}\label{eq:NewtIndNI}
    \perp^{{\hat a}{\hat b}}\dot b_{\hat b}  + D_{\hat b}(v^{\hat b} b^{\hat a}) - D_{\hat b} (v^{\hat a} b^{\hat b})=  \left(\sigma^{{\hat a}{\hat b}} - \omega^{{\hat a}{\hat b}} - \frac{2}{3}\theta\perp^{{\hat a}{\hat b}}\right)b_{\hat b}  \;.
\end{equation}
Focusing on the first term, and using \cref{eq:frameTransport} 
\begin{equation}
    \dot b_{\hat b} = (\nabla_U b)_{\hat b} = u^{\hat c} \partial_{\hat c} b_{\hat b} + \left(a^{\hat c}b_{\hat c}\right)u_{\hat b} + \veps^u_{\hat b \hat e\hat c}W^{\hat e} b^{\hat c}\;.
\end{equation}
The second term vanishes due to the orthogonal projection, while\footnote{Note that we are here identifying the vorticity of the fibration observer with the four-rotation of the local frame chosen.}
\begin{equation}\label{eq:Wdropsout}
    \perp^{\hat a \hat b}\dot b_{\hat b} + \om^{\hat a\hat b}b_{\hat b} = \perp^{\hat a \hat b} \left(u^{\hat c}\partial_{\hat c} b_{\hat b}\right)\;.
\end{equation}
In practice, the term involving the four rotation of the frame drops out of the induction equation. We also note that, because we are now considering the non-inertial equations in the local frame of a single observer, there is no shear or expansion and the induction equation in the Newtonian limit simplifies to 
\begin{equation}
    \partial_t B^{\hat i} + \nabla_{\hat j}\left(v^{\hat j}B^{\hat i} - v^{\hat i}B^{\hat j}\right) = 0 \;.
\end{equation}
At the Newtonian level then, the induction equation in the local frame of a generic observer retains the same form as for an inertial one. 
This is similar to the case of the Lorentz force (entering the Euler equation) and the Ampère law.
Let us nonetheless stress that additional terms involving the four-acceleration of the observer worldline do appear at the special relativistic level, even though working with the ideal MHD induction equation may be somehow controversial in a fully-relativistic regime (cf. discussion in \cref{sec:MHDfibration}). 

When it comes to the non-inertial terms in the Euler equations, these are obviously well known: we have to account for fictitious accelerations. 
We refer to \cite{GourghoulonSR} for a rigorous derivation of the fictitious forces in special relativity, showing also how additional terms involving the observer four-acceleration enter the relativistic expressions.
We also stress that working with a rotating or non-rotating local frame is entirely a matter of choice (see \cite{GravitationMTW}). 
At the local Newtonian level then, we can always get rid of the non-inertial terms associated with the frame rotation. 
As the linear acceleration of the observer drops out of the perturbation equations, this means we can effectively work with the inertial equations. 

We conclude this section by noting that, as previously anticipated, some kind of filtering operation is key to separate between background and fluctuations in a highly dynamical environment. 
Postponing a discussion of this to \cref{sec:MRIsummary}, let us simply note at this point that the notion of local frame discussed here is clearly linked to the covariant filtering procedure discussed in \cref{ch:LES}. 

\section{Going back to hydrodynamics}\label{sec:Back2Hydro}

As briefly hinted at in \cref{sec:WKBlocalbox}, the magneto-rotational instability relies on the \emph{hydrodynamic} stability of axisymmetric modes. The generic instability problem is more involved. 
If we relax the symmetry assumptions on the background, we need to consider the fact that hydrodynamic shear flows tend to be unstable. That is, we expect to find instabilities to appear already at the hydrodynamic level. Clearly, such instabilities would be affected by a magnetic field but not caused by it in the first place. This is an important distiction seeing as the magneto-rotational instability is \emph{due to} the presence of the  magnetic field.

With this observation in mind, let us first consider the fluid problem. 
This will be useful for two reasons: First, it will allow us to get a better grasp on the magnetic field impact on the instability. Second, it will allow us to make contact with the Rayleigh criterion (and ultimately the magneto-rotational instability). 
As the fluid problem is much simpler than the magneto-fluid one, we will study the case where both shear and vorticity gradients are retained, and also discuss the impact of shear viscosity---either of microphysical origin or due to filtering as in the Smagorinsky model (or, more in general, in the so-called eddy-viscosity type models, see \cref{sec:energy_cascade}). 
Shear viscosity is introduced in the usual way (see, for example, \citet{LandauFLuidMechanics}), and the shear viscosity coefficient $\eta$ will be considered constant consistently within the local analysis.

Before we move on to discuss the perturbation equations and the resulting dispersion relation(s) though, it is worth stressing that, in many situations of interest the relevant dynamics is either sub- or supersonic. 
As such, for these problems it is worth considering models that filter out modes that are either faster or slower than the sound waves. 
This can be done starting from a fully compressible dispersion relation and taking either of two limits: either we assume the speed of sound to be very large, in which case the model becomes
sound-proof (we point to \cite{Vasil_2013} for more details), or very small. In the following, we typically work in the sound-proof limit, noting that the MRI is commonly discussed within the so-called Boussinesq approximation \cite{BoussinesqBarletta}, thus removing fast magneto-sonic waves \cite{BalbusHawley1}.

Starting from the continuity equation, perturbing it and introducing the plane-wave expansion we readily obtain
\begin{equation}\label{eq:MHDpertCont}
    \partial_t \delta \rho + \delta\rho \nabla_i v^i + v^i \nabla_i \delta\rho + \rho \nabla_i \delta v^i  = 0  \Longrightarrow -i \omega \delta\rho + i \rho k_i \delta v^i =0 \;, 
\end{equation}
where $\om$ and $k_i$ are defined as in \cref{sec:WKBlocalbox}.
Note that we set $v^i =0$ as we evaluate the relation at the centre of the local box, and assume that  the background expansion rate $\nabla_i v^i$ can be neglected.
In a similar fashion, the perturbed Euler including a shear-viscous term gives
\begin{multline}\label{eq:HydroEulerVisc}
     \partial_t \delta  v_i + \delta v^j \nabla_j v_i + \frac{1}{\rho} \nabla_i \delta P - \delta \left(\eta \nabla^j \tau_{ji}\right) = 0 \\
     \Longrightarrow -i\omega \delta v_i + i \frac{c_s^2 }{\rho} k_i \delta \rho + \sigma_{ij}\delta v^j + \eps_{ijk} W^j \delta v^k - \delta \left(\eta \nabla^j \tau_{ji}\right)=0 \;,
\end{multline}
where $\tau_{ji}$ is the rate-of-strain/shear tensor and $W^i = 1/2 \eps^{ijk}\om_{jk}$.
Working this out we retained gradients in the background flow only, used the velocity gradient decomposition (\cref{subsec:vel_gradients}), introduced the adiabatic speed of sound $c_s^2 = \partial P / \partial \rho$, and considered the gravitational potential to be externally sourced (hence neglecting its perturbations). 

In order to derive the dispersion relation and study the effects of a sheared background, it is convenient to choose a basis that is adapted to it. 
Because the shear is a trace-free symmetric matrix, we know there exists a basis (in the tangent space) whereby 
\begin{equation}\label{eq:ShearBasis}
    \sigma^{ij} = \text{diag} \left(\sigma_1, \sigma_2, - (\sigma_1 + \sigma_2)\right) \;.
\end{equation}
We will make use of this basis to write down the coefficient matrix of the linearized system. 
Before doing so, however, it is reasonable to wonder whether this change of basis has any impact on the perturbation equations.
We are, always free to choose a basis in the tangent space that is not associated with the coordinates chosen, but this (in general) introduces additional terms in the covariant derivative.
Let us spell out why this is not the case here. Working with a non-coordinate basis, we need to account for spin-coefficients when a derivative acts on vectors and tensors. 
The spin coefficients are given by the sum of two terms (see the formula in the Notation chapter, or \cite{GRCarroll} for more details).
The first involves the Christoffel symbols associated with the coordinates chosen, and thus vanish as we are working with a non-rotating Cartesian frame. 
The second term instead stems from the fact that the change of basis matrix (translating the coordinates base into the shear-adapted one) may be different from point to point. 
In the context of this analysis, however, we are looking at scales smaller than those over which background quantities vary.
In essence, also this second term vanishes as the shear matrix is (by construction) constant over the local region of fluid we are zooming on.

Working in the shear adapted basis, we write the coefficients matrix of the linearized system as 
\begin{equation}
    \textbf{M} = \begin{pmatrix}
    -\om & \rho k_1 & \rho k_2 & \rho k_3 \\
    \frac{c_s^2}{\rho}k_1 & - \om - i \sigma_1 - i \eta L_1 & i W^3 - \frac{i}{6} \eta k_1 k_2 & - i W^2 - \frac{i}{6}\eta k_1 k_3\\
    \frac{c_s^2}{\rho}k_2 &- i W^3 - \frac{i}{6}\eta k_2 k_1 &  - \om - i \sigma_2 - i \eta L_2 & i W^1 - \frac{i}{6}\eta k_2 k_3 \\
    \frac{c_s^2}{\rho}k_3 &  i W^2 - \frac{i}{6}\eta k_3 k_1 & - i W^1 - \frac{i}{6}\eta  k_3 k_2 &  - \om + i \sigma_2 + i \sigma_2- i \eta L_3
    \end{pmatrix} \;,
\end{equation}
where $L_1 = \frac{2}{3} k_1 ^2 + \frac{1}{2}k_2 ^2 + \frac{1}{2} k_3^2$ and $L_2,\,L_3$ are similarly defined. 
The dispersion relation is computed taking the determinant of this matrix and equating it to zero.
In order to keep the discussion as general as possible (i.e. without having to refer to a specific background configuration) we will decompose the coefficients of the characteristic polynomial in terms of scalars built from background quantities. 
In the simplest cases this can be done ``by eye'', but the procedure can easily become quite messy.
The logic is nonetheless simple: we group the different terms in each coefficients according to the power of the various background quantities, for example we group all the terms quadratic in the shear and wave-vector components. 
We then build all the possible scalars that are quadratic in the shear and wave-vector, and look for the correct linear combination of them.
This logic can be easily implemented on a computer algebra program such as Mathematica\footnote{\href{https://www.wolfram.com/mathematica/}{https://www.wolfram.com/mathematica/}}. 
We now discuss the dispersion relations obtained by retaining only shear terms, both shear and viscous terms, and lastly shear and vorticity terms. 
Before doing so, we observe that the coefficients will involve scalars constructed from the shear matrix only.
As any $3\cross3$ matrix, the shear matrix $\gvec \sigma$ has three invariants 
\begin{equation}
    I_1 = \text{Tr}(\gvec \sigma) \;, \quad I_2 = \frac{1}{2}\left[ \text{Tr}(\gvec \sigma^2) - \left( \text{Tr}(\gvec \sigma)\right)^2\right]\;, \quad I_3 = \text{det}(\gvec \sigma)\;,
\end{equation}
related via the Cayley-Hamilton theorem as
\begin{equation}
    \gvec \sigma^3 - I_1 \gvec \sigma^2 + I_2 \gvec \sigma - I_3 \mathbb{I} = 0 \;,
\end{equation}
where $\mathbb{I}$ is the $3\cross3$ identity matrix.
Because the shear matrix is trace-free, we will write the coefficients in terms of $\sfrac{1}{2}\text{Tr}(\gvec\sigma^2)$ and $\text{det}(\gvec\sigma)$.

We begin our analysis by considering a background with negligible vorticity, and set the shear viscosity to zero.
The resulting dispersion relation is 
\begin{equation}
    \om^4 + a_2 \om^2 + a_1 \om + a_0 = 0 \;,
\end{equation}
with 
\begin{subequations}
\begin{align}
    a_2 & = -c_s^2 k^2 + \frac{1}{2}\text{Tr}(\gvec \sigma^2) \;,\\
    a_1 &= i \left[c_s^2 \sigma_{ij}k_ik_j - \text{det}(\sigma)\right] \;,\\
    a_0 & = c_s^2 \left[\sigma^2_{ij}k_ik_j -\frac{1}{2}\text{Tr}(\gvec \sigma^2)  k^2\right] \;.
\end{align}
\end{subequations}
We then take the sound-proof limit (i.e. we retain only terms proportional to the speed of sound) and consider the case $\text{det}(\gvec\sigma) =0$, looking for modes such that $\sigma^{ij}k_j =0$. 
Recalling that, as discussed in \cref{subsec:vel_gradients}, a mainly two dimensional flow with negligible vorticity is characterized by having a shear matrix with vanishing determinant---that is $\text{det}(\gvec\sigma) \sim \O(\eps)$---and noting that we can choose the orientation of the local axes in such a way that the background flow is, say, along the $\hat x,\,\hat y$ directions, we can always consider the determinant to be zero. 
This means that there always exists a wave-vector living in the eigen-space corresponding to the zero eigenvalue. 
Then we end up with\footnote{We note that we obtain the same dispersion relation also in the opposite limit where the speed of sound is tiny.}   
\begin{equation}\label{eq:HydroFastestModes}
    \om^2 = -\frac{1}{2}\text{Tr}(\gvec \sigma^2)  \Longrightarrow \om = \pm i \sqrt{\frac{1}{2}\text{Tr}(\gvec \sigma^2) }\;.
\end{equation}
That is such modes are non-propagating, and half of them are unstable with a growth rate independent of the wave-vector. 
Next, we consider wave-numbers such that $\sigma_{ij} k_i  k_j = 0$, noting that such modes will always exist. 
In the shear-adapted basis they are characterized by $k^1 = k^2 = k^3$ if the determinant is not vanishing (that is $s_1 \neq s_2$), and $k_1 = k_2$ when it does.
It follows that for such modes
\begin{equation}\label{eq:N2Ftrick}
    -\frac{1}{2}\text{Tr}(\gvec\sigma^2) + \sigma^2_{ij}\hat k^i \hat k^j  = - \frac{1}{6}\text{Tr}(\gvec\sigma^2)  \;,
\end{equation}
where $\hat k = \vec k / |\vec k|$. 
We then obtain
\begin{equation}\label{eq:HydroNext2Fastest}
    \om^2 = -\frac{1}{6}\text{Tr}(\gvec\sigma^2)\Longrightarrow \om = \pm i \sqrt{\frac{1}{6}\text{Tr}(\gvec \sigma^2) } \;.
\end{equation}
These modes are also non-propagating, and half of them are unstable with a (constant) growth rate about a factor of
$2$ smaller. 
As the dispersion relation is quadratic (in the sound-proof limit), we can explicitly solve
it and confirm the expectation (and well known fact) that shearing flows are generically unstable.

Let us now build on this and discuss how vorticity and shear viscosity impact on the generic instability of sheared flows. 
First we consider the case where the background has negligible vorticity but non-vanishing shear viscosity. 
As sanity check, we observe that if we also set the background shear to be negligible, and take the sound-proof limit\footnote{We have also verified, using the Routh-Hurwitz criterion (see \cite{korn2013mathematical} and \cref{app:RH}) that the same result holds true in general, not only in the sound-proof limit.} we obtain
\begin{equation}
    \om^2 +i \eta k^2 \om - \frac{1}{4}\eta^2 (k^2)^2 = \left(\om + \frac{i}{2}\eta k^2 \right)^2  = 0 \;,
\end{equation}
with stable roots provided $\eta >0$. 
We recall that when viscosity is of microphysical origin, $\eta >0$ follows from the second law of thermodynamics (see \cite{LandauFLuidMechanics,livrev}). 
If viscosity is instead due to filtering, a positive value of $\eta$ corresponds to an eddy-type model where energy is cascading to smaller/unresolved scales (see \cite{SchmidtLES,lesbook,LesieurLES}). 
With this observation in mind, let us go back to the case with both both shear and viscosity, in which case the dispersion relation is
\begin{equation}
    \om^4 + a_3 \om^3 + a_2 \om^2 + a_1 \om + a_0 = 0\;,
\end{equation}
with 
\begin{subequations}
\begin{align}
    a_3 &= \frac{5}{3}i \eta k^2 \;, \\
    a_2 &= -c_s^2 k^2 + \frac{1}{2}\text{Tr}(\gvec\sigma^2) - \frac{1}{12}\eta \left[11 \eta (k^2)^2 -2 \sigma_{ij} k_i  k_j\right] \;, \\
    a_0 &= c_s^2 \left[\sigma^2_{ij} k_i  k_j - \frac{1}{2}\text{Tr}(\gvec\sigma^2) k^2 - \frac{1}{2}\eta k^2 \sigma_{ij} k_i  k_j + \frac{1}{4} \eta^2 (k^2)^3\right]\;,
\end{align}
and 
\begin{multline}
     a_1 = i c_s^2 \left[\sigma_{ij} k_i  k_j - \eta (k^2)^2\right]\\ + i \left\{ -\frac{1}{6} \left[\sigma^2_{ij} k_i k_j -2 \text{Tr}(\gvec \sigma^2)k^2 \right] + \frac{1}{12}\eta^2 k^2 (\sigma_{ij} k_i  k_j) - \frac{1}{6}\eta^3 (k^2)^3 - \text{det}(\gvec \sigma)\right\}\;.
\end{multline}
\end{subequations}
As before, we first consider modes such that $\sigma_{ij}\hat k_j =0$, whose dispersion relation is
\begin{equation}
    \om^2 +i \eta k^2 \om - \frac{1}{4}\left( \eta^2 (k^2)^2 - 2 \text{Tr}(\gvec\sigma^2) \right) =0 \;.
\end{equation}
Assuming $\eta >0$, stability corresponds to 
\begin{equation}
    \eta^2 (k^2)^2 - 2 \text{Tr}(\gvec\sigma^2)  > 0 \;.
\end{equation}
In essence, comparing this to \cref{eq:HydroFastestModes} we see that viscosity tends to stabilize shear-unstable modes, with a larger impact at smaller scales. 
This makes intuitive sense. 
Next, consider modes such that $\sigma_{ij}\hat k_i \hat k_j =0$, whose dispersion relation is
\begin{equation}
     \om^2 + i\eta k^2 \om - \frac{1}{12}\left( 3\eta^2 (k^2)^2 - 2 \text{Tr}(\gvec\sigma^2) \right) =0 \;,
\end{equation}
and we made use of \cref{eq:N2Ftrick}.
As before, these modes---to be compared with their counterparts in \cref{eq:HydroNext2Fastest}---are also stable
(assuming $\eta >0$) provided the last term in the previous equation is negative.
That is, provided the wave-number is sufficiently large. We have verified that the same trend is true for generic wavevectors.
In essence, we learn (as one may have expected) that shear viscosity generically slows the growth rate of unstable shear modes, and stabilises modes with small enough wavelengths.

Turning to the case where the background has non-negligible vorticity and shear, the dispersion relation is 
\begin{equation}\label{eq:S+Whydro}
    \om^4 + a_2 \om^2 + a_1 \om + a_0 = 0 \;,
\end{equation}
with 
\begin{subequations}
\begin{align}
    a_2 &= - c_s^2 k^2 - \vec W^2 + \frac{1}{2}\text{Tr}(\gvec\sigma^2) \;,\\
    a_1 &= i \left[c_s^2 \sigma_{ij}k_i k_j - \text{det}(\gvec\sigma) - \sigma_{ij}W_i W_j\right] \;,\\
    a_0 &= c_s^2 \left[  \sigma^2_{ij}k_i k_j - \frac{1}{2}\text{Tr}(\gvec\sigma^2)k^2 + (\vec k \cdot \vec W)^2\right]\;.
\end{align}
\end{subequations}
Taking the sound-proof limit, we first observe that the fastest growing modes encountered before, namely those characterized by $\sigma_{ij}\hat k_j =0$ are not guaranteed to exist anymore, as the determinant of the shear matrix cannot be assumed to be negligible in general (see \cref{subsec:vel_gradients}). 
Should these modes exist, though, their dispersion relation would be 
\begin{equation}
    \om^2 = -\frac{1}{2}\text{Tr}(\gvec\sigma^2) + (\hat k\cdot \vec W)^2 \Longrightarrow \om = \pm i \sqrt{-\frac{1}{2}\text{Tr}(\gvec\sigma^2) + (\hat k\cdot \vec W)^2 }\;,
\end{equation}
and we see, comparing this to \cref{eq:HydroFastestModes}, that vorticity tends to stabilize them. 
We also observe that---in contrast to shear viscosity---vorticity affects all such modes by reducing their growth rate in a way that does not depend on their wave-number (although the direction of propagation is important).
Next---and also because the modes we just looked at may not exist---we consider modes such that $\sigma_{ij}\hat k_i \hat k_j = 0$ obtaining
\begin{equation}
    \om^2 = -\frac{1}{6}\text{Tr}(\gvec\sigma^2) + (\hat k\cdot \vec W)^2\;.
\end{equation}
Comparing this to \cref{eq:HydroNext2Fastest}, we observe again that vorticity tends to stabilize such modes in a way that does not depend on their wave-number.
We have verified that the same trend is also true for generic wave-vectors. 
As a final point, it is easy to verify that the case with only background vorticity is generally stable (not only in the sound-proof limit). 

In summary, a sheared background flow is generically unstable already at the hydrodynamic level, which is a well-known fact.
However, we have considered the impact that shear viscosity and/or vorticity have on the instability of the possible hydrodynamic modes. The results show that shear viscosity tends to weaken the instability  in general, with larger effects for larger wave-numbers. Meanwhile,   vorticity has a stabilizing effect which does not depend on the wave-number. 
Finally, let us also point to \cref{subapp:RayleighCloser} where we show that the general dispersion relation derived here is shown to encompass the classic Rayleigh stability criterion.

\section{Magneto-shear instability in the local frame}\label{sec:MSinstaLocal}

Having explored the hydrodynamic case, let us  perturb the corresponding MHD equations and study the impact of the magnetic field on the generic shear instabilities we encountered.
We consider a barotropic equation of state and retain gradients in the background velocity only, as we want to focus on the magneto-shear nature of the instability (cf. \cite{BalbusHawley2,shibataNR}). 
The continuity equation is obviously unchanged, while the perturbed Euler equation becomes
\begin{multline}\label{eq:MHDpertEuler}
    \partial_t \delta  v_i + \delta v^j \nabla_j v_i + \frac{1}{\rho}\nabla_i \delta P + \frac{1}{\mu_0\rho} \left[ B_j \nabla_i \delta B^j - B^j \nabla_j \delta B_i\right] = 0 \\ \Longrightarrow -i\omega \delta v_i + i \frac{c_s^2 }{\rho} k_i \delta \rho + \frac{i}{\mu_0\rho} \left[ (B_j \delta B^j)k_i - (B^j k_j)\delta B_i \right] + \sigma_{ij}\delta v^j + \eps_{ijk} W^j \delta v^k =0 \;.
\end{multline}
Finally, the perturbed induction equation is
\begin{multline}\label{eq:MHDpertInd}
    \partial_t \delta B^i + B^i \nabla_j \delta v^j - B^j \nabla_j \delta v^i - \delta B^j \nabla_j v^i + \delta B^i \nabla_j v^j  =0 \\ \Longrightarrow -i\omega \delta B^i + i B^i(k_j \delta v^j ) - i (B^j k_j)\delta v^i  -  \sigma^{ij}\delta B_j -  \eps^{ijk}W_k \delta B_j + \frac{2}{3}\theta \delta B^i= 0 \;.
\end{multline}
We will now discuss the linearized system that follows from these equations. 
In order to keep the discussion tidy, we will first recap the mode analysis for the homogeneous case and then move on to consider a background with non negligible shear and vorticity (separately).  

\subsection{Homogeneous background: a recap}\label{subsec:HomogeneousBG}

In order to derive the fully compressible dispersion relation for the homogeneous case, we first re-scale the magnetic field as 
\begin{equation}\label{eq:RescaleB}
    \vec v_A \doteq \frac{\vec B}{\sqrt{\mu_0\rho}} \;,\quad   \delta \vec v_A \doteq \frac{\delta\vec B}{\sqrt{\mu_0\rho}} \;,
\end{equation}
and introduce a convenient basis $\{\hat v_A,\, \hat q,\,\hat s \}$ where $\hat v_A =\vec v_A /|\vec v_A|$ while $\hat q,\hat s$ complete it to an orthonormal basis.
For instance, assuming $\vec v_A$ is not aligned with $\vec k$ we can construct it as
\begin{equation}
    \vec q = \vec k - \left(\vec k \cdot \hat v_A\right)\hat v_A \,,\qquad \hat q = \frac{\vec q}{|\vec q|} \,, \qquad \hat s = \hat v_A \cross\hat q \;,
\end{equation}
so that\footnote{If the wave-vector is along the background magnetic field we just have to set $k^q =0$ in the following.} 
\begin{equation}
    \vec k = k^{v_A} \hat {v_A} + k^q \hat q  \;.
\end{equation}
The coefficient matrix of the linearized system can then be written as (cf. \cref{eq:MHDpertCont,eq:MHDpertEuler,eq:MHDpertInd} and ignore background vorticity and shear)
\begin{equation}
    \vec M = \begin{pmatrix}
    \vec A & \vec C \\
    \vec C^\top & \vec D
    \end{pmatrix}\;,
\end{equation} 
with 
\begin{subequations}
    \begin{equation}
    \vec A = \begin{pmatrix}
        -\om & \rho k_{v_A} & \rho k_q & 0 \\
        \frac{c_s^2}{\rho} k_{v_A} & - \om & 0 & 0 \\
        \frac{c_s^2}{\rho} k_q & 0 & - \om & 0 \\
        0 & 0 & 0 & -\om 
    \end{pmatrix} \;, \quad 
    \vec C = \begin{pmatrix}
        0 & 0 & 0 \\
        0 & 0 & 0 \\
        b k_q & - b k_{v_A} & 0 \\
        0 & 0 & - b k_{v_A}
    \end{pmatrix} \;, 
\end{equation}
and
\begin{equation}
    \vec D = \begin{pmatrix}
        -\om & 0 & 0 \\
        0 & - \om & 0 \\
        0 & 0 & -\om
    \end{pmatrix} \;.
\end{equation}
\end{subequations}
As $\vec D$ is clearly invertible, we can reduce $\vec M$ into factors via the Schur complement 
\begin{equation}
    \begin{pmatrix}
    \vec A & \vec C \\
    \vec C^\top & \vec D
    \end{pmatrix} = 
    \begin{pmatrix}
    \mathbf{I}_4 &  \vec C \vec D^{-1} \\
    \vec 0_{4 \times 3} & \mathbf{I}_3
    \end{pmatrix} 
    \begin{pmatrix}
    \vec A - \vec C \vec D^{-1} \vec C ^\top & \vec 0_{4\times 3} \\
    \vec 0_{3\times 4} & \vec D
    \end{pmatrix}
    \begin{pmatrix}
    \mathbf{I}_4 &  \vec 0_{4 \times 3} \\
    \vec D^{-1} \vec C ^\top & \mathbf{I}_3
    \end{pmatrix}\;,
\end{equation}
and then compute the determinant as
\begin{equation}
    \text{det}(\vec M) = \text{det}(\vec D)\, \text{det}(\vec A - \vec C \vec D^{-1} \vec C ^\top) \;.
\end{equation}
The resulting dispersion relation is 
\begin{equation}\label{eq:FCMHDwaves}
    -\om \left(\om^2 - \left(\vec v_A \cdot \vec k\right)^2 \right)\left[ \om^4 - \left(v_A^2 + c_s^2 \right) k ^2 \om^2 + c_s^2  k ^2 \left(\vec v_A \cdot \vec k\right)^2 \right] =0 \;,
\end{equation}
where the roots of the quadratic polynomial correspond to Alfv\'{e}n waves, while those of the quartic one in square brackets describe (fast and slow) magneto-sonic waves \cite{galtier2016}. 

Before moving on to discuss the impact of shear and vorticity, let us briefly note what happens to the modes when we take the sound-proof limit---where the speed of sound is large.
From \cref{eq:FCMHDwaves} we see that fast magneto-sonic waves are filtered out, while the slow ones reduce to Alfv\'{e}n waves. 
In the opposite limit---when disturbances are much faster than the sound waves---the dispersion relation describes Alfv\'{e}n waves and the low-$c_s$ limit of fast magneto-sonic waves. 
This limit corresponds to ignoring fluid pressure perturbations while retaining variations in the magnetic pressure. 

\subsection{Sheared Background}
Let us now consider the case where the background vorticity is negligible while shear terms are not.  
Re-scaling the magnetic field as in \cref{eq:RescaleB} and decomposing \cref{eq:MHDpertCont,eq:MHDpertEuler,eq:MHDpertInd} (ignoring vorticity terms) as well as $\delta\vec v $ and $\delta \vec v_A$ in the shear-adapted basis, the coefficients' matrix of the linearized system of equations reads
\begin{equation}\label{eq:ShearedMatrix}
    \begin{pmatrix}
    - \om & \rho k_1 & \rho k_2 & \rho k_3 & 0 & 0 & 0 \\
    \frac{c_s^2}{\rho} k_1 & - \om -i \sigma_1 & 0 & 0 &  I_1 & v_A^2k_1  & v_A^3 k_1 \\
    \frac{c_s^2}{\rho} k_2 & 0 & -\om -i \sigma_2 & 0 & v_A^1k_2 & I_2 & v_A^3 k_2 \\
    \frac{c_s^2}{\rho} k_3 & 0 & 0 & - \om -i \sigma_3  & v_A^1k_3 & v_A^2k_3 & I_3 \\
    0 & I_1 & v_A^1k_2 & v_A^1k_3 & -\om + i \sigma_1 & 0 & 0 \\ 
    0 & v_A^2 k_1 & I_2 & v_A^2k_3 & 0 & -\om + i \sigma_2 & 0 \\
    0 & v_A^3k_1 & v_A^3k_2 & I_3 & 0 & 0 & -\om + i \sigma_3 \\
    \end{pmatrix} \;,
\end{equation}
where
\begin{equation}
    I_1 = v_A^1 k_1 - \left(\vec v_A\cdot\vec k\right) \;, \qquad \sigma_3 = - (\sigma_1 + \sigma_2) \;,
\end{equation}
and $I_2,I_3$ are defined similarly. 

In a similar fashion as for the hydrodynamic case considered above, we will decompose the coefficients of the characteristic polynomial in terms of scalars built from background quantities.
As we might have expected, the resulting dispersion relation is a complicated seventh-degree polynomial (and we sanity-checked it reduces to the homogeneous case when we set to vanish the shear terms). 
In order to learn something useful out of it, we then consider the sound-proof limit and retain only terms proportional to the speed of sound.
We end up with the following dispersion relation
\begin{equation}\label{eq:ShearedCoeffSP}
    a_5 \om^5 + a_4 \om^4 + a_3\om^3 + a_2 \om^2 + a_1 \om + a_0 = 0 \;,
\end{equation}
with
\begin{subequations}
\begin{multline}
    a_0 = - i \bigg\{\text{det}(\gvec \sigma)\Big[\sigma^2_{ij}k^ik^j - \frac{1}{2}\text{Tr}(\gvec{\sigma}^2)\Big] + (\vec v_A\cdot\vec k)^2 \Big[\text{det}(\gvec \sigma)k^2 -\frac{1}{2}(\sigma_{ij}k^ik^j)\text{Tr}(\gvec \sigma^2)  \Big]\\
     + (\vec v_A\cdot\vec k)^4\sigma_{ij}k^ik^j\bigg\} \;,
\end{multline}
\begin{multline}
    a_1 =  \bigg\{(\vec v_A\cdot\vec k)^4 k^2 + \left(\vec v_A \cdot \vec k\right)^2\left[\sigma^2_{ij}k^ik^j - \text{Tr}(\gvec\sigma^2)k^2\right] + \text{det}(\gvec\sigma)\left(\sigma_{ij}k^ik^j\right) +\\ \frac{1}{2}\text{Tr}(\gvec\sigma^2)\left[\frac{1}{2}\text{Tr}(\gvec\sigma^2)k^2 - \sigma^2_{ij}k^ik^j \right]\bigg\} \;,
\end{multline}
\begin{equation}
    a_2 = i \Bigg\{ -\frac{1}{2}(\sigma_{ij}k^ik^j)\text{Tr}(\gvec\sigma^2) + \text{det}(\gvec \sigma)k^2 +2 (\vec v_A\cdot\vec k)^2 (\sigma_{ij}k^ik^j)\Bigg\}\;,  
\end{equation}
\begin{equation}
    a_3 =  \left[\text{Tr}(\gvec\sigma^2)k^2 -2 (\vec v_A\cdot\vec k)^2k^2 - \sigma^2_{ij}k^ik^j\right]\;,
\end{equation}
\begin{equation}
    a_4 = -i (\sigma_{ij}k^ik^j)\;,
\end{equation}
\begin{equation}
    a_5 =  k^2 \;.
\end{equation}
\end{subequations}

As in the hydrodynamic case considered earlier, we first consider the case $\text{det}(\gvec\sigma) =0$, and look for modes such that $\sigma^{ij}k_j =0$. 
It is then easy to see that the general dispersion relation in \cref{eq:ShearedCoeffSP} simplifies to (ignoring a trivial root)
\begin{equation}
    \left[\om^2 - \left(\frac{1}{2}\text{Tr}(\gvec\sigma^2) - (\vec v_A \cdot\vec k)^2\right)\right]^2 = 0\;.
\end{equation}
Comparing to the corresponding hydrodynamic modes in \cref{eq:HydroFastestModes}, we immediately see that the magnetic field tends to have a stabilizing effect (provided it is not orthogonal to the wave-vector, in which case it has no effect whatsoever).

\begin{figure}
\centering
\includegraphics[width=0.99\textwidth]{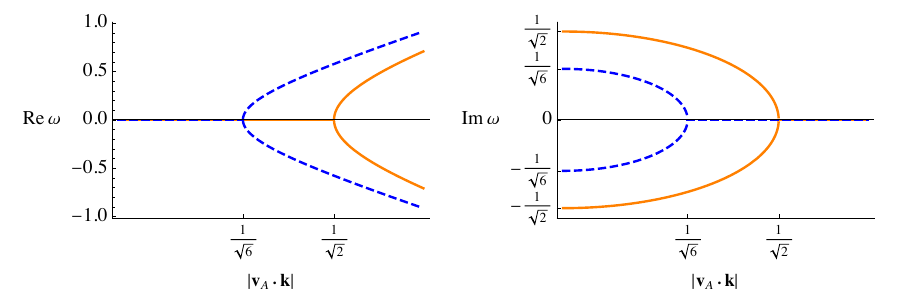}
\caption{Real and imaginary part of the solutions of \cref{eq:N2FMHDdr},with both the frequency and $|\vec v_A\cdot \vec k|$ in units of $\sqrt{\text{Tr}(\gvec \sigma^2)}$. The solutions plotted correspond to the fastest growing modes evolving on top of an MHD sheared background. We see that the magnetic field has a stabilizing effect, as the growth rates are reduced with respect to those of the corresponding hydrodynamic modes. The stabilizing effect is all the more pronounced the more the wave-vector is aligned with the magnetic field lines, and is switched off for modes propagating in the directions perpendicular to the magnetic field lines. In particular, modes corresponding to sufficiently large values of $|\vec v_A\cdot \vec k|$ are turned stable.}
\label{fig:N2FinstaMHD}
\end{figure}

Next, we take (again, as before) $\text{det}(\gvec\sigma) =0$ and consider modes such that $\sigma^{ij}k_ik_j =0$ (but $\sigma^{ij}k_j \neq 0$).
The relevant dispersion relation can then be written (making use of \cref{eq:N2Ftrick})
\begin{equation}\label{eq:N2FMHDdr}
    \om^4 + b_2 \om^2 + b_4 = 0\;,
\end{equation}
with 
\begin{subequations}
\begin{equation}
    b_2 = \frac{2}{3}\text{Tr}(\gvec\sigma^2) - 2 (\vec v_A \cdot\vec k)^2\;,
\end{equation}
and 
\begin{equation}
    b_4 = \frac{1}{12}\text{Tr}(\gvec\sigma^2)^2 - \frac{2}{3}\text{Tr}(\gvec\sigma^2)(\vec v_A \cdot\vec k)^2 + (\vec v_A \cdot\vec k)^4\;.
\end{equation}
\end{subequations}
The same stabilizing effect of the magnetic field is evident from \cref{fig:N2FinstaMHD}, where both the frequency and $|\vec v_A\cdot\vec k|$ are plotted in units of $\sqrt{\text{Tr}(\gvec\sigma^2)}$.

The key point here is that, while the background shear is required for the instability (the vanishing-shear modes are stable Alfvèn waves in the sound-proof limit), the magnetic field is not the main driver.
This is evident from the results as the imaginary part of the unstable modes remains finite in the limit $\vec v_A \to 0$, and the limiting value coincides with the hydrodynamic result (from the previous section).
This observation, possibly unexpected at first sight, deserves a thorough discussion, and we will return to this issue in  \cref{subsec:InstaVsMRI}. 
Before we expand on this, let us stress that the results make intuitive sense.
The magnetic field impacts on the instability in that it breaks the hydrodynamic isotropy and dampens the growth of unstable modes propagating along magnetic field lines. 
This also suggests that shear-instability driven turbulence is isotropic in the hydrodynamic case but inherently anisotropic for magnetized flows, consistently with the overall picture discussed in \cref{sec:IntroMHDTurbulence}. 
Before moving on, it is also worth noting that the background velocity profile considered by Balbus \& Hawley  \citep{BalbusHawley1,BalbusHawleyRev} is characterized by having a shear matrix with vanishing determinant (and expansion rate), and also that for axisymmetric modes $\sigma^{ij}k_ik_j = 0$, while for the fastest growing MRI modes (propagating vertically) $\sigma^{ij}k_j =0$.

\subsection{Background with vorticity}

Before we make contact with the usual MRI and the Rayleigh criterion, let us also consider the case with non-negligible background vorticity only.
We re-scale the magnetic field as in \cref{eq:RescaleB} and introduce a convenient basis $\{\hat W,\, \hat q,\,\hat s \}$, where $\hat W =\vec W /|\vec W|$ while $\hat q,\hat s$ complete it to an orthonormal basis. 
For instance, assuming $\vec v_A$ is not aligned with $\vec W$ we can construct it as
\begin{equation}
    \vec q = \vec v_A - \left(\vec v_A \cdot \hat W\right)\hat W \,,\qquad \hat q = \frac{\vec q}{|\vec q|} \,, \qquad \hat s = \hat W \cross\hat q \;,
\end{equation}
and the magnetic field\footnote{Note that the definition of $\hat q$ changes when the background magnetic field is aligned with the vorticity, even though in what follows we would simply have to set $v_A^q=0$.}  
\begin{equation}
    \vec v_A = v_A^W \hat W + v_A^q\hat q \;.
\end{equation}

The coefficient matrix of the linearized system then is (cf. \cref{eq:MHDpertCont,eq:MHDpertEuler,eq:MHDpertInd} and ignore shear terms)
\begin{equation}
    \begin{pmatrix}
        -\om & \rho k^W & \rho k^q & \rho k^s & 0 & 0 & 0 \\
        \frac{c_s^2}{\rho}k^W & - \om & 0 & 0 & -v_A^q k^q & v_A^q k^W & 0\\
        \frac{c_s^2}{\rho}k^q & 0 & - \om & +i W & v_A^W k^q & - v_A^ W k^W & 0 \\
        \frac{c_s^2}{\rho}k^s & 0 & -i W & - \om & v_A^W k^s & v_A^q k^s & - \left(v_A^W k^W + v_A^q k^q\right) \\
        0 & - v_A^q k^q & v_A^W k^q & v_A^W k^s & - \om & 0 & 0 \\
        0 & v_A^q k^W & -v_A^W k^W & v_A^q k^s & 0 & - \om & -iW \\ 
        0 & 0 & 0 & - \left(v_A^W k^W + v_A^q k^q\right)& 0 & + i W & - \om
    \end{pmatrix}\;.
\end{equation}

As for the sheared case, after having sanity-checked the result by contrasting it against the homogeneous background dispersion relation, we take the sound-proof limit.
The sound-proof dispersion relation can then be written as
\begin{equation}\label{eq:VortDR}
    \om^4 + b_2 \om^2 + b_4 = 0\;,
\end{equation}
with 
\begin{subequations}\label{eq:VortStabQuartic}
\begin{align}
    b_2 &= -\left[W^2 +  (\hat k \cdot\vec W)^2 + 2 (\vec v_A\cdot\vec k)^2 \right]\;, \\
    b_4 &=   \left[(\vec v_A \cdot\vec k)^2 + W^2 \right]  \left[ (\hat k \cdot \vec W)^2 + (\vec v_A\cdot\vec k)^2\right] \;.
\end{align}
\end{subequations}
 As this is a particularly simple quartic polynomial, we can study the stability of its roots analytically. 
Considering \cref{eq:VortDR} as an equation for $\om^2$ and computing the discriminant we obtain %
\begin{equation}
    \left[ W^2 - (\hat k \cdot \vec W)^2\right]^2 \ge 0
\end{equation}
so that $\om^2$-roots are real. As complex roots of a real algebraic polynomials occur in pairs of complex conjugates, complex $\om^2$-roots would correspond to an instability. 
In order to have stable roots though, we also need other conditions to be met. 
We, in fact need $b_2<0$ and $b_4>0$ to make sure that the $\om^2$-roots are real and positive, so that $\om$-roots are real as well. 
As this is evidently the case, we conclude that magnetized flows are generically stable in this case. This feature is unchanged from the corresponding fluid case, so that it is  reasonable to expect that the same trend we discussed for the purely hydrodynamical case will also apply to the magnetized case with both shear and vorticity: Vorticity tends to stabilize shear-unstable modes in a manner independent of the wave number.

\section{Concluding remarks: The MRI in perspective}\label{sec:MRIsummary}

We set out with the intention of discussing the magneto-rotational instability in a general background, relaxing the symmetry constraints associated with the standard analysis and possibly deriving an instability ``criterion'' relevant for (highly) dynamical environments and nonlinear simulations. 
However, having set up the analysis (and the required tools) in an arguably sensible way, we arrived at results which were not in line with the ``naïve'' expectations. Given this, it makes sense to comment  on the implications. Moreover, we need to  highlight an important ``missing ingredient'' in the discussion; the need to involve some suitable filtering operation to make the discussion sensible in the first place.  We will deal with each of these questions in turn, starting with the implications of our results for the MRI. 

\subsection{The  MRI vs the Rayleigh criterion}\label{subsec:InstaVsMRI}

A key  aspect of  the MRI is that adding a weak magnetic field on top of a hydrodynamically stable shearing flow changes the nature of the problem and makes it unstable. 
In discussing this problem, however, it is often ``forgotten'' that the relevant hydrodynamic stability criterion \cite{Rayleigh1917} guarantees stability \underline{only for} axisymmetric modes (cf. the discussion in \cref{subapp:RayleighCloser}). Adding a magnetic field renders such modes unstable---technically, the non-axisymmetric ones are not \citep{BalbusHawley4}. 
Thus it is clear that the MRI is relevant only in situations where we can think of axisymmetric modes being  ``preferred'' in some sense. An immediate example of this is an accretion disk, which involves a  globally axisymmetric background for the perturbations. This then immediately tells us that applying the results to the dynamical context of neutron star mergers is a much more subtle endeavour.
In fact, this exercise is problematic from the outset.

To back up this claim, we show in \cref{app:MRIlocal} that we can reproduce the MRI perturbation equations and dispersion relation through the local frame construction. 
However, for the specific MRI calculation there exists a  preferred local frame.
This local frame is associated with an observer that is co-rotating with the fluid on some orbit, and the coordinate axes  rotate in such a way that one of them always points in the radial direction of the global cylindrical coordinate system. Another coordinate axis  always points in the azimuthal direction. 
This local frame is  ``preferred'' as the axes are (by construction) tied to those of the most natural global coordinate system. In a sense, we could set up different local co-rotating observers and construct the global axes by stitching together the local ones. 
In the case of a general and truly local analysis, however, this additional piece of information is not available. 

Moreover, we show in \cref{subapp:RayleighCloser} how one may set up (for the circular and axisymmetric background flow) a local frame that is ``co-moving but not co-rotating'' with the fluid. 
In doing so, we derived the corresponding dispersion relation, confirmed that the result is consistent with the general formulae, and showed how we can recover the usual Rayleigh criterion (and hence also the MRI criterion) as long as we perform the conversion to the relevant co-rotating frame frequency.

These arguments clarify the sense in which the MRI (and similarly the Rayleigh stability) is a ``global instability analyzed with local tools''. The local analysis needs to be ``augmented'' by pieces of information that cannot be truly local. 
The upshot of this is that, in a merger-like scenario (where assumptions regarding the global properties of the flow are debatable) we should probably not expect the standard instability criteria to provide a faithful indication/diagnostic of what is actually going on. The standard argument will apply, but only if there is a meaningful sense of (Rayleigh stable) flow on a scale larger than that at which the plane-wave analysis is carried out. This complicates the discussion for any given numerical simulation, but so be it.

\subsection{The missing ingredient: Filtering}

Throughout the discussion we have focussed on the analytical development, sweeping issues associated with actual numerical data ``under the carpet''. The key issue here is that we ignored the question of how one would, in practice, construct the background suitable for the perturbation analysis given nonlinear simulation dynamics. In words, the answer is easy: We need to apply some suitable filtering operation to remove small scale fluctuations from a gradually varying ``background''. In a nonlinear setting this split is (obviously) not guaranteed to make sense. Suppose that the instability we are trying to uncover acts on some characteristic scale $L$, say. Then we need a background that varies on a larger scale than this, otherwise the notion of a shear flow that becomes unstable due to smaller scale waves makes no sense.  This argument relies on an explicit filtering step, separating the instability scale $L$ from the variation of the background. The construction of such a filter should be possible, at least in principle, in many situations (see, for example, \citet{fibrLES}). Of course, the scale separation may not apply in actual problems of interest.

Further complicating the discussion is the unavoidable implicit filtering associated with the finite numerical resolution. 
We know from the large body of work on turbulence simulations that sub-grid dynamics may play an important role in a robust description of the dynamics. 
This typically involves a suitable large-eddy scheme to represent the subgrid dynamics. 
Hence, the analysis involves elements of choice (effectively, the closure relations).
Crucially, the effective field theory that is/should be simulated is not that of the ideal theory. 
All current models---both the ones discussed in \cite{carrasco,radice2} as well as the covariant scheme of \cref{ch:LES}---modify the principal part of the equations of motion. 
Therefore the analysis of the model ``that is actually solved'' is fundamentally changed, even when the closure terms are small.
In essence, an instability analysis of numerical simulation data needs to consider the impact of an effective viscosity/resistivity. 
Given the presently available tools, we do not have a particularly good handle on this issue. 
We are forced to conclude that we need to make progress on the development of robust large-eddy models before we can make a sensible attempt to demonstrate the presence of the MRI in a highly dynamical environment.

%% file: Parts/Conclusions.tex

Hydrodynamics is an incredibly useful framework with myriads of applications at all scales, so that even after centuries of research it continues to be extremely fascinating and valuable.
In this thesis, we have studied different and interconnected problems in the modelling of relativistic fluids, from dissipation to turbulence. 
The motivation for this work lies in the (extremely thrilling) promises of gravitational wave astronomy, and the range of exciting physics we can explore with binary neutron star mergers. 

After a brief introduction/review in \cref{ch:DissipationLiteratureReview} of the different modelling strategies for dissipative fluids currently on the market, we explored in \cref{ch:Linearizing} the close to equilibrium regime of the action-based dissipative multi-fluid model of \citet{2015CQAnderssonComer}.  
A first motivation for this lies in the fact that the equations of motion are derived from an action principle, and as such are valid (in principle) in the non-linear regime with no reference to some equilibrium state. 
However, as the close to equilibrium regime is likely to be relevant for much of the applications we have in mind, it is worth exploring how the model behaves in such a limit. 
This also has the additional advantage of facilitating a direct comparison with alternative existing models, all of which are based on an expansion around such a notion of equilibrium. 

In developing complicated dissipative models, however, we need to be pragmatic and keep in mind the extreme computational cost of simulating them. 
Hence, we continued in \cref{ch:LES} focusing on the foundational aspects of performing a ``spatial filtering'' in relativity. 
We do so as this is a common strategy for dealing with computationally demanding turbulent flows, for which direct numerical simulations are often impractical. 
The discussion we provided focuses on the formal underpinnings, as the strategy is complicated by the covariance principle of General Relativity.
We then argued that it is natural to set up the framework in the fibration associated with fluid elements, and showed how one can perform filtering ensuring consistency with the tenets of relativity. 
The framework we put forward has the additional advantage that it allows for a direct link with the underlying thermodynamics, which is ultimately what we aim to constraint with binary neutron star merger observations. 
In the process we also demonstrated how the filtered equations of motion are effectively equivalent to those describing a dissipative fluid, thus leading back to similar (although, as we discuss in \cref{ch:LES}, not equivalent) issues faced when modelling dissipation in relativity. 

As we argued already in this work, for accurate neutron star modelling we are tasked with even more complicated settings involving multiple interpenetrating flows, should this be in the form of a two-fluid plasma and/or superfluid/superconducting mixtures. 
Given this, we continued in \cref{ch:LESMHD} by discussing the first steps towards extending the framework of \cref{ch:LES} to magneto-hydrodynamics. 
We do so as this is the first step towards a multi-fluid LES framework in that it adds the electromagnetic degrees of freedom to the picture, while remaining in the realm of single fluid models. 

In the last part of this work we focused on applications to problems of relevance for binary neutron star mergers. 
In \cref{ch:BVinSIM} we focused on modelling (fast) reactions for neutron star simulations, and the associated bulk viscosity. 
The reason being that reactions are thought to source the dominant dissipative mechanism at play in mergers. 
Our discussion, in particular, focuses on the impact that inevitable numerical limitations have on the way we should frame the modelling. 
We then considered the magneto-rotational instability, which is thought to be a key mechanism for sustaining the development of turbulence in the outer layers of binary neutron star merger remnants. 
Our aim has been to provide an analysis of this mechanism that is well suited for highly-dynamical environments such as mergers, where usual criteria based on rather restricting assumptions may not hold. 

Whilst we have presented our different contributions in a relatively independent fashion, it should have become rather clear by now that they are not. 
This is true from a physics perspective, as all the different aspect we touched upon, from dissipation to turbulence, play a role \emph{``at the same time''} in mergers. 
It is even more true if we consider that similar modelling strategies developed in one ``area'' can find useful applications in another. 
This is demonstrated, for example, by the specific closure scheme we put forward in \cref{ch:LES}, where we adapted some of the ideas developed to address the stability and causality issues of traditional dissipative models to fix the issues encountered with the natural relativistic generalization of the model originally put forward by \citet{Smagorinksy} in a Newtonian setting. 

In terms of future work, there are a number of possible avenues worthwhile pursuing that originate from the analysis presented in this thesis. 
They involve, not surprisingly, both aspects of framework developments as well as more specific applications. 

In terms of framework developments, for example, it would be worthwhile working towards extensions of the framework of \cref{ch:LES} to the case of multiple interpenetrating flows/mixtures.
Moreover, we also mentioned briefly at the end of \cref{ch:LESMHD} how the framework discussed in \cref{ch:LES} suggests---due to the covariance of the resulting model---that we may enhance the role of the ``large-eddy strategy'' to a tool for linking models valid at different scales. 
Clearly, there are several applications we can envisage, from dynamo theory to try and unify the somewhat arbitrarily separated small scale and large scale models, to superfluids to link mesoscopic models where each single vortex line is resolved to coarse-grained descriptions of the kind discussed in \cref{sec:VariationalModels}---for which the two-fluid models were originally developed \cite{livrev}.

Furthermore, as we discussed in \cref{sec:2parEoS}, the scale-gap and fluctuations within a fluid box implies that the equation of state used for large-scale merger simulations may not be trivially linked to the underlying microphysics.
Notably, as any numerical simulation is performed on a finite grid, there is always at least an implicit filtering associated with it. 
As the best grid resolution in large-scale
merger simulations is of order tens of meters, the impact of this is potentially significant.
It would be worth exploring this potential disconnect and try to quantify the impact this may have on neutron star parameters extracted from observations. 

Further developments of the large-eddy strategy in both these directions would obviously be rather incomplete if not provided with suitable closure schemes.
Future developments will also have to involve aspects of developing, testing and calibrating novel/different closure schemes.
As we may also envisage rather different schemes to be better suited for diverse applications, there is plenty of work to be done.

%% file: Parts/Appendix.tex


\chapter{Transporting a tetrad and Fermi Coordinates}\label{app:FermiCoord}

In this appendix we introduce the notion of Fermi coordinates, as these have been used but not derived/discussed in the main body of the thesis. 
We start by discussing the transport of a tetrad along a curve/worldline since the notion of Fermi coordinates builds on this \cite{GourghoulonSR,GravitationMTW}. 
In our discussion we will, in particular, make explicit contact with the notion of spin coefficients that have to be introduced whenever one wants to work with an orthonormal basis or tetrad (see notation section at the beginning of this work). 

\section{Transporting a tetrad along a curve}

We start our analysis by considering a curve/worldline in spacetime, which we here take as time-like with tangent vector $U^a$, and assuming a set of four orthonormal vectors is given along the worldline. 
Namely, we have a set of four vectors $e_{\hat a}$, with $\hat a = 0,1,2,3$ such that 
\begin{equation}
    g(e_{\hat a},e_{\hat b}) = \eta_{\hat a \hat b} \;,
\end{equation}
where we are considering the metric $g$ as a bi-linear form on the tangent space, and $\eta_{\hat a\hat b}$ is the Minkowski metric.
We further assume that we can take the time-like unit vector of the tetrad as the one tangent to the curve, that is $U = e_{\hat 0}$. 
We are interested in the rate-of-change of the tetrad basis vectors as we move along the curve. 
Since $U$ is the tangent to the curve, we can start from 

\begin{equation}\label{eq:OmegaBilinear}
    \nabla_{U}e_{\hat a} = \Omega_{\;\hat a}^{\hat b}\, e_{\hat b} \;,
\end{equation}
where we have introduced the bi-linear form $\Omega_{\;\hat a}^{\hat b}$ to re-write the rate-of-change as a linear combination of the tetrad basis vector themselves.  
Next, we observe that $\Omega_{\hat b\hat a}$ must be an anti-symmetric bi-linear form\footnote{For clarity, let us point out that the bi-linear form is obtained lowering the contravariant index via the metric, namely $\Omega_{\hat a \hat b} = g_{\hat a\hat c}\Omega^{\hat c}_{\;\hat b}$.}, as this follows from the orthonormal character of the basis, that is 
\begin{equation}
    \nabla_{U} \left( g(e_{\hat a},e_{\hat b})\right) = 0 \Longrightarrow \Omega_{\hat a\hat b} = \Omega_{{\hat b}{\hat a}} \;.
\end{equation}
As such, we can decompose $\Om$ as any skew-symmetric bi-linear form \cite{GourghoulonSR}. 
In our case this gives
\begin{equation}\label{eq:OmegaDec}
    \Om_{{\hat a}{\hat b}} = U_{\hat a}a_{\hat b} - a_{\hat a}U_{\hat b} - \veps_{\hat a\hat b\hat c\hat d}U^{\hat c}W^{\hat d} \;,
\end{equation}
where $a_{\hat a}$ is the four-acceleration of the worldline, while $W^{\hat a}$ is a generic vector orthogonal to $U^{\hat a}$. 
To better understand the physical meaning of such decomposition, we introduce the spatial index $\hat i = 1,2,3$ and use it also to denote the three unit vectors orthogonal to $U$. 
We then get
\begin{align}
    \left( \nabla_U U\right)^{\hat a} &= a^{\hat a} \;, \\
    \left(\nabla_U e_{\hat i}\right)_{\hat a} &=a_{\hat i}U_{\hat a} + \veps^{U}_{\;\hat a \hat b\hat i}W^{\hat b} \;.
\end{align}
Hence, when the worldline is a geodesic we have $a^{\hat a} = 0$ and the time-like unit vector is unchanged as we move along the curve, while the spatial ones $e_{\hat i}$ change due to some (spatial) rotation in the subspace orthogonal to $U$. 
In particular, we stress that whilst the four acceleration is an intrinsic property of the worldline, the vector $W$ represents the angular velocity associated with the rotation of the observer frame. 
In essence, given a time-like worldline/curve, there exist an infinite number of compatible tetrads such that $e_{\hat 0} = U$, and these are all related by some rotation in the plane orthogonal to the curve. 
We can then further split the bilinear form $\Omega$ as the sum of two terms
\begin{subequations}
    \begin{equation}
    \Omega_{{\hat a}{\hat b} } = \Omega^{FW}_{{\hat a}{\hat b}} + \Omega^{rot}_{{\hat a}{\hat b}} \;,
\end{equation}
with
\begin{equation}
    \Om^{FW}_{\hat a\hat b} = U_{\hat a}a_{\hat b} - a_{\hat a}U_{\hat b} \;, \quad \text{and }\Om^{rot}_{{\hat a}{\hat b}}  = \veps^U_{\;{\hat a}{\hat b}{\hat c}}W^{\hat c}\;.
\end{equation}
\end{subequations}
This splitting is meaningful because it separates the terms in $\Omega$ we have control over (the rotation part) from those that are given once the curve is specified (the Fermi Walker part). 
In particular, a vector is said to be Fermi-transported (or Fermi-Walker transported) along a worldline if its components change with $\Om^{FW}$. 

Before moving on to discuss Fermi coordinates, we take the opportunity to make contact with the spin connection coefficients, which are normally introduced when working with a tetrad basis. 
Using the spin coefficients \cite{GRCarroll} we would get
\begin{equation}
    \nabla_U e_{\hat a} = U^b\nabla_b e_{\hat a} = U^b \om_{b\;\hat a}^{\;\hat b}e_{\hat b} = \om_{\hat 0\; \hat a}^{\;\hat b}e_{\hat b} \;,
\end{equation}
so that, contrasting this with \cref{eq:OmegaBilinear} we see 
\begin{equation}\label{eq:OmegaTimeSpin}
    \om_{\hat 0\; \hat a}^{\;\hat b} = \Om^{\hat b}_{\;\hat a} \;.
\end{equation}
In essence, the bilinear form $\Omega$ determines the ``time-part'' of the spin connection coefficients. 
This is intuitive given that the spin connection coefficients have to be used when a tetrad is assigned throughout (a region of) the spacetime, and not only along a worldline. 

\section{Fermi coordinates}

We now discuss how to introduce a (local) set of coordinates that is suitable for describing the physics as measured by some local observer \cite{GravitationMTW,GourghoulonSR}, that is in the vicinity of the observer worldline. 
These go under the name of Fermi coordinates, as it was Fermi who originally introduced them in 1922 \cite{fermi1,fermi2}  (the work by \citet{walker_1933} is also relevant).
The concept was then further developed by, in particular \citet{ManasseMisner}, where the authors focus on the particular case where the central curve is a geodesic. 
The same idea was then generalized to the more general case of an accelerated and rotating observer by \citet{GravitationMTW}. 

\begin{figure}
    \centering
    \includegraphics[width = 0.98 \textwidth]{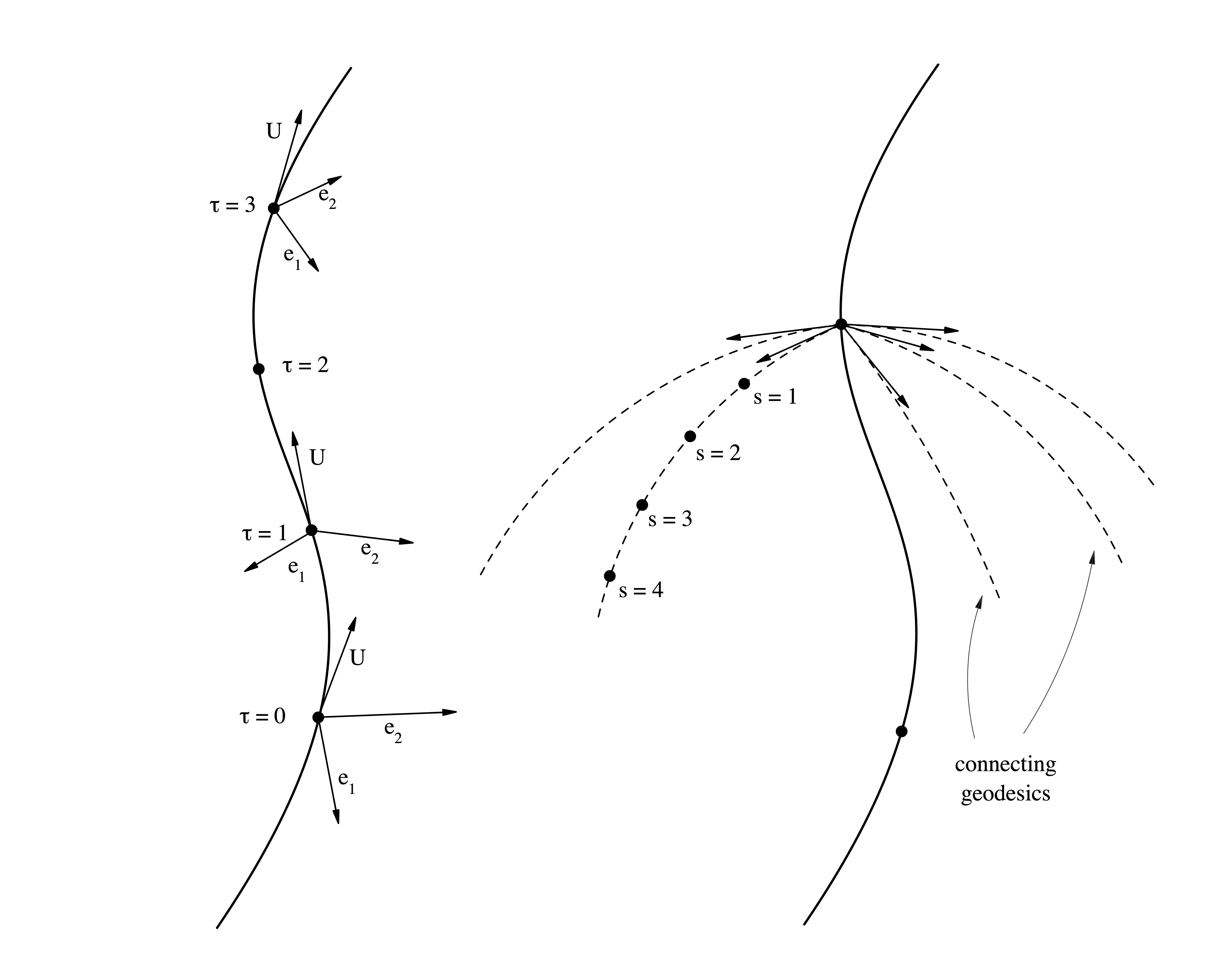}
    \caption{Left: Tetrad transported along the observer worldline. Right: Connecting geodesics starting from and perpendicular to the central curve. Figure adapted from \citet{GravitationMTW}.}
    \label{fig:FermiCoord}
\end{figure}

Given some observer's worldline, the associated Fermi coordinates can be constructed via the following procedure---which is perhaps best described in \cref{fig:FermiCoord}.
First consider the tangent vector to the (time-like) curve $U$, and a tetrad defined over the entire worldline. 
In the case of Fermi coordinates this is given at a point $P_0$ and then Fermi-transported along the curve. 
We can then assign coordinates to spacetime points (or viceversa) in the vicinity of the worldline as follows.
We start with four real numbers $\{x^{\hat 0},\,x^{\hat 1},\,x^{\hat 2},\,x^{\hat 3}\}$, and first move along the central curve by $x^{\hat 0}$ (in a parameterization compatible with $U$).
We then consider the numbers $\alpha^{\hat i} = x^{\hat i}/s$ with $s^2 = \sum_i (x^{\hat i})^2$, and the connecting geodesic starting perpendicular from the central curve with unit tangent vector $v = \alpha^{\hat i} e_{\hat i}$. 
As geodesics satisfy a second order differential equation, they are uniquely defined given a point and a tangent vector (at the point).
Next, move along the connecting geodesic by $s$ (in terms of the connecting geodesic's proper length), and assign to such point the coordinates $(x^{\hat 0},\,x^{\hat 1},\,x^{\hat 2},\,x^{\hat 3})$. 

Let us first of all note that along the central curve the basis vector of the coordinates grid are, by construction, identical to the orthonormal tetrad 
\begin{equation}\label{eq:FermiCoordGridTetrad}
    \pdv{}{x^{\hat a}} = e_{\hat a} \;.
\end{equation}
This is important as we can immediately draw two important conclusions. 
First, the metric in the given coordinates takes the Minkowski form along the entire central curve. 
Second, we can use this simple fact to show that such coordinates are well-defined in some neighbourhood of the central curve \cite{ManasseMisner, GravitationMTW}.
This is intuitively clear as points close enough to the central curve are uniquely connected to the central curve by one (and only one) geodesic.
On larger distances (in terms of the connecting geodesic proper length) the procedure fails because different geodesics can mix and touch. 
This can happen due to both the acceleration of the central curve and the curvature of the spacetime itself---and also due to the tetrad rotation if the basis is not Fermi transported along the curve. 

We can now use such coordinates to write down an expansion for the metric in the neighbourhood of the central curve. 
As we know already that the metric along the central curve takes the Minkowski form, we only need the first derivatives of the metric evaluated on the central curve.
These can be obtained from the Christoffel symbols which are related to the spin coefficients and hence to the bilinear form $\Omega$ introduced above. 
In particular, by inspecting \cref{eq:FermiCoordGridTetrad} one can easily see that 
\begin{equation}
    \Gamma^{\hat a}_{\;{\hat b}{\hat 0 }} = \Omega^{\hat a}_{\;\hat b} \;,
\end{equation}
so that, by means of \cref{eq:OmegaDec}, we readily obtain
\begin{align}\label{eq:ZeroGammaFermi}
    \Gamma_{\hat 0{\hat 0}{\hat 0 }}  & = \Gamma^{\hat 0}_{\;{\hat 0}{\hat 0 }} = 0  \;, \\
    \Gamma_{\hat 0{\hat j}{\hat 0 }} & = - \Gamma^{\hat 0}_{\;{\hat j}{\hat 0 }} = - a_{\hat j} \;, \\
    \Gamma_{\hat i{\hat j}{\hat 0 }} &= \Gamma^{\hat i}_{\;{\hat j}{\hat 0 }} = -\veps_{\hat 0 \hat i \hat j \hat k}W^{\hat k} \;.
\end{align}
The remaining Christoffel symbols can be found using the geodesic equation satisfied by the connecting geodesics
\begin{equation}
    \dv[2]{x^{\hat a}}{s} + \Gamma^{\hat a}_{\;\hat b\hat c} \dv{x^{\hat b}}{s}\dv{x^{\hat c}}{s} = 0 \;.
\end{equation}
As the connecting geodesics are given by $x^{\hat 0} = \text{const}$ and $x^{\hat i }= \alpha^{\hat i} s$, we readily obtain 
\begin{equation}\label{eq:LastGammaFermi}
    \Gamma_{\hat a\hat i \hat j} = \Gamma^{\hat a}_{\;\hat i \hat j} = 0 \;.
\end{equation}
Next, note that by means of the metric compatibility condition (with the connection given by the Christoffel symbols) we have 
\begin{equation}
    g_{\hat a\hat b,\hat c} = 2 \Gamma_{(\hat a\hat b)c} \;,
\end{equation}
and using \cref{eq:ZeroGammaFermi,eq:LastGammaFermi} we obtain
\begin{align}
    g_{\hat 0 \hat 0} &=  g_{\hat 0 \hat 0 } \big|_{G} + g_{\hat 0 \hat 0, \hat a}x^{\hat a } = - \left(1 + 2a_{\hat j}x^{\hat j}\right) + \O(x^{\hat j})^2 \;, \\
    g_{\hat 0 \hat i} &=  g_{\hat 0 \hat i} \big|_{G} + g_{\hat 0 \hat i, \hat a}x^{\hat a } = -\veps_{\hat 0 \hat i \hat j \hat k}W^{\hat k} + \O(x^{\hat j})^2 \;, \\
    g_{\hat i \hat j} &=  g_{\hat i \hat j } \big|_{G} + g_{\hat i \hat j, \hat a}x^{\hat a } = \eta_{\hat i \hat j }+ \O(x^{\hat j})^2 \;.
\end{align}
In essence, we obtained an expansion for the metric away from the central worldline. The first order expansion depends on the worldline acceleration, and also on the (arbitrary) vector describing the angular rotation of the basis vectors as we move along the worldline. 
Notably, no information about the space-time curvature enters at first order in the expansion.

We conclude with some observations/comments. 
First of all, let us observe that we can always choose to work with a non-rotating tetrad, so that the associated expansion for the metric simplifies accordingly (noting that this is precisely the choice made in \cref{ch:LES}, for example). 
In this case, the coordinates are called Fermi-coordinates \cite{fermi1,fermi2,synge}. 
If the observer is also freely-falling, the central curve is a geodesic and the (non-rotating) coordinates are called Fermi normal coordinates \cite{ManasseMisner}---in which case the metric expansion contains no first order terms.
The second order corrections to the Minkowski metric have been computed explicitly by \citet{ManasseMisner} (for Fermi normal coordinates), showing in particular that the second order corrections are uniquely determined by the Riemann tensor (evaluated on the central curve). 
The work by \citet{Rakhmanov_2014}, where the Fermi coordinates expansion of the metric is computed to all orders given a spacetime describing a plane gravitational wave, is also relevant.
Finally, we also note that when a similar scheme is used with light-like connecting geodesics, the resulting coordinates are known as optical coordinates \cite{Rakhmanov_2014}.

\chapter{Multi-scale arguments and the invariant manifold method}\label{app:MultiScale}

Multi-scale methods are useful whenever the (system of) equations to solve contains different scales, so that is physically (and numerically) useful/convenient to solve an approximate system instead.
In this appendix we briefly summarize key results from~\cite{StuartPavliotis,Weinan} that are used in the main body of the thesis, specifically in \cref{ch:BVinSIM}.

\section{Invariant manifold approach}

Assume we have a system of ordinary differential equations written as
\begin{subequations}
    \label{eq:multiscale}
    \begin{align}
        \dot{x} &= f(x, y), \label{eq:multiscale_slow} \\
        \dot{y} &= \epsilon^{-1} g(x, y). \label{eq:multiscale_fast}
    \end{align}
\end{subequations}
The variables $x$ are called \emph{slow}, and the variables $y$ are called \emph{fast}, whilst $\epsilon \ll 1$ is a parameter.

In the \emph{invariant manifold} approach we assume that there exists an \emph{equilibrium (fast) state} $\varphi(x)$ such that
\begin{equation}
    \label{eq:multiscale_inv_manifold_equil}
    g(x, \varphi(x)) \equiv 0.
\end{equation}
We can then write the fast variables $y$ as an expansion in the small parameter $\epsilon$ about the equilibrium state as
\begin{equation}
    \label{eq:multiscale_inv_manifold_expansion}
    y = \varphi(x) + \epsilon y_1 + \mathcal{O}(\epsilon^2).
\end{equation}
Using the equations of motion~\cref{eq:multiscale} we find that the behaviour of the slow variables $x$ is approximated, to second order in $\epsilon$, by the solution to
\begin{subequations}
    \label{eq:multiscale_inv_manifold_solution}
    \begin{align}
    \dot{X} &= F_0(X) + \epsilon F_1(X) \\
    &= f(X, \varphi(X)) + \epsilon \nabla_y f(X, \varphi(X)) \left( \nabla_y g(X, \varphi(X)) \right)^{-1} \nabla_x \varphi(X) f(X, \varphi(X)).
    \end{align}
\end{subequations}
The solution to the simplified system approximates the solution to the full system~\cref{eq:multiscale} to ${\mathcal O}(\epsilon^2)$ up to times ${\mathcal O}(1)$. The first order correction term needs to be applied consistently to all variables in the reduced system for this accuracy result to hold.
\section{Two timescale approach}

Strictly, the invariant manifold approach is only valid for ordinary differential equations. A more general approach that applies to partial differential equations is the two-scale approach. Using the two timescale approach as an example, this introduces the \emph{fast time} $\tau = t / \epsilon$ which is then treated as an independent variable. Applied to~\cref{eq:multiscale} this leads to
\begin{subequations}
    \begin{align}
        \partial_t x + \epsilon^{-1} \partial_\tau x &= f(x, y), \\
        \partial_t y + \epsilon^{-1} \partial_\tau y &= \epsilon^{-1} g(x, y).
    \end{align}
\end{subequations}
By gathering terms in powers of $\epsilon$, the fast behaviour can be integrated out by taking the integral average in $\tau$.

The result of the mathematical calculation is identical to the invariant manifold approach, when applied to ordinary differential equations. The calculation is general enough to include partial differential equations, and illustrates a different interpretation and potential problems. The interpretation is that the reduced system is valid for the \emph{integral average} of the slow variables: the fast scales have been integrated out. The potential problem is the requirement that the integral average of the fast behaviour is assumed to not contribute at leading order in $\epsilon$. This ``resonance'' behaviour cannot be captured by these approaches.

\section{Linear fast dynamics}
\label{sec:multiscale_linear}

A particularly relevant example is where the fast behaviour is linear, or can be linearised. In this case we write the full system~\cref{eq:multiscale} as
\begin{subequations}
    \label{eq:multiscale_linear}
    \begin{align}
        \dot{x} &= f(x, y), \label{eq:multiscale_linear_slow} \\
        \dot{y} &= \epsilon^{-1} (-A y + B), \label{eq:multiscale_linear_fast}
    \end{align}
\end{subequations}
with $A=A(x)$ and $B=B(x)$ are constants in the fast variables $y$. The equilibrium solution is therefore $\varphi(x) = B / A$, and the simplified system is
\begin{equation}
    \dot{X} = f(X, \varphi(X)) \left[ 1 + \epsilon A^{-1} \nabla_y f(X, \varphi(x)) \nabla_x \varphi(X) \right].
\end{equation}

\section{Constructing the fast terms}
\label{sec:appendix_multiscale_construct_fast}

The construction in this section relies on the ratio of scales $\epsilon$ being explicit in the equations of motion. Usually a non-dimensionalisation of the system is needed to make the scales explicit. However, with complex nonlinear terms (such as tabulated net reaction rates which include many reaction channels) the precise form of the terms may not be obvious.

Here we need only the leading order terms and so can proceed as follows. We start from a system of equations which we expect to have fast behaviour
\begin{equation}
    \dot{z} = h(x, z),
\end{equation}
where $x$ are any variables we expect to be slow. We assume that we know how $h$ scales asymptotically with the ratio of scales. That assumption means we can explicitly compute
\begin{equation}
    h^{\text{fast}} = \lim_{\epsilon \to 0} \left( \epsilon h \right).
\end{equation}
This defines the source term for the fast behaviour as the piece that diverges linearly with the ratio of scales in the limit of infinitely fast speeds. We then split the source into fast and slow pieces using
\begin{equation}
    h^{\text{slow}} = h - \epsilon^{-1} h^{\text{fast}},
\end{equation}
and perform the equivalent split on the variables $z$ as
\begin{subequations}
    \begin{align}
        \dot{z}^{\text{fast}} &= \epsilon^{-1} h^{\text{fast}}, \\
        \dot{z}^{\text{slow}} &= h^{\text{slow}}.
    \end{align}
\end{subequations}
We can then identify the fast variables $y$ with ${z}^{\text{fast}}$ and augment the slow variables $x$ with ${z}^{\text{slow}}$.

\chapter{Working with the CompOSE database}\label{app:compOSE}

Our analysis of the reactive problem in \cref{ch:BVinSIM} is---ultimately---aimed at numerical implementations. 
Given this target, it makes sense to consider how the discussion impacts on the matter model that needs to provide the input physics. 
In this appendix we will spell out the connection with the compOSE database, which provides a useful collection of state-of-the-art equation of state models. 
As a result, the arguments draw heavily on the compOSE manual \cite{compOSE}, in particular  section 4.1.2 (``Thermodynamic consistency'') of version 2.0. 
The aim here is to explain how the various thermodynamical coefficients introduced in the main text can be worked out from an actual equation of state table. 
This is obviously a necessary step in the process. 
It also helps highlight to what extent existing tabulated data needs to be augmented in the future.

All equations of state relevant for our work in the compOSE database are provided as tables of $(T,n,Y_\q)$, where $Y_\q$ is the fraction of charged strongly interacting particles, which for a system without muons corresponds to the electron fraction $Y_\q = Y_\e$ (as local charge neutrality is assumed to hold). The central thermodynamical potential is the Helmholtz free energy density $f= \veps - Ts$, and some key quantities in the construction of the tables are 
\begin{equation}
     \left\{ \frac{p}{n},\, \frac{s}{n},\,\frac{\mu_\b}{m_\n}-1,\,\frac{\mu_\q}{m_\n},\,\frac{\mu_\l}{m_\n},\,\frac{f}{nm_\n}-1,\,\frac{\veps}{nm_\n}-1 \right\} \;,
\end{equation}
where $m_\n$ is the neutron mass---also provided in the tables and specific to each model---while $\mu_\b,\,\mu_\q,\,\mu_\l$ are the baryon, charge and lepton ``chemical potentials'' (respectively). The energy cost of adding a neutron, proton or electron to the system is then
\begin{equation}
    \mu_\n = \mu_\b \;,\quad \mu_\p = \mu_\b + \mu_\q \;,\quad \mu_\e = - \mu_\q + \mu_\l
\end{equation}
as follows straightaway from the respective baryon, charge\footnote{This is the total charge, not the charge of the strongly interacting particles corresponding to $n_\q = n Y_\q$.} and lepton number. The  baryon, charge and lepton chemical potentials are (in general) used to build the free energy, even though in the charge-neutral case with leptons this reduces to
\begin{equation}
    f = (\mu_\b  + \mu_\l Y_\e ) n - p \ .
\end{equation}
From this we see that there are only two independent chemical potentials at the thermodynamical level---consistent with a three-parameter equation of state---that can be written as derivatives of the Helmholtz free energy, namely $\mu_\b$ and $\mu_\l$.

Now, we need to connect with the quantities used in the main text. We have  (following from the Gibbs relation \cref{eq:OutEqGibbs})
\begin{equation}
    f = \veps - Ts \;, \quad df = -s dT + \mu_\n dn -\beta dn_\e\;,
\end{equation}
where $\beta = \mu_\n - \mu_\p - \mu_\e$ in the cold equilibrium assumed throughout this paper. From this differential we see that, by definition 
\begin{subequations}
\begin{align}
    \mu_\n &= \left(\frac{\partial f}{\partial n}\right)_{T,n_\e} =  \left(\frac{\partial f}{\partial n}\right)_{T,Y_\e} - Y_\e \left(\frac{\partial f}{\partial n_\e}\right)_{T,n}\;, \\
    -\beta &= \left(\frac{\partial f}{\partial n_\e}\right)_{T,n} = \frac{1}{n} \left(\frac{\partial f}{\partial Y_\e}\right)_{T,n}\;.
\end{align}
\end{subequations}
Contrasting this with the results in section 4.1.2 of the compOSE manual, it is easy to see that\footnote{Note that $\mu_\q$ potential is not an independent thermodynamic quantity as the system cannot create protons alone because of the charge neutrality assumption.} $\mu_\n = \mu_\b$ and $\beta= - \mu_\l$. In practice, the affinity $\beta$ can be either read off from the tables directly, or computed as above. Because the equation of state table is  three-dimensional, the quantity extracted is inevitably a function of $\beta = \beta(n,T,Y_\q)$. This is consistent with the results of \cref{sec:hydro-thermodynamics}, where we accounted for the fact that $\beta$ depends on either the temperature or the energy density.

Now, the coefficients $\A$ and $\B$ introduced in \cref{sec:hydro-thermodynamics} can be obtained as combinations of derivatives of $\beta$ considered as a function of $(n,\veps,Y_\e)$. We need to link these expressions to derivatives that can be computed from the available tables (or, which may be more practical for the future, enhance the table with the required information). That is, we  have to change variables to arrive at
\begin{subequations}\label{eq:fromEtoT}
\begin{align}
    \left(\frac{\partial \beta}{\partial n}\right)_{\veps, Y_\e} &= \left(\frac{\partial \beta}{\partial n}\right)_{T, Y_\e} - \left(\frac{\partial \veps}{\partial n}\right)_{T, Y_\e} \left(\frac{\partial \veps}{\partial T}\right)_{n, Y_\e}^{-1} \left(\frac{\partial \beta}{\partial T}\right)_{n,Y_\e} \;, \\
    \left(\frac{\partial \beta}{\partial \veps}\right)_{n, Y_\e} &=  \left(\frac{\partial \veps}{\partial T}\right)_{n, Y_\e}^{-1} \left(\frac{\partial \beta}{\partial T}\right)_{n,Y_\e} \;, \\
    \left(\frac{\partial \beta}{\partial Y_\e}\right)_{\veps, n} &= \left(\frac{\partial \beta}{\partial Y_\e}\right)_{T, n} - \left(\frac{\partial \veps}{\partial Y_\e}\right)_{T, n} \left(\frac{\partial \veps}{\partial T}\right)_{n, Y_\e}^{-1} \left(\frac{\partial \beta}{\partial T}\right)_{n,Y_\e}  \;.
\end{align}
\end{subequations}
Recalling that these quantities should be evaluated at equilibrium, we see that we also need to construct the corresponding equilibrium table. Operationally, this can be done as follows.
We fix $n,T$ and vary $Y_\e$ until we find a value for which $\beta = 0$. The corresponding value of $Y_\e$ is then what we call $Y_\e^\mathrm{eq}$ and the equilibrium composition will automatically only be a function of $(n,T)$. Evaluating the original three-parameter model at $Y_\e = Y_\e^\mathrm{eq}$  gives the corresponding equilibrium energy density and pressure etc. Using expressions analogous to \cref{eq:fromEtoT}, we can rewrite derivatives of $Y_\e^{\mathrm{eq}}$ with respect to $(n,\veps)$ in terms of derivatives with respect to $(n,T)$, that can be extracted from the tables.  

\begin{figure}
    \centering
    \includegraphics[width=\textwidth]{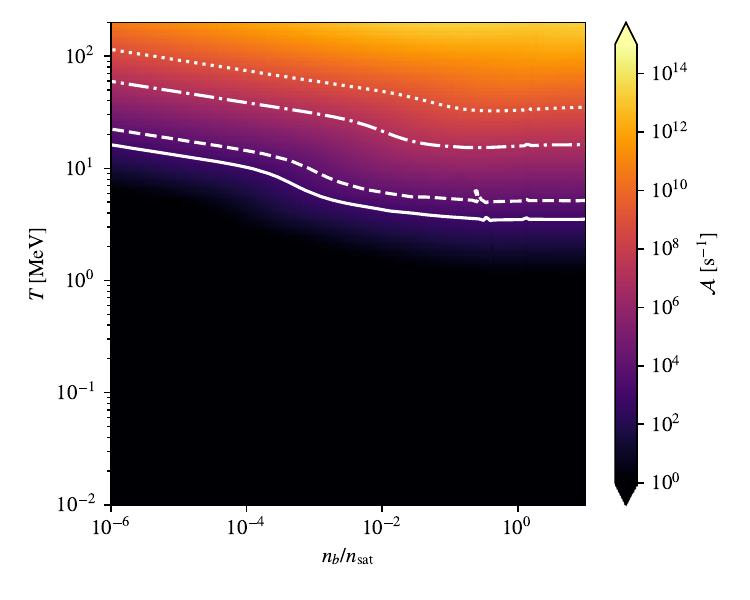}
    \caption{Plot of $\mathcal{A}$ for the APR equation of state used in \cite{PeteThermal}. The restoring term $\gamma$ is calculated assuming the Fermi surface approximation remains valid.  Contours are at $\mathcal{A}=\{10^3,10^4,10^7,10^9\}\mathrm{s^{-1}}$ (solid, dash, dot-dash, dot).}
    \label{fig:A_zoom}
\end{figure}

With the results in \cref{eq:fromEtoT} and evaluating the relevant quantities at equilibrium, we can work out (for a given equation of state) the value of $\B$, as required for \cref{fig:P_test_new}. In order to compute $\A$ though, we also need to evaluate the restoring term $\gamma$ (effectively, a measure of the reaction timescale). In \cref{fig:A_zoom} we show $\A$ as obtained from the modified Urca rates for the APR equation of state \cite{compOSE,APReos} used in~\cite{PeteThermal}. For this figure we have calculated $\gamma$ assuming the Fermi surface approximation, which allows us to use the analytic formulae from \cite{AlfordHarris18}.

Let us stress that the result should be valid for low temperatures ($T \lesssim 1~\mathrm{MeV}$), however the timescales relevant for our purposes occur in the range of (according to $\mathcal{A}$ as calculated above) $2~\mathrm{MeV}\lesssim T\lesssim 20~\mathrm{MeV}$ for the densities relevant to the neutron star core, begging the question of how accurate the approximation is.
Instead of calculating out-of-equilibrium rates without the Fermi surface approximation, we can take a different approach, and estimate the equilibration timescale using neutrino opacities \cite{PetePHD}.
Notably, we find some broad similarities between the two estimates---despite expected qualitative differences---particularly in the regions of interest at relevant temperatures and densities. 

\chapter{Formulating the MRI in the local frame}\label{app:MRIlocal}

In this appendix we show how to formulate the magneto-rotational instability using the local frame construction of \cref{ch:MRI}. 
In particular, we will show that using a co-rotating local frame we can derive the same equations as in \cite{BalbusHawley1}. 
Next in \cref{subapp:RayleighCloser} we focus on the Rayleigh criterion. 
The discussion we provide here has the advantage that it makes explicit an important underlying assumption that is key to the usual Rayleigh and MRI criteria \cite{Rayleigh1917,BalbusHawleyRev}.

Let us start by considering the circular velocity profile assumed in \cite{BalbusHawley1}, $\vec v = v^{\hat \varp} \hat \varp$ with $ v^{\hat \varp} = \Om(R) R$ where we use cylindrical coordinates and an orthonormal basis on the ``tangent space'' (as usual).
Consistently with the notational conventions adopted in the rest of the thesis, we then distinguish between indices with a ``hat'' corresponding to the orthonormal basis, and those without that correspond to the coordinate basis. 
We then pick an orbit at some radial distance $R_0$ and choose an observer that is co-rotating with angular frequency identical to that of the background flow at $R_0$, that is $\vec v_{obs} = \Omega_0 R \hat\varp$ where $\Om_0 = \Om(R_0)$.
The observer is then accelerated with acceleration $\vec a = - \Om_0^2 R \hat R$, and the velocity of the fluid with respect to such an observer then is $\vec v' = (\Om - \Om_0)R\hat\varp$.
We then set up the axes of the observer local frame so that one is pointing in the radial direction ($\hat e_1$), one is pointing in the azimuthal direction ($\hat e_2$) and the third one is aligned with the rotation axis ($\hat e_3$). 
Introducing coordinates associated to this observer, we can then write the background fluid velocity as 
\begin{equation}
    \vec v' = \frac{\text{d}\Om}{\text{dln}R}\bigg|_{R_0} x' \hat e_2  + \O(x'^2)\;.
\end{equation}
We have neglected terms of order $\O(x'^2)$ as we will only need the velocity and its gradients evaluated at the origin of the frame---so that such terms will not enter the perturbation equations anyway.
Computing the gradients we then obtain 
\begin{equation}\label{eq:CRgrad}
    \partial'_iv'_j = \begin{pmatrix}
    0 & s_0 & 0 \\
    0 & 0 & 0 \\
    0 & 0 & 0
    \end{pmatrix} \;,\qquad s_0 = \frac{\text{d}\Om}{\text{dln}R}\bigg|_{R_0} \;.
\end{equation}
As the local frame of the observer is rotating with angular velocity $\Om_0 \hat e_3$, we need to include the Coriolis force into the perturbation equations. 
We then write the perturbed Euler and continuity equations (dropping the primes for clarity, and retaining only fast-gradients in the background velocity) 
\begin{subequations}
\begin{align}
    &\partial_t \delta v_i + 2\Om_0 \eps_{i3k} \delta v^k + \delta v^j \partial_j v_i + \frac{c_s^2}{\rho}\partial_i \delta \rho + \frac{1}{\mu_0\rho} \left[ B_j \partial_i \delta B^j - B^j \partial_j \delta B_i\right] = 0 \;,\\
    &\partial_t \delta \rho + \rho \partial_i \delta v^i = 0 \;,
\end{align}
\end{subequations}
and, introducing a WKB plane-wave expansion (as detailed in \cref{sec:WKBlocalbox}),
\begin{subequations}\label{eq:Euler+contMRI}
\begin{align}
    &-i \om \delta v_i + 2\Om_0 \eps_{i3k}\delta v^k + s_0 \delta_{i2}\delta v^1 + i \frac{c_s^2}{\rho}k_i \delta \rho + \frac{i}{\mu_0 \rho}\left[B_jk_i \delta B^j - B^j k_j \delta B_i\right] =0 \;,\\
    &-i \om \delta \rho + i \rho k_i \delta v^i = 0 \;.
\end{align}
\end{subequations}
Next, focus on the induction equation. As we have discussed \cref{sec:NonInertialEq}, the induction equation in the co-rotating frame retains the inertial form. 
We then have 
\begin{equation}
    \delta \partial_j \left(v^j B^i -v^i b^j\right) = \delta v^j \partial_j B^i + v^j \partial_j \delta B^i - B^j \partial_j \delta v^i - \delta B^j \partial_j v^i \;,
\end{equation}
where we made use of i)the no-monopoles constraint ii)the vanishing expansion  of the background flow iii)the Boussinesq approximation to get rid of the divergence of the perturbed velocity. 
Introducing the WKB plane-wave expansion and evaluating the background quantities at the origin of the local frame we then end up with 
\begin{equation}\label{eq:InductionMRI}
    - i \om \delta B^i - i B^j k_j \delta v^i - \delta B^1 s_0 \delta ^{i2} = 0 \;.
\end{equation}
In \cref{eq:Euler+contMRI,eq:InductionMRI} we recognize the terms entering the perturbation equations in \cite{BalbusHawley1} (with the exception of background gradients in the pressure that we are here neglecting). 
We also note that we here do not need to formally neglect terms of the form $B/ R$ as these terms do not appear in the explicit local frame construction. 
We conclude by noting that, at the special relativistic level, a uniformly rotating observer and the co-rotating one are not the same as the latter is also accelerated (see \citet{GourghoulonSR}, ch. 13). 
However, such a difference is irrelevant at the level of the Newtonian perturbation equations since i) pseudo-acceleration terms drop out of the perturbed Euler equation ii) non-inertial terms in the induction equation involving the four-acceleration are negligible in the Newtonian limit (cf. discussion in \cref{sec:NonInertialEq}).

\section{Another look at the non-inertial MHD equations.}
Before we move on to take a closer look at the Rayleigh criterion, let us here show how the terms involving the local frame rotation drop out of the induction equation. 
Even though we have already argued this happens in general (cf. \cref{eq:Wdropsout}), we here prove this for the specific case of a co-rotating observer. 
We do so as this allows us to appreciate better why the cancellation comes about. 

We show this using the notion of spin coefficients associated with a uniformly rotating observer. 
That this is going to lead us where we want can be anticipated by recalling the result in \cref{eq:OmegaTimeSpin} and that we made use of the bi-linear form $\Omega$ in the argument in \cref{sec:NonInertialEq}.
We then introduce (Born) coordinates associated with a uniformly rotating observer (axes suitably oriented so that the angular velocity is $\Om_0 \hat z$)
\begin{equation}\label{eq:BornCoord}
    t' = t \;,\quad z' = z \;,\quad x' = R \text{cos}(\Om_0 t + \varphi) \;,\quad y' = R \text{sin}(\Om_0 t + \varp) \;,
\end{equation}
where primed coordinates are Cartesian (i.e. non-rotating). Computing the spin coefficients (starting from a flat metric) we then obtain 
\begin{equation}\label{eq:BornSC}
     \om_{\varphi\;\hat \varp}^{\;\hat R} = -1 \;,\quad \om_{\varphi\;\hat R}^{\;\hat \varp} = +1 \;,\quad
     \om_{t\;\hat \varp}^{\;\hat R} = -\Om_0 \;,\quad
     \om_{t\;\hat R}^{\;\hat \varp} = \Om_0 \;,
\end{equation}
showing that, as the coordinates ``mix space and time'' we need to introduce a covariant derivative in the time-direction as well.
We then write the non-inertial induction equation as (cf. \cref{eq:NewtIndNI})
\begin{equation}\label{eq:CorotatingInduction}
    \nabla_tB^{\hat i} +\nabla_{\hat j} \left(v^{\hat j} B^{\hat i} - v^{\hat i} B^{\hat j} \right) + \eps^{\hat i \hat j\hat k}B_{\hat j}\Om^0_{\hat k}  = 0\;.
\end{equation}
It is then easy to verify, by means of \cref{eq:BornSC} that
\begin{equation}
    \nabla_t B^{\hat i } + \eps^{\hat i \hat j\hat k}  B_{\hat j}\Om^0_{\hat k}=  \partial_t  B^{\hat i} \;,
\end{equation}
thus confirming the result in \cref{eq:Wdropsout} and the use of the inertial induction equations. 
The co-rotating frame rotation vector is  $\Om_0\hat e_3$, which happens to be the same (by construction) as the vorticity of the observer. This is why we see the same cancellation as in \cref{eq:Wdropsout} where we assumed the two are equal.

\section{A closer look at the Rayleigh stability criterion}\label{subapp:RayleighCloser}

The key point of the magneto-rotational instability is that the circular velocity background is stable against axisymmetric hydrodynamic perturbations, while adding a (however weak) magnetic field changes completely the nature of the system and makes it unstable to such perturbations. We now revisit the Rayleigh criterion with the aim of bringing to the fore important aspects to be kept in mind when looking at the general results derived in the main text, specifically in \cref{sec:MSinstaLocal,sec:Back2Hydro}. 

Starting from \cref{eq:Euler+contMRI}, and ignoring terms associated with the magnetic field, we write the coefficients matrix  (ordering the perturbed quantities as $\{\delta \rho/\rho, \delta v^{\hat R}, \delta v^{\hat \varp}, \delta v^{\hat \z}\} $)
\begin{equation}\label{eq:CMRayleighCR}
    \begin{pmatrix}
        -\om & k_1 & k_2 & k_3 \\
        c_s^2 k_1 & - \om & 2 i \Om_0 & 0 \\
        c_s^2 k_2 & - i \frac{\kappa^2}{2\Om_0} & - \om & 0 \\
        c_s^2 k_3 & 0 & 0 & - \om 
    \end{pmatrix} \;, \qquad \frac{\kappa^2}{2\Om_0} = 2 \Om_0 + s_0 = 2 \Om_0 + \frac{\text{d}\Om}{\text{dln}R}\bigg|_{R_0} \;,
\end{equation}
and the dispersion relation reads 
\begin{equation}\label{eq:RayleighCoRot}
    \om^4 - \left(c_s^2 \vec k^2 - \kappa^2 \right)\om^2 + i c_s^2 s_0 k_1k_2\om + c_s^2 \kappa^2 (k_3)^2 = 0 \;.
\end{equation}
If we now consider the sound-proof limit, namely $c_s^2 \gg 1$, we then end up with 
\begin{equation}
    \vec k^2 \om^2 - i s_0 k_1 k_2 \om - \kappa^2 k_3^2 = 0\;.
\end{equation}
From this we easily see that, if we assume axisymmetric perturbations, namely $k_2=0$, we obtain the usual Rayleigh stability criterion, that is $\kappa^2 > 0$. We stress that, as is well-known, the criterion does not guarantee that non-axisymmetric modes are stable. In fact, rewriting the dispersion relation in terms of $\Delta = - i\om$ and using the Routh-Hurwitz criterion (cf. \cref{app:RH}) we find that, on top of the Rayleigh criterion we would also need  
\begin{equation}
    s_0 k_1 k_2 \le 0 \;.
\end{equation}
We also note that, the story changes if we take the opposite limit instead, namely $c_s^2 \ll 1$, in which case the Rayleigh criterion is sufficient to guarantee stability of also non-axisymmetric perturbations. This would also be the case had we assumed incompressibility from the start. 

Having discussed the usual Rayleigh criterion using the co-rotating observer, we now re-work through it using an observer that is orbiting with the fluid at a given orbital distance $R_0$ but whose (local frame) axes are non rotating. 
We do this for two reasons. 
First, it will allow for a direct comparison with the general results discussed in the main text, specifically in \cref{sec:Back2Hydro}.
Second, we have argued that choosing to work with a rotating or non-rotating observer is, in general, just a matter of taste.
We then pick up an orbit $R_0$ as before, choose the observer to be co-orbiting with the background flow at the specific orbit
\begin{equation}
    \vec v_{obs} =  - \Om_0 y_0 \hat x + \Om_0 x_0 \hat y\;,
\end{equation}
where we used global Cartesian coordinates and $x_0(t) = R_0 \text{cos}(\Om_0 t), y_0(t) = R_0 \text{sin}(\Om_0 t)$ describe the worldline of the observer (the origin of the axes is suitably chosen so that $z_0(t) = 0$).
The background fluid velocity is then 
\begin{equation}
    \vec v = - \Om x \hat y + \Om y \hat x \;,\qquad \Om = \Om (\sqrt{x^2 + y^2}) 
\end{equation}
so that, considering the relative velocity $\vec v' = \vec v - \vec v_{obs}$ and expanding around $(x_0, y_0)$ we obtain
\begin{multline}
    \vec v' = - \left[s_0 \frac{x_0 y_0}{x_0^2 + y_0^2} x' + \left(\Om_0 + s_0 \frac{y_0^2}{x_0^2 +y_0^2}\right)y'\right]\hat x \\+ \left[ \left(\Om_0 + s_0 \frac{x_0^2}{x_0^2 +y_0^2}\right)x' + s_0 \frac{x_0 y_0}{x_0^2 + y_0^2} x' \right]\hat y \;,
\end{multline}
where $x' = x-x_0,\, y' = y - y_0$.
We can now choose a local region around a specific point $(x_0,y_0,z_0)$ on the orbit and choose to re-orient the axes by a constant rotation so that the observer velocity is moving only in the $y-$direction. 
We then set up the local frame in such a way that the local axis are non-rotating and oriented like the global cartesian ones. 
We can therefore write the gradients as
\begin{equation}\label{eq:NRgrad}
    \partial'_iv'_j = \begin{pmatrix}
    0 & \Om_0 + s_0 & 0 \\
    -\Om_0 & 0 & 0 \\
    0 & 0 & 0
    \end{pmatrix} \;,
\end{equation}
and the coefficients matrix of the linearized Euler plus continuity system is (cf. \cref{eq:Euler+contMRI} and ignore both magnetic field terms and the Coriolis force as the axis are non-rotating)
\begin{equation}\label{eq:CMRayleighNR}
    \begin{pmatrix}
        -\om & k_1 & k_2 & k_3 \\
        c_s^2 k_1 & - \om & i \Om_0 & 0 \\
        c_s^2 k_2 & - i (\Om_0 + s_0) & - \om & 0 \\
        c_s^2 k_3 & 0 & 0 & - \om 
    \end{pmatrix} \;.
\end{equation}
We can then compute the dispersion relation to find 
\begin{equation}
    \om^4 - \left[c_s^2 \vec k^2 + \Om_0(\Om_0 + s_0) \right]\om^2 + i c_s^2 k_1k_2 s_0\, \om + c_s^2 \Om_0(\Om_0 + s_0) (k_3)^2 = 0 \;,
\end{equation}
and observe this is consistent with the general dispersion relation in \cref{eq:S+Whydro} when restricted to the shear and vorticity associated with \cref{eq:NRgrad}.
However, this is not quite the dispersion relation we obtained above (cf. \cref{eq:RayleighCoRot}). 
The reason for this is that the two local observers we have considered measure different frequencies, as the axes of the co-rotating observer rotate with angular velocity $\Om_0 \hat e_3$ with respect to the other. 
To show why this is the resolution to the apparent conflict, let us consider once again the Born coordinates (cf. \cref{eq:BornCoord,eq:BornSC}). Given any vector $a^{\hat i}$ we have 
\begin{equation}
    \nabla_t a^{\hat i} = \partial_t \hat a^i + \Om_0 \eps^{\hat i \hat 3 \hat k} a_{\hat k}\;.
\end{equation}
This relation, when we introduce a plane-wave WKB expansion translates to %
\begin{equation}\label{eq:NRvsCRfrequencyConversion}
    - i \om_{rot} \delta a^{\hat i} = -i \om_{nr} \delta a^{\hat i}  + \Om_0 \eps^{\hat i \hat 3 \hat k} \delta a_{\hat k} \;,
\end{equation}
where $\om_{rot}$ is the frequency measured by the co-rotating observer, while $\om_{nr}$ is the frequency measured by an observer that has the same worldline but uses non-rotating axes. 
Specifying \cref{eq:NRvsCRfrequencyConversion} to the perturbed velocity (noting that it would not apply to the continuity equation as the density is a scalar), and noting that the frequency in \cref{eq:CMRayleighNR} correspons to $\om_{nr}$, we can reconcile the results obtained from \cref{eq:CMRayleighNR} with those from \cref{eq:CMRayleighCR}. 
We also note here that the same logic applies when we consider magnetized flows. 
That is, if we work with the inertial induction equation and compute the background velocity gradients as in \cref{eq:NRgrad}, we also need to take into account of the relation in \cref{eq:NRvsCRfrequencyConversion} for magnetic field disturbances to get back to \cref{eq:InductionMRI} and the MRI dispersion relation.

\chapter{The Routh-Hurwitz criterion}\label{app:RH}
In this appendix we review the Routh-Hurwitz criterion (see \cite{korn2013mathematical}), which gives valuable information about the roots of a polynomial with real coefficients. 
The criterion is often useful for studies of the linear stability of a system of equations, and can also be used prior to a numerical investigation (to inform the numerical study). 

Given a real algebraic equation 
\begin{equation}
    x^n + \ta_1 x^{n-1} + \dots + \ta_{n-1}x + \ta_n =0 \;,
\end{equation}
the Routh-Hurwitz criterion states that the number of roots with positive real part corresponds to the number of sign changes---disregarding vanishing terms---in the following sequence
\begin{equation}
    T_0\;,T_1\;,T_1T_2\;,T_2T_3\;,\dots\;, T_{n-1}T_{n-2},\ta_n \;,
\end{equation}
where
\begin{equation}
    T_0 = 1\;,\quad T_1 = \ta_1 \;,\quad T_2 = \text{det}\begin{pmatrix}
    \ta_1 & 1 \\
    \ta_3 & \ta_2 
    \end{pmatrix} \;, \quad
    T_3 = \text{det}\begin{pmatrix}
    \ta_1 & 1 & 0 \\
    \ta_3 & \ta_2 & \ta_1 \\
    \ta_5 & \ta_4 & \ta_3
    \end{pmatrix} \;,
\end{equation}
and so on. 
Throughout this work we have derived a number of dispersion relations in terms of the frequency $\omega$ (as a function of the wave-vector $\vec k$), whereby linear stability corresponds to its roots having negative imaginary part. 
Rewriting the dispersion relation in terms of $\Delta = - i\omega$, stability corresponds to $\Delta$-roots having negative real part. As such, using of the Routh-Hurwitz criterion we can directly obtain information about the stability of a system without having to explicitly find the solutions to the dispersion relation---which is always possible numerically (but often requires an expensive parameter study) while is viable analytically only in special/simple cases. 

We conclude this appendix by noting a caveat that is not explicitly mentioned in \cite{korn2013mathematical}, namely that Routh-Hurwitz criterion can be used \textit{only} when all the coefficients in a polynomial are non-vanishing. 
We show this with a trivial trivial example:
\begin{equation}
    (x^2 -2)(x^2 -3) = x^4 -5 x^2 + 6 = 0  \Longrightarrow x = \pm \sqrt 2 \;,\pm\sqrt 3 \;.
\end{equation}
Using the Routh-Hurwitz criterion, the sequence we obtain (neglecting vanishing terms) is $1,6$ and we would deduce there are no roots with positive real part. 
This is obviously wrong.